\begingroup
\makeatletter
\newread\arxivbibinput
\newwrite\arxivbiboutput
\immediate\openout\arxivbiboutput=\jobname.bbl
\immediate\write\arxivbiboutput{\@percentchar\space $ biblatex auxiliary file $}
\immediate\write\arxivbiboutput{\@percentchar\space $ biblatex bbl format version 3.3 $}
\openin\arxivbibinput=\jobname.bibdata
\loop\unless\ifeof\arxivbibinput
\readline\arxivbibinput to\arxivbibline
\immediate\write\arxivbiboutput{\arxivbibline}
\repeat
\closein\arxivbibinput
\immediate\closeout\arxivbiboutput
\endgroup

\documentclass[11pt,letterpaper]{article}
\usepackage[T1]{fontenc}
\usepackage{lmodern}
\usepackage[margin=1in]{geometry}
\usepackage{amsmath,amssymb,amsthm,mathtools,microtype}
\usepackage{array,booktabs,xcolor}
\usepackage{xspace}
\usepackage[breakable]{tcolorbox}
\usepackage[n,advantage,operators,sets,adversary,landau,probability,
  notions,logic,ff,mm,primitives,events,complexity,oracles,asymptotics,keys]{cryptocode}
\usepackage{csquotes}
\usepackage[backend=biber,style=numeric,sorting=nyt,giveninits=true,
  maxbibnames=99,doi=false,isbn=false,url=true,eprint=false]{biblatex}
\AtEveryBibitem{\iffieldundef{doi}{}{\clearfield{url}}}
\usepackage{aliascnt}
\usepackage[hidelinks,bookmarksnumbered]{hyperref}
\usepackage[nameinlink]{cleveref}
\hypersetup{pdftitle={Leakage-Resilient Computation via Matrix Refreshing},
 pdfauthor={Stefan Dziembowski and Sebastian Faust},
 pdfsubject={Continual storage, circuit privacy, and efficient simulation under local leakage}}
\newskip\LastLineFill
\parfillskip=\LastLineFill
\makeatletter
\patchcmd{\@trivlist}{\parfillskip \@flushglue}{\parfillskip \LastLineFill}{}{\PackageWarningNoLine{header}{Could not patch the list last-line setting}}
\makeatother
\allowdisplaybreaks[1]
\newtheorem{theorem}{Theorem}[section]
\newaliascnt{lemma}{theorem}
\newtheorem{lemma}[lemma]{Lemma}
\aliascntresetthe{lemma}
\newaliascnt{corollary}{theorem}
\newtheorem{corollary}[corollary]{Corollary}
\aliascntresetthe{corollary}
\newaliascnt{claim}{theorem}
\newtheorem{claim}[claim]{Claim}
\aliascntresetthe{claim}
\newcommand{\restatedlemmaname}{}
\newtheorem*{restatedlemma}{\restatedlemmaname}
\newenvironment{lemmarestatement}[2]{\renewcommand{\restatedlemmaname}{\cref{#1}}\begin{restatedlemma}[#2; restated]}{\end{restatedlemma}}
\theoremstyle{plain}
\newaliascnt{definition}{theorem}
\newtheorem{definition}[definition]{Definition}
\aliascntresetthe{definition}
\numberwithin{equation}{section}
\crefname{equation}{Eq.}{Eqs.}
\Crefname{equation}{Eq.}{Eqs.}
\crefname{theorem}{Thm.}{Thms.}
\Crefname{theorem}{Thm.}{Thms.}
\crefname{lemma}{Lem.}{Lems.}
\Crefname{lemma}{Lem.}{Lems.}
\crefname{corollary}{Cor.}{Cors.}
\Crefname{corollary}{Cor.}{Cors.}
\crefname{claim}{Claim}{Claims}
\Crefname{claim}{Claim}{Claims}
\crefname{definition}{Def.}{Defs.}
\Crefname{definition}{Def.}{Defs.}
\crefname{section}{Sec.}{Secs.}
\Crefname{section}{Sec.}{Secs.}
\crefname{subsection}{Sec.}{Secs.}
\Crefname{subsection}{Sec.}{Secs.}
\crefname{subsubsection}{Sec.}{Secs.}
\Crefname{subsubsection}{Sec.}{Secs.}
\crefname{appendix}{App.}{Apps.}
\Crefname{appendix}{App.}{Apps.}
\crefname{subappendix}{App.}{Apps.}
\Crefname{subappendix}{App.}{Apps.}
\crefname{subsubappendix}{App.}{Apps.}
\Crefname{subsubappendix}{App.}{Apps.}
\crefname{figure}{Fig.}{Figs.}
\Crefname{figure}{Fig.}{Figs.}
\crefname{table}{Tab.}{Tabs.}
\Crefname{table}{Tab.}{Tabs.}
\newcommand{\F}{\mathbb{F}}
\newcommand{\R}{\mathbb{R}}
\NewDocumentCommand{\E}{e{_} g}{\mathbb{E}\IfValueT{#1}{_{#1}}\IfValueT{#2}{\mathopen{}\left[#2\right]\mathclose{}}}
\NewDocumentCommand{\Prb}{e{_} g}{\mathbb{P}\IfValueT{#1}{_{#1}}\IfValueT{#2}{\mathopen{}\left[#2\right]\mathclose{}}}
\newcommand{\LeftParty}{\mathsf{L}}
\newcommand{\RightParty}{\mathsf{R}}
\newcommand{\MatrixPhase}{\mathsf{M}}
\newcommand{\ShortPhase}{\mathsf{S}}
\newcommand{\transpose}{\mathsf{T}}
\newcommand{\cZ}{\mathcal{Z}}
\newcommand{\project}[1]{\mathsf{project}\mathopen{}\left(#1\right)\mathclose{}}
\newcommand{\Circuit}{\Gamma}
\newcommand{\OldData}{\mathcal{A}}
\newcommand{\TableRows}{\mathcal{X}}
\newcommand{\TableCols}{\mathcal{Y}}
\newcommand{\Rev}{R_{\mathrm{ev}}}
\newcommand{\FullRank}{\mathcal{H}}

\newcommand{\DiscBound}{\mathrm{disc}_n}
\newcommand{\UL}{U_{\LeftParty}}
\newcommand{\UR}{U_{\RightParty}}
\newcommand{\InvMatr}{\mathsf{InvMatr}}
\newcommand{\JointRef}{\nu_{\mathrm{joint}}}
\newcommand{\LeakFn}{\mathsf{Leak}}
\renewcommand{\prob}{\mathsf{prob}}
\newcommand{\LongViewDist}[1]{\mathcal{D}_{#1}^{\mathrm{long}}}
\newcommand{\ShortViewDist}[1]{\mathcal{D}_{#1}^{\mathrm{short}}}

\newcommand{\LoneNorm}[1]{\left\lVert#1\right\rVert_1}
\newcommand{\LtwoNorm}[1]{\left\lVert#1\right\rVert_2}
\newcommand{\NuclearNorm}[1]{\left\lVert#1\right\rVert_*}
\newcommand{\MaxNorm}[1]{\left\lVert#1\right\rVert_\infty}
\newcommand{\LtwoNormAt}[2]{\left\lVert#1\right\rVert_{2,#2}}
\newcommand{\LpNormAt}[3]{\left\lVert#1\right\rVert_{#2,#3}}

\newcommand{\VecNorm}[1]{\left\lVert#1\right\rVert}
\newcommand{\EuclNorm}[1]{\left\lVert#1\right\rVert_{\mathrm E}}
\newcommand{\pairing}[2]{\left\langle#1,#2\right\rangle}
\newcommand{\StatDist}[2]{\delta\mathopen{}\left(#1,#2\right)\mathclose{}}
\newcommand{\rank}[1]{\operatorname{rank}\mathopen{}\left(#1\right)\mathclose{}}
\newcommand{\sgn}[1]{\operatorname{sign}\mathopen{}\left(#1\right)\mathclose{}}
\newcommand{\spn}{\operatorname{span}}

\newcommand{\AvSign}{\mathrm{Av}_\Upsilon}
\newcommand{\GapFn}{\mathrm{gap}}
\newcommand{\FourCyc}[1]{\mathrm{Cyc}\mathopen{}\left(#1\right)\mathclose{}}
\newcommand{\Gram}{\mathrm{Gram}}
\newcommand{\Id}{\mathrm{Id}}
\newcommand{\InnerSpace}{\mathbb{V}}
\newcommand{\OldL}{{L^{\circ}}}
\newcommand{\OldR}{{R^{\circ}}}
\newcommand{\vecx}{\mathbf{x}}
\newcommand{\vecs}{\mathbf{s}}
\newcommand{\vecy}{\mathbf{y}}

\newcommand{\cD}{\mathcal{D}}
\newcommand{\cE}{\mathcal{E}}
\newcommand{\cF}{\mathcal{F}}
\newcommand{\cG}{\mathcal{G}}

\newcommand{\cI}{\mathcal{I}}
\newcommand{\cK}{\mathcal{K}}
\newcommand{\cL}{\mathcal{L}}
\newcommand{\cM}{\mathcal{M}}
\newcommand{\cN}{\mathcal{N}}

\newcommand{\cS}{\mathcal{S}}
\newcommand{\cV}{\mathcal{V}}

\newcommand{\MaskL}[1]{f_{\LeftParty,#1}}
\newcommand{\MaskR}[1]{f_{\RightParty,#1}}
\newcommand{\Ipub}{I_{\mathrm{pub}}}
\newcommand{\Ipriv}{I_{\mathrm{priv}}}

\newcommand{\para}[1]{\par\smallskip\noindent\textbf{#1}\enspace\ignorespaces}

\newcommand{\github}{\href{https://github.com/sdziembowski/MatrixRefresh}{\texttt{github.com/\allowbreak sdziembowski/\allowbreak MatrixRefresh}}\xspace}

\newcommand{\githubsub}{\href{https://github.com/sdziembowski/MatrixRefresh/tree/main/version-1.0}{\texttt{version-1.0}}\xspace}

\title{Leakage-Resilient Computation\\ via Matrix Refreshing}
\author{Stefan Dziembowski\\
 University of Warsaw\\ and IDEAS Institute\\
 Warsaw, Poland
 \and
 Sebastian Faust\\
 TU Darmstadt\\
 Darmstadt, Germany} 
\date{}
\begin{document} 
\maketitle

\begin{abstract}
Leakage-resilient cryptography protects secrets against side-channel attacks, in which a physical implementation reveals partial information about its internal state.
A fundamental primitive in this area is a \emph{leakage-resilient storage}, which allows a secret to be stored in a way that remains secure even if an adversary learns some information about the storage's internal state.
A popular approach represents a secret as the inner product of two $n$-bit vectors, called \emph{shares}. The shares are stored and processed separately by two parts of the implementation, called the \emph{left} and \emph{right parties}.
Because leakage accumulates over time, the shares must be \emph{refreshed} repeatedly.
We prove that a simple refresh, in which the left party multiplies its share by a uniform invertible matrix $M$ and the right party multiplies its share by $M^{-1}$, is secure under continual local leakage.
After $T$ refreshes, each allowing at most $\lambda$ bits of leakage per party, the transcripts for two stored secrets are at statistical distance at most $C\cdot T\cdot2^{6\lambda-n/9}$ for an absolute constant~$C$.

We also obtain a Boolean-circuit compiler with statistical privacy error at most $C\cdot S\cdot2^{18\lambda-n/27}$ for $S$ operations and at most $\lambda$ bits of local leakage per party per operation. The guarantee covers repeated stateful evaluations with adaptively chosen public inputs, with $S$ counting all operations across invocations. Thus, for polynomial-size circuits, a constant leakage fraction is tolerated with share length depending only on the security parameter.
We formalize the refreshing bound, the composition and compiler guarantees, and the expected-work bounds in Lean~4.
Our results also provide efficient simulation: the leakage can be reproduced from the public inputs and the circuit outputs alone, with expected polynomial overhead.
\end{abstract}

\section{Introduction}
Physical implementations of cryptographic algorithms leak information about their internal state through side channels such as running time, power consumption, and electromagnetic radiation.
This has motivated a large body of work on \emph{leakage-resilient cryptography}, which extends the traditional black-box security model to account for side-channel leakage and designs schemes that provably withstand it~\cite{DBLP:conf/crypto/IshaiSW03,DBLP:conf/tcc/MicaliR04,DBLP:conf/focs/DziembowskiP08,DBLP:conf/tcc/AkaviaGV09,DBLP:conf/eurocrypt/Pietrzak09,DBLP:conf/crypto/NaorS09,DBLP:conf/crypto/AlwenDW09,DBLP:conf/eurocrypt/FaustRRTV10,DBLP:conf/focs/DodisHLW10,DBLP:conf/focs/BrakerskiKKV10,DBLP:conf/crypto/GoldwasserR10,DBLP:conf/eurocrypt/DucDF14,DBLP:books/acm/19/KalaiR19}.
The central goal of leakage-resilient cryptography is to compute arbitrary programs securely in the presence of side-channel leakage.
This goal is formalized by \emph{leakage-resilient circuit compilers}~\cite{DBLP:conf/crypto/IshaiSW03,DBLP:conf/eurocrypt/FaustRRTV10}, which transform any program, modeled as a Boolean or arithmetic circuit, into a protected circuit that is secure against leakage.
Security means that the leakage is useless to the adversary: anything it learns from the leakage together with the input/output behavior of the protected circuit could also be learned from the input/output behavior alone.

\para{Our leakage model.} Security is impossible if the leakage can be arbitrary, so the admissible leakage functions must be restricted.
In this work, we consider \emph{bounded leakage}: the adversary chooses leakage functions of the internal state with bounded output length and learns their outputs.
Moreover, the leakage is \emph{continual}: the protected computation runs for many rounds, each round leaks, and the adversary chooses the leakage function of every round after seeing the answers of the earlier ones, so the total leakage is unbounded.

Considering bounded leakage alone is not enough, however: by the impossibility of program obfuscation~\cite{DBLP:journals/jacm/BarakGIRSVY12}, there exist programs for which a single bit of leakage, computed by an efficient function of the entire internal state, breaks security.
We therefore also restrict which parts of the state a single leakage function can access. Models differ in how finely the state is split.
At one extreme is the probing model of Ishai, Sahai, and Wagner~\cite{DBLP:conf/crypto/IshaiSW03}, in which the adversary learns the values of a bounded number of individual wires.
At the other extreme is the \emph{split-state model}~\cite{DBLP:conf/focs/DziembowskiP08,DBLP:conf/asiacrypt/DziembowskiF11,DBLP:conf/tcc/DziembowskiF12}, in which the state is divided between two parties that leak separately.
We keep this two-party structure and combine it with the assumption that only computation leaks~\cite{DBLP:conf/tcc/MicaliR04}, as in prior circuit compilers with bounded leakage~\cite{DBLP:conf/crypto/GoldwasserR10,DBLP:conf/crypto/JumaV10,DBLP:conf/tcc/DziembowskiF12,DBLP:conf/focs/GoldwasserR12}.
We call the participants the \emph{left party} $\LeftParty$ and the \emph{right party} $\RightParty$. Each stores its own part of the state and performs its part of the computation.
Each party is thus a processor with its own memory, and the protected computation is a sequence of \emph{operations} that leak in \emph{rounds}: in each round, each party's leakage function reads the data that the party is currently computing on, namely its part of the state involved in the operation, the randomness drawn so far, and the messages it has received so far, and outputs a bounded number of bits.
Memory that is not in use does not leak.
The two leakage functions of a round are chosen before either answer is revealed; across rounds, the adversary may adapt its choices to all earlier answers.
Communication between the parties is private, and each party may permanently erase data it no longer needs; erased data cannot be read by later leakage.

\para{Leakage-resilient encoding.}
A first ingredient of a leakage-resilient compiler is an encoding scheme that splits a secret into two parts, so that separate leakage from each part reveals almost nothing about it.
This notion of leakage-resilient encoding for split-state computation was introduced by Dav{\`\i}, Dziembowski, and Venturi~\cite{DBLP:conf/scn/DaviDV10}, who also proposed a simple scheme based on the inner-product extractor.
For an integer share length $n\ge2$, their scheme encodes a secret bit $s\in\F_2=\{0,1\}$ as two nonzero vectors, a row $\ell\in\F_2^n$ held by the left party $\LeftParty$ and a column $r\in\F_2^n$ held by the right party $\RightParty$, chosen uniformly at random subject to
\[
 s=\ell\cdot r=\sum_{k=1}^n\ell_k\cdot r_k.
\]
Its security rests on the fact that the inner product is a \emph{two-source extractor}~\cite{DBLP:journals/siamcomp/ChorG88}: its output is close to uniform whenever its two inputs are independent and each is sufficiently unpredictable.
This remains true given sufficiently bounded leakage obtained separately from each share, so the encoding hides $s$~\cite{DBLP:conf/scn/DaviDV10,DBLP:conf/tcc/DziembowskiF12}.

\para{Towards continual leakage: refreshing.}
A leakage-resilient encoding protects the secret only when the \emph{total} leakage from each share is less than its length $n$. Otherwise, an adversary can learn both shares and hence $s$.
A device that operates for a long time, however, keeps leaking, so its total leakage eventually exceeds any such bound.
The standard remedy is to \emph{refresh} the encoding periodically, replacing $(\ell,r)$ by a new encoding $(\ell',r')$ of the same secret, so that leakage in different periods concerns different shares~\cite{DBLP:conf/asiacrypt/DziembowskiF11,DBLP:journals/corr/abs-1209-4820}.
Refreshing is a central component of leakage-resilient circuit compilers~\cite{DBLP:conf/crypto/IshaiSW03,DBLP:conf/eurocrypt/FaustRRTV10,DBLP:conf/focs/GoldwasserR12,DBLP:conf/tcc/DziembowskiF12}, and designing it is one of their main technical challenges.
The difficulty is that a refresh must remain secure under composition: the new shares must be correlated with the old ones to encode the same secret, yet the refresh itself leaks, and the adversary may exploit this leakage to combine information from different periods.

The most natural refresh for inner-product encodings adds to each share a random vector that is orthogonal to the other share.
For example, the left party sends a random column $x$ with $\ell\cdot x=0$ to the right party, which replaces $r$ by $r+x$, and the right party then does the same for the left share.
Dziembowski and Faust showed that this protocol is completely insecure under continual leakage~\cite[full version, App.~D]{DBLP:conf/asiacrypt/DziembowskiF11}.
Both parties see the exchanged vectors, so an adversary that leaks two field elements from each party per refresh can accumulate the inner products of longer and longer prefixes of the shares and recover the secret after $n$ refreshes.
One approach to preventing such attacks is to deliver the refreshing randomness \emph{obliviously}: a party learns nothing about the vector added to the other party's share beyond what correctness requires.
Prior works~\cite{DBLP:conf/asiacrypt/DziembowskiF11,DBLP:conf/tcc/DziembowskiF12,DBLP:journals/corr/abs-1209-4820} achieve this with a leak-free component that samples a random pair of orthogonal vectors and gives one vector to each party. The component itself does not leak, so each vector can leak only through the party that receives it.
Leak-free components are a strong assumption since they rely on a piece of hardware that must not leak at all.
This raises the following question: can an inner-product encoding be refreshed securely under continual leakage without any leak-free component, even when the leakage functions may read all the randomness of the refresh?
 
Goldwasser and Rothblum~\cite{DBLP:conf/focs/GoldwasserR12,Goldwasser2015} address this question in the model where only computation leaks: they construct a circuit compiler, including a refreshing procedure, based on the inner-product encoding.
Apart from gaps in the proof of their main theorem (see our detailed discussion in \mbox{\cref{app:gr}}), their refreshing procedure is considerably more complex than the one we present in this work.

\para{Leakage-resilient circuit compilers.}\phantomsection\label{sec:intro-compiler-question}
Given a leakage-resilient refresh, the second ingredient of a compiler is a leakage-resilient \emph{gadget} for each gate of the original circuit: a two-party protocol that takes encodings of the gate's inputs and produces an encoding of the gate's output.
While the gadget runs, each party leaks on the data it uses: its shares of these encodings, its randomness, and the messages it receives.
Composing the gadgets along the circuit yields the protected circuit.
The main difficulty lies in nonlinear gates such as multiplication or NAND: the parties must compute an encoding of $s_1\cdot s_2$ from encodings of $s_1$ and $s_2$, and every intermediate value of this computation may leak.
Since the total leakage over an evaluation grows with the number of gates, the gadgets must refresh encodings as they go, and security must compose across all gates of the circuit.

The compiler of Dziembowski and Faust~\cite{DBLP:conf/tcc/DziembowskiF12}, which works in the split-state model, relies on leak-free components.
The compiler of Goldwasser and Rothblum discussed above avoids them, but its proof has gaps.
This raises a second question: can arbitrary circuits be compiled so that they remain secure when every active gadget leaks locally, without any leak-free internal component?

\subsection{Our contributions}
In this work, we answer both questions affirmatively. We give a refresh for inner-product encodings and a compiler for arbitrary circuits that remain secure under continual bounded leakage from both parties, without any leak-free component. We summarize our contributions in more detail below.

\para{The refreshing protocol.}
Recall that the left party $\LeftParty$ holds the row share $\ell$, the right party $\RightParty$ holds the column share $r$, and the stored secret is their inner product $\ell\cdot r$.
To refresh these shares, the left party samples $M\sample \InvMatr_2^n$, where $\InvMatr_2^n$ denotes the set of invertible $n\times n$ matrices over $\F_2$.
The left party sends $M$ privately to the right party, and the two parties update
\[
 \ell:=\ell\cdot M,\qquad r:=M^{-1}\cdot r,
\]
after which both parties erase $M$ and their old shares.
Correctness is immediate, since $(\ell\cdot M)\cdot\left(M^{-1}\cdot r\right)=\ell\cdot r$.
Each refresh communicates the $n^2$ bits of $M$ and costs an expected $O(n^3)$ bit operations: sampling an invertible $n\times n$ matrix and inverting it by Gaussian elimination takes cubic time in its dimension.
For security, we use the leakage model described above with each refresh as one operation: each party leaks at most $\lambda$ bits through a function of its old share, its new share, and the entire matrix $M$.
Unlike in~\cite{DBLP:conf/asiacrypt/DziembowskiF11}, the refresh randomness is thus not delivered obliviously: both parties know $M$, so each knows exactly how the other party's share is transformed.
The main challenge in our proof is therefore that the two parties' leakage is correlated through the shared randomness $M$: the leakage of a refresh cannot be treated as two independent leakages from the two shares, so the extractor argument for a single encoding does not apply directly.
We use a hybrid argument: starting from the execution with stored secret $0$, we switch the bit encoded by the successive encodings, produced by the refreshes, one at a time until we reach the execution with stored secret $1$, and we bound the change caused by each switch.
The intermediate experiments, called \emph{hybrids}, are not real executions, since in each of them one refresh connects encodings of different bits.
Each switch requires a new estimate that controls the leakage correlation introduced by the shared matrix $M$ (see \mbox{\cref{sec:technical-outline}}).
As in our model, both leakage functions of a refresh are fixed before either answer is revealed; our proof extends to adaptive choices within a refresh, which we sketch in \cref{sec:within-round}.

\para{Secure continual refreshing.}
As a first result, we obtain that the above simple matrix refresh is secure.
Suppose that the parties perform $T$ refreshes, that each share has $n$ bits, and that each party leaks at most $\lambda$ bits per refresh.
Let $P_s^T$ denote the distribution of the complete leakage transcript when the stored secret is $s$, and let $\delta$ denote the statistical distance.
\Cref{thm:main} shows, for arbitrary leakage that is adaptive across refreshes and nonadaptive within each refresh, that
\begin{equation*}
 \StatDist{P_0^T}{P_1^T}
 \le C\cdot T\cdot2^{6\lambda-n/9}
\end{equation*}
for some constant $C$.
For $T$ polynomial in $n$, the bound decays exponentially whenever $\lambda\le\alpha n$ for a fixed $\alpha<1/54$ (see \cref{cor:parameters}).
\Cref{sec:field-size} derives the sufficient share length $n=O(\kappa+\lambda+\log(T+2))$ for target distance $2^{-\kappa}$.

\para{Leakage-resilient circuit compilation.}\phantomsection\label{sec:intro-compiler}
We also use our refresh to protect arbitrary Boolean circuits. We first treat stateless circuits and then extend the result to stateful ones.
A \emph{stateless} circuit $\Circuit$ is evaluated once, on a public input $\vecx$ and a private input $\vecs$, and outputs $\vecy=\Circuit(\vecx,\vecs)$, whereas a \emph{stateful} circuit keeps a secret state, such as a cryptographic key, between invocations and in invocation $t$ computes $(\mathsf{st}_{t+1},\vecy_t)=\Circuit(\mathsf{st}_t,\vecx_t)$ from the current state $\mathsf{st}_t$ and a fresh public input $\vecx_t$ chosen by the adversary~\cite{DBLP:conf/crypto/IshaiSW03}.
We express the circuit to be protected, which we call the \emph{source circuit}, using NAND gates, whose gate function is $\NAND(s_1,s_2)=1+s_1\cdot s_2$ over $\F_2$.
We call the two input bits of a gate its \emph{operands}. A gate of the source circuit may read the same wire twice, as in the negation $\NAND(s,s)=1+s$.
Every circuit over a fixed AND/OR/NOT basis can be rewritten in the NAND basis with constant-factor overhead (see \cref{sec:direct-source-model}).

Each NAND gate of the source circuit is implemented by a two-party \emph{gadget} in three steps, inspired by the multiplication gadget of Dziembowski and Faust~\cite{DBLP:conf/tcc/DziembowskiF12} but without any leak-free component.
First, to multiply the operands, the parties compute the tensor product of their shares locally.
For encodings $(L_1,R_1)$ and $(L_2,R_2)$ of the operands $s_1$ and $s_2$ (refreshed copies, as explained below), the left party forms a long row $a$ and the right party a long column $b$,
\[
 a=(L_1\otimes L_2,1),\qquad b=\binom{R_1\otimes R_2}{1},
\]
where the tensor product $\otimes$ lists all pairwise products of coordinates and a final coordinate equal to one is appended. These long shares have length $m=n^2+1$.
Second, in the \emph{matrix phase}, the parties refresh the long encoding with our refreshing protocol in dimension $m$ and erase the refresh matrix; \cref{sec:technical-outline} explains why this refresh is needed.
Third, in the \emph{shortening phase}, the parties shrink the refreshed long shares $U$ and $V$ back to $n$ bits, so that the share length does not grow along the circuit.
The left party samples a uniform nonzero row $A$ of length $n$ and a uniform matrix $H\in\F_2^{n\times m}$ of full row rank subject to $A\cdot H=U$ and privately sends $H$. The right party then sets $R=H\cdot V$, and $(A,R)$ is the output encoding.
If $R=0$, the parties repeat with fresh $A$ and $H$ within the same leakage budget. The first attempt fails with probability at most $2^{1-n}$.

For correctness, the tensor product multiplies inner products, so $a\cdot b=(L_1\cdot R_1)\cdot(L_2\cdot R_2)+1=1+s_1\cdot s_2$. The refresh preserves this inner product, and so does shortening, since $A\cdot R=A\cdot H\cdot V=U\cdot V$.
Moreover, the completed shortening outputs an exactly uniform nonzero encoding of $U\cdot V$, even conditional on the long shares, a property we call \emph{exact freshness}.
For $m=n$, the matrix $H$ is invertible and shortening is precisely our refresh with $M=H^{-1}$.
The gadget leaks in two rounds, one per phase, each with one function per party and a total budget of $\lambda$ bits per party per gadget: the shortening leakage may depend on the matrix-phase answers, but it cannot read the erased matrix $M$.

For composition, the compiler places a refresh between consecutive gadgets: every operand is refreshed just before it is used.
Without these refreshes, an encoding read by several gates would leak in each of these operations, so its total leakage would grow with the number of times its wire is used.
Concretely, a wire used at $\deg(w)$ \emph{input ports}, that is, operand positions of gates, is refreshed in a chain of $\deg(w)$ copies, each created just before its use.
Each copy then belongs to at most three operations, namely its producing refresh, its gate consumer, and the next refresh, so every encoding leaks in a bounded number of operations.
This handles arbitrary fanout with linear overhead, and a gate that reads the same wire twice receives two independent copies.

Public inputs and constants are encoded directly under leakage, following the public-input step of Dziembowski and Faust~\cite[Figure~3]{DBLP:conf/tcc/DziembowskiF12}.
In the stateless case, a trusted initialization supplies only the private input encodings, and at the end a trusted decoder releases the output $\vecy$ in the clear.
The output shares are never revealed in full; like every encoding, they leak only through the operations that produce and use them.
In the stateful case, the trusted initialization encodes only the initial state; the state encodings are kept between invocations and refreshed before each use like any other wire, and the decoder releases the output $\vecy_t$ of every invocation in the clear.

\para{Efficiency and security of the compiler.}
For $t_0$ NAND gates of the source circuit (after rewriting it into a circuit with only NAND gates in case it had other gates) and $\Ipub$ public-input or constant wires, the compiler has $S=3t_0+\Ipub$ operations: $t_0$ computation gadgets, $2t_0$ input-port refreshes, and $\Ipub$ public-input samplers, which encode the public inputs and constants.
The compiled circuit has three operations per gate of the source circuit, and each gate costs an expected $O(n^4)$ bits of communication and $O(n^6)$ bit operations, both dominated by the $(n^2+1)\times(n^2+1)$ refresh matrix of its gadget.
Since tensoring squares the share length, the matrix phase refreshes shares of length $m=n^2+1$, so its matrix takes $m^2=O(n^4)$ bits to send and $O(m^3)=O(n^6)$ bit operations to sample and invert, as discussed above.

Security is shown by an efficient simulator, described below, that receives the public input $\vecx$ and the output $\vecy=\Circuit(\vecx,\vecs)$ but not the private input $\vecs$, and reproduces the adversary's view of the leakage.
With at most $\lambda$ leaked bits per party per operation, \cref{thm:compiler-direct} bounds the simulation error, the statistical distance between the real and the simulated view, by $C\cdot S\cdot2^{18\lambda-n/27}$ for some constant $C$.
For polynomial-size circuits this is exponentially small whenever $\lambda\le\alpha n$ with a fixed $\alpha<1/486$, so the share length can depend only on the security parameter.
The same bound extends to stateful circuits under continual leakage (see \cref{thm:compiler-stateful}): over $T$ invocations, with the input of each invocation chosen by the adversary after seeing the earlier leakage and outputs, the error is $C\cdot S\cdot2^{18\lambda-n/27}$ with $S$ counting the operations of all invocations, so that $S$ grows linearly in $T$; the share length therefore grows only logarithmically in $T$, while the total leakage grows linearly.
Our simulator is simple and efficient: it runs the compiled circuit on the all-zero private input and the public input $\vecx$, answers the adversary's leakage queries from this execution, and outputs the given $\vecy$, the \emph{authorized output}, instead of decoding its own output encodings.
In particular, it never needs a private input that produces $\vecy$.
\Cref{sec:technical-outline} explains why this dummy execution cannot be distinguished from the real one.

\Cref{tab:results} compares the results using the same parameters: $n\ge2$ is the length of the stored shares, $T$ counts refreshes, $S$ counts all operations of the compiled circuit (gate gadgets, input-port refreshes, and public-input samplers), $\lambda$ is the total leakage per party per operation, and $\delta$ is the statistical distance between the compared views.
The errors accumulate linearly in the number of operations.
For a prescribed finite error $2^{-\kappa}$, this gives the usual additive $\log(S+2)$ term in a sufficient share length. For polynomial-size circuit families, the share length need not depend on the circuit size.

\begin{table}[htbp]
\centering
\small
\renewcommand{\arraystretch}{1.25}
\begin{tabular}{@{}>{\raggedright\arraybackslash}p{0.29\linewidth}>{\raggedright\arraybackslash}p{0.36\linewidth}>{\raggedright\arraybackslash}p{0.27\linewidth}@{}}
\toprule
Application & Statistical-distance bound & Sufficient leakage fraction\\
\midrule
Continual refreshing\newline(see \cref{thm:main}) & $\delta\le C\cdot T\cdot2^{6\lambda-n/9}$ & $\alpha<1/54$\\
\addlinespace
Circuit compilation, hidden output shares\newline(see \cref{thm:compiler-direct}) & $\delta\le C\cdot S\cdot2^{18\lambda-n/27}$ & $\alpha<1/486$\\
\bottomrule
\end{tabular}
\caption{Bounds for continual refreshing and circuit compilation, with a common absolute constant $C$.
For $T$ or $S$ polynomial in $n$ and $\lambda\le\alpha n$ with fixed $\alpha$, the last column guarantees distance $2^{-\Omega(n)}$.
The compiler's output encodings remain hidden.}
\label{tab:results}
\end{table}

\para{Proof techniques.}
In the extractor argument for a single encoding, the leakage splits into a function of the left shares times a function of the right shares; for such split leakage, the probability of each answer is almost the same whether the shares encode $0$ or $1$.
Our leakage does not split in this way, because the parties communicate: a message, such as the refresh matrix $M$ or the shortening matrix $H$, is held by both parties and can be read by both leakage functions, so the probability of an answer depends on the two parties' shares jointly.
Our main tool to handle such dependencies is a \emph{local approximation}: we approximate the probability of each leakage answer of a single operation, up to a small error, by a sum of split functions with bounded total weight, so that the extractor argument applies to each summand.
This sum is obtained from the singular value decomposition (SVD) of the answer probability.

For refreshing, we combine these approximations with the hybrid argument described above: each step changes the bit encoded by one state, which affects only the leakage of the neighboring refreshes, so the local errors add up linearly in the number of refreshes.
For the compiler, the hybrid argument does not extend directly, because the private inputs determine the bits of all wires at once.
Instead, we expand the probability of the whole leakage transcript into a common term that does not depend on the private inputs and correction terms that each contain a local approximation error or the difference between an encoding and independent shares; each correction term can be bounded locally, and the total error is linear in the number of operations.
\Cref{sec:technical-outline} explains both proofs in more detail.

\para{Scope and limitations.}
The storage construction treats one-bit secrets and binary shares; the compiler handles vectors of bits.
The tolerated leakage rate is smaller than for a fixed encoding without refresh, and we keep the absolute constants conservative to simplify the local estimates.
If the initial storage encoding may contain a zero share, the refreshing bound increases by at most $4/2^n$ (see \cref{sec:zero-initialization}).
In the compiler, every internal operation, including the input-port refreshes and the public-input samplers, may leak, so the total leakage grows with the size of the source circuit.
We assume private communication, the specified erasures, and that only the active operation leaks; moreover, the encoding of the private input or initial state and the decoding of the output in every invocation are trusted.
Allowing the decoding to leak would require the simulator to program the output encodings, for instance with selector gates, and we leave this open.
\Cref{sec:formalization} proves the stateful extension and discusses leakage that is adaptive within a refresh and larger finite fields; our quantitative theorems concern the binary field and the stated phase structure.

\para{Lean formalization.}
We provide a Lean~4 formalization of the refreshing and composition arguments for deterministic leakage strategies, the NAND compiler with input-port refreshes, and its counted expected-work simulator; randomized strategies reduce to deterministic ones by averaging, an argument that is not formalized.
\Cref{sec:formalization} describes its scope and simplifying assumptions.
The source files are available in the \githubsub folder of the project repository.

\para{AI usage.}
The authors obtained the results of this paper with assistance from ChatGPT and Claude AI.
The idea for the matrix-based refresh came from the authors, but the security proofs were developed with help from the AI.
The initial proof was first formulated as a result of the authors' interaction with ChatGPT 6.0 on September 5, 2026 (earlier attempts to solve this problem with other versions of ChatGPT were unsuccessful).
The authors spent considerable time iterating with the AI to simplify the proofs.
The AI also helped substantially in writing the paper.
The Lean formalization was done by ChatGPT.

\paragraph{Paper's website and supplementary materials.}
The paper and its supplementary materials (including the Lean formalization) are available \github in the \githubsub folder.

\subsection{Technical outline}\label{sec:technical-outline}
 
We now explain the main ideas behind our two security proofs.
For refreshing, we follow the hybrid argument outlined with our refreshing protocol above: we change the bit encoded by one of the successive encodings at a time and bound the effect of each change.
The main new ingredient is an explicit local approximation of the leakage of one refresh, which we motivate and construct below.
For the compiler, we use the same kind of local approximation for every operation, but instead of a hybrid argument we expand the whole computation into terms that can each be bounded locally (see the last paragraph of this subsection).
We first fix a few terms informally. Precise definitions follow in \crefrange{sec:preliminaries}{sec:compilation}.
A \emph{round} is one refresh, a \emph{state} $X_i=(L_i,R_i)$ is the pair of shares held after round $i$ (see \cref{sec:shares}), the sequence $X_0,\ldots,X_T$ is the \emph{chain}, a \emph{table} is a real function of a left argument and a right argument, and a \emph{leakage answer} $z$ is the pair of outputs of the two leakage functions in one round (see \cref{sec:leakage}).
We write $K:=2^{2\lambda}$ for the number of possible leakage answers in one round. Each factor $K$ in the bounds below comes from enumerating one answer.
 
\para{The challenge and our approach.}
To see where the difficulty lies, consider first a single encoding $(\ell,r)$ that is never refreshed, with leakage $f(\ell)$ from the left party and $g(r)$ from the right party.
The encodings that produce a given answer then form a \emph{rectangle}, a set of left shares times a set of right shares, so the probability of the answer is a product of a function of $\ell$ and a function of $r$.
Lindsey's lemma~\cite[Lemma~14.5]{Jukna2011} bounds the inner-product discrepancy of each such rectangle (see \cref{lem:disc} for the formulation used here).
Summing these bounds over the possible leakage answers shows that, for sufficiently small leakage budgets, the leakage distributions for the two secrets are statistically close.

A refresh destroys this product structure, because both leakage functions also read the matrix~$M$.
For a fixed matrix $M$, the old states that produce a given answer still form a rectangle, since $M$ and the old state determine the new state.
The hybrid argument, however, changes the bit encoded by one state at a time, and each state is the new state of one round and the old state of the next.
We therefore view the probability of an answer as a table whose left argument is the pair of left shares $(\ell,\ell')$ before and after the round and whose right argument is the pair $(r,r')$.
Given both states, the matrix $M$ is still uniformly random among the matrices that map the old state to the new one, and the answer probability is the average of the rectangles over these matrices.
Such an average is in general not a product: through the common matrix, the leakage of the two parties is correlated.
This is the main obstacle, and it is why the extractor argument for a single encoding does not apply directly. 

Our approach is to \emph{approximate} this table by a sum of left/right products $F=\sum_j a_j\cdot b_j^{\transpose}$.
By linearity, the extractor argument applies to each product separately, so it suffices to control the total size of the sum.
The approximation need not be a probability and may take negative values, so we call it a \emph{signed} table.
We measure the size of the sum by its \emph{cost} $\sum_j\LtwoNorm{a_j}\cdot\LtwoNorm{b_j}$, where $\LtwoNorm{\cdot}$ is the root-mean-square value under the uniform measure.
The minimum cost over all such representations of a table $F$ is its \emph{nuclear norm} $\NuclearNorm{F}$ (see \cref{eq:nuclear-factor-cost} in the \cref{sec:preliminaries-proofs}), so every representation satisfies (see \cref{eq:nuclear-product-bound})
\[
 \NuclearNorm{F}\le\sum_j\LtwoNorm{a_j}\cdot\LtwoNorm{b_j}.
\]
Equality is attained by the balanced factors $\sqrt{\sigma_j}\cdot u_j$ and $\sqrt{\sigma_j}\cdot v_j$ of an orthonormal expansion $F=\sum_j\sigma_j\cdot u_j\cdot v_j^{\transpose}$ (see \cref{lem:nuclear-factor-characterization} in the \cref{sec:preliminaries-proofs}).
This expansion is the singular value decomposition of $F$ for the uniform inner products. Its coefficients $\sigma_j$ are the \emph{singular coefficients} of $F$, and their sum is $\NuclearNorm{F}$ (see \cref{sec:linear-algebra}).
The inner-product discrepancy estimate~\cite{DBLP:journals/siamcomp/ChorG88} (see \cref{eq:table-disc}) turns the extractor property into a bound for arbitrary tables:
\[
 \left|\E_{(\ell,r)\sample\mu_0}{F(\ell,r)}
       -\E_{(\ell,r)\sample\mu_1}{F(\ell,r)}\right|
 \le\DiscBound\cdot\NuclearNorm{F},\qquad
 \DiscBound=3/\sqrt{2^n-2},
\]
where $\mu_s$ is the uniform distribution on nonzero encodings of $s$.
Thus changing the bit encoded by one state changes the average of a table by at most $\DiscBound$ times its nuclear norm.
The rest of the refreshing proof constructs, for every answer of a refresh, such an approximation with small nuclear norm and small error.

\para{Independent encodings and conditional leakage.}
To apply the discrepancy bound to one state, we must average over that state with its encoding distribution $\mu_s$, independently of the other states.
Observed leakage, however, correlates the states: given the answers, the successive states are in general not independent.
We therefore never condition the states on the answers.
A uniform invertible matrix sends every fixed encoding to a uniform encoding of the same bit (see \cref{lem:fresh}).
Before leakage, the successive states $X_i=(L_i,R_i)$ are therefore independent samples from $\mu_s$.
We may thus sample all states first and then draw each answer from its conditional leakage probability function $\prob_{i,h,z}\left(X_{i-1},X_i\right)$, the probability that round $i$ yields the answer $z$ given the incoming history $h$ (the answers of the earlier rounds) and the old and new states.
The incoming history only determines which leakage functions are evaluated. The states are not conditioned on the listed answers having occurred.
This description is the exact real execution: the probability of a transcript is the expectation, over independent states, of the product of the answer probabilities of its rounds.
All dependencies created by the leakage are carried by these answer probabilities, each of which involves only the two states of its round, and this is what makes local replacements possible.

\para{Explicit local approximation.}
We now construct the approximation of a single answer.
Fix one answer $z$ and any formal incoming history, that is, any sequence of earlier answers, whether or not it can occur (see \cref{sec:leakage}).
The conditional probability table $\prob_z$ of this answer is a function of the four shares $((L,L'),(R,R'))$ of the old and new state.
We write simply $\prob$ when the answer and history are fixed and clear from context.
Our approximation is the simplest one suggested by the nuclear norm: we take the singular value decomposition of the table and keep only the terms with large singular coefficients.
One point needs care.
In an actual refresh, the old and new state always encode the same bit, and we call a four-share tuple \emph{consistent} if it has this property.
The table $\prob_z$ is thus defined only on consistent tuples, and these do not form a product set, since the condition $\ell\cdot r=\ell'\cdot r'$ couples the two sides.
To obtain a table on all pairs of left and right arguments, we extend $\prob_z$ by zero on inconsistent tuples and take its orthonormal expansion $\prob_z=\sum_j\sigma_j\cdot u_j\cdot v_j^{\transpose}$ with probability-normalized singular coefficients, that is, with inner products and norms taken as expectations under uniform measures (see \cref{lem:orthonormal-expansion}).
Keep precisely the terms whose singular coefficients exceed $\beta$:
\[
 \prob_z^{>\beta}:=\sum_{j:\,\sigma_j>\beta}\sigma_j\cdot u_j\cdot v_j^{\transpose}.
\]
The superscript refers to singular coefficients, not to table entries (see \cref{eq:refresh-explicit}), and our approximation is $2\cdot\prob_z^{>\beta}$.
The factor two compensates for the consistency condition, satisfied by roughly half of independent four-share tuples.
For any table $F$, the consistency projection $\project{F}$ keeps its entries at tuples whose two encoded bits agree and sets the others to zero.
The \emph{residual} $\Delta_z=\prob_z-2\cdot\project{\prob_z^{>\beta}}$ (see \cref{eq:refresh-residual}) measures what the approximation misses. It equals $\prob_z-2\cdot\prob_z^{>\beta}$ on actual refresh tuples.

What do we need from this approximation?
The discrepancy bound needs a small nuclear norm, and the hybrid argument below needs a small error under the actual encoding distributions.
The hybrid argument also multiplies the approximations of two adjacent rounds, which requires control of the individual factors and not only of their total cost.
This control is given by the \emph{factor energies} $\Phi_\beta$ and $\Psi_\beta$ of $\prob_z^{>\beta}$, the sums of squares of its balanced left and right factors (see \cref{def:factor-energy}).
The threshold $\beta$ trades these requirements against each other: a smaller threshold keeps more terms, which decreases the error but increases the energies.
For $q:=2^n-1$ and $q^{-1/3}\le\beta\le1$, the construction has
\begin{align*}
 \NuclearNorm{\prob_z^{>\beta}}&\le2,
 \\
 \Phi_\beta(\ell,\ell')&\le1/\beta
 &&\text{for every }(\ell,\ell')\in\UL^2,
 \\
 \Psi_\beta(r,r')&\le1/\beta
 &&\text{for every }(r,r')\in\UR^2,
 \\
 \E_{(\ell,\ell')\sample\UL^2}{\Phi_\beta(\ell,\ell')}&\le2,
 \\
 \E_{(r,r')\sample\UR^2}{\Psi_\beta(r,r')}&\le2,
 \\
 \LtwoNormAt{\prob_z-2\cdot\prob_z^{>\beta}}{\mu_s\otimes\mu_s}&\le C\cdot\beta^{1/3}
 &&\text{for every }s\in\{0,1\}.
\end{align*}
Here $\UL$ and $\UR$ are the nonzero left and right share spaces.
Multiplying each factor by $\sqrt2$ gives the approximation $2\cdot\prob_z^{>\beta}$, whose energies are $2\cdot\Phi_\beta$ and $2\cdot\Psi_\beta$.
The error norm $\LtwoNormAt{\cdot}{\mu_s\otimes\mu_s}$ is the root-mean-square value when the old and new states are sampled independently from $\mu_s$ (see \cref{sec:copy-local}).
The same $2\cdot\prob_z^{>\beta}$ works for either secret and is defined even on inconsistent tuples, which will let us use it in the hybrids, where a round may connect encodings of different bits.
Under any assignment of the bits encoded by the old and new state, its $L_2$ norm is at most $6$.
All bounds in this display are per answer and uniform over formal histories.

The nuclear bound uses the rectangle structure for a fixed matrix: it combines the masks of the answer, the indicator functions of the left and right sets whose product is its rectangle (see \cref{sec:leakage}), with an exact variance identity for the Gram matrices (matrices of pairwise inner products) under uniform invertible refreshing.
The error bound is the heart of the argument, since it must show that the approximation misses little under the actual encoding distributions.
Two facts combine.
First, cancellation among the binary signs $(-1)^{\ell\cdot r+\ell'\cdot r'}$ that encode the consistency condition controls the largest singular coefficient of $\Delta_z$.
Second, a four-cycle count (an average of products of four table entries over two rows and two columns) and an exact identity for the squared norm of $\Delta_z$ (see \cref{eq:direct-residual-identity}) bound this norm in terms of itself: for $t:=\LtwoNorm{\Delta_z}$, they give the inequality $t^2\le C\cdot\beta^{1/2}\cdot t^{1/2}+C/(q\cdot\beta)$ (see \cref{eq:refresh-error-bootstrap}), and hence $t\le C\cdot\beta^{1/3}$.
\Cref{app:direct-approximation-proof} proves these bounds for refreshing.
\Cref{sec:joint-general} states the extension needed for the long encodings $(a,b)$ that the gadget forms from tensor products (see \cref{sec:intro-compiler}), whose additional proof details are in \cref{app:general-approximation-proof}.

\para{The refreshing hybrid.}
We now combine the approximation with the hybrid argument.
The difficulty is that the intermediate experiments of the hybrid argument are not real executions.
We change the secret from zero to one in steps, labeling the states along the chain by the bits they encode, a prefix of ones followed by zeros (see \cref{fig:chain-hybrid}).
At the boundary between the ones and the zeros, one round connects an old state and a new state that carry different labels.
We call it the \emph{cheating round}, and the other rounds \emph{valid}.
No actual refresh behaves like this, and the true probability function $\prob$ describes only consistent tuples, so it cannot be used at the cheating round.
This is exactly where the approximation helps, since it is defined on inconsistent tuples and is the same for both secrets.
Smaller thresholds make the error smaller but the energies, and hence the discrepancy cost below, larger, and we take the smallest threshold allowed above, $\beta=q^{-1/3}$, giving the per-answer $L_2$ error $\varepsilon=C\cdot q^{-1/9}$.
Write $G$ for the resulting approximation $2\cdot\prob_z^{>\beta}$, omitting the round, history, and answer indices for brevity.
Each hybrid uses $G$ at the cheating round and the true probability functions $\prob$ elsewhere.
A round that uses $G$ instead of $\prob$ is called \emph{signed}, and a round that uses $\prob$ is \emph{unsigned}.
The endpoints, in which all labels are equal, are the actual transcript distributions.
To move the boundary by one state, replace the next $\prob$ by $G$, change the bit of the state between the two signed rounds, and restore the preceding $G$ to $\prob$.
Only two rounds are signed at any time, and true functions are inserted or removed only at valid rounds.

Each such move changes the experiment only slightly.
We must bound the approximation error, the cost of changing a label, and the effect of adaptive leakage.
We start with the approximation error.
For fixed answers, write $e:=\prob-G$ for the local approximation error.
When we replace $\prob$ by $G$ at a valid round, the error $e$ appears multiplied by a signed companion $G'$, the table of the other signed round that shares a state with the replaced round.
The replacement error is therefore $e\cdot G'$ on three independent encoded states $A$, $X$, and $B$.
These states are sampled from the encoding distributions specified by their current hybrid labels, without conditioning on the observed history.
The $L_2$ bounds on $e$ and $G'$ (see \cref{eq:copy-local-error,eq:copy-local-ltwo}) and the Cauchy--Schwarz inequality (see \cref{lem:la-cauchy-schwarz}) give $\E_{A,X,B}{|e\cdot G'|}\le6\varepsilon$, even though the two functions share the middle state.

Next, changing the label of a state, that is, the bit it encodes, is where the discrepancy bound enters.
The state is shared by the two signed rounds adjacent to it.
We multiply their product representations and average the nuclear norm of the result, as a table in that state, over the neighboring states.
This is where we need the factor energies and not only the nuclear norms: we bound one energy pointwise by $2/\beta$ and average the other, whose mean is at most $4$, which gives the bound $8/\beta$.
The discrepancy cost is therefore $8\DiscBound/\beta$.

Finally, the leakage is adaptive: the leakage functions of later rounds depend on earlier answers, so we cannot fix all leakage functions in advance.
We handle this without conditioning on a history.
First, sum the answers of the unsigned rounds after the signed ones in reverse order, which is possible because their true probability functions sum to one.
Then enumerate the answers of the two signed rounds (the \emph{block}) and of the one possible earlier neighboring round, and finally sum the remaining prefix of answers.
The enumeration costs at most $K^3$ and uses no independence conditioned on a history.
Adding $O(T)$ steps gives
\[
 \StatDist{P_0^T}{P_1^T}
 \le C\cdot T\cdot K^3\cdot\left(q^{-1/9}+\DiscBound\cdot q^{1/3}\right)
 \le C'\cdot T\cdot K^3\cdot q^{-1/9}.
\]
The initial and final stored encodings remain private.

This is also where our proof departs from earlier ones.
The security proofs of Dziembowski and Faust~\cite{DBLP:conf/asiacrypt/DziembowskiF11,DBLP:conf/tcc/DziembowskiF12} also switch the encoded bit one encoding at a time.
They simulate the leakage of the refresh at which the bit changes, the counterpart of our cheating round, from leakage of a fixed encoding.
Our comparisons instead use signed approximations common to both secrets.

\para{Why the gadget refreshes before shortening.}
We now turn to the compiler, starting with a design choice in the gadget of \cref{sec:intro-compiler}.
Since shortening already outputs an exactly fresh encoding, one might try to shorten the long encoding $(a,b)$ directly, without the matrix phase.
This is not safe, because the right party's shortening leakage may read the message $H$.
Since $A\cdot H=U$, the row space of $H$ contains $U$.
Without the refresh, $U$ would be the long row $a=(L_1\otimes L_2,1)$, which is typically the only vector of this tensor form in that row space, so the right party's leakage could recover $L_1$ and $L_2$ from $H$ and leak $s_1=L_1\cdot R_1$.
After the refresh, $U$ is a uniformly random long share, and once $M$ is erased, the row space of $H$ no longer identifies the operand shares.

\para{Transport to a computation gadget.}
For circuits, we need a local approximation of the same kind for every operation.
For the input-port refreshes, this is the local approximation behind \cref{thm:main}, which controls one refresh with its input and output encodings retained.
A computation gadget, however, has two new features.

First, each party's leakage may also read its two operand shares, so the approximation must keep these as additional arguments.
The possible operand pairs, $(L_1,L_2)$ on the left and $(R_1,R_2)$ on the right, form sets of size $d=q^2$.
The augmented tensor maps $(L_1,L_2)\mapsto(L_1\otimes L_2,1)$ and $(R_1,R_2)\mapsto(R_1\otimes R_2,1)$ take this set into nonzero vectors of length $m=n^2+1$ and are injective.
Thanks to this injectivity, the approximation lemma with retained local data (see \cref{lem:joint-map}) applies to the matrix phase for $q^{-2/3}\le\beta\le1$.

Second, the gadget ends by shortening, so the approximation, which is a table in the long shares, must be transported to the short output encoding.
A random rectangular matrix shortens the fully refreshed long encoding after the refresh matrix is erased.
A successful shortening attempt can also be sampled backward, starting from its output $(A,R)$: the shortening matrix $H$ is uniform of full row rank, $V$ is uniform on $H\cdot V=R$, and $U=A\cdot H$.
This backward description transports each product representation by averaging over $H$ and its fibers, the sets of long shares consistent with a given short output.
For each $H$, averaging a product over a fiber again gives a product, so the transported table is again a sum of left/right products.
Jensen's inequality preserves the pointwise energy bound $2/\beta$, and since $U$ and $V$ are uniform in the backward description, the mean energy bounds $4$ of the left and right factors are preserved.
The Cauchy--Schwarz inequality (see \cref{lem:la-cauchy-schwarz}) contracts the $L_2$ error under valid labelings, those in which the three wire bits satisfy the gate equation.
The rare retry changes each answer probability by at most $4/2^n$.
Thus the complete gadget (see \cref{lem:compiler-local-interface}) retains the per-answer nuclear and energy bounds and error $C\cdot\beta^{1/3}$, so the composition argument below can treat gadgets and refreshes alike.
Under any assignment of bits to its three wires (the two operands and the output), the approximation has $L_2$ norm at most $C_0:=4\sqrt{27}$.

\para{Composition with hidden output shares.}
It remains to compose the local approximations of all operations of a compiled circuit.
As for refreshing, we first make the wire encodings independent.
For fixed public and private circuit inputs, exact freshness of each completed operation permits a canonical experiment in which we sample one independent encoding for each wire before drawing leakage answers from their conditional probability functions (see \cref{lem:compiler-canonical} in the \cref{sec:matrix-estimates}).
These are unconditional samples, and observing the answers can create dependencies between them.

Two features make the composition harder than for a chain of refreshes.
First, a wire can occur in three operations, so the errors of three local approximations may meet on one wire, and products of local errors call for an $L_3$ bound instead of an $L_2$ bound.
Second, the wires form a circuit rather than a chain.
Instead of moving a boundary through the circuit, we compare the executions for all private inputs with one common main term.

To obtain the $L_3$ bound, we use the same threshold $\beta=q^{-1/3}$ for each computation gadget and each input-port refresh, and we \emph{trim} each approximation $G$: we remove its rows and columns whose factor energy exceeds $B:=\beta^{-1/3}$.
The separate mean-energy bounds and the valid-label $L_2$ error imply that the trimmed approximation has $L_3$ error $O(\beta^{1/9})$, while both pointwise energies are at most $B$.
Write $e:=\prob-G$ for this trimmed error, and write $D_s:=\mu_s-\nu$ for the signed difference between an encoding distribution and the distribution $\nu$ of two independent uniform nonzero shares.
The density of $D_s$ relative to $\nu$ has largest singular coefficient $O(q^{-1/2})$.

To obtain the main term, we expand the product of the local functions using $\prob=G+e$.
For each resulting set of error operations, keep the true encoding distributions on their incident wires and expand $\mu_s=\nu+D_s$ only on the other non-public wires.
The term with no errors and no differences is the main term, called the \emph{empty term} in \cref{sec:direct-composition}, and every other term contains at least one error $e$ or one difference $D_s$.
Bounding these terms one at a time does not work: each term still contains approximations $G$ at the remaining operations, and since these are not probabilities, their answers do not sum to one, so we would have to enumerate the answers of all operations, at a cost of up to a factor $K$ per operation.
We therefore group the terms.
Fix a set $A_\star$ of \emph{marked} errors and a set $W_\star$ of marked wires, and sum all terms containing these marked differences before taking absolute values.
This sum restores the true probability functions everywhere except at the marked errors and at the operations touching marked wires, whose joint factor energies are at most $B^{3|W_\star|}$.
Tensoring the discrepancy estimate, applying H\"older's inequality to the marked errors, and enumerating the answers of at most $9|A_\star|+3|W_\star|$ neighboring operations gives the bound
\[
 \bigl(C\cdot K^9\cdot q^{-1/27}\bigr)^{|A_\star|}\cdot
 \bigl(C\cdot K^3\cdot q^{-1/6}\bigr)^{|W_\star|}.
\]
The remaining true answer probabilities sum to one in reverse time, even under adaptive leakage.
Inclusion--exclusion combines these marked bounds and controls the sum of all terms other than the main term by $O(S\cdot K^9\cdot q^{-1/27})$ in the nontrivial range.
Finally, the main term is a \emph{signed measure} on leakage transcripts: it assigns each transcript a real weight that may be negative.
It is independent of every private wire label and is therefore common to all private inputs.
It cancels when the real and dummy-input executions are compared, and it never needs to be sampled.
\Cref{sec:direct-composition} gives the full argument.

\subsection{Related work}\label{sec:related-work}

\para{Leakage models.}
Micali and Reyzin~\cite{DBLP:conf/tcc/MicaliR04} proposed the axiom that only computation leaks information.
Dziembowski and Pietrzak~\cite{DBLP:conf/focs/DziembowskiP08} combined it with bounded leakage per round, and constructed a stream cipher whose memory is divided into parts that leak independently.
Akavia, Goldwasser, and Vaikuntanathan~\cite{DBLP:conf/tcc/AkaviaGV09} studied bounded leakage from memory, and~\cite{DBLP:conf/focs/BrakerskiKKV10,DBLP:conf/focs/DodisHLW10} introduced continual memory leakage, in which secret keys are updated periodically.
Tolerating leakage during the update itself is a recurring difficulty in that line of work. See~\cite{DBLP:conf/focs/BrakerskiKKV10,DBLP:conf/stoc/LewkoLW11} and the references therein.
Our refresh plays the role of such an update, and our leakage functions see all of its randomness.
Our model combines the two-party split state of~\cite{DBLP:conf/focs/DziembowskiP08,DBLP:conf/asiacrypt/DziembowskiF11} with the axiom of~\cite{DBLP:conf/tcc/MicaliR04} at the granularity of one operation, as in the compilers of~\cite{DBLP:conf/crypto/GoldwasserR10,DBLP:conf/tcc/DziembowskiF12}.
Kalai and Reyzin~\cite{DBLP:books/acm/19/KalaiR19} survey this area.

Other models are closer to the practice of side-channel attacks.
An early formal treatment of power analysis is due to Chari, Jutla, Rao, and Rohatgi~\cite{DBLP:conf/crypto/ChariJRR99}, who proposed an abstract model of power consumption and a generic technique for provably resistant implementations.
In the probing model of Ishai, Sahai, and Wagner~\cite{DBLP:conf/crypto/IshaiSW03}, the adversary learns the values on a bounded number of wires of a circuit.
In the noisy leakage model, leakage is inherently noisy, and Prouff and Rivain~\cite{DBLP:conf/eurocrypt/ProuffR13} proved the security in it of \emph{masking}, a common countermeasure that splits each sensitive value into random shares and computes on these shares.
Duc, Dziembowski, and Faust~\cite{DBLP:conf/eurocrypt/DucDF14} gave a proof that avoids the leak-free gates of~\cite{DBLP:conf/eurocrypt/ProuffR13} by reducing noisy leakage to probing, and Brian, Dziembowski, and Faust~\cite{DBLP:conf/eurocrypt/BrianDF24} removed the dependence of the noise parameter on the field size.
Our leakage is bounded and local, and it is neither probing nor noisy.

\para{Leakage-resilient storage and its refreshing.}
Dav{\`\i}, Dziembowski, and Venturi~\cite{DBLP:conf/scn/DaviDV10} introduced leakage-resilient storage based on two-source extractors, including the inner product.
Dziembowski and Faust~\cite{DBLP:conf/asiacrypt/DziembowskiF11} showed how to refresh inner-product encodings under continual leakage, and used the refreshed encodings to protect standard schemes such as the Okamoto identification scheme and an ElGamal-based encryption scheme.
As explained above, their refresh delivers its randomness through a leak-free component that outputs pairs of orthogonal vectors.
For fields of size $\Omega(n)$, their refreshing protocol tolerates $0.15\cdot n\cdot\log_2(|\F|)-1$ bits of leakage per refresh~\cite[full version, Corollary~2]{DBLP:conf/asiacrypt/DziembowskiF11}.
Their circuit compiler~\cite{DBLP:conf/tcc/DziembowskiF12} uses a variant of this refresh whose output encoding is independent of the input encoding apart from the secret, again with a leak-free component, and Andrychowicz~\cite{DBLP:journals/corr/abs-1209-4820} gave a more efficient refreshing protocol under the same type of assumption.
Refreshing of encoded secrets has also been studied for masking schemes.
For circuits with only affine gates, the compilers of Andrychowicz, Dziembowski, and Faust~\cite{DBLP:conf/eurocrypt/AndrychowiczDF16} give constant-rate refreshing of an encoded secret against probing.
Dziembowski, Faust, and {\.Z}ebrowski~\cite{DBLP:conf/asiacrypt/DziembowskiFZ19} prove that a simple and efficient refreshing of masked values is secure in the noisy leakage model.
In our protocol, the randomness that correlates the old and new encodings is exposed in full to the leakage functions, and no component is leak-free.
The price is a smaller leakage rate: over $\F_2$, our refreshing bound is exponentially small for rates below $1/54$.
We do not know whether this loss is inherent to the protocol or only to our analysis.

\para{Local leakage from shares.}
Benhamouda, Degwekar, Ishai, and Rabin~\cite{DBLP:journals/joc/BenhamoudaDIR21} study linear secret sharing under local leakage, where a function with a bounded output is applied to each share separately.
They show that additive sharing and high-threshold Shamir sharing withstand such leakage over large prime fields when the number of parties is sufficiently large, whereas every linear scheme over a field of small characteristic is insecure, regardless of the number of parties.
The inner-product encoding is bilinear rather than linear in its shares, and its security against local leakage over $\F_2$ rests on the extractor property~\cite{DBLP:journals/siamcomp/ChorG88,DBLP:conf/scn/DaviDV10}.

\para{Fixed encodings and two-source extractors.}
For an encoding that is never refreshed, much more leakage can be tolerated.
Dziembowski and Faust~\cite[Lemma~1 and Corollary~1]{DBLP:conf/tcc/DziembowskiF12} show that a variant of the $k$-coordinate inner-product encoding, with the first left coordinate required to be nonzero, over a field of size $q=\Omega(k)$ tolerates $0.49\cdot k\cdot\log_2(q)-1$ bits of leakage from each share, with error negligible in $k$.
For a discrete random variable $X$, its \emph{min-entropy} is $H_\infty(X):=-\log_2\max_x\Prb{X=x}$. For an $n$-bit source its min-entropy rate is $H_\infty(X)/n$.
The inner product is an extractor for two independent sources whose min-entropy rate exceeds $1/2$~\cite{DBLP:journals/siamcomp/ChorG88}.
Other explicit two-source extractors need less: Bourgain~\cite{doi:10.1142/S1793042105000108} reaches min-entropy rate $0.499$, and Chattopadhyay and Zuckerman~\cite{ChattopadhyayZuckerman2019} reach polylogarithmic min-entropy.
We use the inner product because the refresh multiplies the shares by $M$ and $M^{-1}$, which preserves their inner product.

\para{Circuit compilers.}
Ishai, Sahai, and Wagner~\cite{DBLP:conf/crypto/IshaiSW03} introduced circuit compilers that protect computations on randomly masked shares against probing attacks, and Andrychowicz, Dziembowski, and Faust~\cite{DBLP:conf/eurocrypt/AndrychowiczDF16} construct compilers with perfect security against probing and leakage rate $O(1/\log n)$, where $n$ is the security parameter.
Other compilers restrict the class of leakage functions.
Faust et al.~\cite{DBLP:conf/eurocrypt/FaustRRTV10} treat computationally bounded and noisy leakage.
Rothblum~\cite{DBLP:conf/crypto/Rothblum12} compiles circuits that resist $\mathrm{AC}^0$ leakage without secure hardware, under a computational assumption about inner products.
Miles and Viola~\cite{DBLP:conf/stoc/MilesV13} use products over the alternating group to resist leakage by $\mathrm{NC}^1$ functions under the assumption $\mathrm{L}\ne\mathrm{NC}^1$, and by further classes for which average-case lower bounds are known.
Ishai and Song~\cite{DBLP:conf/eurocrypt/IshaiS24} construct leakage-tolerant circuits, whose leakage can be simulated by the same class of leakage functions applied only to the input and output, for global leakage by parities, disjunctions, or conjunctions of wires.
Ajtai~\cite{DBLP:conf/stoc/Ajtai11} protects a RAM computation against an adversary who observes the memory accesses and, at chosen times, the memory contents.

Closest to our setting are compilers that allow arbitrary bounded leakage functions on parts of the computation that leak independently.
Juma and Vahlis~\cite{DBLP:conf/crypto/JumaV10} rely on fully homomorphic encryption and a single leak-free token that samples from a fixed distribution independent of the key.
Goldwasser and Rothblum~\cite{DBLP:conf/crypto/GoldwasserR10} use public-key encryption and leak-free hardware that outputs encryptions of zero; their later compiler~\cite{DBLP:conf/focs/GoldwasserR12,Goldwasser2015} avoids leak-free hardware, but its proof has gaps (see \cref{app:gr}).
The compiler of Dziembowski and Faust~\cite{DBLP:conf/tcc/DziembowskiF12} is information-theoretic and works in a model of independently leaking parts similar to ours, but it requires leak-free components.
Later works seek to remove this hardware requirement by adding computing components or using computational assumptions.
Bitansky, Dachman-Soled, and Lin~\cite{BitanskyDachmanSoledLin2014} propose using two main computing components and $O(1)$ auxiliary components to emulate Goldwasser and Rothblum's ciphertext banks and replace the leak-free hardware of the Dziembowski--Faust compiler.
The security argument for this construction~\cite[full version, Sec.~C.2.4]{BitanskyDachmanSoledLin2014} relies on Goldwasser and Rothblum's bank analysis and is therefore also subject to the proof gap involving Claim~6.24 described in \cref{app:gr}.
Our counterexample to that claim does not establish an attack on the Bitansky--Dachman-Soled--Lin construction.
Dachman-Soled, Liu, and Zhou~\cite{DachmanSoledLiuZhou2015} obtain computationally secure compilers with only two computing components and no leak-free hardware, using universal deniable encryption and receiver non-committing encryption under assumptions including indistinguishability obfuscation.

In comparison, our compiler is statistically secure without leak-free hardware inside the computation, but it trusts the encoding of the secret and the decoding of the outputs, and each leakage function reads only its party's data for the active operation, whereas the two-component guarantee of~\cite{DachmanSoledLiuZhou2015} allows leakage on each component's full evaluation state.
Our leakage model also differs from the probing models above, so their numerical leakage rates are not directly comparable.

\subsection{Organization}
\begin{figure}[htbp]
\centering
\begingroup
\setlength{\fboxrule}{0.3pt}
\setlength{\fboxsep}{8pt}
\fcolorbox{black!25}{white}{\begin{minipage}{\dimexpr\linewidth-2\fboxsep-2\fboxrule\relax}
\centering
\begin{tikzpicture}[x=1cm,y=1cm,font=\footnotesize,
 sec/.style={draw=black!40,fill=black!2,rounded corners=2pt,align=center,text width=3.2cm,minimum height=1.45cm,inner sep=4pt},
 res/.style={draw=black!60,fill=black!6,rounded corners=2pt,align=center,text width=3.7cm,minimum height=1.45cm,inner sep=4pt},
 app/.style={draw=black!30,dashed,rounded corners=2pt,align=center,text width=3.2cm,minimum height=1.15cm,inner sep=4pt},
 flow/.style={->,draw=black!65,semithick}]
 \node[sec] (pre) at (0,0) {\crefrange{sec:preliminaries}{sec:leakage}\\[2pt]encodings, tables, discrepancy, refresh leakage};
 \node[res] (refresh) at (4.6,0) {\cref{sec:refreshing}: refreshing\\[2pt]explicit approximation, moving-boundary hybrid\\\cref{thm:main}};
 \node[res] (circ) at (9.4,0) {\cref{sec:compilation}: compiler\\[2pt]shortening, composition, simulation\\\cref{thm:compiler-direct}};
 \node[app] (appA) at (0,-2.5) {\cref{sec:preliminaries-proofs}\\[2pt]preliminary proofs};
 \node[app] (appB) at (9.4,-2.5) {\cref{sec:matrix-estimates}\\[2pt]compiler approximation and shortening};
 \draw[flow] (pre)--(refresh);
 \draw[flow] (refresh)--(circ);
 \draw[flow,dashed,black!45] (appA)--(pre);
 \draw[flow,dashed,black!45] (appB)--(circ);
\end{tikzpicture}
\end{minipage}}
\endgroup
\caption{The proof is first presented for continual refreshing and then extended to circuits.
Dashed boxes are appendices. The arrow to the compiler represents an extension of the method, not a deduction from refreshing security alone.}
\label{fig:roadmap}
\end{figure}
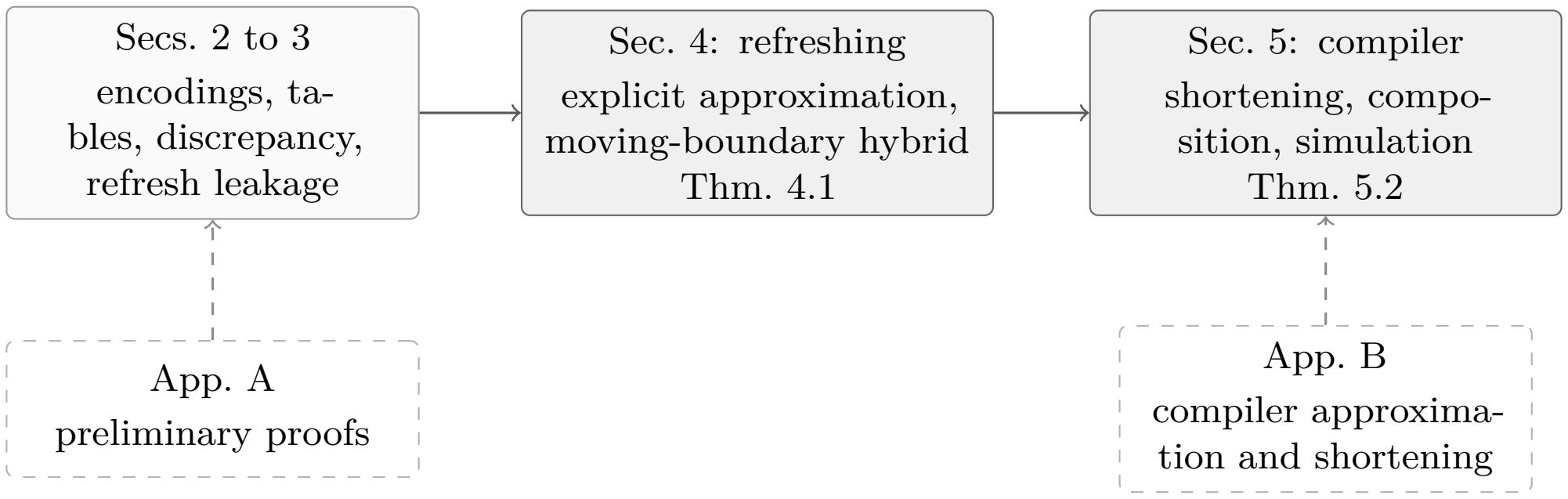

\Cref{sec:preliminaries} introduces the encodings, the linear algebra of real tables, and the two-secret discrepancy bound.
\Cref{sec:leakage} defines leakage from a refresh, its answer masks, and histories of successive refreshes.
\Cref{sec:refreshing} gives the complete refreshing argument: the security statement, the independent-encoding experiment, the explicit local approximation, and the hybrid proof that moves a boundary along the chain.
It also derives the leakage rate and share length and treats initialization with zero shares.
\Cref{sec:compilation} gives the compiler, with its input-port refreshes and trusted output decoding, the computation gadget and its local approximation, the composition proof, and the simulator.
\Cref{sec:formalization} proves the extension to stateful circuits, records the formalization status, and discusses further extensions of the leakage model and field.
The appendices contain the remaining proofs and supporting material.
\Cref{sec:preliminaries-proofs} proves the preliminary lemmas.
\Cref{sec:matrix-estimates} proves the extension of the approximation lemma to retained local data (the operand shares that the leakage may also read), its tensor-encoding instance, and the shortening results.
\Cref{app:gr} gives a counterexample to an auxiliary claim in the proof of Goldwasser and Rothblum.
\Cref{fig:roadmap} shows the progression from refreshing to computation.

\section{Preliminaries}\label{sec:preliminaries}
The refreshing and circuit compilation proofs use the inner-product encoding and describe local leakage by real-valued tables.
We first fix notation in \cref{sec:conventions} and define the encoding and matrix refresh in \cref{sec:shares}.
We then develop the linear algebra of tables, including the definition of the nuclear norm, in \cref{sec:linear-algebra}, and establish the counting, freshness, and discrepancy estimates in \cref{sec:counting} that connect these tables to encoded secrets.
Proofs and auxiliary definitions are collected in the \cref{sec:preliminaries-proofs}.

\subsection{Conventions}\label{sec:conventions}
The share and matrix spaces are finite, and all real vector spaces used below are finite-dimensional.
Share arithmetic takes place in $\F_2$, whereas the tables (real functions of a left share and a right share, defined in \cref{def:table-norms} below) and their factor functions are real.
The factor functions, introduced below in \cref{def:product-factors}, depend only on one party's data.
We write $\log$ for the base-two logarithm and $\ln$ for the natural logarithm.
The letter $C$ denotes an absolute positive constant whose value may increase between inequalities.
On a finite or countable set, a probability distribution also denotes its point-probability function.
We write $x:=f(y)$ for deterministic assignment.
For a nonempty finite set $S$, the notation $X\sample S$ means that $X$ is sampled uniformly from $S$.
If $\rho$ is a probability distribution, $X\sample\rho$ means sampling according to $\rho$.
The notation $\rho\otimes\rho'$ denotes independent sampling from the two distributions.
The marginal distribution of one component of a random tuple is its distribution when the other components are ignored.
For real tables, $F\cdot H$ denotes entrywise multiplication, and inequalities between tables are pointwise.
Each expectation indicates the variables over which it averages, and its argument is enclosed in square brackets.

\begin{definition}[Finite signed weights]\label{def:measure-norm}
Let $V$ be a finite set.
A \emph{signed weight function} on $V$, also called a \emph{signed measure}, is a function $P:V\to\R$ whose values may be negative.
We use the unnormalized $L_1$ norm
\[
 \LoneNorm{P}:=\sum_{v\in V}|P(v)|,
\]
and evaluate a real function $F:V\to\R$ against $P$ by the finite weighted sum
\[
 \sum_{v\in V}F(v)\cdot P(v).
\]
\end{definition}
Every probability distribution, such as the distribution of the transcripts of a real execution, is a signed measure with nonnegative values.
The security proofs also need signed measures with negative values.
These arise when the probability of a leakage answer is replaced by a signed approximation, and when an encoding distribution is written as independent uniform shares plus a signed correction.
Such intermediate measures need not have total mass one and are never sampled: they only serve to compare the genuine transcript distributions at the endpoints of an argument.
Keeping the signs in a weighted sum allows positive and negative contributions to cancel, and the later discrepancy estimates bound such a sum as a whole before taking its absolute value.
Note that the $L_1$ norm above is an unnormalized sum of absolute values, unlike the uniform averages that define the table norms in \cref{def:table-norms}.
\begin{definition}[Statistical distance]\label{def:statistical-distance}
Let $V$ be a finite set and let $P$ and $Q$ be probability distributions on $V$.
Their \emph{statistical distance} is
\[
 \StatDist{P}{Q}:=\frac12\cdot\LoneNorm{P-Q}
 =\max_{\cM:V\to\{0,1\}}
 \left|\Prb_{v\sample P}{\cM(v)=1}-\Prb_{v\sample Q}{\cM(v)=1}\right|.
\]
\end{definition}
We reserve $\delta$ for probability distributions and use the $L_1$ norm for arbitrary signed weights.
\Cref{app:prelim-probability} justifies the two expressions for $\delta$.
\Cref{tab:notation} in the \cref{sec:preliminaries-proofs} lists the main symbols for reference.
Each symbol is defined where it is introduced.

\subsection{Encodings and the matrix refresh}\label{sec:shares}
We first fix the encoding and the refresh protocol whose security we analyze.
Throughout the main text, arithmetic on shares and refresh matrices is over $\F_2=\{0,1\}$, and $n\ge2$ is an integer.
\begin{definition}[Shares, states, and encoding distributions]\label{def:joint-state}\phantomsection\label{def:encoding-distribution}
Put $N:=2^n$ and $q:=N-1$.
Let $\UL$ be the set of nonzero row vectors and $\UR$ the set of nonzero column vectors in $\F_2^n$.
Each set has $q$ elements.
A \emph{left share} is a row $\ell\in\UL$, a \emph{right share} is a column $r\in\UR$, and a \emph{joint state}, or simply \emph{state}, is a pair $(\ell,r)\in\UL\times\UR$ of shares held at the same time.
The state \emph{encodes} the secret bit
\[
 \ell\cdot r=\sum_{k=1}^n\ell_k\cdot r_k\in\F_2.
\]
For $s\in\F_2$ define
\[
 \Omega_s:=\{(\ell,r)\in\UL\times\UR:\ell\cdot r=s\},
\]
and let $\mu_s$ be the uniform distribution on $\Omega_s$.
For a map $f:X\to Y$ and $y\in Y$, the \emph{fiber} of $f$ at $y$ is the set $\{x\in X:f(x)=y\}$.
We call $\Omega_s$ the class of the secret $s$, or the fiber of the inner-product map at $s$: it is the set of all encodings with that secret.
\end{definition}
Uppercase letters denote random shares and states.
The corresponding lowercase letters denote their values.
Requiring both shares to be nonzero is a convenience.
\Cref{sec:zero-initialization} shows that allowing zero shares at initialization increases the refreshing bounds by at most $4/N$.

\begin{definition}[Matrix refresh]\label{def:protocol}
Let $s\in\F_2$ and let $T\ge0$ be an integer.
Initialize $(L_0,R_0)\sample\mu_s$ and write $X_0:=(L_0,R_0)$.
In each round $i=1,\ldots,T$:
\begin{enumerate}
\item The left party samples $M_i\sample \InvMatr_2^n$, where $\InvMatr_2^n$ is the set of invertible $n\times n$ matrices over $\F_2$, independently of the entire preceding execution, and sends it to the right party over a private channel.
\item The parties update their shares:
\[
 L_i:=L_{i-1}\cdot M_i,\qquad R_i:=M_i^{-1}\cdot R_{i-1}.
\]
\item Each party erases its old share and the matrix, and keeps only its new share.
\end{enumerate}
The state after round $i$ is $X_i=(L_i,R_i)$.
\end{definition}

Every refresh preserves the secret and keeps both shares nonzero (see \cref{app:prelim-counting}).
Because of the erasure in step~3, the data held by the parties at the start of round $i+1$ are exactly the state $X_i$.
Sending the square matrix costs $n^2$ bits per refresh.

\subsection{Linear algebra for tables}\label{sec:linear-algebra}\label{sec:la-shared}
Having fixed the protocol, we turn to the tool we use to analyze its leakage: the linear algebra of real functions on a product of two finite sets, which we call tables.
In the security proofs, a table records the probability of a leakage answer as a function of the left party's data and the right party's data.
The simplest tables are products $a(\ell)\cdot b(r)$. For these tables, the extractor property of the inner product bounds the difference between their means under $\mu_0$ and $\mu_1$ by $\frac{3}{\sqrt{2^n-2}}\cdot\LtwoNorm{a}\cdot\LtwoNorm{b}$ (see \cref{eq:disc}).
Here the $L_2$ norms, defined below in \cref{def:table-norms}, use the uniform measures on nonzero shares, so the means are close when these norms are bounded independently of $n$.
For a fixed refresh matrix, the Boolean function indicating whether a given answer occurs is such a product, namely of the two masks defined below in \cref{def:rectangle}, because each party's output depends only on its own share and the matrix.
The security proof, however, must also retain the new shares, since they feed the next round.
It therefore uses a table on both the old and the new shares, obtained by averaging over the matrices compatible with those endpoints (see \cref{sec:technical-outline}).
This table differs from the ordinary average of the fixed-matrix masks on the old shares alone, and in general it is not a product.
By linearity, this bound extends to sums of products, with a cost proportional to the total size of their terms.
The \emph{nuclear norm} is the minimum such cost, so a table with small nuclear norm has a mean that is nearly independent of the secret.
The averaged table above need not have a small nuclear norm, and this is why the proofs approximate it.

We define the nuclear norm below in \cref{def:nuclear-norm} as the sum of the singular coefficients in the orthonormal expansion of \cref{lem:orthonormal-expansion}.
The uniform inner products and orthonormality used here are defined in \cref{def:table-norms,def:orthonormal-families}.
This expansion is the singular value decomposition for those inner products.
For sets of equal size $q$, the normalized singular coefficients are $1/q$ times the ordinary singular values of the matrix $(F(\ell,r))$.
We call this convention, in which inner products and norms are expectations under the uniform measures, the \emph{probability normalization}.
The analysis also uses product factorizations whose factors need not be orthogonal.
\Cref{app:prelim-linear-algebra} proves the properties we need, including the characterization of the nuclear norm as a minimum over product representations and the extension to tables on sets of different sizes.
Both security proofs use product representations to control local comparisons, and the local approximation in \cref{sec:joint-refresh} also separates an orthonormal expansion into its large and small coefficients.
Every expectation in this section is uniform on the indicated set or product of sets.
Although $\UL$ and $\UR$ both have size $q$, all the arguments also apply to two finite sets of different sizes, each equipped with its own uniform measure.
In \cref{sec:joint-refresh}, we use the sets $\UL^2$ and $\UR^2$.

\subsubsection{Tables and the Cauchy--Schwarz inequality}
\begin{definition}[Tables, inner products, and norms]\label{def:table-norms}
A table is a real function $F:\UL\times\UR\to\R$.
We call $\ell\in\UL$ the \emph{row} and $r\in\UR$ the \emph{column} of the entry $F(\ell,r)$.
Let $U$ be one of the two sets $\UL$ or $\UR$.
For a real function $g:U\to\R$,
\[
 \E_{x\sample U}{g(x)}:=\frac1{|U|}\sum_{x\in U}g(x).
\]
For real functions $a$ and $b$ from $U$ to $\R$, put
\[
 \pairing{a}{b}:=\E_{x\sample U}{a(x)\cdot b(x)},\qquad
 \LtwoNorm{a}:=\left(\E_{x\sample U}{a(x)^2}\right)^{1/2}.
\]
For tables $F$ and $W$, put
\[
 \pairing{F}{W}:=\E_{\ell,r}{F(\ell,r)\cdot W(\ell,r)},\qquad
 \LtwoNorm{F}:=\left(\E_{\ell,r}{F(\ell,r)^2}\right)^{1/2},
\]
where $\ell$ and $r$ are independent and uniform, so $\E_{\ell,r}$ is $1/q^2$ times the sum over all entries.
We call $\pairing{F}{W}$ the \emph{pairing} and $\LtwoNorm{F}$ the \emph{$L_2$ norm}, or root-mean-square value.
We write $\E_\ell$ for $\E_{\ell\sample\UL}$ and $\E_r$ for $\E_{r\sample\UR}$ when the set is clear.
\end{definition}

\begin{definition}[Product tables]\label{def:product-table}
A \emph{product table} is a table of the form $(\ell,r)\mapsto a(\ell)\cdot b(r)$, where $a:\UL\to\R$ and $b:\UR\to\R$.
We also write it as $a\cdot b^{\transpose}$.
A function with $L_2$ norm one is a \emph{unit function}, and $a\cdot b^{\transpose}$ is a \emph{unit product table} when both factors are unit functions.
\end{definition}
Here ``unit'' refers to the norms of the factors, not to their individual entries.

\begin{lemma}[Cauchy--Schwarz inequality]\label{lem:la-cauchy-schwarz}
For every finite index set $I$, every pair of real families $(x_i)_{i\in I}$ and $(y_i)_{i\in I}$, and all nonnegative real weights $(p_i)_{i\in I}$,
\[
 \left|\sum_{i\in I} p_i\cdot x_i\cdot y_i\right|^2
 \le\left(\sum_{i\in I} p_i\cdot x_i^2\right)
      \cdot\left(\sum_{i\in I} p_i\cdot y_i^2\right).
\]
\end{lemma}
The inequality includes the sum and expectation versions.
See Axler~\cite[Sec.~6A, Thm.~6.14, p.~189]{axler2023linear} for the Cauchy--Schwarz inequality.
The inequality is used throughout the proofs, starting in \cref{sec:copy-local,sec:refresh-replacement,sec:refresh-wire-change}.

\subsubsection{Orthonormal expansions and the nuclear norm}
\begin{definition}[Orthonormal families]\label{def:orthonormal-families}
Functions on the same finite set are \emph{orthogonal} when their inner product is zero.
A family $(u_j)_j$ is \emph{orthonormal} if each $u_j$ has $L_2$ norm one and $\pairing{u_i}{u_j}=0$ whenever $i\ne j$.
The inner products and norms use the uniform measure of \cref{def:table-norms}.
\end{definition}

\begin{lemma}[Pythagoras' theorem and Bessel's inequality]\label{lem:la-orthogonal-sums}
Let $m\ge0$ be an integer, let $U$ be a finite nonempty set, and let $(w_j)_{j=1}^m$ be an orthonormal family of functions $U\to\R$ for the uniform inner product.
For every coefficient vector $(c_1,\ldots,c_m)\in\R^m$, Pythagoras' theorem gives
\begin{equation}\label{eq:orthonormal-sum-norm}
 \LtwoNorm{\sum_{j=1}^m c_j\cdot w_j}^2=\sum_{j=1}^m c_j^2.
\end{equation}
For every function $f:U\to\R$, Bessel's inequality gives
\begin{equation}\label{eq:bessel-inequality}
 \sum_{j=1}^m\pairing{f}{w_j}^2\le\LtwoNorm{f}^2.
\end{equation}
Both statements also hold for the empty family.
\end{lemma}
See Axler~\cite[Sec.~6B, Thms.~6.24 and~6.26, p.~198]{axler2023linear}.
We use \cref{eq:orthonormal-sum-norm,eq:bessel-inequality} in \cref{sec:copy-local} and in the \cref{app:prelim-linear-algebra}.

\begin{lemma}[Orthonormal expansion]\label{lem:orthonormal-expansion}
For every real table $F:\UL\times\UR\to\R$, there exists an expansion
\begin{equation}\label{eq:orthonormal-expansion}
 F(\ell,r)=\sum_{j=1}^{m}\sigma_j\cdot u_j(\ell)\cdot v_j(r).
\end{equation}
Here $m$ is the rank of the matrix $(F(\ell,r))_{\ell,r}$, that is, the dimension of its column span.
The coefficients satisfy $\sigma_j>0$, and $(u_j)_j$ and $(v_j)_j$ are orthonormal families on $\UL$ and $\UR$, respectively.
The coefficients $\sigma_j$, called the \emph{singular coefficients} of $F$, are uniquely determined up to ordering.
For $F=0$, the expansion is the empty sum.
\end{lemma}
The expansion~\cref{eq:orthonormal-expansion} is the singular value decomposition for the uniform inner products.
See Axler~\cite[Sec.~7E, Thm.~7.70, p.~273]{axler2023linear}.
Truncation means discarding the expansion terms whose singular coefficients are at or below a chosen threshold.
We use it in \cref{sec:copy-local,sec:compiler-local}.

\begin{definition}[Nuclear norm]\label{def:nuclear-norm}
For an orthonormal expansion as in~\cref{eq:orthonormal-expansion}, define the \emph{nuclear norm} by
\[
 \NuclearNorm{F}:=\sum_{j=1}^{m}\sigma_j.
\]
In particular, $\NuclearNorm{0}=0$.
\Cref{lem:orthonormal-expansion} ensures that this value does not depend on the chosen expansion.
\end{definition}
The orthonormal expansion gives a convenient way to choose factors, but multiplying tables or averaging them during the shortening step introduced in \cref{sec:compiler-gadget} below need not preserve orthogonality.
We therefore also keep track of product representations whose factors are not orthogonal.

\begin{definition}[Product factors]\label{def:product-factors}
In a representation
\[
 F(\ell,r)=\sum_{j=1}^m a_j(\ell)\cdot b_j(r),
\]
the functions $a_j:\UL\to\R$ are the \emph{left factors}, and $b_j:\UR\to\R$ are the \emph{right factors}.
The number $m$ is finite and may depend on the representation.
The empty representation is allowed for the zero table.
\end{definition}

The next lemma collects basic properties of the nuclear norm, including a bound for arbitrary product representations.

\begin{lemma}[Basic nuclear-norm bounds]\label{lem:factor-minimum}
The function $\NuclearNorm{\cdot}$ is a norm on the real tables on $\UL\times\UR$.
For every such table $F$,
\begin{equation}\label{eq:nuclear-dominates-ltwo}
 \LtwoNorm{F}\le\NuclearNorm{F}.
\end{equation}
For every integer $m\ge0$ and all functions $a_j:\UL\to\R$ and $b_j:\UR\to\R$, indexed by $j=1,\ldots,m$,
\begin{equation}\label{eq:nuclear-product-bound}
 \NuclearNorm{\sum_j a_j\cdot b_j^{\transpose}}
 \le\sum_j\LtwoNorm{a_j}\cdot\LtwoNorm{b_j}.
\end{equation}
For every pair of functions $a:\UL\to\R$ and $b:\UR\to\R$,
\begin{equation}\label{eq:nuclear-product-identity}
 \NuclearNorm{a\cdot b^{\transpose}}=\LtwoNorm{a}\cdot\LtwoNorm{b}.
\end{equation}
\par\nopagebreak[4]\smallskip\noindent
\textnormal{\emph{Proof: see \cref{app:prelim-factor-minimum}.}}
\end{lemma}
We use~\cref{eq:nuclear-product-bound} whenever approximations are multiplied, in \cref{sec:refresh-wire-change,sec:direct-marked-expansion} and in the \cref{sec:compiler-shortening}.
We also use~\cref{eq:nuclear-dominates-ltwo} in the \cref{sec:compiler-shortening}.

\subsubsection{Largest coefficients and factor energies}
The nuclear norm summarizes a table by one number. We next define two quantities that give finer information: the largest singular coefficient (see \cref{def:largest-singular-coefficient} below) and factor energies (see \cref{def:factor-energy} below).
The largest singular coefficient bounds the pairing of a table with every product of functions of unit $L_2$ norm, so it shows how little the terms discarded by a truncation can correlate with a product test.
Factor energies record where, among the rows and columns, a chosen product representation is large.
We need this local information when two approximations that share an encoding are multiplied, because their sizes at the same row or column multiply.
\begin{definition}[Largest singular coefficient]\label{def:largest-singular-coefficient}
Let $\sigma_{\max}(F)$ denote the largest coefficient of an orthonormal expansion of $F$, with $\sigma_{\max}(0):=0$.
\end{definition}
\begin{lemma}[Largest singular coefficient]\label{lem:largest-singular-coefficient}
For every real table $F:\UL\times\UR\to\R$ and all functions $u:\UL\to\R$ and $v:\UR\to\R$,
\begin{equation}\label{eq:largest-singular-coefficient}
 \left|\pairing{F}{u\cdot v^{\transpose}}\right|
 \le\sigma_{\max}(F)\cdot\LtwoNorm{u}\cdot\LtwoNorm{v}.
\end{equation}
For every nonzero $F$, there exist unit functions $u$ and $v$ attaining equality in~\cref{eq:largest-singular-coefficient}.
For every real table $F$,
\begin{equation}\label{eq:largest-singular-ltwo}
 \sigma_{\max}(F)\le\LtwoNorm{F}.
\end{equation}
\par\nopagebreak[4]\smallskip\noindent
\textnormal{\emph{Proof: see \cref{app:prelim-largest-coefficient}.}}
\end{lemma}
The local approximation proof in \cref{sec:copy-local} uses~\cref{eq:largest-singular-coefficient} to control the part of an expansion with small coefficients, and \cref{sec:direct-encoding-difference} uses it for encoding differences.

\begin{definition}[Factor energies]\label{def:factor-energy}
Fix a finite product representation $F=\sum_j a_j\cdot b_j^{\transpose}$.
Its left and right \emph{factor energies} are
\[
 \Phi_F(\ell):=\sum_j a_j(\ell)^2,\qquad
 \Psi_F(r):=\sum_j b_j(r)^2.
\]
They refer to the chosen representation, not only to the table $F$, and have means
\[
 \E_\ell{\Phi_F(\ell)}=\sum_j\LtwoNorm{a_j}^2,\qquad
 \E_r{\Psi_F(r)}=\sum_j\LtwoNorm{b_j}^2.
\]
\end{definition}

\begin{definition}[Balanced factors]\label{def:balanced-factors}
Fix an orthonormal expansion $F=\sum_j\sigma_j\cdot u_j\cdot v_j^{\transpose}$.
Its \emph{balanced factors} are
\[
 a_j:=\sqrt{\sigma_j}\cdot u_j,\qquad
 b_j:=\sqrt{\sigma_j}\cdot v_j,
\]
so that $F=\sum_j a_j\cdot b_j^{\transpose}$.
Their energies are those of \cref{def:factor-energy}, for this chosen representation.
\end{definition}
For these balanced factors, $\LtwoNorm{a_j}^2=\LtwoNorm{b_j}^2=\sigma_j$, so the mean energies are
\begin{equation}\label{eq:balanced-mean-energies}
 \E_\ell{\Phi_F(\ell)}=\E_r{\Psi_F(r)}=\sum_j\sigma_j=\NuclearNorm{F}.
\end{equation}
These factors also attain the minimum in the characterization of the nuclear norm (see \cref{eq:nuclear-factor-cost} in the \cref{app:prelim-factor-minimum}).
At a row $\ell$, the energy $\Phi_F(\ell)$ is the squared Euclidean length of the vector $(a_j(\ell))_j$: it measures how large the left factors are at that row.
Similarly, $\Psi_F(r)$ measures the right factors at column $r$.
The energies bound each entry (see \cref{app:prelim-factor-covariance}): for every $\ell\in\UL$ and $r\in\UR$,
\begin{equation}\label{eq:factor-energy-entry-bound}
 |F(\ell,r)|\le\sqrt{\Phi_F(\ell)\cdot\Psi_F(r)}.
\end{equation}
Thus energy bounds also control the entries of the table.
For balanced factors, each energy has mean $\NuclearNorm{F}$, while its values record where this size is concentrated.
For example, on three rows the energy profiles $(1,1,1)$ and $(3,0,0)$ both have mean $1$, but the second concentrates all the energy in one row.
The local approximation supplies both pointwise and mean energy bounds: when multiplying the approximations of neighboring rounds or gates, the proofs bound some energies pointwise and average another.
The representations used for rectangular shortening in \cref{sec:compilation} need not be balanced, so there the two mean energies are bounded separately.
\Cref{fig:orthogonal-expansion} works out these notions for a $3\times3$ table.

\begin{figure}[htbp]
\centering
\begingroup
\setlength{\fboxrule}{0.3pt}
\setlength{\fboxsep}{8pt}
\fcolorbox{black!25}{white}{\begin{minipage}{\dimexpr\linewidth-2\fboxsep-2\fboxrule\relax}
\centering
\setlength{\abovedisplayskip}{2.5pt}
\setlength{\belowdisplayskip}{2.5pt}
\[
 F=\bordermatrix{
  & {\scriptstyle r=\mathtt{01}} & {\scriptstyle r=\mathtt{10}} & {\scriptstyle r=\mathtt{11}}\cr
  {\scriptstyle\ell=\mathtt{01}} & -1 & -4 & 13\cr
  {\scriptstyle\ell=\mathtt{10}} & 13 & 4 & -1\cr
  {\scriptstyle\ell=\mathtt{11}} & 4 & 10 & -4\cr
 }
   =a_1\cdot b_1^{\transpose}+a_2\cdot b_2^{\transpose}+a_3\cdot b_3^{\transpose},
\]
\[
 F=\underbrace{\begin{pmatrix}-6&-6&6\\6&6&-6\\6&6&-6\end{pmatrix}}_{a_1\cdot b_1^{\transpose}}
  +\underbrace{\begin{pmatrix}6&0&6\\6&0&6\\0&0&0\end{pmatrix}}_{a_2\cdot b_2^{\transpose}}
  +\underbrace{\begin{pmatrix}-1&2&1\\1&-2&-1\\-2&4&2\end{pmatrix}}_{a_3\cdot b_3^{\transpose}}.
\]
\[
 a_1=\begin{pmatrix}\sqrt6\\-\sqrt6\\-\sqrt6\end{pmatrix},\qquad
 a_2=\begin{pmatrix}\sqrt6\\\sqrt6\\0\end{pmatrix},\qquad
 a_3=\begin{pmatrix}1\\-1\\2\end{pmatrix},
\]
\[
 b_1^{\transpose}=\begin{pmatrix}-\sqrt6&-\sqrt6&\sqrt6\end{pmatrix},\qquad
 b_2^{\transpose}=\begin{pmatrix}\sqrt6&0&\sqrt6\end{pmatrix},\qquad
 b_3^{\transpose}=\begin{pmatrix}-1&2&1\end{pmatrix}.
\]
Here $(\sigma_1,\sigma_2,\sigma_3)=(6,4,2)$.
Dividing $a_j$ and $b_j$ by $\sqrt{\sigma_j}$ gives the unit functions
\[
 u_1=\begin{pmatrix}1\\-1\\-1\end{pmatrix},\qquad
 u_2=\begin{pmatrix}\sqrt{3/2}\\\sqrt{3/2}\\0\end{pmatrix},\qquad
 u_3=\begin{pmatrix}1/\sqrt2\\-1/\sqrt2\\\sqrt2\end{pmatrix},
\]
\[
 v_1^{\transpose}=\begin{pmatrix}-1&-1&1\end{pmatrix},\qquad
 v_2^{\transpose}=\begin{pmatrix}\sqrt{3/2}&0&\sqrt{3/2}\end{pmatrix},\qquad
 v_3^{\transpose}=\begin{pmatrix}-1/\sqrt2&\sqrt2&1/\sqrt2\end{pmatrix}.
\]
These form two orthonormal families under uniform averaging over three coordinates,
with $\LtwoNorm{u_j}=\LtwoNorm{v_j}=1$.
Thus $F=\sum_{j=1}^3\sigma_j\cdot u_j\cdot v_j^{\transpose}$.

\smallskip
\begin{minipage}[t]{0.49\linewidth}
\centering
\small
\textbf{First contribution to $F$}
\[
\begin{aligned}
 a_1\cdot b_1^{\transpose}
 &=\begin{pmatrix}\sqrt6\\-\sqrt6\\-\sqrt6\end{pmatrix}
   \cdot\begin{pmatrix}-\sqrt6&-\sqrt6&\sqrt6\end{pmatrix}\\[3pt]
 &=\begin{pmatrix}-6&-6&6\\6&6&-6\\6&6&-6\end{pmatrix}.
\end{aligned}
\]
\end{minipage}\hfill
\begin{minipage}[t]{0.49\linewidth}
\centering
\small
\textbf{First unit product table}
\[
\begin{aligned}
 u_1\cdot v_1^{\transpose}
 &=\begin{pmatrix}1\\-1\\-1\end{pmatrix}
   \cdot\begin{pmatrix}-1&-1&1\end{pmatrix}\\[3pt]
 &=\begin{pmatrix}-1&-1&1\\1&1&-1\\1&1&-1\end{pmatrix}.
\end{aligned}
\]
\end{minipage}
\[
 \underbrace{a_1\cdot b_1^{\transpose}}_{\text{balanced factors}}
 =\underbrace{6}_{\sigma_1}\cdot
 \underbrace{u_1\cdot v_1^{\transpose}}_{\text{unit functions}}.
\]
\begingroup
\small
\setlength{\arraycolsep}{3.5pt}
\renewcommand{\arraystretch}{1.2}
\[
 \begin{array}{c|ccc|c}
  \ell & a_1(\ell)^2 & a_2(\ell)^2 & a_3(\ell)^2
       & \Phi_F(\ell)\\\hline
  \mathtt{01} & 6 & 6 & 1 & 13\\
  \mathtt{10} & 6 & 6 & 1 & 13\\
  \mathtt{11} & 6 & 0 & 4 & 10\\\hline
  \text{mean} & 6 & 4 & 2 & 12
 \end{array}
 \qquad
 \begin{array}{c|ccc|c}
  r & \mathtt{01} & \mathtt{10} & \mathtt{11} & \text{mean}\\\hline
  b_1(r)^2 & 6 & 6 & 6 & 6\\
  b_2(r)^2 & 6 & 0 & 6 & 4\\
  b_3(r)^2 & 1 & 4 & 1 & 2\\\hline
  \Psi_F(r) & 13 & 10 & 13 & 12
 \end{array}
\]
\endgroup
Averaging either energy gives the nuclear norm:
\[
 \E_\ell{\Phi_F(\ell)}=\E_r{\Psi_F(r)}
 =6+4+2=12=\NuclearNorm{F}.
\]
\end{minipage}}
\endgroup
\caption{An orthonormal expansion with different left and right factors.
The functions $a_j$ and $u_j$ depend on $\ell$, and $b_j$ and $v_j$ depend on $r$, with coordinate order $\mathtt{01}$, $\mathtt{10}$, and $\mathtt{11}$.
The energy tables sum squared factors across rows on the left and down columns on the right.}
\label{fig:orthogonal-expansion}
\end{figure}

\subsubsection{Example: a small nuclear norm need not bound every energy}
The following example shows why the mean energies, which equal the nuclear norm for balanced factors, do not suffice on their own.
Consider the table
\[
 F=\begin{pmatrix}1&0&0\\1&0&0\\1&0&0\end{pmatrix}
 =\begin{pmatrix}1\\1\\1\end{pmatrix}\begin{pmatrix}1&0&0\end{pmatrix}.
\]
Multiplying the left factor by $3^{-1/4}$ and the right factor by $3^{1/4}$ balances their norms.
The resulting energies are
\[
 \Phi_F=\left(1/\sqrt3,1/\sqrt3,1/\sqrt3\right),\qquad
 \Psi_F=\left(\sqrt3,0,0\right),
\]
and both means equal $\NuclearNorm{F}=1/\sqrt3$.
The energy at the first column is nevertheless three times its mean, and for the analogous $d\times d$ table the ratio is $d$.
Thus a nuclear-norm bound alone does not control every energy, or the nuclear norm of a product of several tables.
Our explicit approximation therefore supplies additional pointwise energy bounds, which are used together with the mean bounds.

\subsection{Counting and discrepancy}\label{sec:counting}
We now collect the binary counting estimates used to analyze a refresh.
The first two facts describe the distributions $\mu_s$ and how a fresh matrix acts on them.
The discrepancy bound then compares the means of real tables under two secrets.
The refreshing proof uses this bound to change the bit encoded by one state.

\begin{definition}[Independent reference distribution]\label{def:product-distribution}
Let $\nu$ be the uniform distribution on $\UL\times\UR$.
\phantomsection\label{def:secret-probability}
For each $s\in\F_2$, put $p_s:=\Prb_{(L,R)\sample\nu}{L\cdot R=s}$.
\end{definition}
Under $\nu$ the two shares are independent, whereas under $\mu_s$ they are not.

\para{Counting encodings.}
The fiber sizes and class probabilities are
\begin{align}
 |\Omega_0|&=q\cdot(N-2)/2,
 \nonumber\\
 |\Omega_1|&=q\cdot N/2,
 \nonumber\\
 p_0&=(N-2)/(2q),
 \label{eq:encoding-zero-probability}\\
  p_1&=N/(2q).
 \label{eq:encoding-one-probability}
\end{align}
Both $\mu_0$ and $\mu_1$ have uniform left and right marginals, and conditioning $\nu$ on $\Omega_s$ gives $\mu_s$.
For $n\ge2$,
\begin{equation}\label{eq:encoding-probability-lower}
 p_1\ge p_0=\frac{N-2}{2(N-1)}\ge\frac13.
\end{equation}
The counting argument is in the \cref{app:prelim-counting}.

\begin{lemma}[Freshness]\label{lem:fresh}
For every $s\in\F_2$ and every $(\ell,r)\in\Omega_s$, if $M\sample\InvMatr_2^n$, then the refreshed pair $\left(\ell\cdot M,M^{-1}\cdot r\right)$ has distribution $\mu_s$.
\par\nopagebreak[4]\smallskip\noindent
\textnormal{\emph{Proof: see \cref{app:prelim-counting}.}}
\end{lemma}

Freshness also holds conditionally on the entire preceding execution, but it need not hold after conditioning on the leakage answers of the current round.
For example, an answer that reveals one coordinate of a new share restricts the possible values of that share.
Thus exact freshness describes the sampling before the answer is observed.
Hence, the security proof must still account for what the answer reveals.
Freshness gives the exact experiment with independent encodings in \cref{sec:copy-canonical}, and its argument is reused in \cref{sec:joint-setup} and in the \cref{sec:compiler-shortening}.

The next lemma says that a small nuclear norm makes the mean of a table nearly independent of the secret.
\begin{lemma}[Two-secret discrepancy]\label{lem:disc}
For every real table $F$,
\begin{equation}\label{eq:table-disc}
 \left|\E_{(\ell,r)\sample \mu_0}{F(\ell,r)}-\E_{(\ell,r)\sample \mu_1}{F(\ell,r)}\right|
 \le\frac3{\sqrt{N-2}}\cdot\NuclearNorm{F}.
\end{equation}
In particular, for real functions $u:\UL\to\R$ and $v:\UR\to\R$,
\begin{equation}\label{eq:disc}
 \left|\E_{(\ell,r)\sample \mu_0}{u(\ell)\cdot v(r)}-\E_{(\ell,r)\sample \mu_1}{u(\ell)\cdot v(r)}\right|
 \le\frac3{\sqrt{N-2}}\cdot\LtwoNorm{u}\cdot\LtwoNorm{v}.
\end{equation}
\par\nopagebreak[4]\smallskip\noindent
\textnormal{\emph{Proof: see \cref{app:prelim-discrepancy}.}}
\end{lemma}
We use~\cref{eq:table-disc} in \cref{sec:refresh-wire-change} to change the bit encoded by one state.
In \cref{sec:direct-encoding-difference}, we prove the analogous estimate for the differences $\mu_s-\nu$.
\section{Leakage in continual refreshing}\label{sec:leakage}
Before analyzing the storage protocol of \cref{def:protocol}, we must specify what an adversary can learn from it.
Recall that each refresh replaces $(L,R)$ by $(L',R')=\left(L\cdot M,M^{-1}\cdot R\right)$.
We first specify, for one refresh, which data a leakage function may read, when the adversary must choose it, and what the adversary then observes.
We then introduce histories and answer rectangles in \cref{def:history,def:rectangle} below, before reducing randomized strategies to deterministic ones for the proof in \cref{sec:refreshing}.

\subsection{Leakage in a matrix refresh}
During a refresh, each party's leakage function, an arbitrary function whose output the adversary observes, may read everything the party holds: its old share, its new share, and the entire refresh matrix.
It cannot read the other party's shares. The only protocol data common to the two functions are the entries of the matrix, which both parties hold.

The timing of the queries is a separate restriction.
The adversary chooses both parties' leakage functions using the answers of earlier refreshes and its random tape.
Both functions are fixed before either output of the current refresh is revealed.
Thus leakage is adaptive across refreshes and nonadaptive within a refresh: an answer may influence the functions of every later refresh, but neither output can influence the other party's function in the same refresh.
\Cref{sec:within-round} sketches how the proof extends to adaptivity within a refresh.

\begin{definition}[Leakage queries]\label{def:leakage-budget}
Fix an integer $\lambda\ge0$, the leakage budget of each party per refresh.
For every round $i\in\{1,\ldots,T\}$, let $(L,R)$ be its old state and let $M$ be its refresh matrix.
\phantomsection\label{def:leakage-query}
For a preceding history $h$, that is, the sequence of answers already observed in refreshes $1,\ldots,i-1$ (defined formally in \cref{def:history}), the two leakage functions are arbitrary maps
\[
 \begin{aligned}
 \LeakFn^{\LeftParty}_{i,h}&:\UL\times\InvMatr_2^n\longrightarrow\{0,1\}^{\lambda},\\
 \LeakFn^{\RightParty}_{i,h}&:\UR\times\InvMatr_2^n\longrightarrow\{0,1\}^{\lambda}.
 \end{aligned}
\]
\end{definition}
The functions may also depend on the adversary's tape, which we omit from the notation.
The left function is evaluated on $(L,M)$ and the right function on $(R,M)$.
These inputs determine the party's new share, so the function may read that share as well.
Returning exactly $\lambda$ bits is a convention: a function with a smaller fixed output length can be padded with zeros.
The adversary observes no other information about the shares, and in particular does not observe the private transmission of the matrix.

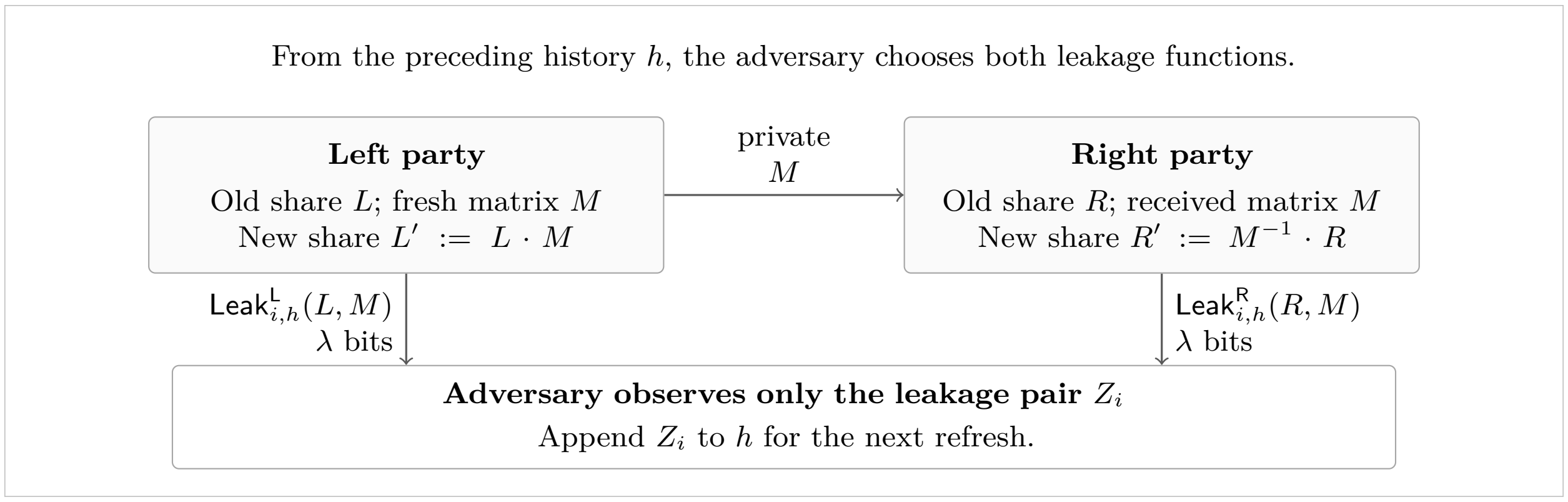
\begin{figure}[htbp]
\centering
\begingroup
\setlength{\fboxrule}{0.3pt}
\setlength{\fboxsep}{8pt}
\fcolorbox{black!25}{white}{\begin{minipage}{\dimexpr\linewidth-2\fboxsep-2\fboxrule\relax}
\centering
\begin{tikzpicture}[x=1cm,y=1cm,font=\footnotesize,
 party/.style={draw=black!35,fill=black!2,rounded corners=2pt,
   align=center,text width=5.1cm,minimum height=1.65cm,inner sep=5pt},
 flow/.style={->,draw=black!65,semithick}]
 \node[align=center] at (0,1.45)
   {From the preceding history $h$, the adversary chooses both leakage functions.};
 \node[party] (left) at (-4,0)
   {\textbf{Left party}\\[3pt]Old share $L$; fresh matrix $M$\\New share $L':=L\cdot M$};
 \node[party] (right) at (4,0)
   {\textbf{Right party}\\[3pt]Old share $R$; received matrix $M$\\New share $R':=M^{-1}\cdot R$};
 \draw[flow] (left.east)--(right.west)
   node[midway,above,align=center] {private\\$M$};
 \node[draw=black!35,rounded corners=2pt,align=center,
   text width=12.6cm,minimum height=0.85cm,inner sep=5pt] (adv) at (0,-2.35)
   {\textbf{Adversary observes only the leakage pair $Z_i$}\\[2pt]
    Append $Z_i$ to $h$ for the next refresh.};
 \draw[flow] (left.south)--(left.south |- adv.north)
   node[midway,left,align=right] {$\LeakFn^{\LeftParty}_{i,h}(L,M)$\\$\lambda$ bits};
 \draw[flow] (right.south)--(right.south |- adv.north)
   node[midway,right,align=left] {$\LeakFn^{\RightParty}_{i,h}(R,M)$\\$\lambda$ bits};
\end{tikzpicture}
\end{minipage}}
\endgroup
\caption{Leakage in one matrix refresh.
Both functions are fixed before either answer is revealed.
After leakage the parties erase their old shares and $M$, keeping $L'$ and $R'$.}
\label{fig:refresh-leakage}
\end{figure}

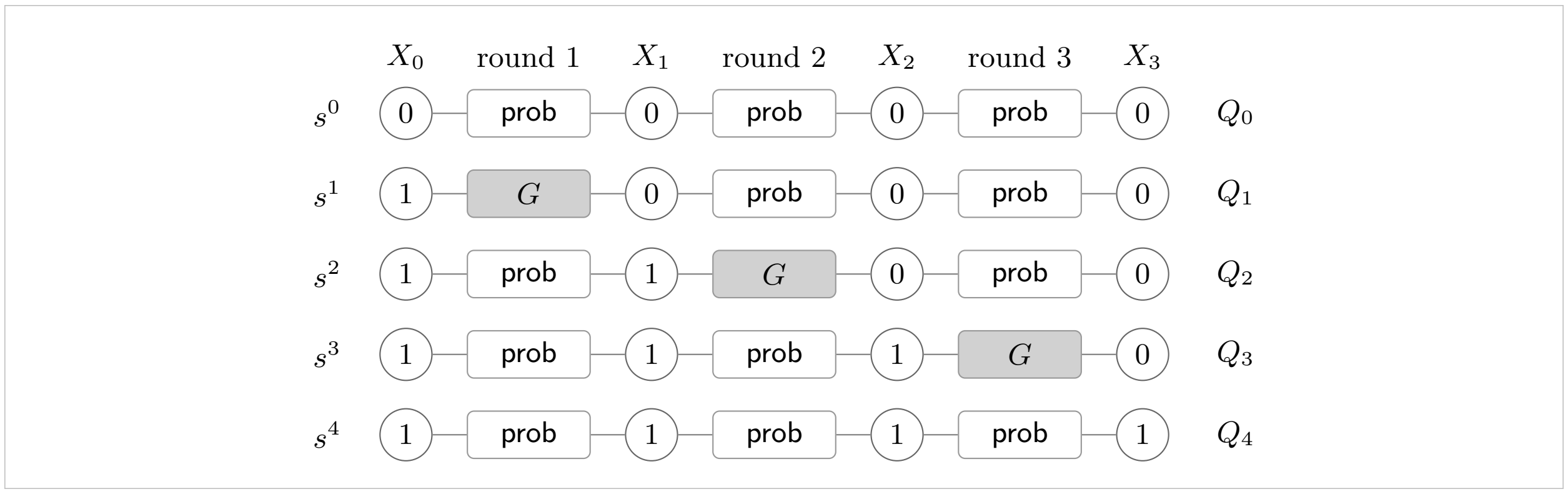
\begin{figure}[htbp]
\centering
\begingroup
\setlength{\fboxrule}{0.3pt}
\setlength{\fboxsep}{8pt}
\fcolorbox{black!25}{white}{\begin{minipage}{\dimexpr\linewidth-2\fboxsep-2\fboxrule\relax}
\centering
\begin{tikzpicture}[x=1cm,y=1cm,font=\footnotesize,
 state/.style={draw=black!60,circle,inner sep=1pt,minimum size=0.55cm},
 kern/.style={draw=black!40,rounded corners=2pt,minimum width=1.3cm,minimum height=0.5cm,inner sep=2pt},
 flow/.style={draw=black!50}]
 \foreach \i in {0,1,2,3}{\node at (2.6*\i,0.6) {$X_\i$};}
 \foreach \i in {1,2,3}{\node at (2.6*\i-1.3,0.6) {round $\i$};}
 \foreach \j in {0,1,2,3,4}{
  \node[anchor=east] at (-0.55,-0.85*\j) {$s^{\j}$};
  \foreach \i in {0,1,2,3}{
   \pgfmathtruncatemacro{\lab}{ifthenelse(\i<\j,1,0)}
   \node[state] (w\j\i) at (2.6*\i,-0.85*\j) {\lab};
  }
  \foreach \i in {1,2,3}{
   \pgfmathtruncatemacro{\h}{\i-1}
   \ifnum\i=\j
    \node[kern,fill=black!18] (g\j\i) at (2.6*\i-1.3,-0.85*\j) {$G$};
   \else
    \node[kern] (g\j\i) at (2.6*\i-1.3,-0.85*\j) {$\prob$};
   \fi
   \draw[flow] (w\j\h)--(g\j\i)--(w\j\i);
  }
  \node[anchor=west] at (8.45,-0.85*\j) {$Q_{\j}$};
 }
\end{tikzpicture}
\end{minipage}}
\endgroup
\caption{The labelings $s^0,\ldots,s^{T+1}$ for a chain of $T=3$ rounds.
Each intermediate labeling has exactly one cheating round, the signed round $j$ (shaded), which uses its signed table $G$.
Every other round uses its true leakage probability function $\prob$.}
\label{fig:chain-hybrid}
\end{figure}

\begin{definition}[Complete answers]\label{def:complete-answer}
The \emph{complete answer}, or \emph{answer}, of the refresh is the pair of leakage outputs
\[
 Z_i:=\left(\LeakFn^{\LeftParty}_{i,h}(L,M),
           \LeakFn^{\RightParty}_{i,h}(R,M)\right).
\]
It belongs to the fixed answer alphabet
\[
 \cZ:=\{0,1\}^{\lambda}\times\{0,1\}^{\lambda},
 \qquad K:=|\cZ|=2^{2\lambda}.
\]
\end{definition}
Some answer pairs may have probability zero.
\Cref{fig:refresh-leakage} shows the local data and the answers observed in one refresh.

\subsection{Histories}
To specify an adaptive query, we must record which answers the adversary has already seen.
We also retain sequences that have probability zero in the real execution, because the intermediate expressions in the proof can assign them nonzero weight.
\begin{definition}[Histories]\label{def:history}
For an execution of $T$ refreshes and each $i\in\{0,\ldots,T\}$, the \emph{history} after refresh $i$ is
$H_i:=(Z_1,\ldots,Z_i)\in\cZ^i$, and $H_0:=\varnothing$ is the empty history.
The history $H_{i-1}$ is the \emph{incoming history} of refresh $i$.
For $h=(z_1,\ldots,z_{i-1})$, write $h\|z:=(z_1,\ldots,z_{i-1},z)$ for the history obtained by appending $z$.
A \emph{formal history} is any element of $\cZ^i$, that is, any sequence of $i$ answers, including those with probability zero.
\end{definition}
Together with the adversary's tape, the incoming history determines the two leakage functions of the refresh.
The \emph{leakage strategy} is the rule by which the adversary chooses each pair of leakage functions from the preceding history.
The experiments for the two secrets use the same rule.
The strategy is defined on every formal history, including those that no execution can reach.
Its choices there may be arbitrary functions allowed by \cref{def:leakage-query}, with the same output lengths.
Every sum over $z$ ranges over $\cZ$, unless stated otherwise, and history sums include all formal histories.

Fixing a formal history selects the leakage functions.
It does not condition the shares or matrices on the listed answers having occurred.
Conditioning on an observed history can instead change the state distributions and make them dependent.
The proofs keep these operations apart by sampling states without conditioning on the transcript and accounting for the answers through their probabilities (see \cref{sec:copy-canonical,sec:refresh-adaptivity}).

\subsection{Answer rectangles and masks}
Histories tell us which functions are evaluated in a refresh.
We now ask what one answer of these functions says about the old state.
The new share is determined by the old share and $M$, so, once the incoming history and the adversary's tape are fixed, each leakage function depends only on its party's old share and $M$.
Consequently, for a fixed matrix, the old states that produce a given answer form a rectangle: the product of a set of left shares and a set of right shares.
In \cref{def:rectangle} below, we describe each answer rectangle by two masks, Boolean functions equal to one on its left and right sets, respectively, and zero elsewhere.
For each fixed matrix $M$, their product is a product table (see \cref{def:product-table}), so \cref{lem:disc} controls how its mean depends on the encoded bit.
We also record that the rectangles of all answers partition the set of old states.
\begin{definition}[Answer rectangles and masks]\label{def:rectangle}
Fix $i\in\{1,\ldots,T\}$, a formal incoming history $h\in\cZ^{i-1}$, the adversary's tape, a matrix $M\in\InvMatr_2^n$, and a complete answer $z=(z_{\LeftParty},z_{\RightParty})\in\cZ$.
Define
\[
 \begin{aligned}
 U_{\LeftParty,z}(M)&:=\{\ell\in\UL:\LeakFn^{\LeftParty}_{i,h}(\ell,M)=z_{\LeftParty}\},\\
 U_{\RightParty,z}(M)&:=\{r\in\UR:\LeakFn^{\RightParty}_{i,h}(r,M)=z_{\RightParty}\}.
 \end{aligned}
\]
The old states producing $z$ are exactly the rectangle $U_{\LeftParty,z}(M)\times U_{\RightParty,z}(M)$.
These rectangles partition $\UL\times\UR$, because each old state produces exactly one answer pair.
The left and right \emph{masks} of the answer $z$ are functions $\MaskL{z}:\UL\times\InvMatr_2^n\to\{0,1\}$ and $\MaskR{z}:\UR\times\InvMatr_2^n\to\{0,1\}$, defined by
\begin{equation}\label{eq:answer-masks}
 \MaskL{z}(\ell,M):=
 \begin{cases}
  1,&\ell\in U_{\LeftParty,z}(M),\\
  0,&\ell\notin U_{\LeftParty,z}(M),
 \end{cases}
\end{equation}
and
\begin{equation}\label{eq:answer-right-mask}
 \MaskR{z}(r,M):=
 \begin{cases}
  1,&r\in U_{\RightParty,z}(M),\\
  0,&r\notin U_{\RightParty,z}(M).
 \end{cases}
\end{equation}
\end{definition}
We omit the fixed indices $i$ and $h$ for brevity.
The rectangle partition is the identity
\begin{equation}\label{eq:partition}
 \sum_{z\in\cZ}\MaskL{z}(\ell,M)\cdot \MaskR{z}(r,M)=1
 \quad\text{for every }\ell\in\UL,\ r\in\UR,\ M\in\InvMatr_2^n.
\end{equation}
The identity in~\cref{eq:partition} says that every fixed old state and matrix produce exactly one answer.
Averaging over the matrix then shows that the answer probabilities of a refresh sum to one.
The bound $K$ on the number of answers enters when estimates for single answers are summed.

The rectangle describes a set of inputs for each fixed matrix.
It does not assert that the two parties' outputs are independent.
The encoding correlates the two shares, and averaging over the common matrix can introduce further dependence.
The approximation constructed in \cref{sec:joint-refresh} below uses the masks separately for each matrix before controlling the average.
The leakage strategy enters \cref{sec:joint-refresh} only through the masks in~\cref{eq:answer-masks,eq:answer-right-mask}.

\subsection{Removing randomness}\label{sec:removing-randomness}
For a fixed tape, the leakage functions above are deterministic.
An adversary may, however, use randomness both when choosing the functions and inside them. We now explain why it suffices to prove security for deterministic leakage functions.
The encodings and refresh matrices remain random throughout the proof.
We fix only the adversary's tape and the presampled auxiliary choices described below while proving the bound, and then average over those choices.
This lets us work with deterministic leakage functions without changing the randomized security claim.

\para{Randomized adversaries.}
The adversary's random tape is independent of the secret, the initial encodings, and every refresh matrix.
We fix an arbitrary tape throughout the proofs.
Since the bound holds uniformly for every tape, taking the ordinary expectation over the possibly infinite tape preserves it, also when the tape is included in the adversary's view (see \cref{thm:main}).

\para{Leakage functions with private randomness.}
The two leakage functions of a refresh may also use fresh auxiliary randomness.
The parties' auxiliary values may be correlated, with a joint distribution depending on $(i,h,M)$ but, conditional on this triple, independent of the shares.
For the proof, we may sample all auxiliary randomness before the execution as follows.
For every possible triple $(i,h,M)$, sample a separate pair of auxiliary values, one for each party, from the joint distribution prescribed for that triple.
Sample these pairs independently across triples and independently of the secret, initial encodings, and refresh matrices.
In round $i$, the incoming history $h$ and the sampled matrix $M$ determine which presampled pair the two leakage functions use.
Because the indexing triple includes the round number, none of the pairs indexed by round $i$ is used in an earlier round.
Consequently, conditional on the current triple $(i,h,M)$, the selected pair still has the prescribed joint distribution and is independent of the current shares.
Thus presampling leaves the execution distribution unchanged.
For fixed presampled values, the leakage functions are deterministic functions of the local data and matrix, selected according to the preceding history.
Averaging the deterministic security bounds therefore also covers these randomized functions.
\section{Continual refreshing}\label{sec:refreshing}
With the leakage model in place, we can now prove that the matrix refresh of \cref{def:protocol} is secure.
We first state the theorem and then prove it, working from a single refresh up to an entire execution.
In round $i$, the matrix $M_i$ takes the state $X_{i-1}=(L_{i-1},R_{i-1})$ to $X_i=(L_i,R_i)$, where
\[
 L_i:=L_{i-1}\cdot M_i,\qquad R_i:=M_i^{-1}\cdot R_{i-1}.
\]
When we analyze a single round, we drop the index and write $(L,R)$ and $(L',R')$ for its old and new shares and $M$ for its matrix.
The adversary observes only the leakage answers.
The initial and final states remain private.
The bit encoded by a state will be called its label.
\Cref{def:labels} below makes this terminology precise.
The proof has three steps.
First, it approximates the leakage of one round by a \emph{signed} table that retains both the old and the new state (see \cref{lem:copy-local}). Such a table is real-valued and may take negative values.
Second, it bounds local replacement errors by the Cauchy--Schwarz inequality (see \cref{lem:la-cauchy-schwarz}) and label changes by averaged factor energies (see \cref{lem:refresh-replacement,lem:refresh-wire-change}).
Third, it compares the executions for the two secrets by a hybrid argument: we introduce intermediate transcript weight functions, assign a bit to every state, and move the boundary between the states labeled $1$ and those labeled $0$ along the sequence of states (see \cref{sec:copy-comparison} below).

\subsection{Security statement}\label{sec:model}
\phantomsection\label{sec:security}
Security means that the complete leakage transcript is almost independent of the stored secret.
\phantomsection\label{def:transcript-distribution}
When the stored secret is $s$, let $P_s^T$ denote the distribution of the complete transcript $H_T=(Z_1,\ldots,Z_T)$ of $T$ refreshes.
This distribution depends on the adversary's strategy for choosing leakage functions, which we omit from the notation for brevity.
Quantitatively, we bound the statistical distance $\StatDist{P_0^T}{P_1^T}$ (see \cref{def:statistical-distance}), which bounds the advantage of every distinguisher.

\begin{theorem}[Security of continual refreshing]\label{thm:main}
There exist absolute constants $C\ge1$ and $C'\ge1$ such that, for all integers $n\ge2$, $T\ge0$, and $\lambda\ge0$, with $q:=2^n-1$ and $K:=2^{2\lambda}$, every adversary in the leakage model of \cref{sec:leakage} satisfies
\begin{equation}\label{eq:copy-rate}
 \StatDist{P_0^T}{P_1^T}\le C\cdot T\cdot K^3\cdot q^{-1/9}.
\end{equation}
The bound also satisfies
\begin{equation}\label{eq:copy-rate-binary}
 C\cdot T\cdot K^3\cdot q^{-1/9}\le C'\cdot T\cdot2^{6\lambda-n/9}.
\end{equation}
The same bound holds when the adversary's independent random tape is included in its view.
\end{theorem}
The theorem applies to all $T$ and $\lambda$, and it gives a nontrivial guarantee exactly when the right-hand side is below one.
For $T=0$, the transcript is empty, so the distance is zero.
The proof is given at the end of \cref{sec:copy-comparison}.

\subsection{An exact experiment with independent encodings}\label{sec:copy-canonical}
Our analysis rests on an exact description of an execution in which the states are sampled first, independently of one another, and the answers are drawn afterward.
To set it up, we fix the adversary's tape as in \cref{sec:leakage} and define the conditional probabilities of the answers.

\begin{definition}[Leakage probability functions]\label{def:leakage-probability}
For every round $i\in\{1,\ldots,T\}$, formal incoming history $h\in\cZ^{i-1}$, answer $z\in\cZ$, secret $s\in\F_2$, and states $x=(\ell,r)$ and $x'=(\ell',r')$ in $\Omega_s$, define
\begin{equation}\label{eq:refresh-conditional-probability}
 \prob_{i,h,z}\left(x,x'\right)
 :=\Prb_{M\sample \InvMatr_2^n}{Z_i=z\mid \ell\cdot M=\ell',\ M^{-1}\cdot r=r'}.
\end{equation}
Here $Z_i$ is the answer that the two leakage functions selected by $i$ and $h$ produce from the old state $x$ and the matrix $M$.
We call $\prob_{i,h,z}:\Omega_s\times\Omega_s\to[0,1]$ a \emph{leakage probability function}.
When the round, history, and answer are fixed, we may abbreviate it to $\prob$.
\end{definition}
The conditioning event in~\cref{eq:refresh-conditional-probability} is nonempty, since $x'$ is a possible refresh of $x$ by \cref{lem:fresh}.
For every $(x,x')\in\Omega_s\times\Omega_s$, we have $\sum_{z\in\cZ}\prob_{i,h,z}(x,x')=1$.
The value of $\prob_{i,h,z}$ for states that encode different bits is not defined, because a refresh never produces such a pair.

For each secret $s\in\F_2$, sample the states $X_0,\ldots,X_T$ independently from $\mu_s$, and then draw the answers $z_1,\ldots,z_T$ in order, drawing $z_i$ with probability $\prob_{i,(z_1,\ldots,z_{i-1}),z_i}\left(X_{i-1},X_i\right)$.
This procedure gives exactly the real transcript distribution, for the following reason.
A fresh matrix sends every fixed state to a uniform state that encodes the same bit (see \cref{lem:fresh}), so the real states are independent samples from $\mu_s$.
For a fixed sequence of states, call the set of matrices that take $X_{i-1}$ to $X_i$ the \emph{transition set} of round $i$.
The matrix tuples compatible with the sequence form the Cartesian product of the transition sets, one set per round.
Conditioning the initially independent uniform matrices on this product makes them independent and uniform on their respective transition sets.
Choosing each round's leakage functions from the preceding answers then reproduces the real execution.
In particular, for every complete history $h=(z_1,\ldots,z_T)\in\cZ^T$,
\begin{equation}\label{eq:copy-canonical}
 P_s^T(h)=
 \E_{(X_0,\ldots,X_T)\sample \mu_s^{\otimes(T+1)}}{\prod_{i=1}^T \prob_{i,h_{i-1},z_i}\left(X_{i-1},X_i\right)},
 \qquad h_{i-1}:=(z_1,\ldots,z_{i-1}).
\end{equation}
The independence in~\cref{eq:copy-canonical} concerns the states before any answer is observed.
The states need not remain independent after conditioning on the observed history.
This distinction will matter when we apply the approximation error bound in \cref{eq:copy-local-error} below, whose average uses independent endpoint encodings.

The hybrids of the proof also use states that encode different bits, which we describe by labels.
\begin{definition}[Labels, valid rounds, and cheating rounds]\label{def:labels}
A \emph{label vector}, or \emph{labeling}, $(s_0,\ldots,s_T)\in\F_2^{T+1}$ specifies independent samples $X_i\sample \mu_{s_i}$ for $i=0,\ldots,T$.
The \emph{label} $s_i$ is the bit encoded by state $X_i$.
Round $i$ is \emph{valid} for these labels when $s_{i-1}=s_i$.
Otherwise, we call it a \emph{cheating round}, because the encoded bit changes across that round in the hybrid.
\end{definition}
This terminology describes the proof, not malicious behavior in the real protocol.
The leakage probability functions $\prob_{i,h,z}$ are used only at valid rounds, and the signed approximations introduced below are defined even at cheating rounds.
The real execution for secret $s$ uses the constant label vector $(s,\ldots,s)$, all of whose rounds are valid.

\subsection{Approximating one refresh}\label{sec:joint-refresh}\label{sec:copy-local}
We start with the first step of the proof, which concerns a single round: we replace the leakage probability function $\prob_{i,h,z}$ of an answer (see \cref{def:leakage-probability}) by a real table $G$ that may take negative values.
This table is a sum of products of a left-party function and a right-party function, with a bound on the total size of these products.
We need this replacement because the hybrid argument changes the bit encoded by a state using the discrepancy bound of \cref{lem:disc}, whose cost grows with the nuclear norm of the table being averaged.
For a fixed matrix, the answer mask is a product (see \cref{def:rectangle}), but the true function averages over matrices compatible with the four shares, and the result need not have a small nuclear norm.
The table keeps both the old and the new state: its left argument is the pair $(L,L')$ and its right argument is the pair $(R,R')$.
Keeping the new shares is essential, because they become the old shares of the next round.

\subsubsection{The four shares and their reference distribution}\label{sec:refresh-joint-setup}
Orthonormal expansions and the nuclear norm treat the row and the column of a table as independent uniform coordinates, whereas the four shares of a refresh are related: the old and new pairs encode the same bit.
We therefore define our tables on the full product of all left pairs and all right pairs, where these tools apply, and use a density to convert averages over this product into averages in a refresh.
A refresh preserves the inner product, so tuples whose old and new pairs encode different bits cannot arise from a refresh.
We therefore also introduce an operation that restricts a table to tuples whose two pairs encode the same bit.
\begin{definition}[Four-share tuples and the one-round experiment]\label{def:one-round}
A \emph{four-share tuple} is an element $\omega=((\ell,\ell'),(r,r'))$ of $\UL^2\times\UR^2$, where $\UL^2=\UL\times\UL$ and $\UR^2=\UR\times\UR$.
The tuple is \emph{consistent} when $\ell'\cdot r'=\ell\cdot r$.
Let $\JointRef$ be the uniform distribution on $\UL^2\times\UR^2$, under which all four shares are independent and uniform on the nonzero shares.
The \emph{one-round experiment} samples $L$ and $R$ as independent uniform nonzero shares and $M$ as an independent uniform invertible matrix, and sets $L':=L\cdot M$ and $R':=M^{-1}\cdot R$.
\end{definition}
We treat $(L,L')$ and $(R,R')$ as two \emph{grouped} arguments: a table $F$ on $\UL^2\times\UR^2$ is written $F((\ell,\ell'),(r,r'))$, and its norms are those of \cref{sec:linear-algebra} for the two sets $\UL^2$ and $\UR^2$, each with its uniform measure.
The one-round experiment produces only consistent tuples, and its tuple $\omega$ has density $\varphi$ relative to $\JointRef$: for each tuple, $\varphi(\omega)$ is its probability in the one-round experiment divided by its probability under $\JointRef$.
Explicitly,
\begin{equation}\label{eq:refresh-density}
 \varphi\left((\ell,\ell'),(r,r')\right) =
 \begin{cases}
  1/p_{\ell\cdot r},&\ell'\cdot r'=\ell\cdot r,\\
  0,&\ell'\cdot r'\ne\ell\cdot r.
 \end{cases}
\end{equation}
Recall from \cref{sec:counting} that $p_s$ is the probability that two independent uniform nonzero shares encode the bit $s$.
\Cref{eq:refresh-density} holds because the old shares are independent and uniform, and for fixed old shares freshness (see \cref{lem:fresh}) makes the new pair a uniform encoding of their bit.

\begin{definition}[Consistency projection]\label{def:consistency-projection}
For every real table $F:\UL^2\times\UR^2\to\R$, its \emph{consistency projection} is the table on the same domain that discards the entries a refresh cannot produce:
\begin{equation*}
 \bigl(\project{F}\bigr)\left((\ell,\ell'),(r,r')\right):=
 \begin{cases}
 F\left((\ell,\ell'),(r,r')\right),&\ell'\cdot r'=\ell\cdot r,\\
 0,&\ell'\cdot r'\ne\ell\cdot r.
 \end{cases}
\end{equation*}
\end{definition}
For example, the projection discards an entry whose old shares encode $0$ and whose new shares encode $1$.
Applying the projection twice changes nothing.

\subsubsection{An explicit approximation to one answer}\label{sec:refresh-masks}
We now express the conditional answer probability as a table on the full reference space, so that we can use its singular coefficients to construct an approximation.
Fix a round $i\in\{1,\ldots,T\}$, a formal incoming history $h\in\cZ^{i-1}$, the adversary's tape, and an answer $z\in\cZ$.
We omit the fixed round and history indices in $\prob$.
Let $\MaskL{z}(\ell,M)$ and $\MaskR{z}(r,M)$ be the Boolean masks of $z$ from~\cref{eq:answer-masks,eq:answer-right-mask}.
For a consistent tuple $\omega=((\ell,\ell'),(r,r'))$, put
\begin{equation}\label{eq:refresh-answer-probability}
 \prob_z\left(\omega\right):=\Prb_{M\sample \InvMatr_2^n}{
 \MaskL{z}(\ell,M)=\MaskR{z}(r,M)=1
 \mid \ell\cdot M=\ell',\ M^{-1}\cdot r=r'},
\end{equation}
and set $\prob_z\left(\omega\right):=0$ on inconsistent tuples.
By \cref{lem:fresh}, every consistent tuple can arise from a refresh, so the conditional probability in~\cref{eq:refresh-answer-probability} is well defined.
The result is a table on the full product space with $0\le \prob_z\le1$ and $\project{\prob_z}=\prob_z$.
At a cheating round, whose endpoint states encode different bits, the true probability function is not used.
Instead, we use the approximation constructed below.

The discrepancy estimate is useful only if the table has a controlled nuclear norm.
We obtain such a table by discarding small singular coefficients and then correcting for the fact that the true answer table is zero off consistent tuples.
The example following the construction explains the correction and why we project only when measuring the error.
Recall that $q=2^n-1$ is the number of nonzero shares.
Take an orthonormal expansion
$\prob_z=\sum_j\sigma_j\cdot u_j\cdot v_j^{\transpose}$ (see \cref{lem:orthonormal-expansion}) with the probability normalization of \cref{sec:linear-algebra}: orthonormality refers to the uniform measures on $\UL^2$ and $\UR^2$, so each $\sigma_j$ is $q^{-2}$ times an ordinary singular value of the $q^2\times q^2$ matrix of entries of $\prob_z$.
For $q^{-1/3}\le\beta\le1$, keep only the coefficients strictly greater than $\beta$:
\begin{equation}\label{eq:refresh-explicit}
 \prob_z^{>\beta}:=\sum_{j:\,\sigma_j>\beta}\sigma_j\cdot u_j\cdot v_j^{\transpose}.
\end{equation}
The superscript $>\beta$ refers to the singular coefficients, not to the entries of the table.
Our approximation is $2\cdot\prob_z^{>\beta}$.
If no coefficient exceeds $\beta$, the retained family is empty and $\prob_z^{>\beta}=0$.
The construction depends only on the masks of the answer, not on the encoded bit.
Define the \emph{residual table} on the full four-share space by
\begin{equation}\label{eq:refresh-residual}
 \Delta_z:=\prob_z-2\cdot\project{\prob_z^{>\beta}}.
\end{equation}
Thus $\Delta_z$ is the approximation error on consistent tuples, extended by zero elsewhere.

\para{Why the factor two.}
Consider the case with no leakage: the answer alphabet has a single element $z$, so its probability is $1$ for every actual refresh.
The table $\prob_z$ also includes tuples that cannot arise from a refresh, and those entries are $0$.
Roughly half the entries describe possible refreshes, so the table contains roughly half $1$s and half $0$s.
Replacing all entries by their average gives about $1/2$, although the desired value on possible refreshes is $1$.
Doubling brings that value back to approximately $1$.
The projection in~\cref{eq:refresh-residual} ensures that we measure the error only on possible refreshes.
For a general answer, the kept singular terms need not form a constant table, and \cref{lem:refresh-rectangle} below proves the error bound for $2\cdot\prob_z^{>\beta}$.
The approximation itself remains defined on inconsistent tuples, where it is used later in the hybrid argument.

\begin{lemma}[Approximation of an answer under refreshing]\label{lem:refresh-rectangle}
There exists an absolute constant $C>0$ such that, for every dimension $n\ge2$, every fixed formal incoming history and adversary tape, every leakage answer $z\in\cZ$, and every threshold $\beta\in[q^{-1/3},1]$, the truncation in~\cref{eq:refresh-explicit} and its residual satisfy
\begin{align}
 \NuclearNorm{\prob_z^{>\beta}}&\le2,
 \label{eq:refresh-rectangle-nuclear}\\
 \LtwoNorm{\prob_z^{>\beta}}&\le1,
 \label{eq:refresh-rectangle-ltwo}\\
 \LtwoNorm{\Delta_z}&\le C\cdot\beta^{1/3}.
 \label{eq:refresh-rectangle-error}
\end{align}
The balanced factors $\sqrt{\sigma_j}\cdot u_j$ and $\sqrt{\sigma_j}\cdot v_j$ of $\prob_z^{>\beta}$ have energies $\Phi_\beta$ and $\Psi_\beta$ bounded at every input:
\begin{align}
 \Phi_\beta(\ell,\ell')&\le1/\beta
 &&\text{for every }(\ell,\ell')\in\UL^2,
 \label{eq:refresh-left-energy}\\
 \Psi_\beta(r,r')&\le1/\beta
 &&\text{for every }(r,r')\in\UR^2.
 \label{eq:refresh-right-energy}
\end{align}
Since the factors are balanced, both mean energies equal $\NuclearNorm{\prob_z^{>\beta}}$ by~\cref{eq:balanced-mean-energies}, and hence are at most $2$ by~\cref{eq:refresh-rectangle-nuclear}.
The same $C$ applies to every dimension, threshold, answer, formal incoming history, and admissible pair of leakage functions.
\end{lemma}

We now explain informally why the bounds of \cref{lem:refresh-rectangle} hold.
\Cref{app:direct-approximation-proof} gives the full proof.

The $L_2$ bound in~\cref{eq:refresh-rectangle-ltwo} and the energy bounds in~\cref{eq:refresh-left-energy,eq:refresh-right-energy} follow row by row.
By Pythagoras' theorem (see \cref{eq:orthonormal-sum-norm}), the row of $\prob_z$ indexed by $(\ell,\ell')$ has squared $L_2$ norm $\sum_j\sigma_j^2\cdot u_j(\ell,\ell')^2$, which is at most one because $0\le\prob_z\le1$.
Truncation only drops terms of this sum, so the rows of $\prob_z^{>\beta}$ also have squared norm at most one, which gives~\cref{eq:refresh-rectangle-ltwo}.
Each kept coefficient exceeds $\beta$, so $\Phi_\beta(\ell,\ell')=\sum_{j:\,\sigma_j>\beta}\sigma_j\cdot u_j(\ell,\ell')^2\le1/\beta$.
The argument for columns is analogous.

The main step is the nuclear-norm bound in~\cref{eq:refresh-rectangle-nuclear}.
Let $W$ be the sum of the kept unit product tables $u_j\cdot v_j^{\transpose}$ (see \cref{def:product-table}).
By orthonormality, $\pairing{\prob_z}{W}$ is the sum of the kept coefficients, that is, $\NuclearNorm{\prob_z^{>\beta}}$.
Since $\prob_z$ is the conditional probability of the answer given the four shares, this pairing is, up to a weight $p_{\ell\cdot r}$ within $1/(2q)$ of $1/2$, the average over $\ell$, $r$, and $M$ of $\MaskL{z}(\ell,M)\cdot \MaskR{z}(r,M)$ times $W$ at the refreshed tuple $((\ell,\ell\cdot M),(r,M^{-1}\cdot r))$.
For fixed $M$, the average over $\ell$ and $r$ splits as $\sum_jc_j\cdot d_j$, where $c_j$ is the inner product of the left mask $\ell\mapsto \MaskL{z}(\ell,M)$ with $\ell\mapsto u_j(\ell,\ell\cdot M)$, and $d_j$ is its right-side counterpart.
If these functions of $\ell$ were orthonormal, Bessel's inequality (see \cref{eq:bessel-inequality}) would give $\sum_jc_j^2\le1$, as the mask takes values in $\{0,1\}$, and with the same bound for the $d_j$ the sum would be at most $1$.
The functions are only approximately orthonormal: their Gram matrix (of pairwise inner products) equals the identity on average over $M$, and its mean squared deviation is at most the number of kept terms divided by $(q-1)\cdot\beta^2$, because $M$ maps any two distinct nonzero shares to a uniform pair of distinct ones (see \cref{lem:refresh-Gram-variance}).
The number of kept terms is at most $\NuclearNorm{\prob_z^{>\beta}}/\beta$, so with $q\cdot\beta^3\ge1$ we get an inequality with the nuclear norm on both sides,
\begin{equation}\label{eq:refresh-nuclear-bootstrap}
 2\cdot\NuclearNorm{\prob_z^{>\beta}}\le2+\sqrt{2\cdot\NuclearNorm{\prob_z^{>\beta}}},
\end{equation}
which forces~\cref{eq:refresh-rectangle-nuclear} (see \cref{lem:direct-nuclear}).
This argument relies on the mask structure of $\prob_z$, so it does not apply to an arbitrary table $F$ with $0\le F\le1$ and $\project{F}=F$.

Finally, consider the error bound in~\cref{eq:refresh-rectangle-error}, with $t:=\LtwoNorm{\Delta_z}$.
The argument combines two properties of the residual.
Its largest singular coefficient is controlled, limiting its correlation with product tests, and its squared norm is nearly its pairing with $\prob_z$.
The answer-mask identity, derived below in \cref{eq:kernel-pairing}, expresses this pairing as a weighted average over $M$ of the two masks multiplied by the residual evaluated at refreshed tuples.
Together, these properties and the answer-mask identity prevent $\Delta_z$ from having a large norm.
The mask products appear only after the refresh equations are substituted, however, so the argument must also carry the information about small coefficients from independent grouped coordinates to refreshed tuples.
The four-cycle average, defined below in \cref{eq:four-cycle-definition}, provides this control.
The consistency condition is encoded by the signs $(-1)^{\ell\cdot r+\ell'\cdot r'}$, which form a restriction of a binary sign matrix whose rows are orthogonal on the full bit-vector domain.
The corresponding near-orthogonality to product tables is proved in \cref{lem:consistency-sign} below.
Hence $2\cdot\project{\prob_z^{>\beta}}$ differs from $\prob_z^{>\beta}$ by a table nearly orthogonal to product tables.
Since the discarded coefficients are at most $\beta$, the largest singular coefficient of $\Delta_z$ is at most $3\beta$ (see \cref{lem:consistency-sign}).
A small largest coefficient alone does not make $t$ small, as many small coefficients can have a large sum of squares.
But truncation is an orthogonal projection, so the approximation is nearly orthogonal to its error, and $t^2$ equals $\pairing{\prob_z}{\Delta_z}$ up to a term of order $1/(q\cdot\beta)$.
As for the nuclear norm, this pairing averages, over $M$, correlations of the two masks with $\Delta_z$ at refreshed tuples.
Each correlation is at most the fourth root of the \emph{four-cycle average} of the table: the mean product of its entries at the corners of a random rectangle (two random rows and two random columns).
For $\Delta_z$ this average is at most $\sigma_{\max}(\Delta_z)^2\cdot t^2\le9\beta^2\cdot t^2$, and refreshing raises it, on average over $M$, by at most a constant factor plus small corrections (see \cref{lem:four-cycle-refresh}).
Together these estimates give
\begin{equation}\label{eq:refresh-error-bootstrap}
 t^2\le C\cdot\beta^{1/2}\cdot t^{1/2}+C/(q\cdot\beta),
\end{equation}
and since $q\cdot\beta^3\ge1$, \cref{eq:refresh-error-bootstrap} implies the error bound in~\cref{eq:refresh-rectangle-error}.

\subsubsection{Proof of the refreshing approximation}\label{app:direct-approximation-proof}
We prove \cref{lem:refresh-rectangle} in two steps:
\begin{itemize}
 \item Bound the nuclear norm of $\prob_z^{>\beta}$ using the answer masks and the approximate orthonormality of the singular functions after refreshing.
 \item Bound the $L_2$ norm of the residual $\Delta_z$ by relating its squared norm to a correlation with the answer masks, then controlling that correlation through a four-cycle average.
\end{itemize}
The row and column projection argument also gives the pointwise energy bounds and the $L_2$ bound on $\prob_z^{>\beta}$.
Throughout, fix a formal incoming history, the adversary's tape, an answer $z$, and a threshold $q^{-1/3}\le\beta\le1$.
Recall that $q=N-1\ge3$ is the number of nonzero shares, where $N=2^n$.
We use the grouped coordinates $x=(\ell,\ell')\in\UL^2$ and $y=(r,r')\in\UR^2$.
As in \cref{sec:la-shared}, all averages and norms use the uniform measures on their respective row and column sets unless another distribution is specified.

\para{Relating a table to the leakage masks.}
For any table $F$ on $\UL^2\times\UR^2$ and fixed $M\in\InvMatr_2^n$, substituting the refresh equations gives a table on the old shares:
\begin{equation*}
 F^M(\ell,r):=F\bigl((\ell,\ell\cdot M),(r,M^{-1}\cdot r)\bigr).
\end{equation*}
Since a refresh produces only consistent tuples, $\bigl(\project{F}\bigr)^M=F^M$.
Write $\MaxNorm{F}:=\max_{\ell,\ell',r,r'}|F((\ell,\ell'),(r,r'))|$ for its largest absolute entry.
For any real table $W$ on the four-share space, conditioning on the four shares yields the \emph{answer-mask identity}
\begin{equation}\label{eq:kernel-pairing}
 \pairing{\prob_z}{W}
 =\E_{\ell,r,M}{p_{\ell\cdot r}\cdot
 \MaskL{z}(\ell,M)\cdot \MaskR{z}(r,M)\cdot W^M(\ell,r)}.
\end{equation}
Here $\ell$ and $r$ are independent uniform nonzero shares, and $M$ is independent of them and uniform in $\InvMatr_2^n$.
The factor $p_{\ell\cdot r}$ cancels the density $\varphi=1/p_{\ell\cdot r}$ of the one-round experiment on consistent tuples (see \cref{eq:refresh-density}).
Recall from~\cref{eq:encoding-zero-probability,eq:encoding-one-probability} that $p_0=(q-1)/(2q)$ and $p_1=(q+1)/(2q)$, and hence $|2p_s-1|=1/q$.
We will use~\cref{eq:kernel-pairing} twice: first to bound the nuclear norm, and later with $W=\Delta_z$ to bound the approximation error.

\para{Projection bounds.}
Write $I:=\{j:\sigma_j(\prob_z)>\beta\}$ for the indices of the kept coefficients.
Each row of $\prob_z^{>\beta}$ is the orthogonal projection of the corresponding row of $\prob_z$ onto the span of the kept right singular functions.
Each column is similarly projected onto the span of the kept left singular functions.
Pythagoras' theorem (see \cref{eq:orthonormal-sum-norm}) and $0\le\prob_z\le1$ give, for every row $x$,
\begin{equation}\label{eq:direct-row-projection}
 \E_y{\bigl(\prob_z^{>\beta}(x,y)\bigr)^2}
 =\sum_{j\in I}\sigma_j^2\cdot u_j(x)^2
 \le\E_y{\bigl(\prob_z(x,y)\bigr)^2}\le1.
\end{equation}
The same argument applies to every column, and averaging~\cref{eq:direct-row-projection} over $x$ gives~\cref{eq:refresh-rectangle-ltwo}.
Since each kept $\sigma_j$ exceeds $\beta$, we obtain the following bounds at every $x$ and $y$:
\begin{align}
 e_u(x)&:=\sum_{j\in I}u_j(x)^2\le\beta^{-2},
 \label{eq:direct-left-unit-energy}\\
 e_v(y)&:=\sum_{j\in I}v_j(y)^2\le\beta^{-2},
 \label{eq:direct-right-unit-energy}\\
 \Phi_\beta(x)&:=\sum_{j\in I}\sigma_j\cdot u_j(x)^2\le\beta^{-1},
 \nonumber\\
 \Psi_\beta(y)&:=\sum_{j\in I}\sigma_j\cdot v_j(y)^2\le\beta^{-1}.
 \label{eq:direct-right-factor-energy}
\end{align}
Orthonormality also gives $\E_x{e_u(x)}=\E_y{e_v(y)}=|I|$.
Together with the pointwise bounds in~\cref{eq:direct-left-unit-energy,eq:direct-right-unit-energy}, this implies that both $\E_x{e_u(x)^2}$ and $\E_y{e_v(y)^2}$ are at most $|I|/\beta^2$.
The unweighted sums $e_u$ and $e_v$ will control the Gram matrices below, that is, the matrices of pairwise inner products of the functions involved.
The weighted sums $\Phi_\beta$ and $\Psi_\beta$ are the factor energies required by \cref{lem:refresh-rectangle}.

\para{Bounding the nuclear norm.}
The kept singular functions are orthonormal on the full grouped space, but their restrictions along a fixed refresh need not remain orthonormal.
We must control this change before applying the mask bounds.
The next lemma controls how much an average along a random refresh can differ from its mean.
We will apply it to matrices of products $u_j(x)\cdot u_k(x)$, whose averages form the Gram matrix of the kept functions after refreshing.

\begin{lemma}[Gram variance for an invertible refresh]\label{lem:refresh-Gram-variance}
Let $\InnerSpace$ be any finite-dimensional real inner-product space with induced norm $\VecNorm{\cdot}$.
For each $\ell\in\UL$, let $g_\ell:\UL\to\InnerSpace$, and put
\[
 \overline g_\ell:=\E_{\ell'\sample \UL}{g_\ell(\ell')},\quad
 \widehat g_\ell(\ell'):=g_\ell(\ell')-\overline g_\ell,\quad
 Y_M:=\frac1q\sum_{\ell\in\UL}g_\ell(\ell\cdot M).
\]
For independent uniform $M\in\InvMatr_2^n$ and $M'\in\InvMatr_2^n$,
\begin{align}
 \E_M{\VecNorm{Y_M-\E_{M'}{Y_{M'}}}^2}
 &=\frac1{q(q-1)}\sum_\ell\E_{\ell'}{\VecNorm{\widehat g_\ell(\ell')}^2}\notag\\
 &\quad{}-\frac1{q^2(q-1)}\E_{\ell'}{\VecNorm{\sum_{\ell\in\UL}\widehat g_\ell(\ell')}^2},
 \label{eq:invertible-Gram-identity}\\
 \E_M{\VecNorm{Y_M-\E_{M'}{Y_{M'}}}^2}
 &\le\frac1{q-1}\E_{\ell,\ell'}{\VecNorm{g_\ell(\ell')}^2}.
 \label{eq:invertible-Gram-variance}
\end{align}
All $\ell'$ averages are uniform on the nonzero rows and $\ell$ is uniform on $\UL$.
The same statement holds for column images under $M^{-1}$.
\end{lemma}
\begin{proof}
Any two distinct nonzero binary rows are linearly independent, so for $\ell\ne \tilde\ell$ their images under $M$ are a uniform ordered pair of distinct nonzero rows.
Since every $\widehat g_\ell$ has mean zero, summing first over all ordered pairs and removing the equal pairs gives
\begin{equation}\label{eq:invertible-offdiagonal-covariance}
 \E_{(\ell',\tilde\ell')\sample \{(u,u')\in\UL^2:u\ne u'\}}{\pairing{\widehat g_\ell(\ell')}{\widehat g_{\tilde\ell}(\tilde\ell')}}
 =-\frac1{q-1}\E_{\ell'}{\pairing{\widehat g_\ell(\ell')}{\widehat g_{\tilde\ell}(\ell')}}.
\end{equation}
Expand the squared norm of $q^{-1}\sum_{\ell\in\UL}\widehat g_\ell(\ell\cdot M)$ and use~\cref{eq:invertible-offdiagonal-covariance} for its off-diagonal terms.
Adding and subtracting the diagonal terms inside the final square gives~\cref{eq:invertible-Gram-identity}.
Discard its nonpositive last term and use
$\E_{\ell'}{\VecNorm{g_\ell(\ell')-\overline g_\ell}^2}\le\E_{\ell'}{\VecNorm{g_\ell(\ell')}^2}$ to obtain~\cref{eq:invertible-Gram-variance}.
Uniformity of $M^{-1}$ and the same pair-transitivity argument prove the column version.
\end{proof}

\begin{lemma}[Nuclear norm of the truncated answer table]\label{lem:direct-nuclear}
Fix any formal incoming history and adversary tape.
For every answer $z$ and every $q^{-1/3}\le\beta\le1$, the truncation defined in~\cref{eq:refresh-explicit} satisfies $\NuclearNorm{\prob_z^{>\beta}}\le2$.
\end{lemma}
\begin{proof}
If $I$ is empty, the claim is immediate.
Otherwise, put $\tau:=2\NuclearNorm{\prob_z^{>\beta}}$ and $W:=\sum_{j\in I}u_j\cdot v_j^{\transpose}$.
Then $\pairing{\prob_z}{W}=\tau/2$ and, by~\cref{eq:direct-left-unit-energy,eq:direct-right-unit-energy}, $\MaxNorm{W}\le\beta^{-2}$.
Write $u(x):=(u_j(x))_{j\in I}$ and $v(y):=(v_j(y))_{j\in I}$ as column vectors of coefficients, and define
\begin{equation}\label{eq:refresh-gram-matrix}
 \Gram_{\LeftParty,M}:=\E_{\ell\sample \UL}{u(\ell,\ell\cdot M)\cdot u(\ell,\ell\cdot M)^{\transpose}}.
\end{equation}
Define $\Gram_{\RightParty,M}$ analogously with $v(r,M^{-1}\cdot r)$.
The entries of $\Gram_{\LeftParty,M}$ in~\cref{eq:refresh-gram-matrix} are the pairwise inner products of the functions $\ell\mapsto u_j(\ell,\ell\cdot M)$, so it is their \emph{Gram matrix}.
The same interpretation applies on the right.
For coefficient vectors and matrices only, $\EuclNorm{\cdot}$ denotes the square root of the sum of squared entries, without averaging.
Apply \cref{lem:refresh-Gram-variance} in this inner-product space of coefficient matrices, with
$g_\ell(\ell'):=u(\ell,\ell')\cdot u(\ell,\ell')^{\transpose}$.
The mean Gram matrix is the $|I|\times |I|$ identity matrix $\Id$ by orthonormality, and
$\EuclNorm{u\cdot u^{\transpose}}^2=e_u^2$.
Consequently, on each side,
\begin{align}
 \E_M{\EuclNorm{\Gram_{\LeftParty,M}-\Id}^2}
 &\le \frac{|I|}{(q-1)\cdot\beta^2},
 \label{eq:left-Gram-variance}\\
 \E_M{\EuclNorm{\Gram_{\RightParty,M}-\Id}^2}
 &\le \frac{|I|}{(q-1)\cdot\beta^2}.
 \label{eq:right-Gram-variance}
\end{align}
For each fixed $M$, define coefficient vectors $c=(c_j)_{j\in I}$ and $d=(d_j)_{j\in I}$ by
\begin{align*}
 c_j&:=\E_\ell{\MaskL{z}(\ell,M)\cdot u_j(\ell,\ell\cdot M)},
 \\
 d_j&:=\E_r{\MaskR{z}(r,M)\cdot v_j(r,M^{-1}\cdot r)}.
\end{align*}
For any coefficient vector $\xi$ of ordinary Euclidean norm one, the Cauchy--Schwarz inequality (see \cref{lem:la-cauchy-schwarz}) and $0\le \MaskL{z}\le1$ give
\begin{equation}\label{eq:refresh-mask-coefficient-bound}
 |\xi^{\transpose}\cdot c|^2
 \le \xi^{\transpose}\cdot \Gram_{\LeftParty,M}\cdot \xi
 \le1+\EuclNorm{\Gram_{\LeftParty,M}-\Id}.
\end{equation}
Maximizing~\cref{eq:refresh-mask-coefficient-bound} over $\xi$ gives the same bound on $\EuclNorm{c}^2$, and likewise for $d$.
Since the average of $\MaskL{z}\cdot \MaskR{z}\cdot W^M$ at fixed $M$ is $c^{\transpose}\cdot d$, the Cauchy--Schwarz inequality (see \cref{lem:la-cauchy-schwarz}) over $M$ and~\cref{eq:left-Gram-variance,eq:right-Gram-variance} yield
\begin{equation}\label{eq:refresh-mask-pairing-bound}
 \E_{\ell,r,M}{\MaskL{z}(\ell,M)\cdot \MaskR{z}(r,M)\cdot W^M(\ell,r)}
 \le1+\sqrt{\frac{|I|}{(q-1)\cdot\beta^2}}.
\end{equation}
This step uses no independence between the two Gram matrices.
Using~\cref{eq:kernel-pairing,eq:refresh-mask-pairing-bound}, $|2p_s-1|=1/q$, and $\MaxNorm{W}\le\beta^{-2}$, we get
\begin{equation}\label{eq:direct-nuclear-bootstrap}
 \tau\le1+\sqrt{\frac{|I|}{(q-1)\cdot\beta^2}}+\frac1{q\cdot\beta^2}
 \le1+\sqrt{\frac1{2(q-1)\cdot\beta^3}}\cdot\sqrt\tau+\frac1{q\cdot\beta^2},
\end{equation}
where $|I|\cdot\beta\le\NuclearNorm{\prob_z^{>\beta}}=\tau/2$.
Since $q\cdot\beta^3\ge1$ and $q\ge3$, the coefficient of $\sqrt\tau$ and the last term in~\cref{eq:direct-nuclear-bootstrap} are at most one.
Thus~\cref{eq:direct-nuclear-bootstrap} gives $\tau\le2+\sqrt\tau$, the inequality in~\cref{eq:refresh-nuclear-bootstrap}, implying $\tau\le4$.
The balanced factors of $\prob_z^{>\beta}$ therefore have mean energies $\tau/2\le2$, while their pointwise bounds $1/\beta$ and the bound $\LtwoNorm{\prob_z^{>\beta}}\le1$ follow from the projection bounds.
\end{proof}

\para{Cancellation from consistency.}
The remaining task is to bound the residual $\Delta_z=\prob_z-2\cdot\project{\prob_z^{>\beta}}$.
To separate the effect of the consistency projection from that of truncation, define the real table
\begin{equation}\label{eq:refresh-consistency-sign-table}
 J\bigl((\ell,\ell'),(r,r')\bigr):=
 \begin{cases}
  1/2,&\ell\cdot r=\ell'\cdot r',\\
  -1/2,&\ell\cdot r\ne\ell'\cdot r'.
 \end{cases}
\end{equation}
For any real table $F$ on the same domain,
\begin{equation}\label{eq:refresh-projection-sign}
 \project{F}=F/2+J\cdot F,
\end{equation}
where $J\cdot F$ denotes entrywise multiplication.
The following lemma shows that the positive and negative entries of $J$ in~\cref{eq:refresh-consistency-sign-table} almost cancel when paired with a product of a left function and a right function.

\begin{lemma}[Cancellation of the consistency sign]\label{lem:consistency-sign}
For arbitrary real functions $u$ on $\UL^2$ and $v$ on $\UR^2$,
\begin{equation}\label{eq:consistency-sign}
 \left|\pairing{J}{u\cdot v^{\transpose}}\right|
 \le\frac{N}{2q^2}\LtwoNorm{u}\LtwoNorm{v}
 \le\frac1q\LtwoNorm{u}\LtwoNorm{v}.
\end{equation}
Fix any formal incoming history and adversary tape.
For every answer $z$, every $\beta\in[q^{-1/3},1]$, and all functions $u:\UL^2\to\R$ and $v:\UR^2\to\R$ with $\LtwoNorm{u}=\LtwoNorm{v}=1$,
\begin{equation}\label{eq:projected-product-bound}
 \left|\pairing{J\cdot \prob_z^{>\beta}}{u\cdot v^{\transpose}}\right|\le\frac1{q\cdot\beta}.
\end{equation}
\end{lemma}
\begin{proof}
The sign table $2J$ is the submatrix of
$(-1)^{\ell\cdot r+\ell'\cdot r'}$ indexed by $(\ell,\ell')$ and $(r,r')$.
Extend functions by zero to all pairs of $n$-bit vectors.
The grouped coordinates form subsets of this full domain.
On the full $2n$-bit domain, distinct binary sign rows are orthogonal: the sum of a nontrivial binary sign is zero by pairing vectors that differ in a coordinate where its coefficient is one.
For the full sign matrix $\Upsilon^{\mathrm{full}}$, this gives $\Upsilon^{\mathrm{full}}\cdot(\Upsilon^{\mathrm{full}})^{\transpose}=N^2\cdot\Id$.
The Cauchy--Schwarz inequality (see \cref{lem:la-cauchy-schwarz}) therefore bounds the unnormalized pairing by $N$ times the two Euclidean norms.
Each grouped domain here has $q^2$ points.
Dividing the pairing by $q^4$ and converting each Euclidean norm to its probability-normalized $L_2$ norm gives~\cref{eq:consistency-sign}.
Finally $N/(2q)\le1$.

For the last assertion, write $\prob_z^{>\beta}=\sum_j a_j\cdot b_j^{\transpose}$ with its balanced factors.
Apply~\cref{eq:consistency-sign} to $u\cdot a_j$ and $v\cdot b_j$ and then the Cauchy--Schwarz inequality (see \cref{lem:la-cauchy-schwarz}) over $j$:
\[
 \left|\pairing{J\cdot \prob_z^{>\beta}}{u\cdot v^{\transpose}}\right|
 \le\frac1q
 \sqrt{\E_x{u(x)^2\cdot\Phi_\beta(x)}}\cdot
 \sqrt{\E_y{v(y)^2\cdot\Psi_\beta(y)}}\le\frac1{q\cdot\beta}.
\]
\end{proof}

\para{The residual and its exact identity.}
Applying~\cref{eq:refresh-projection-sign} to the table $\prob_z^{>\beta}$ rewrites the residual in~\cref{eq:refresh-residual} as
\begin{equation}\label{eq:direct-residual-decomposition}
 \Delta_z=\prob_z-2\cdot\project{\prob_z^{>\beta}}
        =(\prob_z-\prob_z^{>\beta})-2\cdot J\cdot\prob_z^{>\beta}.
\end{equation}
Put $t:=\LtwoNorm{\Delta_z}$.
The residual is supported on consistent tuples and equals $\prob_z-2\cdot\prob_z^{>\beta}$ there.
Recall that $\sigma_{\max}$ denotes the largest singular coefficient (see \cref{def:largest-singular-coefficient}).
The discarded singular coefficients are at most $\beta$, so applying~\cref{eq:largest-singular-coefficient,eq:projected-product-bound} to~\cref{eq:direct-residual-decomposition} gives
\begin{align}
 \sigma_{\max}(\Delta_z)&\le\beta+\frac2{q\cdot\beta}\le3\beta,
 \label{eq:residual-largest-coefficient}\\
 \MaxNorm{\Delta_z}&\le3/\beta.
 \label{eq:residual-entry-bound}
\end{align}
To obtain~\cref{eq:residual-largest-coefficient}, check the pairing bound on every unit product table and then on a pair of singular functions of $\Delta_z$.
The entry bound~\cref{eq:residual-entry-bound} follows from $|\prob_z^{>\beta}(x,y)|\le\sqrt{\Phi_\beta(x)\cdot\Psi_\beta(y)}\le1/\beta$ and $|\prob_z|\le1$.
Every row and column of $\Delta_z$ has $L_2$ norm at most three: $\prob_z$ has norm at most one, $\prob_z^{>\beta}$ has norm at most one on each row and column, and the consistency projection contracts each such norm.
These bounds alone do not yet show that $t$ is small: a table can have many small singular coefficients whose squares have a large sum.
We therefore use an identity specific to this residual.
The support of $\prob_z$ and orthogonality of the kept terms give
\begin{equation}\label{eq:direct-retained-pairing}
 \pairing{\prob_z}{\project{\prob_z^{>\beta}}}
 =\pairing{\prob_z}{\prob_z^{>\beta}}
 =\LtwoNorm{\prob_z^{>\beta}}^2.
\end{equation}
Also, because the consistency projection leaves each entry unchanged or sets it to zero,
\begin{equation}\label{eq:direct-projected-norm}
 \LtwoNorm{\project{\prob_z^{>\beta}}}^2
 =\frac12\cdot\LtwoNorm{\prob_z^{>\beta}}^2
   +\pairing{\prob_z^{>\beta}}{J\cdot\prob_z^{>\beta}}.
\end{equation}
Expanding $\LtwoNorm{\Delta_z}^2$ and using~\cref{eq:direct-retained-pairing,eq:direct-projected-norm} yields
\begin{equation}\label{eq:direct-residual-identity}
 t^2=\pairing{\prob_z}{\Delta_z}+4\pairing{\prob_z^{>\beta}}{J\cdot \prob_z^{>\beta}}.
\end{equation}
Indeed, both sides expand to $\LtwoNorm{\prob_z}^2-2\LtwoNorm{\prob_z^{>\beta}}^2+4\pairing{\prob_z^{>\beta}}{J\cdot \prob_z^{>\beta}}$.
Applying~\cref{eq:projected-product-bound} to each normalized product in the expansion of $\prob_z^{>\beta}$ yields
\begin{equation}\label{eq:direct-cross-term}
 \left|\pairing{\prob_z^{>\beta}}{J\cdot \prob_z^{>\beta}}\right|\le\frac{\NuclearNorm{\prob_z^{>\beta}}}{q\cdot\beta}\le\frac2{q\cdot\beta}.
\end{equation}
It remains to bound $\pairing{\prob_z}{\Delta_z}$.
The answer-mask identity~\cref{eq:kernel-pairing} expresses this pairing through $\Delta_z^M$, the residual evaluated along an actual refresh.
The largest-coefficient bound concerns independent grouped coordinates.
It does not directly control the correlation with masks after those coordinates are restricted along a fixed refresh.
We instead bound this correlation on average over the matrix.
The four-cycle average introduced next bounds correlation with two masks at each fixed matrix, admits a counting argument when averaged over matrices, and is controlled by the largest singular coefficient under independent sampling.

\para{Four-cycle averages.}
For a real table $F$ on a product of two finite nonempty sets, let $\nu_{\LeftParty}$ and $\nu_{\RightParty}$ denote the uniform measures on its row and column sets.
Its \emph{four-cycle average} is
\begin{equation}\label{eq:four-cycle-definition}
 \FourCyc{F}:=\E_{(x_1,x_2,y_1,y_2)\sample\nu_{\LeftParty}^{\otimes2}\otimes\nu_{\RightParty}^{\otimes2}}
 {F(x_1,y_1)\cdot F(x_1,y_2)\cdot F(x_2,y_1)\cdot F(x_2,y_2)}.
\end{equation}
The four entries in~\cref{eq:four-cycle-definition} are the corners of a rectangle formed by two independently sampled rows and two independently sampled columns.
Expanding an orthonormal decomposition from \cref{lem:orthonormal-expansion} gives $\FourCyc{F}=\sum_j\sigma_j(F)^4$.
In particular, the average is nonnegative even when individual products are negative, and
\begin{equation}\label{eq:four-cycle-norms}
 0\le \FourCyc{F}\le\sigma_{\max}(F)^2\cdot\LtwoNorm{F}^2.
\end{equation}
For arbitrary real functions $f$ and $g$ on the row and column sets, two applications of the Cauchy--Schwarz inequality (see \cref{lem:la-cauchy-schwarz}) give
\begin{equation}\label{eq:four-cycle-test}
 \left|\E_{x,y}{f(x)\cdot F(x,y)\cdot g(y)}\right|
 \le\LtwoNorm{f}\cdot\LtwoNorm{g}\cdot \FourCyc{F}^{1/4}.
\end{equation}
Indeed, first remove $f$ by the Cauchy--Schwarz inequality (see \cref{lem:la-cauchy-schwarz}) over $x$.
The remaining squared expression is the pairing of $g(y)\cdot g(y')$ with $\E_x{F(x,y)\cdot F(x,y')}$, and the Cauchy--Schwarz inequality over $(y,y')$ bounds it by $\LtwoNorm{g}^2\cdot\sqrt{\FourCyc{F}}$.
For Boolean masks, both function norms are at most one, so the fourth root of $\FourCyc{F}$ alone bounds their correlation with $F$.

\para{Uniform binary matrices.}
The four-cycle calculation is simpler for an unrestricted uniform matrix, whose entries are independent.
A singular matrix has no inverse, so we will first express the refresh constraints by equations on an enlarged table.
The complete four-cycle of that table is nonnegative, allowing us to compare its average over invertible matrices with its average over all matrices.
The next two lemmas supply this comparison and count the linear constraints in the unrestricted average.

\begin{lemma}[Comparison with unrestricted matrices]\label{lem:invertible-comparison}
Let $A_0$ be uniform on $\F_2^{n\times n}$ and $M$ uniform on $\InvMatr_2^n$.
Then
\begin{equation}\label{eq:invertible-probability}
 \Prb{A_0\text{ is invertible}}=\prod_{j=1}^n\left(1-2^{-j}\right)>\frac14,
\end{equation}
and for every nonnegative function $g$ on all binary $n\times n$ matrices,
\begin{equation}\label{eq:invertible-comparison}
 \E_{M}{g(M)}\le4\E_{A_0}{g(A_0)}.
\end{equation}
\end{lemma}
\begin{proof}
The $j$th row of an invertible matrix must avoid the span of the first $j-1$ rows, which has $2^{j-1}$ elements, so the probability that a uniform matrix is invertible is $\prod_{j=1}^n\left(1-2^{-n+j-1}\right)=\prod_{j=1}^n\left(1-2^{-j}\right)$.
Since $\prod_{j=2}^n\left(1-2^{-j}\right)\ge1-\sum_{j=2}^n2^{-j}>1/2$, this probability exceeds $1/4$.
Conditioned on invertibility, $A_0$ is uniform on $\InvMatr_2^n$, and $g\ge0$ gives $\E_{A_0}{g(A_0)}\ge\Prb{A_0\text{ is invertible}}\cdot\E_M{g(M)}$.
\end{proof}

\begin{lemma}[Prescribed rows and columns]\label{lem:prescribed-entries}
Let $x_1$ and $x_2$ be rows in $\F_2^{1\times n}$, and let $v_1$ and $v_2$ be columns in $\F_2^{n\times1}$.
Let $r_x$ and $r_v$ be the ranks of the lists $(x_1,x_2)$ and $(v_1,v_2)$, respectively.
Let $u_1$ and $u_2$ be prescribed rows in $\F_2^{1\times n}$, and let $c_1$ and $c_2$ be prescribed columns in $\F_2^{n\times1}$.
For uniform $A_0\in\F_2^{n\times n}$,
\begin{equation}\label{eq:prescribed-entries}
 \Prb_{A_0}{x_i\cdot A_0=u_i\text{ and }A_0\cdot v_j=c_j\text{ for all }i,j\in\{1,2\}}
 =\begin{cases}2^{r_x r_v}N^{-(r_x+r_v)},&\text{if consistent},\\0,&\text{otherwise.}\end{cases}
\end{equation}
The prescriptions are \emph{consistent} when every linear relation among the $x_i$ holds among the $u_i$, every linear relation among the $v_j$ holds among the $c_j$, and $u_i\cdot v_j=x_i\cdot c_j$ for all $i,j\in\{1,2\}$.
\end{lemma}
\begin{proof}
Choose a maximal independent sublist of the rows and one of the columns, with $r_x$ and $r_v$ elements.
The conditions for the remaining indices follow from the relation conditions.
An invertible change of basis on each side maps a uniform matrix to a uniform matrix and turns the chosen rows and columns into standard basis vectors.
The row conditions then fix $r_x$ full rows of $A_0$ and the column conditions fix $r_v$ full columns.
The $r_x r_v$ entries where a fixed row meets a fixed column are prescribed twice, and the two prescriptions agree exactly when $u_i\cdot v_j=x_i\cdot c_j$ for the chosen indices.
All other pairings follow from the relation conditions.
Hence consistent prescriptions fix exactly $r_x n+r_v n-r_x r_v$ independent binary entries.
The probability of satisfying them is $2^{-(r_x n+r_v n-r_x r_v)}=2^{r_x r_v}N^{-(r_x+r_v)}$, as stated in~\cref{eq:prescribed-entries}.
Inconsistent prescriptions have probability zero.
\end{proof}

The bound on $\sigma_{\max}(\Delta_z)$ controls $\FourCyc{\Delta_z}$ under independent sampling.
The next lemma transfers this control to $\FourCyc{\Delta_z^M}$, as required by the answer-mask identity.
The bound in \cref{eq:four-cycle-refresh} below includes correction terms for repeated rows or columns in the four-cycle expansion.

\begin{lemma}[Four cycles after a refresh]\label{lem:four-cycle-refresh}
Let $\Delta:\UL^2\times\UR^2\to\R$ be any real table supported on consistent tuples.
Put $t:=\LtwoNorm{\Delta}$ and $B_\infty:=\MaxNorm{\Delta}$, and let $B_2\ge0$ be a real number bounding the $L_2$ norm of every row and every column.
Then
\begin{equation}\label{eq:four-cycle-refresh}
 \E_{M\sample \InvMatr_2^n}{\FourCyc{\Delta^M}}
 \le64\FourCyc{\Delta}+\frac{32B_2^2}{q}\cdot t^2
             +\frac{72B_\infty^2}{q^2}\cdot t^2.
\end{equation}
\end{lemma}
\begin{proof}
\emph{Allowing singular matrices.}
Extend $\Delta$ by zero whenever $\ell'$ or $r'$ is zero, obtaining $\widehat{\Delta}$.
In the extended averages, $\ell$ and $r$ remain uniform on $\UL$ and $\UR$, while $\ell'$ and $r'$ range independently over all $N$ binary rows and columns.
Then
\begin{align}
 \widehat{t}^2&:=\E_{\ell,\ell',r,r'}{\widehat{\Delta}((\ell,\ell'),(r,r'))^2}=(q/N)^2\cdot t^2,
 \label{eq:zero-extension-ltwo}\\
 \FourCyc{\widehat{\Delta}}&=(q/N)^4\cdot \FourCyc{\Delta}.
 \label{eq:zero-extension-four-cycle}
\end{align}
The rows and columns of $\widehat{\Delta}$, with these measures, still have $L_2$ norm at most $B_2$.
For a possibly singular matrix we cannot use $A_0^{-1}\cdot r$.
Instead, we keep $r'$ as part of the column coordinate and impose the equation $A_0\cdot r'=r$.
For an unrestricted uniform binary matrix $A_0$, define
\begin{equation}\label{eq:unrestricted-refresh-table}
 \widetilde\Delta_{A_0}(\ell,(r,r')):=
 \begin{cases}
 \sqrt N\cdot\widehat{\Delta}((\ell,\ell\cdot A_0),(r,r')),&A_0\cdot r'=r,\\
 0,&\text{otherwise}.
 \end{cases}
\end{equation}
The table in~\cref{eq:unrestricted-refresh-table} has row measure uniform on $\UL$ and column measure uniform on $\UR\times\F_2^{n\times1}$.
For invertible $M$, each $r$ permits exactly one $r'=M^{-1}\cdot r$.
Expanding the complete four-cycle shows that the two $r'$ averages contribute $N^{-2}$, exactly canceled by the four factors $\sqrt N$.
Hence $\FourCyc{\widetilde\Delta_M}=\FourCyc{\Delta^M}$.
By nonnegativity of $\FourCyc{\widetilde\Delta_{A_0}}$ and~\cref{eq:invertible-comparison},
\begin{equation}\label{eq:joint-four-cycle-unrestricted}
 \E_M{\FourCyc{\Delta^M}}\le4\E_{A_0}{\FourCyc{\widetilde\Delta_{A_0}}}.
\end{equation}
We apply~\cref{eq:invertible-comparison} to the complete four-cycle, which is nonnegative.
An individual product of four entries may be negative.

\emph{Separating the four rank cases.}
Expanding the four-cycle and inserting independent uniform rows $\ell'_1$ and $\ell'_2$ in $\F_2^{1\times n}$ gives
\begin{equation}\label{eq:four-cycle-prescribed-expansion}
\begin{aligned}
 &\E_{A_0}{\FourCyc{\widetilde\Delta_{A_0}}}\\
 &\quad=N^4\E_{\ell_1,\ell_2,\ell'_1,\ell'_2,r_1,r_2,r'_1,r'_2}
 {\begin{aligned}
 &\prod_{i,j=1}^2\widehat{\Delta}((\ell_i,\ell'_i),(r_j,r'_j))\\[-1mm]
 &{}\cdot\Prb_{A_0}{\ell_i\cdot A_0=\ell'_i,\ A_0\cdot r'_j=r_j\ (i,j=1,2)}
 \end{aligned}}.
\end{aligned}
\end{equation}
The factor $N^4$ in~\cref{eq:four-cycle-prescribed-expansion} consists of $N^2$ from the four factors $\sqrt N$ in $\widetilde\Delta_{A_0}$ and $N^2$ compensating for the two newly sampled rows $\ell'_1$ and $\ell'_2$.
The pairing constraints in \cref{lem:prescribed-entries} are enforced by the support of $\widehat{\Delta}$.
Terms with $\ell'_i=0$ or $r'_j=0$ vanish by the zero extension.
Over $\F_2$, rank one for the two old rows means $\ell_1=\ell_2$, which forces $\ell'_1=\ell'_2$.
Rank one for the two nonzero columns means $r'_1=r'_2$, which forces $r_1=r_2$ by the prescribed column equations.
For the next display, $\E{g;\cE}$ means the unnormalized restricted average: the quantity being averaged is $g$ on $\cE$ and zero otherwise.
All listed variables are still independent and uniform on the full spaces specified above.
Distinct old rows and distinct new columns give the main four-cycle term.
Equal rows or columns give products of squares.
Applying~\cref{eq:prescribed-entries} to the four rank cases in~\cref{eq:four-cycle-prescribed-expansion} gives the exact decomposition
\begin{equation}\label{eq:joint-four-rank-expansion}
\begin{aligned}
 &\E_{A_0}{\FourCyc{\widetilde\Delta_{A_0}}}\\
 &=16\E_{\ell_1,\ell_2,\ell'_1,\ell'_2,r_1,r_2,r'_1,r'_2}
 {\prod_{i,j=1}^2\widehat{\Delta}((\ell_i,\ell'_i),(r_j,r'_j));\ \ell_1\ne \ell_2,\ r'_1\ne r'_2}\\
 &+\frac4q\E_{\ell,\ell',r_1,r_2,r'_1,r'_2}
 {\widehat{\Delta}((\ell,\ell'),(r_1,r'_1))^2\widehat{\Delta}((\ell,\ell'),(r_2,r'_2))^2;\ r'_1\ne r'_2}\\
 &+\frac4q\E_{\ell_1,\ell_2,\ell'_1,\ell'_2,r,r'}
 {\widehat{\Delta}((\ell_1,\ell'_1),(r,r'))^2\widehat{\Delta}((\ell_2,\ell'_2),(r,r'))^2;\ \ell_1\ne \ell_2}\\
 &+\frac2{q^2}\E_{\ell,\ell',r,r'}{\widehat{\Delta}((\ell,\ell'),(r,r'))^4}.
\end{aligned}
\end{equation}
For example, rank $(1,2)$ has weight $4N$ after multiplication by $N^4$.
The equalities $\ell_1=\ell_2$ and $\ell'_1=\ell'_2$ cost $1/q$ and $1/N$, giving $4/q$.
Rank $(2,1)$ costs $1/N$ and $1/q$ in the same way, and rank $(1,1)$ has weight $2N^2$ and four equalities of joint probability $1/(q^2N^2)$.
The main term excludes only $\ell$- and $r'$-collisions.
Collisions in $\ell'$ and $r$ remain allowed.

\emph{Bounding the four contributions.}
Apply inclusion--exclusion to the two restrictions in the main term.
The contribution with $\ell_1=\ell_2$ is nonnegative: for each fixed $\ell$, it is a complete four-cycle of the sliced table with row coordinate $\ell'$ and column coordinate $(r,r')$, multiplied by $1/q$.
Likewise the contribution with $r'_1=r'_2$ is a nonnegative complete four-cycle with row coordinate $(\ell,\ell')$ and column coordinate $r$, averaged over $r'$ and multiplied by $1/N$.
Their subtraction therefore only lowers the upper bound.
The intersection is bounded by $B_\infty^2\widehat{t}^2/(q\cdot N)$: for four entries $e_i$ bounded by $B_\infty$,
\[
 |e_1\cdot e_2\cdot e_3\cdot e_4|\le\frac{B_\infty^2}{4}\sum_{i=1}^4e_i^2,
\]
and conditioned on both collisions every individual entry retains its full product marginal.
Thus the first term on the right-hand side of~\cref{eq:joint-four-rank-expansion} is at most
$16\FourCyc{\widehat{\Delta}}+16B_\infty^2\widehat{t}^2/(q\cdot N)$.

Drop the remaining restrictions in the two mixed-rank terms, whose summands are nonnegative.
For a fixed row, its mean squared entry is at most $B_2^2$, so the square of this mean is at most $B_2^2$ times the mean itself.
Averaging over rows bounds the first mixed-rank term by $(4B_2^2/q)\widehat{t}^2$.
The other has the same bound using columns.
The final term is at most $(2B_\infty^2/q^2)\widehat{t}^2$.
Combining the four estimates, multiplying by four in~\cref{eq:joint-four-cycle-unrestricted}, and substituting~\cref{eq:zero-extension-ltwo,eq:zero-extension-four-cycle} with $q/N\le1$ and $N>q$ gives~\cref{eq:four-cycle-refresh}: the coefficients are $4\cdot16=64$, $4\cdot8=32$, and $4\cdot(16+2)=72$ after bounding $1/(q\cdot N)$ by $1/q^2$.
\end{proof}

\begin{proof}[Completion of the proof of \cref{lem:refresh-rectangle}]
Apply \cref{lem:four-cycle-refresh} to the residual $\Delta_z$, with $t=\LtwoNorm{\Delta_z}$, $B_2=3$, and $B_\infty\le3/\beta$.
By~\cref{eq:four-cycle-norms,eq:residual-largest-coefficient}, $\FourCyc{\Delta_z}\le9\beta^2\cdot t^2$.
Since $q\cdot\beta^3\ge1$ and $\beta\le1$, both $q^{-1}$ and $q^{-2}\cdot\beta^{-2}$ are at most $\beta^2$.
It follows that
\begin{equation}\label{eq:direct-refreshed-four-cycle-bound}
 \E_M{\FourCyc{\Delta_z^M}}\le C\cdot\beta^2\cdot t^2.
\end{equation}
For each $M$, the test bound~\cref{eq:four-cycle-test} bounds the pairing with the two masks by $\FourCyc{\Delta_z^M}^{1/4}$.
Averaging over $M$ and applying Jensen's inequality with~\cref{eq:direct-refreshed-four-cycle-bound} gives
\begin{equation}\label{eq:direct-residual-mask-bound}
 \left|\E_{\ell,r,M}{\MaskL{z}(\ell,M)\cdot \MaskR{z}(r,M)\cdot\Delta_z^M(\ell,r)}\right|
 \le C\cdot\beta^{1/2}\cdot t^{1/2}.
\end{equation}
Apply the answer-mask identity in~\cref{eq:kernel-pairing} with $W=\Delta_z$ and replace $p_{\ell\cdot r}$ by $1/2$.
The additional absolute error is at most $\MaxNorm{\Delta_z}/(2q)\le3/(2q\cdot\beta)$.
Consequently, \cref{eq:direct-residual-mask-bound} gives
\begin{equation}\label{eq:direct-residual-pairing-bound}
 \left|\pairing{\prob_z}{\Delta_z}\right|
 \le C\cdot\beta^{1/2}\cdot t^{1/2}+\frac3{2q\cdot\beta}.
\end{equation}
Combining~\cref{eq:direct-residual-pairing-bound,eq:direct-residual-identity,eq:direct-cross-term} gives
\begin{equation}\label{eq:direct-error-bootstrap}
 t^2\le C\cdot\beta^{1/2}\cdot t^{1/2}+\frac C{q\cdot\beta}
 \le C\cdot\beta^{1/2}\cdot t^{1/2}+C\cdot\beta^2.
\end{equation}
If the second term on the final right-hand side of~\cref{eq:direct-error-bootstrap} is at least $t^2/2$, then $t\le C\cdot\beta\le C\cdot\beta^{1/3}$.
Otherwise, the first term is at least $t^2/2$.
For $t>0$, divide by $t^{1/2}$ to obtain $t\le C\cdot\beta^{1/3}$.
The case $t=0$ is immediate.
This argument covers the case where no coefficient is kept and every threshold in the stated range, including $\beta=1$, for every $n\ge2$.
\end{proof}

\subsubsection{One family of approximations for both secrets}\label{sec:joint-instances}
The estimates just proved use independent uniform grouped coordinates.
To compare executions, we instead sample encodings of prescribed bits and need the same approximation while those bits change.
We now transfer the estimates to these encoding distributions.
For states $X=(\ell,r)$ and $X'=(\ell',r')$, write
$G(X,X')$ for $G((\ell,\ell'),(r,r'))$.
For any distribution $\rho$, use
$\LtwoNormAt{F}{\rho}:=\left(\E_{x\sample\rho}{|F(x)|^2}\right)^{1/2}$ for the root-mean-square value of $F$ under $\rho$.
The unqualified norms continue to use independent uniform grouped coordinates.

\begin{lemma}[Local approximation for a refresh]\label{lem:copy-local}
There exists an absolute constant $C>0$ such that the following holds with $\beta:=q^{-1/3}$ and $\varepsilon:=C\cdot q^{-1/9}$.
For every round $i\in\{1,\ldots,T\}$, formal incoming history $h\in\cZ^{i-1}$, and answer $z\in\cZ$, there exists a single table $G_{i,h,z}:\UL^2\times\UR^2\to\R$, common to all endpoint labels, with a chosen product representation whose energies (see \cref{def:factor-energy}) satisfy
\begin{align}
 \NuclearNorm{G_{i,h,z}}&\le4,
 \label{eq:copy-local-nuclear}\\
 \Phi_{i,h,z}(\ell,\ell')&\le2/\beta
 &&\text{for every }(\ell,\ell')\in\UL^2,
 \label{eq:copy-local-left-energy}\\
 \Psi_{i,h,z}(r,r')&\le2/\beta
 &&\text{for every }(r,r')\in\UR^2,
 \label{eq:copy-local-right-energy}\\
 \E_{(\ell,\ell')\sample \UL^2}{\Phi_{i,h,z}(\ell,\ell')}&\le4,
 \label{eq:copy-local-left-mean}\\
 \E_{(r,r')\sample \UR^2}{\Psi_{i,h,z}(r,r')}&\le4,
 \label{eq:copy-local-right-mean}\\
 \LtwoNormAt{\prob_{i,h,z}-G_{i,h,z}}{\mu_s\otimes\mu_s}&\le\varepsilon
 &&\text{for every }s\in\{0,1\},
 \label{eq:copy-local-error}\\
 \LtwoNormAt{G_{i,h,z}}{\mu_s\otimes\mu_{s'}}&\le6
 &&\text{for every }(s,s')\in\{0,1\}^2.
 \label{eq:copy-local-ltwo}
\end{align}
The same table and product representation satisfy these bounds simultaneously for every $s\in\{0,1\}$ and every $(s,s')\in\{0,1\}^2$.
All bounds are per answer and uniform over formal histories, including histories that are never reached.
\end{lemma}
Before we prove the lemma, we explain its meaning and how it is used in the hybrid argument.
The two leakage functions of a round share the refresh matrix $M$, so the true leakage probability function $\prob_{i,h,z}$ need not split into a function of the left pair $(\ell,\ell')$ times a function of the right pair $(r,r')$.
Such a \emph{product} would reduce the analysis to one function of each party's grouped shares.
The nuclear norm of a table is the least product cost $\sum_j\LtwoNorm{a_j}\cdot\LtwoNorm{b_j}$ over its decompositions $\sum_j a_j\cdot b_j^{\transpose}$ into products (see \cref{eq:nuclear-factor-cost} in the \cref{sec:preliminaries-proofs}).
The approximation bound in~\cref{eq:copy-local-error} says that $\prob_{i,h,z}$ is nevertheless close to a signed sum of products, and~\cref{eq:copy-local-nuclear} bounds their total cost independently of $q$.
Closeness is required only where the later comparison of executions uses it: for an old and a new state that encode the same bit $s$ and are sampled from $\mu_s\otimes\mu_s$, as in a real execution.
The same table $G_{i,h,z}$ serves every assignment of labels to the endpoint states, so executions with different labels multiply identical tables and differ only in the distributions of the states.
At a cheating round, the old and new states encode different bits, so the true leakage probability function is undefined.
In that case, $G_{i,h,z}$ is only a signed table of controlled size, with $L_2$ norm at most $6$ under $\mu_s\otimes\mu_{s'}$ by~\cref{eq:copy-local-ltwo} and the same product-representation bounds as at every other round.

The bounds of \cref{lem:copy-local} serve different later steps.
The nuclear bound~\cref{eq:copy-local-nuclear} and the mean-energy bounds~\cref{eq:copy-local-left-mean,eq:copy-local-right-mean} show that, averaged over the pairs $(\ell,\ell')$ and $(r,r')$ respectively, the products have cost at most $4$.
The mean bounds alone would allow the energy at a single pair to be as large as $4q^2$, since each side has $q^2$ pairs.
The pointwise bounds~\cref{eq:copy-local-left-energy,eq:copy-local-right-energy}, each $2/\beta=2q^{1/3}$, rule this out, although they are far above the mean.
A nuclear-norm bound alone would not give such a pointwise bound (see \cref{sec:linear-algebra}), and \cref{lem:refresh-wire-change} needs it: when two neighboring rounds are multiplied over their common state, it bounds one energy pointwise and averages the other.
The $L_2$ bounds~\cref{eq:copy-local-error,eq:copy-local-ltwo}, with constants $\varepsilon$ and $6$, are used through the Cauchy--Schwarz inequality (see \cref{lem:la-cauchy-schwarz}) in \cref{lem:refresh-replacement}, which shows that replacing the true table of one round by $G_{i,h,z}$ costs at most $6\varepsilon$.
Finally, $\beta=q^{-1/3}$ is the smallest threshold for which \cref{lem:refresh-rectangle} applies, and it gives the smallest error $\varepsilon=C\cdot q^{-1/9}$.
The price is the large pointwise energy $2q^{1/3}$, which the discrepancy bound of order $q^{-1/2}$ (see~\cref{eq:table-disc}) more than compensates for when one encoding is changed.

\begin{proof}[Proof of \cref{lem:copy-local}]
Apply \cref{lem:refresh-rectangle} with $\beta=q^{-1/3}$ to the masks of the answer $z$, and set $G_{i,h,z}:=2\cdot\prob_{i,h,z}^{>\beta}$.
The nuclear norm doubles, so~\cref{eq:refresh-rectangle-nuclear} gives~\cref{eq:copy-local-nuclear}.
Multiplying each balanced factor by $\sqrt2$ doubles both energies: they become $2\cdot\Phi_\beta$ and $2\cdot\Psi_\beta$.
Thus~\cref{eq:refresh-left-energy,eq:refresh-right-energy} give~\cref{eq:copy-local-left-energy,eq:copy-local-right-energy}, respectively.
The factors remain balanced, so each mean energy equals the nuclear norm by~\cref{eq:balanced-mean-energies}.
This proves~\cref{eq:copy-local-left-mean,eq:copy-local-right-mean}.
By~\cref{eq:encoding-probability-lower}, each fixed-bit encoding distribution $\mu_s$ has density at most three relative to independent uniform shares, so $\mu_s\otimes\mu_{s'}$ has density at most nine relative to $\JointRef$.
On the support of $\mu_s\otimes\mu_s$, the residual $\Delta_z$ equals $\prob_z-2\cdot\prob_z^{>\beta}$.
\Cref{eq:refresh-rectangle-error} therefore gives~\cref{eq:copy-local-error} after enlarging $C$.
For arbitrary endpoint labels, the same density bound and the $L_2$ bound in~\cref{eq:refresh-rectangle-ltwo} give $\LtwoNormAt{2\cdot\prob_z^{>\beta}}{\mu_s\otimes\mu_{s'}}\le3\cdot2=6$, proving~\cref{eq:copy-local-ltwo}.
For an unreachable history, fix any admissible leakage functions (see \cref{sec:leakage}) and use the same construction.
\end{proof}
The tables may be signed, and their sum over answers need not be one.
Only the true leakage probability functions have the positivity and normalization used below.
\subsection{Comparing two neighboring rounds}\label{sec:refresh-replacement}
We now carry out the second step of the proof: bounding what it costs to replace a true leakage probability function by its approximation and, in \cref{sec:refresh-wire-change}, to change the label of a state.
We use the tables of \cref{lem:copy-local} with $\beta=q^{-1/3}$ and $\varepsilon=C\cdot q^{-1/9}$.
Recall from \cref{def:labels} that the label of a state is the bit it encodes in the hybrid. A round is valid when its old and new states have equal labels. Otherwise, it is a cheating round.
Recall also that $G$ comes with a chosen product representation whose factor energies are $\Phi$ and $\Psi$ (see \cref{def:factor-energy}).
All tables for every formal history are fixed before any labels are changed.

\begin{definition}[Signed rounds and signed measures]\label{def:signed-round}
A round is \emph{signed} when its leakage probability function $\prob$ is replaced by its approximation $G$, and \emph{unsigned} otherwise.
Every unsigned round uses its true probability function, called its \emph{true factor}, and must have equal endpoint labels.
For a fixed labeling, the resulting \emph{signed measure} assigns to a complete history the expected product of these local factors over the independently sampled states, as in~\cref{eq:copy-canonical}, using $G$ at signed rounds and $\prob$ elsewhere.
\end{definition}
The total mass of a signed measure need not be one.

The hybrid argument of \cref{sec:copy-comparison} changes the bit encoded by one state at a time.
An interior state is shared by the round producing it and the round reading it.
One of these rounds has unequal endpoint labels before the change, and the other has unequal endpoint labels after it.
Both rounds are therefore signed when the label changes, and replacing or restoring a true function can put its approximation error next to another signed factor.
Because the factors share a state, we bound their product using their $L_2$ norms under the joint distribution of the three states.

\begin{lemma}[Replacing one round]\label{lem:refresh-replacement}
Fix $j\in\{1,\ldots,T-1\}$, formal histories and answers in rounds $j$ and $j+1$, and labels $s_{j-1},s_j,s_{j+1}\in\F_2$ for the states
$A=X_{j-1}$, $X=X_j$, and $B=X_{j+1}$.
Sample $A\sample\mu_{s_{j-1}}$, $X\sample\mu_{s_j}$, and $B\sample\mu_{s_{j+1}}$ independently.
If round $j$ is valid, $e(A,X):=\prob(A,X)-G(A,X)$ is its error, and round $j+1$ is signed with approximation $G'(X,B)$ (its \emph{signed companion}), then
\begin{equation}\label{eq:refresh-replacement-error}
 \E_{A,X,B}{|e(A,X)\cdot G'(X,B)|}\le6\varepsilon.
\end{equation}
The same estimate holds when the valid error depends on $(X,B)$ and its signed companion on $(A,X)$.
For every valid round with no signed companion, the corresponding bound is $\varepsilon$.
\end{lemma}
\begin{proof}
By~\cref{eq:copy-local-error}, the $L_2$ norm of $e$ under its two-state marginal is at most $\varepsilon$.
By~\cref{eq:copy-local-ltwo}, that of $G'$ is at most $6$, even if its endpoint labels differ.
The Cauchy--Schwarz inequality (see \cref{lem:la-cauchy-schwarz}) under the original joint distribution proves~\cref{eq:refresh-replacement-error}:
\[
 \E_{A,X,B}{|e(A,X)\cdot G'(X,B)|}
 \le\bigl(\E_{A,X,B}{e(A,X)^2}\bigr)^{1/2}\cdot\bigl(\E_{A,X,B}{G'(X,B)^2}\bigr)^{1/2}\le6\varepsilon.
\]
The two tables may share the state $X$, since the inequality is applied under the joint distribution of all three states.
With no companion, the Cauchy--Schwarz inequality (see \cref{lem:la-cauchy-schwarz}) applied to $e$ and the constant function $1$ gives the bound $\varepsilon$.
\end{proof}

\subsection{Changing one encoding}\label{sec:refresh-wire-change}
Changing the label of a state from $0$ to $1$ replaces its encoding distribution $\mu_0$ by $\mu_1$.
We call this a \emph{label change}.
Put $\DiscBound:=3/\sqrt{q-1}$, the one-encoding discrepancy bound of~\cref{eq:table-disc}.
Only the two incident rounds depend on the state whose label changes.
After fixing the encodings at the other endpoints of those rounds, their signed tables form the product to which we apply the one-encoding discrepancy estimate.
The nuclear norm of this product can be large at exceptional outer encodings.
We bound its average using the pointwise bound for one energy and the mean bound for the other.

\begin{lemma}[Changing one encoding between signed rounds]\label{lem:refresh-wire-change}
For fixed formal histories and answers, suppose both rounds incident to $X_j$ are signed.
With $A=X_{j-1}$, $X=X_j$, and $B=X_{j+1}$ as in \cref{lem:refresh-replacement}, and $G_j$ and $G_{j+1}$ the approximations at rounds $j$ and $j+1$, let $F_{A,B}(X):=G_j(A,X)\cdot G_{j+1}(X,B)$.
For every pair of labels $(s_A,s_B)\in\F_2^2$, sample $A\sample\mu_{s_A}$ and $B\sample\mu_{s_B}$ independently.
Then
\begin{equation}\label{eq:refresh-wire-nuclear}
 \E_{A,B}{\NuclearNorm{F_{A,B}}}\le8/\beta,
\end{equation}
where the nuclear norm is in the two shares of $X$.
Consequently,
\begin{equation}\label{eq:refresh-wire-discrepancy}
 \E_{A,B}{\left|
   \E_{X\sample\mu_0}{F_{A,B}(X)}
   -\E_{X\sample\mu_1}{F_{A,B}(X)}
 \right|}\le8\cdot\DiscBound/\beta.
\end{equation}
At a chain endpoint, with just one incident signed round, the bounds are $4$ and $4\cdot\DiscBound$, respectively.
\end{lemma}
\begin{proof}
Write $\Phi_j$, $\Psi_j$, $\Phi_{j+1}$, and $\Psi_{j+1}$ for the energies of their chosen product representations (see \cref{lem:copy-local}).
Multiply the chosen product representations of the two signed tables.
At fixed $A=(L_A,R_A)$ and $B=(L_B,R_B)$, the product-representation bound~\cref{eq:nuclear-product-bound} and the Cauchy--Schwarz inequality (see \cref{lem:la-cauchy-schwarz}) over factor indices give
\begin{equation}\label{eq:refresh-wire-product-cost}
 \NuclearNorm{F_{A,B}}\le\sqrt{\Phi_\times(L_A,L_B)\cdot\Psi_\times(R_A,R_B)},
\end{equation}
where
\[
\begin{aligned}
 \Phi_\times(L_A,L_B)&:=\E_{L_X\sample \UL}{
  \Phi_j(L_A,L_X)\cdot\Phi_{j+1}(L_X,L_B)},\\
 \Psi_\times(R_A,R_B)&:=\E_{R_X\sample \UR}{
  \Psi_j(R_A,R_X)\cdot\Psi_{j+1}(R_X,R_B)}.
\end{aligned}
\]
Apply the Cauchy--Schwarz inequality (see \cref{lem:la-cauchy-schwarz}) to~\cref{eq:refresh-wire-product-cost} when averaging over $A$ and $B$.
Their left marginals are independent uniform shares, as are their right marginals, for every assignment of their labels.
For $\Phi_\times$, bound $\Phi_j$ pointwise by $2/\beta$ using~\cref{eq:copy-local-left-energy} and average $\Phi_{j+1}$ using~\cref{eq:copy-local-left-mean}.
This gives $\E_{A,B}{\Phi_\times(L_A,L_B)}\le(2/\beta)\cdot4=8/\beta$.
The same argument with~\cref{eq:copy-local-right-energy,eq:copy-local-right-mean} gives $\E_{A,B}{\Psi_\times(R_A,R_B)}\le8/\beta$.
This proves~\cref{eq:refresh-wire-nuclear}.
Applying~\cref{eq:table-disc} at each fixed $A$ and $B$ and then averaging proves~\cref{eq:refresh-wire-discrepancy}.
With one signed round, the same calculation uses just its two mean energies, each at most $4$ by~\cref{eq:copy-local-left-mean,eq:copy-local-right-mean}.
\end{proof}

\subsection{Accounting for adaptive histories}\label{sec:refresh-adaptivity}
The bounds in~\cref{eq:refresh-replacement-error,eq:refresh-wire-discrepancy} concern fixed formal histories and answers, with states that are not conditioned on an observed transcript.
To use them in the full execution, we enumerate the answers near a \emph{block} of two consecutive signed rounds that are changed together, at a cost of at most $K^3$.
Conditioning the states on an observed prefix could destroy the independent sampling used in the local estimates.
Instead, we keep the state samples unconditioned and explicitly sum over the answers near the changed block.
The calculation proceeds in the following order.
\begin{itemize}
\item For a replacement, move the absolute value inside the state average.
For a label change, first take the difference of the two averages over the changed state, and only then move the absolute value inside the average over the outer states.
All unsigned factors are nonnegative.
\item Sum unsigned answers after the block backward.
Their true probability functions sum to one pointwise at every incoming history.
These functions do not depend on the changing state, which is incident only to the signed block.
This summation removes the entire suffix before any answers are enumerated.
\item Enumerate the two block answers, costing at most $K^2$.
If round $j-1$ exists, enumerate its answer too, costing at most $K$, and drop its true factor using $0\le \prob\le1$.
Round $j-1$ is the only earlier round touching $\{X_{j-1},X_j,X_{j+1}\}$.
\item The remaining prefix factors depend only on states outside this three-state set.
At each fixed formal prefix, average over the three states of this set, sampled independently from their respective encoding distributions, using \cref{lem:refresh-replacement} or~\cref{lem:refresh-wire-change}, whose bounds are uniform in that prefix.
Then sum the remaining prefix answers backward, again using normalization.
\end{itemize}
Thus a replacement costs at most $6K^3\cdot\varepsilon$, and a label change at most $8K^3\cdot\DiscBound/\beta$, in the $L_1$ norm of signed history measures: the sum of absolute masses over complete histories (see \cref{def:measure-norm}).
A missing round simply removes one enumeration.
These same bounds cover the endpoints.
The order matters: the suffix is summed first because enumerating the later neighbor of a block as well would give an unnecessary fourth factor of $K$.
The earlier neighbor cannot be removed in the same way.
Its answer belongs to the incoming histories of the block, so the block's tables change with this answer, and normalization of its true probability function applies only once those tables are gone.
The tables cannot be summed away first, because unlike true probability functions they need not sum to one over their answers.
They are instead bounded by averaging over the three states, but the true factor of the earlier neighbor depends on one of these states, $X_{j-1}$.
Enumerating its answer and discarding this factor resolves the difficulty, at the cost of the third factor of $K$.

\subsection{Moving one boundary along the chain}\label{sec:copy-comparison}
We can now combine the local replacement and label-change bounds to compare the executions for the two secrets.
Assume $T\ge1$.
For $j=0,\ldots,T+1$, give each state $X_i$ the label
\[
 s_i^j:=\begin{cases}1,&i<j,\\0,&i\ge j.\end{cases}
\]
Sample the states independently from these encoding distributions.
For $1\le j\le T$, use $G$ at round $j$ and the true leakage probability function $\prob$ elsewhere.
Call the resulting signed history measure $Q_j$.
The labeling has exactly one cheating round, round $j$, where the global table $G$ is defined even though the encoded bit changes.
At the endpoints no round is signed:
$Q_0=P_0^T$ and $Q_{T+1}=P_1^T$ are the actual transcript distributions.
The intermediate measures may be signed and need not be normalized.

For $1\le j<T$, move from $Q_j$ to $Q_{j+1}$ as follows.
\begin{itemize}
\item Replace $\prob$ by $G$ at round $j+1$.
Its endpoint labels are both zero.
\item Change the label of $X_j$ from zero to one.
Both incident rounds are signed.
\item Restore $G$ to $\prob$ at round $j$.
Its endpoint labels are now both one.
\end{itemize}
Every true factor is inserted or removed only at a valid round, and at most two rounds are signed at any time.
For $Q_0\to Q_1$, sign round $1$ and change $X_0$.
For $Q_T\to Q_{T+1}$, change $X_T$ and restore round $T$.
For $T=1$, these two endpoint moves constitute the entire argument.

Consecutive labelings differ in the label of a single state, which the adversary never sees.
The triangle inequality, the comparisons from~\cref{eq:refresh-replacement-error,eq:refresh-wire-discrepancy} with the answer enumeration in~\cref{sec:refresh-adaptivity}, and the factor $1/2$ between the $L_1$ norm and statistical distance give
\begin{equation}\label{eq:copy-finite}
 \StatDist{P_0^T}{P_1^T}
 \le C\cdot T\cdot K^3\cdot
 \left(\varepsilon+\DiscBound/\beta\right).
\end{equation}
The factor $1/2$ applies because the endpoints $Q_0$ and $Q_{T+1}$ are actual probability distributions.

\begin{proof}[Proof of \cref{thm:main}]
For $q=2^n-1\ge3$, take $\beta=q^{-1/3}$ and $\varepsilon=C\cdot q^{-1/9}$.
Since $\DiscBound=3/\sqrt{q-1}\le C\cdot q^{-1/2}$,
\begin{equation}\label{eq:refresh-discrepancy-rate}
 \DiscBound/\beta\le C\cdot q^{-1/6}\le C\cdot q^{-1/9}.
\end{equation}
Substituting~\cref{eq:refresh-discrepancy-rate} and $\varepsilon=C\cdot q^{-1/9}$ in~\cref{eq:copy-finite} proves the first inequality in~\cref{eq:copy-rate}.
Also $q\ge2^{n-1}$ and $K^3=2^{6\lambda}$, which gives~\cref{eq:copy-rate-binary} after increasing an absolute constant.
For $T=0$, the transcript is empty and the statistical distance is zero.
All estimates are uniform in the adversary's fixed tape.
Averaging proves the assertion with that tape included in the view.
\end{proof}

\subsection{Leakage rate and share length}\label{sec:field-size}
\phantomsection\label{def:leakage-rate}
We now translate \cref{thm:main} into a leakage rate and a sufficient share length.
The per-party leakage rate is $\lambda/n$, since each stored share has $n$ bits.
The temporary refresh matrix is not counted.
\begin{corollary}[Binary leakage rate]\label{cor:parameters}
If $T$ is polynomial in $n$ and $\lambda\le\alpha n$ for a fixed $\alpha<1/54$, then $\StatDist{P_0^T}{P_1^T}$ is exponentially small in $n$.
More generally, for all integers $T\ge1$ and $\lambda\ge0$ and every real target parameter $\kappa\ge0$, distance at most $2^{-\kappa}$ is obtained by every integer $n\ge2$ satisfying
\begin{equation}\label{eq:rate-bound}
 n\ge9\cdot\left(\kappa+6\lambda+\log_2(C'\cdot T)\right),
\end{equation}
where $C'$ is the absolute constant of \cref{thm:main}.
\end{corollary}
\begin{proof}
Since $K^3=2^{6\lambda}$, \cref{eq:copy-rate,eq:copy-rate-binary} bound the distance by
$C'\cdot T\cdot2^{6\lambda-n/9}$.
Under the first assumption, the exponent is at most $-(1-54\alpha)\cdot n/9$.
Multiplying by a polynomial factor in $n$ preserves exponential decay.
Substituting~\cref{eq:rate-bound} proves the second assertion.
\end{proof}
In particular, the sufficient share length depends logarithmically on the number of refreshes.
Since the proof leaves $C'$ unspecified, this estimate does not give a concrete numerical share length.

\subsection{Initialization with zero shares}\label{sec:zero-initialization}
The protocol of \cref{def:protocol} requires both initial shares to be nonzero.
The restriction is only a convenience: allowing zero shares in the initial encoding changes every refreshing bound by at most $4/N$, where $N=2^n$ (see \cref{sec:shares}).
Suppose that the initial state is instead uniform on all pairs $(\ell,r)\in\F_2^n\times\F_2^n$ with $\ell\cdot r=s$, so that a share may be zero.
Extend each leakage function to zero-share inputs in any fixed way.
For $s=1$, neither share can be zero, so the transcript distribution is unchanged.
For $s=0$, the number of such pairs is $N(N+1)/2$, and $2N-1$ of them contain a zero share.
The probability of a zero share is therefore
\begin{equation}\label{eq:zero-share-probability}
 \gamma_0:=\frac{2(2N-1)}{N(N+1)}\le\frac4N,
\end{equation}
and conditioned on its complement the initial state has distribution $\mu_0$.
The new transcript distribution for $s=0$ is thus $(1-\gamma_0)\cdot P_0^T+\gamma_0\cdot\widetilde P$, where $\widetilde P$ is the transcript distribution conditioned on the initial state containing a zero share, and its statistical distance from $P_0^T$ is at most $\gamma_0$.
By the triangle inequality and~\cref{eq:zero-share-probability}, every refreshing bound therefore increases by at most $4/N$ under this initialization.
\section{Circuit compilation}\label{sec:compilation}
In the previous section we addressed the problem of refreshing an encoded secret under leakage.
Let us now turn to the more complex task of protecting an arbitrary computation against leakage.
Recall from \cref{sec:intro-compiler-question} that a leakage-resilient circuit compiler transforms a circuit into a protected circuit such that anything the adversary learns from the leakage together with the input/output behavior could also be learned from the input/output behavior alone.
We asked there whether arbitrary circuits can be compiled so that they remain secure when every active gadget, the two-party protocol that implements one gate on encodings, leaks locally, without any leak-free internal component.
\Cref{thm:compiler-direct} answers this question for a single evaluation: a simulator that receives the output of the circuit but not the private input reproduces the adversary's view within a small statistical distance.
\Cref{thm:compiler-stateful} in \cref{sec:stateful} extends this to stateful circuits that are invoked repeatedly with inputs chosen by the adversary during the execution, which is the continual-leakage setting for computation.
The guarantees for circuit compilation do not follow from \cref{thm:main} as a black box: they reuse the matrix refresh and its local approximation, but composing the gadgets requires a new argument.

The construction, previewed in \cref{sec:intro-compiler}, compiles NAND source circuits into two-party protocols using the matrix refresh of \cref{def:protocol}.
Chains of refreshes give every gate input a fresh encoding, which handles arbitrary fanout and repeated operands.
The private-input encoder and the final output decoder are trusted.
Every internal operation has its stated leakage budget. Communication is private, and queries access only the permitted data of the active operation.
The trusted decoder releases only the output in the clear. The output shares stay hidden.
\Cref{sec:compiler-direct} gives the compiler and its security statement.
\Cref{sec:compiler-gadget} specifies the NAND gadget, \cref{sec:compiler-local} extends the local approximation to it, \cref{sec:direct-composition} proves the composition bound, and \cref{sec:direct-simulator} gives the concrete simulator.

\subsection{Compilation with refreshing between gadgets}\label{sec:compiler-direct}
We first give the compiler for stateless circuits with a trusted input encoder and output decoder.
The encoder supplies only the private input encodings.
The decoder reveals only the output in the clear, after all leakage queries, and the output share pairs remain hidden.
Every internal operation has its stated leakage budget: no refresh, sampling operation, or phase of a computation gadget is assumed to be unobserved.

\subsubsection{NAND sources and preprocessing}\label{sec:direct-source-model}
We begin with the circuits to be compiled.
Let $\Circuit:\F_2^{\Ipub}\times\F_2^{\Ipriv}\to\F_2^k$ be a Boolean circuit with public input $\vecx$, private input $\vecs$, and a fixed topological order of its $t_0\ge0$ \emph{source gates}.
Every source gate computes $\NAND(s_1,s_2):=1+s_1\cdot s_2$ over $\F_2$.
An \emph{input port} is one of the two operand positions of a gate, and a \emph{consumer} of a wire is a gate that reads it.
Repeated operand wires and arbitrary fanout are allowed.
The ordered list of $k\ge0$ \emph{designated outputs} may repeat a wire or designate an input wire directly.
Public constants are public input wires.
An input is \emph{unused} if it has neither a consumer nor an output position.
Unused private inputs may be deleted, or retained without affecting the leakage transcript.
Unused public inputs may also be removed during preprocessing. The compiled execution and operation counts then refer to the resulting source.

This source basis represents every Boolean circuit given in a fixed AND/OR/NOT basis with constant-factor overhead.
Negation is $\NAND(s,s)$.
Replace an AND gate with operands $s_1$ and $s_2$ by the two NAND gates
\begin{align}
 w'&:=\NAND(s_1,s_2),\label{eq:direct-basis-and-first}\\
 w&:=\NAND(w',w').\label{eq:direct-basis-and-second}
\end{align}
Here $w'$ is one intermediate wire feeding both input ports of the second gate, and \cref{eq:direct-basis-and-first,eq:direct-basis-and-second} give $w=s_1\cdot s_2$ for every pair of input bits.
An OR gate is $\NAND(\NAND(s_1,s_1),\NAND(s_2,s_2))$, using three NAND gates.
Each replacement preserves the old output value, hence preserves the circuit function and the designated output list.
These replacements need no new constant wire, and any constants present in the source description are counted among the $\Ipub$ public input wires.
All counts below refer to the source after preprocessing.
We impose no fanout bound and do not require distinct source operands or a positive gate count: the refresh chains provide distinct encoding copies for repeated source operands, and a source with no gates needs no such copies.

\begin{definition}[Operations, wire encodings, scopes, and labelings]\label{def:main-wires}
An \emph{operation} of the compiled protocol is a computation gadget, an input-port refresh, or a public-input sampler, all specified below.
A stored wire with plaintext label $s(w)\in\F_2$ has an encoding $X_w=(L_w,R_w)$ with $L_w\in\UL$, $R_w\in\UR$, and $L_w\cdot R_w=s(w)$.
The \emph{scope} $S_i$ of operation $i$ is the set of encoding locations it reads or produces.
For a set $A$ of locations, write $X_A:=(X_w)_{w\in A}$.
A \emph{labeling} assigns a bit to every location. It is \emph{valid} when it satisfies the equations of all computation and refresh operations and the prescribed public labels.
\end{definition}
The \emph{persistent share length}, the length of the shares that wires carry between operations, is $n\ge2$. The sets $\UL$ and $\UR$ and the encoding distributions $\mu_s$ are those of \cref{def:joint-state}.

\begin{figure}[htbp]
\centering
\begingroup
\setlength{\fboxrule}{0.3pt}
\setlength{\fboxsep}{8pt}
\fcolorbox{black!25}{white}{\begin{minipage}{\dimexpr\linewidth-2\fboxsep-2\fboxrule\relax}
\centering
\begin{tikzpicture}[x=1cm,y=1cm,font=\footnotesize,
 op/.style={draw=black!50,fill=black!4,rounded corners=2pt,align=center,minimum height=0.8cm,inner sep=4pt},
 enc/.style={draw=black!60,circle,inner sep=1pt,minimum size=0.75cm},
 flow/.style={->,draw=black!60,semithick}]
 \node[op] (p) at (0,0) {gadget\\producing $w$};
 \node[enc] (w0) at (2.2,0) {$w_0$};
 \node[op] (r1) at (4.1,0) {input-port\\refresh};
 \node[enc] (w1) at (6.0,0) {$w_1$};
 \node[op] (r2) at (7.9,0) {input-port\\refresh};
 \node[enc] (w2) at (9.8,0) {$w_2$};
 \node[op] (g1) at (6.0,-1.6) {first consumer\\of $w$};
 \node[op] (g2) at (9.8,-1.6) {second consumer\\of $w$};
 \draw[flow] (p)--(w0);
 \draw[flow] (w0)--(r1);
 \draw[flow] (r1)--(w1);
 \draw[flow] (w1)--(r2);
 \draw[flow] (r2)--(w2);
 \draw[flow] (w1)--(g1);
 \draw[flow] (w2)--(g2);
\end{tikzpicture}
\end{minipage}}
\endgroup
\caption{The input-port refreshes of a wire $w$ with $\deg(w)=2$.
The copy $w_1$ lies in the scopes of three operations: its producing refresh, its consumer, and the next refresh.}
\label{fig:port-chain}
\end{figure}
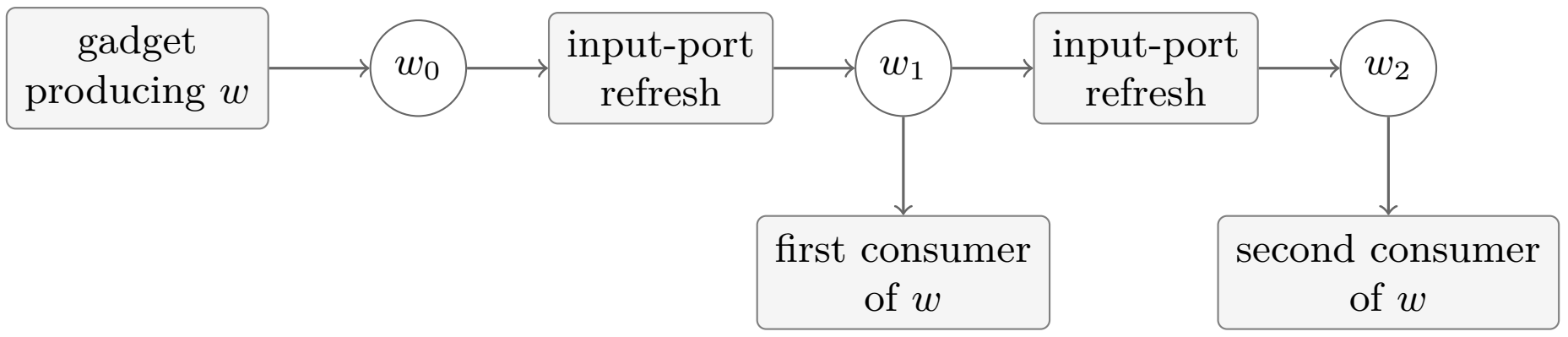

\para{A refreshed encoding for each input port.}
The compiler gives each input port its own refreshed copy of the encoding of its source wire.
These copies control how many operations can inspect any one encoding, even when a source wire has many consumers.
The composition proof accounts for operations touching a changed encoding through their answer alphabets and factor energies (see \cref{sec:direct-composition}). Bounding the number of operations touching an encoding prevents those costs from growing with the source fanout.
Refreshing every copy directly from the original would still read the original once per consumer.
We therefore arrange the copies in a chain so that generating each subsequent copy does not require reading the original encoding again.
For each source wire $w$, enumerate its $\deg(w)$ input-port occurrences in a fixed public order, namely topological gate order with the left input port before the right.
Introduce a chain of dimension-$n$ refreshes, the \emph{input-port refreshes} of $w$,
\[
 w=w_0\longrightarrow w_1\longrightarrow\cdots\longrightarrow w_{\deg(w)},
\]
and route occurrence $j$ to $w_j$.
A refresh from $w_{j-1}$ to $w_j$ samples a fresh matrix $M\sample\InvMatr_2^n$, sends it privately, and sets
\[
 L_{w_j}:=L_{w_{j-1}}\cdot M,\qquad R_{w_j}:=M^{-1}\cdot R_{w_{j-1}}.
\]
Execute the next link immediately before its assigned input port is consumed.
The current chain endpoint is retained until the next link is generated, and the original source-wire encoding is also retained whenever it is a designated output.
An older copy may be erased only after all of its assigned uses have finished.
After this refresh, erase $M$, its inverse, its \emph{sampling tape} (the random bits used to sample it, including rejected proposals), and the temporary workspace.
Retaining an old encoding in inactive memory does not enlarge the active view: each party sees precisely its old share and the matrix, which determine its new share.
An input-port refresh is thus the local refresh analyzed in \cref{lem:copy-local}. Only the lifetime of the stored old share differs from continual storage.
\Cref{fig:port-chain} shows the chain of a wire with two consumers.

\begin{tcolorbox}[breakable,colback=white,colframe=black!25,
 boxrule=0.3pt,arc=0pt,boxsep=0pt,left=8pt,right=8pt,top=8pt,bottom=8pt,
 before skip=8pt,after skip=8pt,fontupper=\small]
\phantomsection\label{alg:direct-compiler}
\textbf{Execution of the compiler.}
\begin{enumerate}
\setlength{\itemsep}{3pt}
\item Independently encode every private input bit using the trusted encoder.
For each public input or constant bit $c$, run a \emph{public-input sampler}: the left party samples $(L,R)\sample\mu_c$, sends $R$ privately, and retains $L$.
During this sampler, the left party's leakage may read the entire pair and sampling tape, and the right party's leakage may read its received share.
The left party then erases $R$ and the sampling tape.
\item Process source gates in topological order.
Immediately before a gate, refresh the current endpoint of its left operand's chain, then refresh the current endpoint of its right operand's chain, updating the endpoint after each refresh.
Evaluate the protected NAND gadget of \cref{sec:compiler-gadget} on those two fresh copies.
If both source operands are the same wire, the right refresh starts from the newly updated endpoint, giving two consecutive, distinct copies.
The gadget tensors its operands, refreshes in dimension $n^2+1$, erases that matrix, its inverse, and its sampling tape, and shortens back to length $n$.
Its output is the original encoding of the new source wire and starts that wire's chain.
\item Erase each operation's temporary workspace when it finishes.
Erase a wire encoding once its consumer, its next refresh, and any designated output use have all finished.
Unrelated retained encodings are inaccessible to the active operation's leakage.
\item After all leakage answers, the trusted decoder reads the designated original source-wire encodings and releases their inner products, namely $\vecy=\Circuit(\vecx,\vecs)$.
It releases no shares or other information.
\end{enumerate}
\end{tcolorbox}

This schedule gives every original input port a preceding refresh, and a gate that uses the same source wire twice receives two distinct refreshed copies.
Each copy $w_j$ belongs to at most three operation scopes: its producing refresh, its sole source-gate consumer, and the next refresh.
An original gate output belongs only to its producer and its first refresh, and an original public input belongs only to its sampler and its first refresh.
Thus every scope contains at most three distinct wires, and every wire belongs to at most three scopes.
Decoding is outside the leakage experiment.
The number of input-port refreshes is exactly $\sum_w\deg(w)=2t_0$, so the operation counts are
\begin{align}
 \text{NAND computation gadgets}&=t_0,\label{eq:direct-computation-count}\\*
 \text{input-port refreshes}&=2t_0,\label{eq:direct-refresh-count}\\*
 \text{public-input samplers}&=\Ipub,\label{eq:direct-public-count}\\*
 S&=3t_0+\Ipub.\label{eq:direct-size}
\end{align}
\Cref{eq:direct-size} counts all internal operations, including the public constants supplied after preprocessing.

\para{Leakage model.}
All communication is private, and each query reads only its party's data for the active operation, with the erasures above.
Each computation gadget has the matrix and shortening phases of \cref{sec:compiler-gadget}, with budgets $\lambda_{\MatrixPhase}$ and $\lambda_{\ShortPhase}$ per party and $\lambda=\lambda_{\MatrixPhase}+\lambda_{\ShortPhase}$.
Each input-port refresh and public-input sampler has one phase with at most $\lambda$ bits per party.
The functions in a phase are fixed before either answer is returned, and they may depend on all previous answers, including those of the earlier phase of the same gadget.
All rejected proposals and shortening retries belong to their phase's local tape and budget, and the adversary observes no separate timing or retry count.
Randomized leakage obeys the erasure-respecting convention of \cref{sec:compiler-phase-randomness}.
Consequently the complete answer of each operation has at most $K:=2^{2\lambda}$ possible values.

An \emph{auxiliary input} is a fixed string supplied to the adversary before the execution. It is the same in the compared executions and is also supplied to the simulator.
The randomness used for initialization and the fresh protocol randomness are independent of this string.
We fix this auxiliary input and the adversary's independent random tape. All bounds hold uniformly for these choices and can afterward be averaged over them.
The \emph{authorized output} of an evaluation is $\vecy:=\Circuit(\vecx,\vecs)$, the output that the evaluation is meant to reveal.
An adversary that receives $\vecy$ before its queries can include it in its auxiliary input.
The comparison fixes that same auxiliary string in the real and dummy executions. Uniformity over auxiliary strings covers this choice.

\begin{theorem}[Security of the compiler]\label{thm:compiler-direct}
The compiler correctly evaluates every NAND source circuit specified above.
Moreover, there exist absolute constants $C\ge1$ and $C'\ge1$ with the following property.
For every source circuit, dimension $n\ge2$, pair of nonnegative integer phase budgets, public input $\vecx$, and admissible leakage strategy, there exists a single simulator, independent of the private input and of any target error, that is given $\vecx$, the authorized output, and the adversary's auxiliary input.
For every private input $\vecs$, let $P_{\vecx,\vecs}$ be the distribution of the real view and let $\widehat P_{\vecx,\vecy}$ be the distribution of the simulator's view for $\vecy=\Circuit(\vecx,\vecs)$. Each view consists of the complete leakage transcript and the output $\vecy$, and contains no output shares.
Then
\begin{align}
 \StatDist{P_{\vecx,\vecs}}{\widehat P_{\vecx,\vecy}}
 &\le \min\left\{1,\ C\cdot S\cdot K^9\cdot(2^n-1)^{-1/27}\right\}\label{eq:direct-simulation}\\
 &\le C'\cdot S\cdot2^{18\lambda-n/27}.\label{eq:direct-simulation-binary}
\end{align}
The simulator has polynomial expected primitive work in the circuit and input sizes and $n$.
It makes $8t_0+2\Ipub$ local-answer evaluations, with adversary and leakage computation charged separately (see \cref{sec:direct-simulator} for both notions).

For every fixed $\alpha<1/486$, choosing $\lambda\le\alpha\cdot n$ gives error $S\cdot2^{-\Omega(n)}$.
Thus for polynomial-size circuit families the share length can be chosen as a function of the security parameter alone, independently of the circuit size.
\end{theorem}

\begin{proof}
\emph{Correctness.}
Each refresh preserves its bit, and the computation gadget outputs an encoding of its gate value by \cref{lem:compiler-shortening-freshness} in the \cref{sec:matrix-estimates}.

\emph{The canonical experiment.}
Exact freshness of a refresh (see \cref{lem:fresh}) and of a computation gadget implies that, conditional on the preceding state before its fresh randomness is drawn, the output of the operation has distribution $\mu_s$, where $s$ is its prescribed bit.
Public-input samplers also have this property.
The \emph{canonical experiment} therefore samples all wire encodings independently and then samples the leakage answers sequentially from their conditional probability functions.
The argument of \cref{lem:compiler-canonical} in the \cref{sec:matrix-estimates} applies to the mixture of these three operation types, including retained old encodings in refresh chains.
The independence concerns the encodings before any answer is observed, not their distribution after conditioning on leakage.

\emph{Comparing private inputs.}
For every formal history and answer, the computation gadget has the common approximation of \cref{lem:compiler-local-interface}, and the dimension-$n$ refresh has the corresponding approximation of \cref{lem:copy-local} at $\beta=q^{-1/3}$, with $q=2^n-1$.
Their valid-label $L_2$ errors are at most $C_{\mathrm{err}}\cdot\beta^{1/3}$ for an absolute constant $C_{\mathrm{err}}$ (see \cref{eq:compiler-local-error,eq:copy-local-error}), and their separate mean factor energies are at most four (see \cref{eq:compiler-local-left-mean,eq:compiler-local-right-mean,eq:copy-local-left-mean,eq:copy-local-right-mean}).
A public-input sampler retains its true normalized conditional answer functions throughout the comparison, and its encoded bit never changes.
The scope and incidence bounds established above allow \cref{thm:direct-composition} to compare the leakage of any two private-input executions with error $C\cdot S\cdot K^9\cdot q^{-1/27}$ (see \cref{eq:direct-composition-bound}).

\emph{The simulator.}
The simulator encodes the all-zero private input and executes all internal operations of the actual compiled protocol against the supplied adversary.
It stops before the output decoder, discards the dummy output shares, and returns the supplied $\vecy$ as the final authorized output.
Apply the comparison with that dummy input, fixing the real $\vecy$ as auxiliary data if the strategy uses it.
Appending the same $\vecy$ to both leakage distributions preserves their distance.
The argument also covers private inputs designated directly as outputs and repeated output positions, and unused private encodings have no effect on leakage.
If $S=0$, the transcript is empty and simulation is exact.
This proves~\cref{eq:direct-simulation}.

\emph{Efficiency and parameters.}
The concrete samplers and their polynomial expected-work argument are given in \cref{sec:direct-simulator}.
Each of the $t_0$ computation gadgets requires four local-answer evaluations, each of the $2t_0$ refreshes requires two, and each public-input sampler requires two.
If adversary selection and leakage evaluation cost at most $\Rev$ per call, expected total work is at most a polynomial in the stated parameters plus $(8t_0+2\Ipub)\cdot\Rev$.
The geometric-tail truncation described there gives a simulator with polynomially bounded primitive work and any prescribed additional negligible statistical error.
Efficiency of total simulation requires the corresponding bound on the adversary's and leakage functions' computation, whereas the statistical comparison imposes no computational restriction.
Finally, $q\ge2^{n-1}$ gives~\cref{eq:direct-simulation-binary}, and $18\alpha<1/27$ gives the stated leakage fraction.
\end{proof}

The guarantee is for one stateless evaluation with a trusted output boundary. \Cref{thm:compiler-stateful} in \cref{sec:stateful} extends it to repeated invocations of a stateful circuit with persistent private registers and inputs chosen adaptively by the adversary.
The view also excludes the final encodings, because their labels can differ between a real and a dummy execution: a dummy output pair may encode a bit different from the corresponding entry of $\vecy$, and revealing its shares would expose this difference at once.
Because these pairs stay hidden, the composition proof can treat output wires like other internal wires, whose labels may change between the compared executions (see \cref{sec:direct-composition-setup}).
The signed intermediate measures used in \cref{sec:direct-composition} are solely proof devices.

\subsection{The protected NAND gadget}\label{sec:direct-gadget}\label{sec:compiler-gadget}
Having described how the compiler connects its operations, we now turn to the gadget that implements a source gate: a two-party protocol that computes NAND on encodings.

\subsubsection{Local views and phase order}\label{sec:direct-phase-model}
We start by fixing the leakage model for one NAND computation gadget.
Recall from \cref{def:main-wires} that the scope of an operation is the set of wire encodings it reads or produces. The two refreshed operand encodings of a computation gadget and its new output are distinct.
All protocol communication is private.
Leakage is \emph{gate-local}: a query can access only the active gadget's local data, comprising its party's encodings in $X_{S_i}$, fresh local randomness, and private messages.
This is the only-computation-leaks granularity of the leakage model in the introduction, with each phase as one leakage round.
The encodings of unrelated \emph{live wires}, those still needed by later gates, are inaccessible.
The restriction applies separately to each computation gadget, input-port refresh, and public-input sampler.
After a gadget finishes, its workspace is erased, and only the encodings still needed by the circuit remain.
After the last leakage query, only the trusted decoder reads the designated output encodings and releases their bits in the clear, so the leakage transcript contains no output shares.

Each gadget has two phases.
Its \emph{matrix phase} tensors the operand shares, appends a final coordinate equal to one, and applies the refresh in dimension $n^2+1$.
The matrix is then erased, and the \emph{shortening phase} reduces the refreshed shares to length $n$ using a random full-row-rank matrix.
Fix nonnegative integer budgets $\lambda_{\MatrixPhase}$ and $\lambda_{\ShortPhase}$, and set
\[
 \lambda:=\lambda_{\MatrixPhase}+\lambda_{\ShortPhase},\qquad
 K:=2^{2\lambda}.
\]
Each party leaks at most $\lambda_{\MatrixPhase}$ bits in the matrix phase and at most $\lambda_{\ShortPhase}$ bits in shortening.
The functions may depend on all earlier answers, including the matrix-phase answer when shortening begins, so the adversary may adapt across phases and gates (see \cref{sec:within-round} for an extension beyond this model).
Within each phase, both functions are fixed before either answer is revealed.
All shortening attempts belong to that single phase: each function may read its party's entire local sampling tape and private messages from the attempts, within the original phase budget.
The right party privately reports whether another attempt is needed. No leakage answer is released between attempts, and the adversary observes no separate retry count, timing, or protocol messages.
\Cref{fig:gate} illustrates the two local views and their leakage to the adversary.

The \emph{answer} $Z_i$ of a gate concatenates the two parties' outputs from both phases.
Padding gives a fixed alphabet $\cZ$ of size at most $K$.
As for refreshing, $H_i=(Z_1,\ldots,Z_i)$ is the history after gate $i$, and $H_0=\varnothing$.
The incoming history and the adversary's tape determine the matrix-phase functions. The shortening functions may additionally depend on that gate's matrix-phase answer.
We use the concatenation notation $h\|z$ and the convention of fixing admissible functions at all formal histories (see \cref{def:history}), so that every quantity below is defined for every formal history.

For a fixed matrix and matrix-phase answer, the operand pairs producing that answer form a \emph{rectangle}: each party's half of the answer depends only on that party's own data, so the indicator of the answer is the product of a Boolean \emph{mask} for the left party and one for the right party.
These masks are defined exactly as in \cref{sec:leakage}, with $\ell$ and $r$ replaced by $(\ell_1,\ell_2)$ and $(r_1,r_2)$, respectively, and with a matrix of dimension $n^2+1$.
The mask partition is taken separately for the matrix phase. The shortening answer is handled after the matrix has been erased.

\para{Randomness of the leakage functions.}\phantomsection\label{sec:compiler-phase-randomness}
Leakage functions are allowed to be randomized, and as in \cref{sec:removing-randomness} their randomness is removed by presampling it and averaging the resulting deterministic bounds.
Splitting a gadget into two phases adds one condition, because the refresh matrix is erased between them.
The matrix-phase functions may use correlated auxiliary randomness whose distribution depends on the gate index, incoming history, and matrix, but is conditionally independent of the operand shares.
Presampling these values gives the deterministic reduction of \cref{sec:removing-randomness}.
In particular, the matrix sampler of \cref{sec:direct-simulator} draws uniform matrices until one is invertible, and its rejected proposals are independent of both the accepted matrix and the operand shares.
These rejected proposals can therefore be presampled as part of the correlated matrix-phase auxiliary randomness, fixed when defining the Boolean masks, and averaged afterward.
Thus the analysis also covers functions of the complete matrix-sampling tape, without additional protocol messages.
The entire tape is erased before shortening.
In shortening, auxiliary randomness must respect erasure: conditional on the gate index and all preceding answers, its joint distribution is independent of the shares and the erased matrix.
This randomness may therefore depend on the matrix only through those answers.
Presample one auxiliary pair for each gate index, incoming history, and matrix-phase answer, independently across these choices and of the initial encodings and matrices.
Fixing these pairs gives deterministic phase functions. Averaging restores their original distribution.

\subsubsection{Tensor, refresh, and shorten}
\begin{figure}[htbp]
\centering
\begingroup
\setlength{\fboxrule}{0.3pt}
\setlength{\fboxsep}{8pt}
\fcolorbox{black!25}{white}{\begin{minipage}{\dimexpr\linewidth-2\fboxsep-2\fboxrule\relax}
\centering
\begin{tikzpicture}[x=1cm,y=1cm,font=\footnotesize,
 party/.style={draw=black!35,fill=black!2,rounded corners=2pt,
   align=center,text width=4.2cm,minimum height=1.7cm,inner sep=5pt},
 answer/.style={draw=black!35,rounded corners=2pt,align=center,
   text width=2.4cm,minimum height=1.05cm,inner sep=4pt},
 flow/.style={->,draw=black!65,semithick}]
 \node at (-4.7,0.85) {\textbf{Left party: active gadget}};
 \node at (4.7,0.85) {\textbf{Right party: active gadget}};
 \node at (0,0.85) {\textbf{Adversary}};
 \node[party] (lm) at (-4.7,-0.4)
   {\textbf{Matrix phase}\\[3pt]Local data: $L_1$, $L_2$, and $M$\\$U:=a\cdot M$};
 \node[party] (rm) at (4.7,-0.4)
   {\textbf{Matrix phase}\\[3pt]Local data: $R_1$, $R_2$, and $M$\\$V:=M^{-1}\cdot b$};
 \node[answer] (am) at (0,-0.4) {Matrix-phase\\answer pair};
 \draw[flow] (lm.east)--(am.west) node[midway,above] {$\lambda_{\MatrixPhase}$};
 \draw[flow] (rm.west)--(am.east) node[midway,above] {$\lambda_{\MatrixPhase}$};
 \node[party] (ls) at (-4.7,-3.7)
   {\textbf{Shortening phase}\\[3pt]Sample $A$ and $H$ with $A\cdot H=U$.\\Privately send $H$.};
 \node[party] (rs) at (4.7,-3.7)
   {\textbf{Shortening phase}\\[3pt]Compute $R:=H\cdot V$.\\Privately request a retry if $R=0$.};
 \node[answer] (as) at (0,-3.7) {Shortening\\answer pair};
 \draw[flow] (ls.east)--(as.west) node[midway,above] {$\lambda_{\ShortPhase}$};
 \draw[flow] (rs.west)--(as.east) node[midway,above] {$\lambda_{\ShortPhase}$};
 \draw[flow] (lm.south)--(ls.north);
 \draw[flow] (rm.south)--(rs.north);
 \node[fill=white,align=center,text width=4.3cm,inner sep=4pt] at (-4.7,-2.05)
   {Erase $M$, $M^{-1}$ and sampling tape.\\Retain operands and $U$.};
 \node[fill=white,align=center,text width=4.3cm,inner sep=4pt] at (4.7,-2.05)
   {Erase $M$, $M^{-1}$ and sampling tape.\\Retain operands and $V$.};
 \draw[flow] (am.south)--(as.north)
   node[midway,fill=white,align=center,text width=2.6cm,inner sep=3pt]
   {Choose new\\functions using\\earlier answers.};
 \node (lo) at (-4.7,-5.15) {Output share $L=A$: $n$ bits};
 \node (ro) at (4.7,-5.15) {Output share $R$: $n$ bits};
 \draw[flow] (ls.south)--(lo.north);
 \draw[flow] (rs.south)--(ro.north);
\end{tikzpicture}
\end{minipage}}
\endgroup
\caption{Local leakage in one circuit gadget.
The matrix is erased between the two phases, while the adversary retains its answers.
All retries belong to the shortening phase.}
\label{fig:gate}
\end{figure}
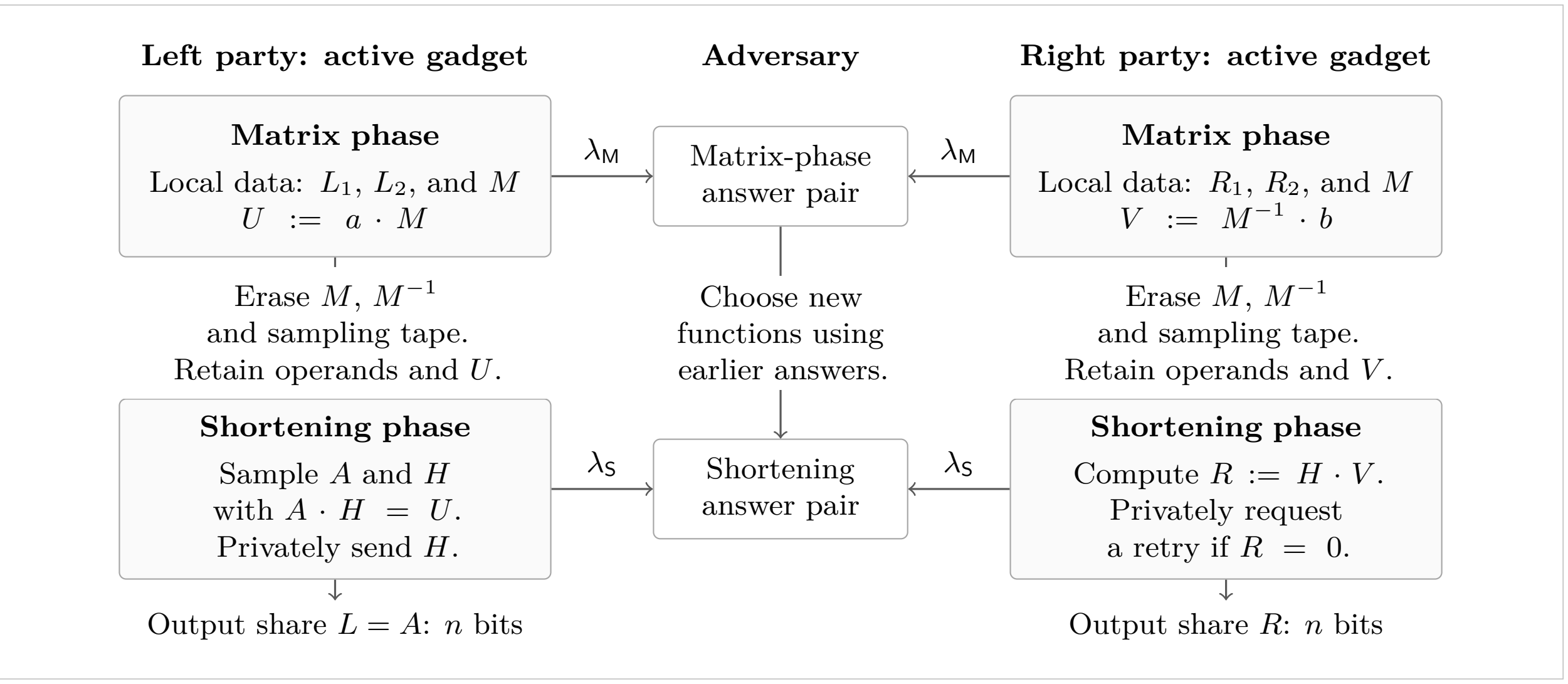

Every source NAND gate is implemented by the following three-step gadget, which uses the refresh of \cref{sec:refreshing}. First, tensoring the operands makes the gate output available as a single inner product. The refresh then randomizes the resulting long encoding, and a random projection brings it back to the persistent length $n$.
\Cref{fig:gate} summarizes the gadget and its two leakage phases.
Let $(L_1,R_1)$ and $(L_2,R_2)$ encode the two operands.
Each $L_j$ is a row in $\F_2^{1\times n}$, and each $R_j$ is a column in $\F_2^{n\times1}$.
Even when the source gate repeats an operand, its two input ports receive distinct refreshed encoding pairs.
The tensor product $\otimes$ of two shares lists all pairwise products of their coordinates.
Set $m:=n^2+1$ and form the nonzero long shares
\[
 a:=(L_1\otimes L_2,1),\qquad
 b:=\begin{pmatrix}R_1\otimes R_2\\1\end{pmatrix}.
\]
The tensor $L_1\otimes L_2$ is a row, so $a$ is a row obtained by appending one coordinate.
The displayed $b$ is a column formed by vertical concatenation.
These long shares already encode the desired NAND output:
\[
 a\cdot b=1+(L_1\cdot R_1)\cdot(L_2\cdot R_2).
\]
The equality follows because the inner product of the two tensor products is $(L_1\cdot R_1)\cdot(L_2\cdot R_2)$. The two appended coordinates contribute one.
Every later operation preserves this bit.
The matrix phase applies \cref{def:protocol} in dimension $m$:
\[
 U:=a\cdot M,\qquad V:=M^{-1}\cdot b,
 \qquad M\sample \InvMatr_2^m.
\]
The left leakage function may read $L_1$, $L_2$, and $M$, and the right function may read $R_1$, $R_2$, and $M$. These data determine the respective long shares.
Before shortening, each party retains only its old operand shares and its refreshed long share from the matrix-phase workspace.
Each party erases the rest of that workspace, including $M$, its inverse, and its sampling tape.
The adversary retains the answers already obtained.

\para{Shortening by a random projection.}
The refreshed shares have length $m$, while the wires of the circuit carry shares of length $n$, so the gadget must compress them without disturbing the bit they encode.
The gadget does so by projecting onto $n$ random independent directions.
Given the nonzero long shares $U$ and $V$ of length $m\ge n$, perform the following attempt.
\begin{enumerate}
\item The left party samples a nonzero row $A\sample \UL$ and then samples
\[
 H\sample \{H'\in\F_2^{n\times m}:\rank{H'}=n,\ A\cdot H'=U\}.
\]
It privately sends $H$.
\item The right party computes $R:=H\cdot V$ and privately reports whether $R=0$.
If $R=0$, repeat with fresh $A$ and $H$ and the same $U$ and $V$.
Otherwise, output
\begin{equation*}
 L:=A,\qquad R:=H\cdot V.
\end{equation*}
\end{enumerate}
Correctness follows from one identity:
\[
 L\cdot R=A\cdot H\cdot V=U\cdot V
 =1+(L_1\cdot R_1)\cdot(L_2\cdot R_2).
\]
The rows of $H$ are independent and their span contains $U$.
The new left share gives the coefficients expressing $U$ in these rows. The new right share lists their inner products with $V$.

For every fixed pair of shares $U$ and $V$, the final output is exactly a fresh encoding of $U\cdot V$ (see \cref{lem:compiler-shortening-freshness} in the \cref{sec:matrix-estimates}).
Here ``fixed'' means that we may condition on any particular pair produced by the preceding refresh. The fresh shortening randomness still gives the same output distribution.
If $U\cdot V=1$, the first attempt always succeeds.
If $U\cdot V=0$, its failure probability is
\[
 \frac{2^{m-n}-1}{2^{m-1}-1}\le 2^{1-n}.
\]
For $n\ge2$, at most two attempts are needed in expectation.
The protocol never explicitly evaluates $U\cdot V$, but a failed attempt implies $U\cdot V=A\cdot(H\cdot V)=0$, so the private failure acknowledgment lets both parties infer that the encoded bit is zero.
The retry-error bound below accounts for this possibility.
Exact freshness explains why shortening produces a suitable encoding for the next operation. It does not by itself control leakage from the shortening transcript.
In particular, the message $H$ reveals a subspace containing the long left share.
Our proof first refreshes the long encoding built from $L_1\otimes L_2$, erases the refresh matrix, and then analyzes the shortening transcript together with the retained operands.
The refresh gives this analysis its starting point: for fixed operands and before any answer is observed, the refreshed pair $(U,V)$ is uniform among all nonzero long encodings of the output bit, whereas the unrefreshed long shares would be determined by the operands.
This uniformity allows us to view a successful first shortening attempt in reverse, starting from its short output and sampling the long shares that could have produced it (see \cref{eq:compiler-inverse-shortening} in the \cref{sec:matrix-estimates}).

Each shortening leakage function may read its own old operands, its long share, the full shortening sampling tape and private transcript held by that party, and its final output share.
All retries remain in the same phase with the same two leakage functions and total budget. The adversary observes no additional answers or timing information.
The erased matrix is unavailable, so the proof does not cover a leakage function that reads both $M$ and the later projection matrix $H$.

\para{Relation to refreshing and cost.}
For $m=n$, every full-row-rank $H$ is invertible, and the same sampling rule is equivalent to choosing $H$ uniformly in $\InvMatr_2^n$ and setting
\[
 A:=U\cdot H^{-1},\qquad R:=H\cdot V.
\]
This is the matrix refresh of \cref{def:protocol} with $M=H^{-1}$. It never retries and communicates $n^2$ bits.
The storage protocol of \cref{sec:refreshing} uses this square case directly, with no preceding refresh or additional shortening stage.
A rectangular shortening attempt communicates $n\cdot m$ matrix bits and one private success bit. Thus the complete shortening procedure communicates at most $2\cdot(n\cdot m+1)$ bits in expectation.
The NAND gadget has $m=n^2+1$ and communicates an expected $O\left(n^4\right)$ bits in total, dominated by its full $m\times m$ refresh matrix.
All sampling and local operations run in expected polynomial time (see \cref{sec:direct-simulator}).

\subsection{Extending the local approximation to a computation gadget}\label{sec:compiler-local}
We now extend the local approximation of \cref{sec:joint-refresh} from a refresh to a computation gadget, which will allow us to compose the operations in \cref{sec:direct-composition}.
Recall that a \emph{table} is a real function of a left argument and a right argument (see \cref{def:table-norms}), and its \emph{nuclear norm} is the least total size $\sum_j\LtwoNorm{a_j}\cdot\LtwoNorm{b_j}$ of a representation $\sum_j a_j\cdot b_j^{\transpose}$ as a sum of products (see \cref{eq:nuclear-factor-cost} in the \cref{sec:preliminaries-proofs}).
For a chosen representation, the left and right \emph{energies} $\Phi$ and $\Psi$ are the sums of the squares of the factors (see \cref{def:factor-energy}).
A \emph{formal incoming history} is any sequence of earlier answers, whether or not it can occur (see \cref{sec:leakage}).
The \emph{leakage probability function} $\prob$ of an answer is its probability given the encodings involved, and a round or gate is \emph{signed} when $\prob$ is replaced by its signed approximation $G$ (see \cref{def:leakage-probability,def:signed-round}).

\subsubsection{Retaining the operand shares}\label{sec:joint-general}\label{sec:joint-setup}
The refreshing proof retains the same shares to which the random matrix is applied.
A computation gadget instead retains its two operand shares while refreshing their tensor encoding.
The table must keep those operands because the leakage functions may read them and because the gadget shares them with the input-port refreshes that produce them, so the composition proof compares executions in which their labels may differ.
Since the tensor map is injective, an operand pair carries the same information as its old long share, which suggests applying \cref{lem:refresh-rectangle} in the long dimension.
That lemma, however, bounds an error averaged over all nonzero old shares, whereas the tensors of operand shares occupy only a thin subset of the long-share space. A small average over the whole space need not be small on that subset.
We therefore extend the approximation lemma of \cref{lem:refresh-rectangle} to distinguish the \emph{old local data}, over which the estimates are averaged, from the \emph{share to be refreshed}, which is a function of those data.
The old data no longer range over the full space of shares: they range over an arbitrary finite set that is mapped injectively into that space, which is what makes the lemma apply to tensors.
The analytic argument remains the one from \cref{app:direct-approximation-proof}, because the Gram and four-cycle estimates require only the corresponding counts for pairs of old inputs, and a uniform invertible matrix sends any two distinct nonzero vectors to a uniform pair of distinct nonzero vectors, whether or not they are tensors.
Beyond injectivity, the tensor structure therefore enters the estimates only through the number of old inputs, which determines the admissible thresholds.

As in \cref{sec:shares}, for a share length $n'\ge2$ let $\UL^{(n')}$ and $\UR^{(n')}$ be the nonzero row and column vectors of length $n'$ over $\F_2$, and let $\mu_s^{(n')}$ be the uniform distribution on pairs with inner product $s$.
At $n'=n$ we omit the superscript.
Independent uniform nonzero shares have inner product $s$ with probability
\[
 p_s^{(n')}
 =\begin{cases}
 2^{n'-1}/\left(2^{n'}-1\right),&s=1,\\
 \left(2^{n'-1}-1\right)/\left(2^{n'}-1\right),&s=0.
 \end{cases}
\]
In particular, $p_s^{(n)}=p_s$, and $p_s^{(n')}\ge1/3$.

Let $\OldData$ be a finite nonempty set indexing the possible old local data of either party, equipped with the uniform distribution.
In our applications, we identify a row and its transposed column with the same bit string, so both parties use the same set $\OldData$. A party's role determines the orientation of its shares.
Fix $m\ge2$ and injective maps
\[
 a:\OldData\longrightarrow\UL^{(m)},\qquad
 b:\OldData\longrightarrow\UR^{(m)}.
\]
Thus every $a(\OldL)$ is a nonzero row and every $b(\OldR)$ is a nonzero column. The maps need not be linear.
Here $\OldL\in\OldData$ and $\OldR\in\OldData$ are the old local data of the left and right parties, respectively, while $a(\OldL)$ and $b(\OldR)$ are the shares of length $m$ to which the matrix refresh is applied:
\begin{itemize}
\item \emph{Refreshing:} $\OldL=L$, $\OldR=R$, $a(\OldL)=L$, $b(\OldR)=R$, and $m=n$.
\item \emph{NAND computation:} $\OldL=(L_1,L_2)$, $\OldR=(R_1,R_2)$, $a(\OldL)=(L_1\otimes L_2,1)$, $b(\OldR)=\binom{R_1\otimes R_2}{1}$, and $m=n^2+1$.
\end{itemize}
Both choices are injective: this is immediate for refreshing, and \cref{sec:joint-tensor-instance} explains it for the tensor maps.
Injectivity ensures that distinct old inputs give distinct nonzero vectors, which over $\F_2$ are linearly independent in pairs. The extension of the matrix estimates in the \cref{app:general-approximation-proof} uses this fact.
Put
\[
 d:=|\OldData|,\qquad N_m:=2^m,\qquad q_m:=N_m-1,\qquad
 \TableRows:=\OldData\times\UL^{(m)},\qquad
 \TableCols:=\OldData\times\UR^{(m)}.
\]
Injectivity gives $1\le d\le q_m$.
Let $\JointRef$ be the uniform reference distribution on $\TableRows\times\TableCols$.
Under this distribution, $\OldL$, $\OldR$, $U$, and $V$ are independent and uniform in their respective spaces. The refresh equations are not imposed.
All unqualified norms of tables on $\TableRows\times\TableCols$ use this distribution.
In the actual refresh experiment, sample the coordinates as follows.
Let $\OldL$ and $\OldR$ be independent uniform elements of $\OldData$, choose an independent matrix $M\sample\InvMatr_2^m$, and set
\[
 U:=a(\OldL)\cdot M,\qquad V:=M^{-1}\cdot b(\OldR).
\]
Write $x:=(\OldL,U)$, $y:=(\OldR,V)$, and $\omega:=(x,y)=((\OldL,U),(\OldR,V))$ for the grouped coordinates, which keep both the old local data and the refreshed shares, as needed when composing gadgets.
We also write $\MaxNorm{F}:=\max_{x,y}|F(x,y)|$ for the largest absolute entry of a table.
In averages indexed by $\OldL$, $\OldR$, $U$, and $V$, each named variable is sampled independently and uniformly from its indicated space, and $M$ is uniform in $\InvMatr_2^m$.
For any table $F:\TableRows\times\TableCols\to\R$ and fixed invertible matrix $M$, let $F^M:\OldData\times\OldData\to\R$ denote the table obtained by substituting the refresh equations:
\[
 F^M(\OldL,\OldR):=F\left((\OldL,a(\OldL)\cdot M),(\OldR,M^{-1}\cdot b(\OldR))\right).
\]
The inner product $a(\OldL)\cdot b(\OldR)\in\F_2$ is the single bit encoded by the shares being refreshed.
For refreshing, it equals $L\cdot R$. For a NAND gate, it equals $1+(L_1\cdot R_1)\cdot(L_2\cdot R_2)$, its prescribed output bit.

\begin{definition}[Consistent retained data and consistency projection]\label{def:consistent-retained}
Call $\omega\in\TableRows\times\TableCols$ \emph{consistent} when $U\cdot V=a(\OldL)\cdot b(\OldR)$.
For every real table $F:\TableRows\times\TableCols\to\R$, its \emph{consistency projection} removes the entries at inconsistent tuples:
\[
 (\project{F})\left((\OldL,U),(\OldR,V)\right):=
 \begin{cases}
  F\left((\OldL,U),(\OldR,V)\right),&U\cdot V=a(\OldL)\cdot b(\OldR),\\
  0,&U\cdot V\ne a(\OldL)\cdot b(\OldR).
 \end{cases}
\]
\end{definition}
This projection has the same properties as in~\cref{sec:refresh-joint-setup,app:direct-approximation-proof}: applying it twice changes nothing, and $(\project{F})^M=F^M$ for every invertible $M$.
The refresh experiment, that is, the distribution of $\omega$ when $\OldL$, $\OldR$, and $M$ are sampled as above, has density
\[
 \varphi\left((\OldL,U),(\OldR,V)\right)=
 \begin{cases}
  1/p_{a(\OldL)\cdot b(\OldR)}^{(m)},&U\cdot V=a(\OldL)\cdot b(\OldR),\\
  0,&U\cdot V\ne a(\OldL)\cdot b(\OldR)
 \end{cases}
\]
relative to $\JointRef$: its probability at each tuple is $\varphi$ times the reference probability.
Indeed, for each fixed $\OldL$ and $\OldR$, invertible matrices act transitively on all nonzero encodings of $a(\OldL)\cdot b(\OldR)$, so by the argument of \cref{lem:fresh} in dimension $m$ the refreshed pair $(U,V)$ is uniform on them.
Since $p_s^{(m)}\ge1/3$, this density is at most three.

For later use, define the sign table
\[
 J(x,y):=\begin{cases}1/2,&(x,y)\text{ is consistent},\\-1/2,&\text{otherwise}.\end{cases}
\]
Then $\project{F}=F/2+J\cdot F$, where $J\cdot F$ is entrywise multiplication.

\subsubsection{The extended approximation lemma}\label{sec:joint-masks}
We now apply the singular-coefficient truncation of \cref{eq:refresh-explicit}, retaining the old local data as part of each grouped coordinate.
The resulting table must remain independent of the plaintext labels, so that later comparisons can change the encoding distributions without changing the approximation.
Fix any formal incoming history and the adversary's random tape.
Let $\MaskL{z}(\OldL,M)$ and $\MaskR{z}(\OldR,M)$ be the Boolean masks for one answer $z$ (see \cref{sec:leakage}).
On consistent tuples, let $\prob_z\left((\OldL,U),(\OldR,V)\right)$ be their product averaged over the uniform matrices satisfying $a(\OldL)\cdot M=U$ and $M^{-1}\cdot b(\OldR)=V$. Extend $\prob_z$ by zero elsewhere.
Thus $\project{\prob_z}=\prob_z$ and $0\le \prob_z\le1$.
For any real table $W$, conditioning on the old and refreshed shares and then averaging gives the \emph{answer-mask identity}
\begin{equation}\label{eq:general-kernel-pairing}
 \pairing{\prob_z}{W}
 =\E_{\OldL,\OldR,M}{p_{a(\OldL)\cdot b(\OldR)}^{(m)}\cdot
 \MaskL{z}(\OldL,M)\cdot \MaskR{z}(\OldR,M)\cdot W^M(\OldL,\OldR)}.
\end{equation}
Here $\OldL$ and $\OldR$ are independent uniform elements of $\OldData$, and $M\sample\InvMatr_2^m$ is independent of them.
Indeed, the density of the refresh tuples is $1/p_{a(\OldL)\cdot b(\OldR)}^{(m)}$ on their support, so the factor $p^{(m)}$ changes the refresh average back to the reference average.
We also record that
\begin{align}
 p_0^{(m)}&=\frac{q_m-1}{2q_m},\nonumber\\
 p_1^{(m)}&=\frac{q_m+1}{2q_m},\nonumber\\
 |2p_s^{(m)}-1|&=1/q_m\qquad\text{for every }s\in\F_2.\label{eq:general-encoding-gap}
\end{align}

\begin{lemma}[Approximation of an answer with retained local data]\label{lem:joint-map}
There exists an absolute constant $C>0$ such that the following holds for every formal incoming history, adversary tape, answer $z$, and threshold $\beta\in[d^{-1/3},1]$.
Take a probability-normalized orthonormal expansion (see \cref{lem:orthonormal-expansion})
$\prob_z=\sum_j\sigma_j\cdot u_j\cdot v_j^{\transpose}$ and put
\begin{equation}\label{eq:joint-direct-construction}
 I:=\{j:\sigma_j>\beta\},\qquad
 \prob_z^{>\beta}:=\sum_{j\in I}\sigma_j\cdot u_j\cdot v_j^{\transpose}.
\end{equation}
Then
\begin{align}
 \NuclearNorm{\prob_z^{>\beta}}&\le2,\label{eq:joint-map-budget}\\
 \LtwoNorm{\prob_z^{>\beta}}&\le1.\label{eq:joint-map-L2}
\end{align}
The balanced factors $\sqrt{\sigma_j}\cdot u_j$ and $\sqrt{\sigma_j}\cdot v_j$ of $\prob_z^{>\beta}$ (see \cref{def:balanced-factors}) have energies $\Phi_\beta$ and $\Psi_\beta$ satisfying, for every grouped row $x\in\TableRows$ and column $y\in\TableCols$,
\begin{align}
 \Phi_\beta(x)&\le1/\beta,\label{eq:joint-direct-energies}\\
 \Psi_\beta(y)&\le1/\beta,\label{eq:joint-direct-right-energy}\\
 \E_{x\sample \TableRows}{\Phi_\beta(x)}&=\NuclearNorm{\prob_z^{>\beta}}\le2,\label{eq:joint-direct-left-mean}\\
 \E_{y\sample \TableCols}{\Psi_\beta(y)}&=\NuclearNorm{\prob_z^{>\beta}}\le2.\label{eq:joint-direct-right-mean}
\end{align}
Finally,
\begin{equation}\label{eq:joint-map-error}
 \LtwoNorm{\prob_z-2\cdot\project{\prob_z^{>\beta}}}\le C\cdot\beta^{1/3}.
\end{equation}
The constant $C$ is independent of the dimension, threshold, answer, formal history, and admissible masks.
The construction depends on the masks and public maps, not on any plaintext labels.
\end{lemma}

\Cref{app:general-approximation-proof} proves this extension.
By~\crefrange{eq:joint-direct-energies}{eq:joint-direct-right-mean}, the approximation $2\cdot\prob_z^{>\beta}$ has balanced energies $2\cdot\Phi_\beta$ and $2\cdot\Psi_\beta$, bounded pointwise by $2/\beta$ and in mean by~$4$.
The factor two compensates for the consistency condition, which keeps roughly half of the independent tuples.
The approximation $2\cdot\prob_z^{>\beta}$ is a global table and may be nonzero on inconsistent tuples. Only the residual $\prob_z-2\cdot\project{\prob_z^{>\beta}}$ is zero there.
On consistent tuples this residual equals $\prob_z-2\cdot\prob_z^{>\beta}$, so~\cref{eq:joint-map-error} also bounds its root-mean-square value under the actual refresh distribution by $\sqrt3\cdot C\cdot\beta^{1/3}$.
The same bound holds for its mean absolute value by the Cauchy--Schwarz inequality (see \cref{lem:la-cauchy-schwarz}).
The nuclear bound~\cref{eq:joint-map-budget} relies on the answer-mask identity~\cref{eq:general-kernel-pairing}, so it applies to leakage probability functions and not to arbitrary bounded consistent tables.
If no coefficient is kept, then $\prob_z^{>\beta}=0$, with all energies zero.

\begin{proof}[Proof outline]
The projection and factor-energy bounds are unchanged from the refreshing proof.
Three estimates must account for the number $d$ of old inputs, which may differ from the number $q_m$ of nonzero long shares.
Injectivity makes the images of two distinct old inputs a uniform pair of distinct nonzero vectors under the random matrix. This gives the Gram-variance bound of \cref{lem:refresh-Gram-variance}, which controls the nuclear norm.
Injectivity also allows the consistency sign to be treated as a submatrix of the full binary sign matrix.
Finally, in the four-cycle calculation, coinciding old inputs have probability $1/d$.
With these changes, the proof of \cref{lem:refresh-rectangle} in \cref{app:direct-approximation-proof} gives~\cref{eq:joint-map-budget} and, for $t:=\LtwoNorm{\prob_z-2\cdot\project{\prob_z^{>\beta}}}$,
\begin{equation}\label{eq:general-error-bootstrap}
 t^2\le C\cdot\beta^{1/2}\cdot t^{1/2}+C/(d\cdot\beta).
\end{equation}
Since $d\cdot\beta^3\ge1$, \cref{eq:general-error-bootstrap} yields $t\le C\cdot\beta^{1/3}$ exactly as for refreshing.
\Cref{app:general-approximation-proof} gives the changed estimates and the complete deductions.
\end{proof}

\subsubsection{Transporting the approximation through shortening}
The counterpart of \cref{lem:copy-local} for a computation gadget retains three encodings: the two operands and the output.
The approximation of \cref{lem:joint-map} covers the matrix phase in dimension $n^2+1$: it retains the operands and accounts for the thin tensor support through the size $d$ of the old data.
Its table still has the refreshed long shares among its coordinates, however, and two steps remain.
The table must instead depend on the short output encoding, because neighboring operations share that encoding, whereas the long shares are temporary and no other operation reads them.
The answer must also include the shortening phase, whose leakage functions may read the message $H$ and may depend on the matrix-phase answer.
We therefore carry the matrix-phase approximation through the shortening step, for each pair of matrix-phase and shortening answers, and so obtain an approximation of the complete local view of the gadget.

The compiled execution has an exact \emph{canonical} description: sample all wire encodings first, independently, and then sample gate answers sequentially from leakage probability functions conditioned on each gate's three encodings.
This unconditional description supplies the independent wire variables in the composition argument. Conditioning on observed answers can create dependencies and is not used to justify independence.
The description is exact because of the freshness of \cref{lem:compiler-shortening-freshness} in the \cref{sec:matrix-estimates}: even conditional on its operand encodings and all internal state preceding the gadget, a completed gadget has a uniformly random encoding of its output bit.
This conditioning does not include the fresh randomness drawn within the gadget.
The answer at that gadget may depend on its output encoding, and the conditional leakage probability function records this dependence.
\Cref{lem:compiler-canonical} in the \cref{sec:matrix-estimates} proves equality of the joint encoding-and-answer distributions.
Tables below group the three left shares into one coordinate and the three right shares into the other: write $\UL^{S_i}$ and $\UR^{S_i}$ for the sets of triples of nonzero left shares and of nonzero right shares indexed by the wires of $S_i$. Their norms use the corresponding product uniform measures.

\begin{definition}[Local labelings and their norms]\label{def:local-labeling}
Put $q:=2^n-1$ and fix a computation gadget $i$ with scope $S_i$.
A \emph{local labeling} $s$ assigns a bit to each of the three wires of $S_i$. It is \emph{valid} when these bits satisfy the equation of gate $i$.
For a local labeling $s$ and a table $F$ on $S_i$, that is, a real function of $X_{S_i}$, put
\[
 \LtwoNormAt{F}{s}:=\left(\E_{X_{S_i}\sample\bigotimes_{w\in S_i}\mu_{s(w)}}{|F(X_{S_i})|^2}\right)^{1/2}.
\]
\end{definition}
Unqualified norms still use independent uniform left and right shares.

\begin{lemma}[A gate with its inputs and output retained]\label{lem:compiler-local-interface}
There exists an exact canonical experiment with conditional leakage probability functions $\prob_{i,h,z}\left(X_{S_i}\right)$, defined on valid local labelings, with $\prob_{i,h,z}\ge0$ and $\sum_z \prob_{i,h,z}=1$.
Its joint distribution of generated wire encodings and answers equals that of the actual execution.
There exists an absolute constant $C>0$ such that, for every threshold $\beta\in[q^{-2/3},1]$, computation operation $i$, formal incoming history $h\in\cZ^{i-1}$, and answer $z\in\cZ$, there exists a real table $G_{i,h,z}^{\beta}$, common to all wire labelings, with a product representation whose left and right energies $\Phi$ and $\Psi$ (the sums of squared factors; see \cref{def:factor-energy}) satisfy, for all grouped shares $x\in\UL^{S_i}$ and $y\in\UR^{S_i}$,
\begin{align}
 \NuclearNorm{G_{i,h,z}^{\beta}}&\le4,\label{eq:compiler-local-interface}\\
 \Phi(x)&\le2/\beta,\label{eq:compiler-local-left-energy}\\
 \Psi(y)&\le2/\beta,\label{eq:compiler-local-right-energy}\\
 \E_{x\sample \UL^{S_i}}{\Phi(x)}&\le4,\label{eq:compiler-local-left-mean}\\
 \E_{y\sample \UR^{S_i}}{\Psi(y)}&\le4,\label{eq:compiler-local-right-mean}\\
 \LtwoNormAt{\prob_{i,h,z}-G_{i,h,z}^{\beta}}{s}
 &\le C\cdot\beta^{1/3}\quad\text{for every valid local labeling }s,\label{eq:compiler-local-error}\\
 \LtwoNormAt{G_{i,h,z}^{\beta}}{s}
 &\le C_0\quad\text{for every local labeling }s.\label{eq:compiler-local-rms}
\end{align}
Here $C_0:=4\cdot\sqrt{27}$, and $C$ is independent of the dimension, threshold, answer, formal history, and admissible leakage functions.
\end{lemma}
The energies refer to the chosen product representation, which shortening can leave unbalanced, so~\cref{eq:compiler-local-left-mean,eq:compiler-local-right-mean} state the two mean-energy bounds separately.
The same approximation families are fixed for all formal histories before any wire labels are changed.
Unlike true leakage probabilities, these tables may be negative and need not sum to one over the answers.

\begin{proof}[Proof outline]
The matrix-phase argument applies the local approximation of \cref{lem:joint-map} to the injective maps
$(L_1,L_2)\mapsto(L_1\otimes L_2,1)$ and $(R_1,R_2)\mapsto(R_1\otimes R_2,1)$.
Over $\F_2$, a nonzero rank-one tensor determines both nonzero factors, so the set $\OldData$ of old local data has size $d=q^2$.
Consequently the admissible threshold range is $d^{-1/3}=q^{-2/3}\le\beta\le1$.
The long approximation is twice the part of the answer table, extended by zero to the inconsistent tuples, whose singular coefficients exceed $\beta$.
It has nuclear norm at most $4$ and energies at most $2/\beta$ (see \cref{cor:joint-tensor} in the \cref{sec:matrix-estimates}), and valid-label $L_2$ error at most $C\cdot\beta^{1/3}$ (see \cref{eq:joint-tensor-error} in the \cref{sec:matrix-estimates}).

To reach the short output encoding, we describe a successful first shortening attempt in reverse (see \cref{eq:compiler-inverse-shortening} in the \cref{sec:matrix-estimates}): given the output $(A,R)$, the matrix $H$ is uniform of full row rank, $V$ is uniform in the fiber $\{v:H\cdot v=R\}$, and $U=A\cdot H$.
These reverse sampling statements precede conditioning on either phase's leakage answer. Those answers enter through the leakage probability function and shortening masks.
For fixed $H$, the left factor uses $A$ and the retained left operands, while the right factor averages over the fiber determined by $R$ and uses the retained right operands.
The shortening masks use only these respective local data and the shortening tape. Neither may read the erased refresh matrix.
Averaging over the fiber and inserting the masks therefore turns each product of the long table into one product for each $H$, yielding a product representation of the short table.
A uniform short encoding followed by these reverse steps also gives exactly a uniform long encoding of the same bit, so the $L_2$ error is still averaged under the distribution for which it was bounded.
Jensen's inequality and masks bounded by one control the $L_2$ error and pointwise energies. The uniform reverse marginals preserve each separate mean-energy bound (see \cref{eq:compiler-shortening-energies,eq:compiler-output-right-energy,eq:compiler-output-left-mean,eq:compiler-output-right-mean,eq:compiler-shortening-L2} in the \cref{sec:matrix-estimates}).
The actual retry procedure adds at most $4/2^n$ per answer pointwise, which is absorbed by $C\cdot\beta^{1/3}$ throughout the stated range.
Under any three wire labels, the canonical density relative to the independent-share reference measure is at most $27$.
Hence~\cref{eq:compiler-local-rms} follows from $\LtwoNormAt{G^\beta}{s}\le\sqrt{27}\cdot\LtwoNorm{G^\beta}\le4\cdot\sqrt{27}$.
Exact freshness gives the canonical execution (see \cref{lem:compiler-canonical} in the \cref{sec:matrix-estimates}).
\Cref{sec:matrix-estimates} proves these statements with the required erasure and retry conditions.
\end{proof}

\subsection{Composition with hidden output shares}\label{sec:direct-composition}
So far we have approximated the leakage of each computation gadget and each input-port refresh separately.
We now combine these local approximations into a composition bound for the entire execution.
The argument compares the complete leakage transcripts of two executions in which the private, internal, and hidden output wires may carry different bits. The adversary sees only leakage from their encodings.
The bits may also be chosen during the execution as functions of the preceding leakage answers. This covers public inputs that the adversary chooses between the invocations of a stateful circuit (see \cref{sec:stateful}).
Signed tables appear only inside the proof: the actual computation and the simulator continue to sample from ordinary probability distributions.

For continual refreshing we changed one label at a time along a chain, keeping at most two rounds signed (see \cref{sec:copy-comparison}).
For the compiled circuit, we compare each execution with one common signed term that does not depend on the changing labels.
Obtaining this term requires two replacements.
At operations covered by the local approximation, we replace the true leakage probability function, which is defined only on tuples satisfying the operation's equation, by its approximation, whose table is common to all labelings.
Dependence on the changing labels then remains only in the encoding distributions, so on those wires we also replace the fixed-bit encoding distribution by independent uniform shares.
The main idea is therefore to expand two kinds of differences together: the error of a local approximation, and the difference between a fixed-bit encoding and independent uniform shares.
After marking a set of these differences, we sum over all the unmarked differences, which restores the true leakage probabilities and encoding distributions away from the marked differences.
The only remaining signed factors are then the marked differences themselves and the approximations at operations touching marked wires, so the local bounds accumulate linearly in the circuit size.

In this subsection, write $\mathcal X:=\UL\times\UR$ for the encoding space.
For any finite wire set $W$, a tuple $x=(x_w)_{w\in W}\in\mathcal X^W$ contains one share pair $x_w=(\ell_w,r_w)$ per wire. A table on grouped left and right shares is evaluated at $x$ by grouping its left coordinates and its right coordinates.
For intuition, consider a single operation that touches a single wire, with true factor $\prob=G+e$ and encoding distribution $\mu_s$.
Then
\[
 \begin{aligned}
 \E_{X\sample\mu_s}{\prob(X)}
 &=\E_{X\sample\nu}{G(X)}\\
 &\quad+\sum_{x\in\mathcal X}G(x)\cdot\left(\mu_s(x)-\nu(x)\right)
       +\E_{X\sample\mu_s}{e(X)}.
 \end{aligned}
\]
The first term is a common reference term, independent of $s$.
The second is a discrepancy term: a signed sum whose absolute value is bounded as a whole, so that its positive and negative weights can cancel.
The third is the approximation error averaged under $\mu_s$, the distribution for which its local bound holds.
The expansion below organizes such terms for all operations at once.

\subsubsection{The composition statement}\label{sec:direct-composition-setup}
We first state the composition bound in an abstract form that covers both the stateless compiler and its stateful extension in \cref{sec:stateful}.
Fix the adversary's tape and its auxiliary input (see \cref{sec:compiler-direct}).
The protocol has $S$ operations in a fixed order, each with a nonempty wire scope $S_i$ of size at most three and an answer alphabet of size at most $K$.
Each wire belongs to at most three scopes.
Wires that lie in no scope may be omitted, since they do not affect the leakage transcript, so at most $3S$ wires are relevant.
Let $U$ be the full ambient set of wire locations in this experiment after that omission, including the public wires.

Let $\cI\subseteq\{1,\ldots,S\}$ be the set of operations to which the local approximation applies. In the compiler these are the computation gadgets and the input-port refreshes.
Every other operation has its scope entirely among public wires, whose labels are the same in the executions being compared. In the compiler these are the public-input samplers, which remain true factors throughout the proof.
Let $\cV\subseteq U$ be the set of remaining wires, including private input wires, internal wires, and hidden output wires.
We call these the \emph{changing wires}, because the comparison does not assume that their labels agree.

A valid labeling $s$ assigns a bit to every wire and satisfies every operation's equation.
We also allow labels that are chosen during the execution, as in \cref{sec:stateful}, where the adversary chooses each public input after seeing earlier leakage.
A \emph{history-dependent labeling} $s$ assigns to every wire $w$ a bit $s_h(w)$ that is a function of the formal history $h$ of the operations preceding the operation that produces $w$. The private input wires, which the trusted encoder supplies, have constant labels.
For a complete formal history $z$, the \emph{induced labeling} $s_z$ assigns to $w$ the bit $s_h(w)$ for the corresponding prefix $h$ of $z$.
The labeling $s$ is \emph{valid} if $s_z$ is valid for every $z$, and two history-dependent labelings \emph{agree on the public wires} if their induced labelings do so for every $z$.
A labeling in the previous sense is the special case whose labels do not depend on the history.
The canonical experiment runs the operations in order: before operation $i$, with incoming history $h_{i-1}$, it samples the encodings of the wires that operation $i$ produces independently, $X_w\sample\mu_{s_{h_{i-1}}(w)}$, and then draws the answer $z_i$ from the true conditional probabilities $\prob_{i,h_{i-1},z_i}(X_{S_i})$. The private input encodings are sampled first.
These probabilities are nonnegative and sum to one for every valid local tuple and every formal incoming history.
Denote the resulting transcript distribution by $P_s$.
Because the label of a wire is fixed before the wire is produced, the encoding weights accumulated along a complete transcript $z=(z_1,\ldots,z_S)$ are exactly $\bigotimes_{w\in U}\mu_{s_z(w)}$, so with $h_{i-1}:=(z_1,\ldots,z_{i-1})$,
\begin{equation}\label{eq:direct-canonical}
 P_s(z)=\E_{X_U\sample\bigotimes_{w\in U}\mu_{s_z(w)}}{
       \prod_{i=1}^S\prob_{i,h_{i-1},z_i}(X_{S_i})}.
\end{equation}
For labels that do not depend on the history, this is the experiment that samples all encodings before any answer is drawn.
The wire encodings are independent in this unconditional expectation. Conditioning on a transcript can create dependencies.
The two shares of a fixed-bit encoding are generally correlated, whereas $\nu$ samples the two shares of a wire independently.
The true tables, like the approximation families below, are fixed for all formal histories before labelings are compared. They depend on the labeling only through the wire encodings supplied to them.

\begin{theorem}[Composition with hidden output shares]\label{thm:direct-composition}
There exists an absolute constant $C$ with the following property.
Let $n\ge2$ and $q:=2^n-1$.
Suppose the canonical experiment above is exact, and every $i\in\cI$ has the local approximation and separate mean-energy bounds of \cref{lem:copy-local,lem:compiler-local-interface} at $\beta=q^{-1/3}$.
Then any two valid history-dependent labelings $s$ and $s'$ that agree on the public wires satisfy
\begin{equation}\label{eq:direct-composition-bound}
 \StatDist{P_s}{P_{s'}}\le C\cdot S\cdot K^9\cdot q^{-1/27}.
\end{equation}
The right-hand side may be replaced by its minimum with one, and the same bound holds when the adversary's independent random tape is included in the view.
\end{theorem}
The threshold $q^{-1/3}$ is allowed for a dimension-$n$ refresh by \cref{lem:copy-local}, and for a computation gadget by \cref{lem:compiler-local-interface}, whose allowed range starts at $q^{-2/3}$.
The proof below uses only the valid-label $L_2$ errors and the separate mean-energy bounds of these approximations. They need be neither positive nor normalized over the answers.

\subsubsection{Trimming and products of local errors}\label{sec:direct-trimming}
A wire can occur in three operations, so a product of local errors is controlled by H\"older's inequality with exponent three (see \cref{clm:direct-product-holder}), which requires $L_3$ errors.
Trimming each side's factor energy separately turns the available $L_2$ error into the required $L_3$ error.
Trimming also gives the uniform energy cap used to bound the product of tables at operations touching a marked wire (see \cref{eq:direct-marked-wire-integration}).

All wire and answer spaces in this section are finite, so the encoding and answer sums below are finite sums.
The adversary's possibly infinite tape is fixed, and unbounded sampler records have already been averaged into the true local probabilities.
For a local real table $F$, a local labeling $s$, and real $p\ge1$, write
\[
 \LpNormAt{F}{p}{s}
 :=\left(\E_{X_{S_i}\sample\bigotimes_{w\in S_i}\mu_{s(w)}}{|F(X_{S_i})|^p}\right)^{1/p}.
\]

\begin{claim}[Trimming the factor energies]\label{clm:direct-trimming}
Let $0<\beta\le1$ and $C_{\mathrm{err}}\ge0$.
On a fixed local scope, let $\prob$ be a table taking values in $[0,1]$ on every valid local tuple.
Suppose $G$ has a chosen product representation with energies $\Phi$ and $\Psi$ (see \cref{def:factor-energy}) such that $\E_x{\Phi(x)}\le4$ and $\E_y{\Psi(y)}\le4$ under the independent uniform grouped-share reference measures, and $\LtwoNormAt{\prob-G}{s}\le C_{\mathrm{err}}\cdot\beta^{1/3}$ for every valid local labeling $s$.
Set $B:=\beta^{-1/3}$ and define
\begin{equation}\label{eq:direct-trimmed-table}
 \overline G(x,y):=\begin{cases}
 G(x,y),&\Phi(x)\le B\text{ and }\Psi(y)\le B,\\
 0,&\text{otherwise}.
 \end{cases}
\end{equation}
This table has a product representation whose two pointwise energies are at most $B$.
Moreover, for a constant $C_1$ depending only on $C_{\mathrm{err}}$ and every valid local labeling $s$,
\begin{equation}\label{eq:direct-trimmed-error}
 \LpNormAt{\prob-\overline G}{3}{s}
 \le C_1\cdot\beta^{1/9}.
\end{equation}
\end{claim}
\begin{proof}
Set every left factor to zero on rows with $\Phi(x)>B$ and every right factor to zero on columns with $\Psi(y)>B$, keeping the other factors.
This gives~\cref{eq:direct-trimmed-table}.
The new energies are at most $B$, and hence $|\overline G|\le B$ by~\cref{eq:factor-energy-entry-bound}.

Fix a valid local labeling $s$, and write $\E_s$ for the expectation over $X_{S_i}\sample\bigotimes_{w\in S_i}\mu_{s(w)}$.
Under this distribution the grouped left shares are uniform, as are the grouped right shares.
Markov's inequality therefore bounds the probability of the event ``$\Phi(x)>B$ or $\Psi(y)>B$'' by $8/B$.
On this event $|\prob-\overline G|=\prob\le1$.
On its complement, $\overline G=G$ and $|G|\le B$, so
\[
 \begin{aligned}
 \E_s{|\prob-\overline G|^3}
 &\le (1+B)\cdot\E_s{|\prob-G|^2}+8/B\\
 &\le C_{\mathrm{err}}^2\cdot(1+B)\cdot\beta^{2/3}+8/B
 \le C\cdot\beta^{1/3},
 \end{aligned}
\]
where $C$ depends only on $C_{\mathrm{err}}$.
Taking the cube root proves~\cref{eq:direct-trimmed-error}.
\end{proof}

\begin{claim}[H\"older's inequality for scopes of incidence three]\label{clm:direct-product-holder}
Let $(X_w)_{w\in U}$ be independent random variables on finite spaces, indexed by a finite set $U$.
Let $I$ be a finite index set and, for each $i\in I$, let $f_i$ be a nonnegative function of the variables in a scope $A_i\subseteq U$.
Suppose each variable occurs in at most three scopes.
Then, with expectations over all the variables,
\begin{equation}\label{eq:direct-product-holder}
 \E{\prod_{i\in I} f_i(X_{A_i})}
 \le\prod_{i\in I}\left(\E{f_i(X_{A_i})^3}\right)^{1/3}.
\end{equation}
\end{claim}
\begin{proof}
Induct on the number of variables. The case of no variables is an equality.
At most three factors depend on a given variable $X_w$.
Holding the other variables fixed, apply H\"older's inequality with exponent three to these factors, filling unused positions with the constant one.
Replace each affected factor $f_i$ by $\left(\E_{X_w}{f_i^3}\right)^{1/3}$, now a function of the remaining variables in its scope.
The incidence bound is unchanged.
Applying the induction hypothesis to the remaining variables gives~\cref{eq:direct-product-holder}, because the cube of each replaced factor is $\E_{X_w}{f_i^3}$.
\end{proof}

Apply \cref{clm:direct-trimming} separately to each operation, formal history, and answer, fixing all the resulting tables once.
From now on set
\begin{align}
 \beta&:=q^{-1/3},\label{eq:direct-parameters}\\
 B&:=q^{1/9},\nonumber\\
 \varepsilon&:=C_1\cdot q^{-1/27},\label{eq:direct-error-parameter}\\
 e_{i,h,z}&:=\prob_{i,h,z}-\overline G_{i,h,z}.\nonumber
\end{align}
Every $e_{i,h,z}$ is used only on valid local tuples, where its $L_3$ norm is at most $\varepsilon$ by~\cref{eq:direct-trimmed-error}.
If a value outside that domain is needed to make it a table, choose any extension. The proof never evaluates an error factor there.

\subsubsection{A signed difference for each encoding}\label{sec:direct-encoding-difference}
The trimmed approximations handle the leakage probabilities, and the second expansion concerns the encoding distributions.
Recall from \cref{def:product-distribution} that $\nu$ is the uniform distribution on $\UL\times\UR$: it samples the two shares independently and does not encode a fixed bit.
Define the signed weight function
\begin{equation*}
 D_b(x):=\mu_b(x)-\nu(x),\qquad b\in\{0,1\},\quad x\in\mathcal X.
\end{equation*}
The weights $D_b(x)$ can be negative and are never sampled.
In the expansion of \cref{sec:direct-marked-expansion}, the term with $\nu$ on every changing wire and no approximation error does not depend on the private input. Every other term contains a difference $D_b$ or an approximation error.

\begin{claim}[Tensor discrepancy for encoding differences]\label{clm:direct-encoding-discrepancy}
Put $\gamma:=3/\sqrt q$.
For every set $W$ of wires, all bits $(b_w)_{w\in W}$, and every real table $F$ on the grouped left and right shares of $W$,
\begin{equation}\label{eq:direct-tensor-discrepancy}
 \left|\sum_{x\in\mathcal X^W}F(x)\cdot\prod_{w\in W}D_{b_w}(x_w)\right|
 \le\gamma^{|W|}\cdot\NuclearNorm{F},
\end{equation}
where the nuclear norm uses independent uniform grouped shares.
\end{claim}
\begin{proof}
By~\cref{eq:encoding-zero-probability,eq:encoding-one-probability}, $p_0=(q-1)/(2q)$ and $p_1=(q+1)/(2q)$.
For $x=(\ell,r)\in\mathcal X$, the reference probability $\nu(x)$ is positive.
The density of $D_b$ relative to $\nu$ is therefore well-defined and equals
\[
 \varphi_b(x):=\frac{D_b(x)}{\nu(x)}=
 \frac{(-1)^b\cdot(-1)^{\ell\cdot r}+1-2p_b}{2p_b}.
\]
The signed sum in~\cref{eq:direct-tensor-discrepancy} is equivalently the ordinary expectation
\[
 \E_{X_W\sample\bigotimes_{w\in W}\nu}
 {F(X_W)\cdot\prod_{w\in W}\varphi_{b_w}(X_w)},
\]
where the $X_w$ are sampled independently from $\nu$. The function being averaged can be negative.
By~\cref{eq:binary-sign-product-bound} in the \cref{app:prelim-sign-discrepancy}, the largest singular coefficient of the sign table $(-1)^{\ell\cdot r}$ is at most $\sqrt{q+1}/q$.
The constant table $1$ has largest singular coefficient one, $p_b\ge1/3$, and $|1-2p_b|=1/q$.
Consequently
\[
 \sigma_{\max}(\varphi_b)
 \le\frac32\left(\frac{\sqrt{q+1}}q+\frac1q\right)
 \le\frac3{\sqrt q}=\gamma,
\]
where the last inequality uses $q\ge3$.

Orthonormal expansions of the individual densities tensor to an orthonormal expansion on the grouped coordinates, whose coefficients are products of the individual coefficients.
The largest coefficient of $\prod_{w\in W}\varphi_{b_w}$ is therefore at most $\gamma^{|W|}$.
The weighted sum in~\cref{eq:direct-tensor-discrepancy} is the pairing of $F$ with this product density.
Applying~\cref{eq:largest-singular-coefficient} to each term in an orthonormal expansion of $F$ and summing its singular coefficients gives~\cref{eq:direct-tensor-discrepancy}.
\end{proof}

\subsubsection{Expanding and summing marked terms}\label{sec:direct-marked-expansion}
We use two expansions: leakage probabilities are written as approximations plus errors, and fixed-bit encoding distributions as the independent uniform reference distribution plus signed differences.
These expansions must be coordinated to use the local error bounds.
An approximation error contains the true leakage probability function, which is defined only on tuples whose bits satisfy the equation of its operation, and~\cref{eq:direct-trimmed-error} bounds the error only when these bits come from the fixed-bit encoding distributions of a valid labeling.
Independent uniform shares on a wire of its scope can produce tuples outside this domain, where the true function is undefined and no bound is available.
To use a local error bound, we must therefore preserve the original fixed-bit encoding distributions on the entire scope of its error factor.
We choose the error operations first, keep those wire distributions unchanged, and expand encoding differences only on the remaining changing wires, so the set of wires we may expand depends on the chosen error operations.
The following sets record which wires must be kept unchanged and which operations can inspect a selected wire.
For $A\subseteq\cI$ and a set $W$ of wires, let $\cV_A$ be the set of wires of the operations in $A$, and let $\cN(W)$ be the set of operations touching $W$:
\[
 \cV_A:=\bigcup_{i\in A}S_i,
 \qquad \cN(W):=\{i:S_i\cap W\ne\varnothing\}.
\]
Fix a complete formal history $z$. Abbreviate $\prob_{i,h_{i-1},z_i}$ to $\prob_i$, and use analogous abbreviations for $\overline G_i$ and $e_i$. Also abbreviate the induced label $s_z(w)$ to $s(w)$.
This fixes the tables and the labels without conditioning the encoding distributions on an observed transcript. The identities below hold for each $z$ separately.
Operation indices range over $\{1,\ldots,S\}$, including the public-input samplers.
In~\cref{eq:direct-canonical}, expand $\prob_i=\overline G_i+e_i$ for $i\in\cI$.
In a term whose error operations are $A$, keep the original encoding weights on $\cV_A$ and expand $\mu_{s(w)}=\nu+D_{s(w)}$ only on $\cV\setminus\cV_A$.
This gives a finite identity
\begin{equation}\label{eq:direct-expansion}
 P_s(z)=\sum_{A\subseteq\cI}
          \ \sum_{W\subseteq\cV\setminus\cV_A}T^s_{A,W}(z).
\end{equation}
Explicitly, $T^s_{A,W}$ uses $e_i$ at operations in $A$, $\overline G_i$ at operations in $\cI\setminus A$, and the true factor $\prob_i$ at operations outside $\cI$.
Its coordinate weights are $D_{s(w)}$ on $W$, $\nu$ on $\cV\setminus(\cV_A\cup W)$, and $\mu_{s(w)}$ on every other wire of $U$.
Thus every error factor keeps the original independent fixed-bit encoding distributions on its entire scope, where its $L_3$ bound holds.
Error factors with overlapping scopes need not be independent, so their product is averaged under the joint encoding distribution.

Bounding the identity in~\cref{eq:direct-expansion} term by term would discard the normalization that the proof needs.
A single term $T^s_{A,W}$ has an approximation $\overline G_i$ at every operation of $\cI\setminus A$, and these tables may be negative and need not sum to one over the answers.
Bounding its absolute mass over complete histories directly would therefore enumerate the answers of all those operations, at a cost of up to a factor $K$ for each.
To recover the normalization of the true probabilities, we instead group the expanded terms before bounding their absolute mass.
For any chosen marks, we add every term that contains those differences.
Summing the additional differences will restore true probability factors away from the marked errors and the operations touching marked wires.
The bound on the absolute mass of each such sum depends on the numbers of marks rather than on the size of the circuit, and inclusion--exclusion in \cref{sec:direct-final-comparison} recovers all the nonempty terms from these sums.
For marked sets $A_\star\subseteq\cI$ and $W_\star\subseteq\cV$, let
\begin{equation}\label{eq:direct-marked-definition}
 F^s_{A_\star,W_\star}
 :=\sum_{\substack{A\supseteq A_\star,\ W\supseteq W_\star\\
                    W\subseteq\cV\setminus\cV_A}}T^s_{A,W}.
\end{equation}
This is the signed sum of all terms containing the marked differences, including any additional unmarked differences.
If $W_\star\cap\cV_{A_\star}\ne\varnothing$, the sum is empty and equals zero.

\begin{claim}[Summing all unmarked differences]\label{clm:direct-marked-sum}
If $W_\star\cap\cV_{A_\star}=\varnothing$, then
\begin{equation}\label{eq:direct-marked-resummation}
 \begin{aligned}
 F^s_{A_\star,W_\star}(z)
 =\sum_{x\in\mathcal X^U}
 &\left(\prod_{i\in A_\star}e_i(x_{S_i})\right)\cdot
       \left(\prod_{i\in\cN(W_\star)}\overline G_i(x_{S_i})\right)\\
 &{}\cdot\left(\prod_{i\notin A_\star\cup\cN(W_\star)}\prob_i(x_{S_i})\right)\\
 &{}\cdot\left(\prod_{w\in W_\star}D_{s(w)}(x_w)\right)\cdot
       \left(\prod_{w\in U\setminus W_\star}\mu_{s(w)}(x_w)\right).
 \end{aligned}
\end{equation}
The last product includes all fixed public wires, and the third product includes every operation outside $A_\star\cup\cN(W_\star)$, including public-input samplers.
Every error and true factor on the right is used only on valid local tuples with nonzero original encoding weight.
Empty products equal one.
\end{claim}
\begin{proof}
A term with $A\supseteq A_\star$ and $W\supseteq W_\star$ occurs in~\cref{eq:direct-marked-definition} only if $W_\star\cap\cV_A=\varnothing$, that is, only if $A\cap\cN(W_\star)=\varnothing$.
For a fixed such $A$, sum first over all additional wire differences outside $W_\star$.
Every unmarked changing-wire weight becomes $\nu+D_{s(w)}=\mu_{s(w)}$, and wires in $\cV_A$ and fixed public wires already have this weight.
The resulting coordinate weights no longer depend on $A$.
Now sum over the additional error operations.
At each operation in $\cI\setminus(A_\star\cup\cN(W_\star))$, the sum is $\overline G_i+e_i=\prob_i$.
The operations in $\cN(W_\star)$ cannot be error operations, so they keep $\overline G_i$, and they all lie in $\cI$ because all other operations have fixed public scopes.
This proves~\cref{eq:direct-marked-resummation}.

No true factor touches $W_\star$.
Every wire in the scope of a true factor thus keeps its original fixed-bit encoding weight $\mu_{s(w)}$, so the labeling is valid wherever that factor is used.
\end{proof}

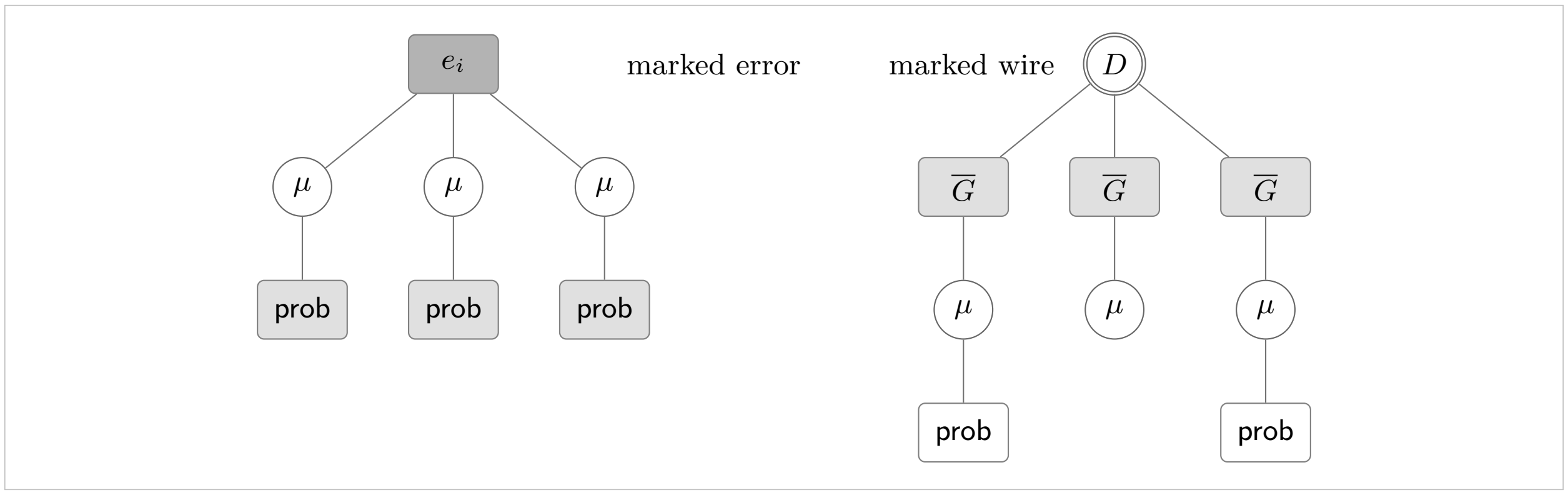
\begin{figure}[htbp]
\centering
\begingroup
\setlength{\fboxrule}{0.3pt}
\setlength{\fboxsep}{8pt}
\fcolorbox{black!25}{white}{\begin{minipage}{\dimexpr\linewidth-2\fboxsep-2\fboxrule\relax}
\centering
\begin{tikzpicture}[x=1cm,y=1cm,font=\footnotesize,
 op/.style={draw=black!50,rounded corners=2pt,minimum width=0.95cm,minimum height=0.62cm,inner sep=2pt},
 guard/.style={op,fill=black!12},
 err/.style={op,fill=black!30},
 enc/.style={draw=black!60,circle,inner sep=1pt,minimum size=0.62cm},
 line/.style={draw=black!55}]
 \node[err] (e) at (1.0,0) {$e_i$};
 \node[enc] (a1) at (-0.6,-1.3) {$\mu$};
 \node[enc] (a2) at (1.0,-1.3) {$\mu$};
 \node[enc] (a3) at (2.6,-1.3) {$\mu$};
 \node[guard] (ga) at (-0.6,-2.6) {$\prob$};
 \node[guard] (gb) at (1.0,-2.6) {$\prob$};
 \node[guard] (gc) at (2.6,-2.6) {$\prob$};
 \draw[line] (e)--(a1) (e)--(a2) (e)--(a3) (a1)--(ga) (a2)--(gb) (a3)--(gc);
 \node[enc,double] (d) at (8.0,0) {$D$};
 \node[guard] (g1) at (6.4,-1.3) {$\overline G$};
 \node[guard] (g2) at (8.0,-1.3) {$\overline G$};
 \node[guard] (g3) at (9.6,-1.3) {$\overline G$};
 \node[enc] (b1) at (6.4,-2.6) {$\mu$};
 \node[enc] (b2) at (8.0,-2.6) {$\mu$};
 \node[enc] (b3) at (9.6,-2.6) {$\mu$};
 \draw[line] (d)--(g1) (d)--(g2) (d)--(g3) (g1)--(b1) (g2)--(b2) (g3)--(b3);
 \node[op] (f2) at (6.4,-3.9) {$\prob$};
 \node[op] (f3) at (9.6,-3.9) {$\prob$};
 \draw[line] (b1)--(f2) (b3)--(f3);
 \node[anchor=west] at (2.7,0) {marked error};
 \node[anchor=east] at (7.5,0) {marked wire};
\end{tikzpicture}
\end{minipage}}
\endgroup
\caption{The structure of one marked term.
Shaded operations are guards, whose answers are enumerated.
Discrepancy removes the tables $\overline G$ around the marked wire, H\"older's inequality bounds the marked errors, the true factors at guards are bounded by one, and the remaining true factors sum to one.}
\label{fig:marked-term}
\end{figure}

\begin{claim}[Bound for jointly marked differences]\label{clm:direct-marked-bound}
For all $A_\star\subseteq\cI$ and $W_\star\subseteq\cV$,
\begin{equation}\label{eq:direct-marked-bound}
 \LoneNorm{F^s_{A_\star,W_\star}}
 =\sum_z|F^s_{A_\star,W_\star}(z)|
 \le (K^9\cdot\varepsilon)^{|A_\star|}\cdot
      (K^3\cdot B^3\cdot\gamma)^{|W_\star|}.
\end{equation}
This bound is uniform in the valid labeling and in all formal histories.
\end{claim}
\begin{proof}
We may suppose that $W_\star\cap\cV_{A_\star}=\varnothing$, since otherwise $F^s_{A_\star,W_\star}=0$.
Fix a complete history and all encodings outside $W_\star$.
The product of the approximation factors in~\cref{eq:direct-marked-resummation} is a table in the grouped left and right shares of $W_\star$:
\[
 \Pi_\star:=\prod_{i\in\cN(W_\star)}\overline G_{i}.
\]
Multiply the chosen product representations of its $|\cN(W_\star)|\le3|W_\star|$ factors.
The energy of the resulting left factors is the product of their left energies, and hence is at most $B^{3|W_\star|}$ pointwise. The same holds on the right.
By~\cref{eq:nuclear-product-bound} and the Cauchy--Schwarz inequality (see \cref{lem:la-cauchy-schwarz}) over the factor indices,
\[
 \NuclearNorm{\Pi_\star}\le
 \sqrt{\E_{L_{W_\star}\sample\UL^{W_\star}}{\Phi_{\Pi_\star}(L_{W_\star})}}\cdot
 \sqrt{\E_{R_{W_\star}\sample\UR^{W_\star}}{\Psi_{\Pi_\star}(R_{W_\star})}}\le B^{3|W_\star|}.
\]
Consequently, pointwise in all outside encodings, the tensor discrepancy~\cref{eq:direct-tensor-discrepancy} gives
\begin{equation}\label{eq:direct-marked-wire-integration}
 \left|\sum_{x\in\mathcal X^{W_\star}}\Pi_\star(x)\cdot
       \prod_{w\in W_\star}D_{s(w)}(x_w)\right|
 \le B^{3|W_\star|}\cdot\gamma^{|W_\star|}.
\end{equation}
Every other factor in~\cref{eq:direct-marked-resummation} is independent of $W_\star$.
The absolute value in~\cref{eq:direct-marked-wire-integration} encloses the whole signed sum, preserving cancellation.
This bound removes every signed approximation factor before any true factor is discarded.

We still need to sum the adaptive answers while averaging the remaining error factors jointly.
These factors may share wire variables, and each depends on its own incoming answer prefix.
The normalization of true probabilities cannot be used at every operation.
A marked error and an operation touching a marked wire have no true factor, and their tables need not sum to one over the answers. The discrepancy bound above removes an approximation table, but its answer still has to be summed.
A true factor that reads a wire of a marked error does sum to one, but it may depend on encodings over which the error bound must average, so it cannot in general be moved outside that average.
If such a factor precedes a marked error, its answer can also help select that error's table through the incoming history, preventing us from summing it out first in general.
We therefore enumerate the answers of these operations, which form a bounded set, so that the other true factors can be summed in reverse time order without changing the encoding distributions used for the error bound.
We treat earlier and later neighbors alike, unlike in \cref{sec:refresh-adaptivity}. This is simpler and leads to the factor $K^9$ for each marked error.
Let
\[
 \cG_\star:=\cN(\cV_{A_\star})\cup\cN(W_\star).
\]
We call this set of operations the \emph{guards}: they are the operations that share a wire with a marked error or touch a marked wire (see \cref{fig:marked-term}).
It contains $A_\star$ and has size at most $9|A_\star|+3|W_\star|$, because every scope has at most three wires and every wire lies in at most three scopes.
Operations outside $\cG_\star$ have scopes disjoint from $\cV_{A_\star}\cup W_\star$.

Write $X_{\mathrm{out}}:=X_{U\setminus(\cV_{A_\star}\cup W_\star)}$ for the outside encodings.
Enumerate the answers of the guards, at a cost of at most $K^{9|A_\star|+3|W_\star|}$, and discard the remaining true factors among the guards using $0\le\prob_i\le1$.
Fix one such answer tuple $z_\star$ on $\cG_\star$.
For a complete history $z$ extending $z_\star$, put
\[
 \Pi_z(X_{\mathrm{out}}):=
       \prod_{i\notin\cG_\star}\prob_{i,h_{i-1},z_i}(X_{S_i}).
\]
Since operations outside $\cG_\star$ have scopes disjoint from $\cV_{A_\star}\cup W_\star$, the factor $\Pi_z$ depends only on $X_{\mathrm{out}}$, whereas the error factors depend only on $X_{\cV_{A_\star}}$.
After~\cref{eq:direct-marked-wire-integration} it therefore remains to bound
\[
 \sum_{z:\,z|_{\cG_\star}=z_\star}\ \sum_{X_{\mathrm{out}}}
 \Bigl(\prod_{w\notin\cV_{A_\star}\cup W_\star}\mu_{s_z(w)}(X_w)\Bigr)\cdot\Pi_z(X_{\mathrm{out}})\cdot
 \E_{X_{\cV_{A_\star}}\sample\bigotimes_{w\in\cV_{A_\star}}\mu_{s_z(w)}}{
       \prod_{i\in A_\star}|e_{i,h_{i-1},z_i}(X_{S_i})|}.
\]
For each complete history $z$, \cref{clm:direct-product-holder} and the valid-label $L_3$ error~\cref{eq:direct-trimmed-error}, applied to the valid labeling $s_z$, give
\begin{equation}\label{eq:direct-error-product}
 \E_{X_{\cV_{A_\star}}\sample\bigotimes_{w\in\cV_{A_\star}}\mu_{s_z(w)}}{
       \prod_{i\in A_\star}|e_{i,h_{i-1},z_i}(X_{S_i})|}
 \le\varepsilon^{|A_\star|}.
\end{equation}
The encodings in this expectation are sampled independently from their original fixed-bit distributions, and the bound is uniform in $z$.
It remains to show that
\begin{equation}\label{eq:direct-unguarded-normalization}
 \sum_{z:\,z|_{\cG_\star}=z_\star}\ \sum_{X_{\mathrm{out}}}
 \Bigl(\prod_{w\notin\cV_{A_\star}\cup W_\star}\mu_{s_z(w)}(X_w)\Bigr)\cdot\Pi_z(X_{\mathrm{out}})=1.
\end{equation}
Sum the answers of the operations outside $\cG_\star$ and the outside encodings in reverse time order, taking the answer of an operation before the encodings of the wires that it produces.
Each true factor sums to one over its answer for every formal prefix and every valid local tuple, and the tuples with nonzero weight are valid because $s_z$ is valid. Each encoding weight sums to one whatever its label is.
When a variable is summed, nothing else depends on it: all later factors and weights have already been summed, earlier factors depend only on earlier answers and encodings, and the label of a wire is fixed by the answers preceding the operation that produces it.
The enumerated guard answers remain in the prefixes but have no remaining factors of their own.
This proves~\cref{eq:direct-unguarded-normalization}. For labels that do not depend on the history, it says that $\Pi_z$ sums to one over the histories for every fixed $X_{\mathrm{out}}$ of nonzero weight.
Combining~\cref{eq:direct-marked-wire-integration,eq:direct-error-product,eq:direct-unguarded-normalization} with the count of guard answers proves~\cref{eq:direct-marked-bound}.
The original independent encoding distributions are never conditioned on an observed history.
\end{proof}

\subsubsection{The common term and the final comparison}\label{sec:direct-final-comparison}
It remains to treat the one term without any difference, which is common to the compared executions, and to combine all bounds.
The term $T^s_{\varnothing,\varnothing}$ in~\cref{eq:direct-expansion} uses $\nu$ on every wire in $\cV$.
Its remaining coordinate weights at a complete history $z$ are the encoding distributions of the public bits $s_z(w)$, and all its tables are common to the labelings.
Write $T_\circ:=T^s_{\varnothing,\varnothing}$ for this common term.
Thus $T_\circ$ is the same signed weight function for all valid history-dependent labelings that agree on the public wires.
It need not be nonnegative or have total mass one, and we do not sample it.

Put
\begin{align*}
 \rho_A&:=K^9\cdot\varepsilon,\\
 \rho_W&:=K^3\cdot B^3\cdot\gamma.
\end{align*}
Inclusion--exclusion sums all nonempty terms through the marked sums:
\begin{equation*}
 P_s-T_\circ
 =\sum_{\substack{A_\star\subseteq\cI,\ W_\star\subseteq\cV\\
                    (A_\star,W_\star)\ne(\varnothing,\varnothing)}}
     (-1)^{|A_\star|+|W_\star|+1}\cdot F^s_{A_\star,W_\star}.
\end{equation*}
To verify this finite identity, fix a nonempty term $T^s_{A,W}$.
Its coefficient on the right is the sum of $(-1)^{|Y|+1}$ over all nonempty subsets $Y$ of the disjoint union of its operation and wire indices, and this sum is one.
The empty term has coefficient zero.

Taking norms only after this regrouping and using~\cref{eq:direct-marked-bound} gives
\begin{equation}\label{eq:direct-common-term-error}
 \LoneNorm{P_s-T_\circ}
 \le (1+\rho_A)^{|\cI|}\cdot(1+\rho_W)^{|\cV|}-1.
\end{equation}
For the parameters in~\crefrange{eq:direct-parameters}{eq:direct-error-parameter}, we have
\begin{align}
 \rho_A&\le C_1\cdot K^9\cdot q^{-1/27},\label{eq:direct-activity-values}\\
 \rho_W&\le3\cdot K^3\cdot q^{-1/6}.\label{eq:direct-wire-activity-value}
\end{align}

\begin{proof}[Proof of \cref{thm:direct-composition}]
Subtract the same signed weight function $T_\circ$ from both real distributions.
The triangle inequality, the factor $1/2$ in statistical distance, and~\cref{eq:direct-common-term-error} imply
\[
 \StatDist{P_s}{P_{s'}}
 \le (1+\rho_A)^{|\cI|}\cdot(1+\rho_W)^{|\cV|}-1.
\]
Let $\tau:=|\cI|\cdot\rho_A+|\cV|\cdot\rho_W$.
If $\tau\le1$, the right-hand side is at most $\exp(\tau)-1\le2\tau$.
If $\tau>1$, use instead $\StatDist{P_s}{P_{s'}}\le1\le\tau$.
Hence the distance is at most $2\tau$ in all cases.
Now $|\cI|\le S$, $|\cV|\le3S$, $K\ge1$, and $q^{-1/6}\le q^{-1/27}$.
Substituting~\cref{eq:direct-activity-values,eq:direct-wire-activity-value} proves~\cref{eq:direct-composition-bound}, after enlarging the absolute constant.
The case $S=0$ has an empty transcript.
Finally, all estimates are uniform in the adversary's fixed tape, so taking expectation over its independent random tape gives the same bound for the view that includes the tape.
\end{proof}

\subsection{Concrete simulation and expected work}\label{sec:direct-simulator}
It remains to justify the claim of polynomial expected primitive work in \cref{thm:compiler-direct}.
We do so by giving concrete sampling algorithms and accounting for every rejected proposal and retry.
Recall from its proof that the simulator executes the compiled protocol on zero private inputs, keeps the public input, discards its dummy output encodings, and reports the supplied authorized output $\vecy$.
The dummy computation need not produce $\vecy$, since the trusted decoder hides its output shares. \Cref{thm:direct-composition} supplies the statistical comparison.
The simulator's algorithm is fixed independently of the private input and target error, and $\vecy$ is available before its first simulated leakage answer.

\para{Primitive work.}
\emph{Primitive work} counts binary arithmetic, single-bit sampling, and indexed reads, writes, and comparisons of circuit addresses. Addresses are machine words.
We use the same concrete samplers for real and simulated executions, retaining every accepted or rejected proposal in the prescribed local phase tape.

\para{Concrete samplers.}
To initialize a uniform encoding of $s$, sample two independent uniform vectors in $\F_2^n$ and reject the pair unless both vectors are nonzero and their inner product is $s$.
The accepted pair is uniform on the desired encoding fiber, and for $n\ge2$ each proposal is accepted with probability at least $1/8$.
The private-input encoder is trusted, so its initialization proposals are excluded from leakage.
In a public-input sampler, which is not trusted, the entire local proposal tape is included in that operation's permitted leakage and budget.
To sample a uniform invertible $n'\times n'$ binary matrix, sample a uniform matrix and reject it if singular. The acceptance probability exceeds $1/4$ by~\cref{eq:invertible-probability}.
To sample the nonzero row $A$, sample a uniform vector and reject zero.
To sample the shortening matrix given $A$ and $U$, fix the first index $j$ with $A_j\ne0$, sample all $n-1$ rows $H_i$ with $i\ne j$ independently and uniformly in $\F_2^m$, and set
\[
 H_j:=U-\sum_{i\ne j}A_i\cdot H_i.
\]
Here $A_j=1$, so the filled matrix satisfies $A\cdot H=U$.
Reject the entire free-row proposal unless $H$ has full row rank.
Filling row $j$ is a bijection from the free-row choices to all matrices satisfying the affine constraint, so the accepted matrix is uniform among the required full-row-rank completions.
The acceptance probability is
\[
 \prod_{i=1}^{n-1}\left(1-2^{i-m}\right)>1/4.
\]
Indeed, after an invertible row transformation whose first row is $A$, the matrix has full row rank exactly when the $n-1$ uniform rows extend the fixed nonzero row $U$ to a linearly independent set.
All rejected and accepted shortening proposals, together with all outer attempts, remain in the appropriate party's local tape, as allowed by the leakage model.
Each nonzero-row proposal and each outer shortening attempt succeeds with probability at least $1/2$.

\para{Expected work.}
Each completed sampler record is a finite sequence of proposals over finite alphabets, but its length is unbounded, so these records have countable support.
The uniform success bounds imply almost-sure termination.
For a nonnegative integer work counter $J$, its expectation is the nonnegative series $\E{J}=\sum_{j=0}^{\infty}j\cdot\Prb{J=j}$. Nonnegativity justifies rearranging these series when combining costs.
Each trial, rank test, inversion, tensor product, and projection uses polynomially many binary operations.
Every operation produces one fresh encoding, so together with the private input encodings at most $\Ipriv+S$ encoding locations are needed before erasures.
The circuit makes polynomially many top-level sampler calls.
Each nested rejection loop has a uniformly bounded expected number of attempts, conditional on the state entering that loop.
Applying these conditional bounds from the inner loops outward and summing the resulting nonnegative expected costs gives a polynomial bound $W$ on total expected primitive work.

A \emph{local-answer evaluation} consists of the adversary's selection of one leakage function and its evaluation on the local state.
A computation gadget requires four local-answer evaluations, an input-port refresh two, and a public-input sampler two, so by~\cref{eq:direct-computation-count,eq:direct-refresh-count,eq:direct-public-count} their total number is $8t_0+2\Ipub$.
If one local-answer evaluation costs at most $\Rev$, expected total work is at most $W+(8t_0+2\Ipub)\cdot\Rev$.
More generally, expected total work is at most $W$ plus the extended nonnegative expectation of the cost of the actual additional computation. That computation may depend on the same private sampling tapes, and its expected cost may be infinite.
Bounded answer length alone does not bound this evaluation cost, and we assume no finite expected evaluation cost for arbitrary functions on unbounded tapes.

\para{Capping the running time.}
The same simulator has strictly polynomial primitive work if we cap the number of shortening attempts and the trials of every rejection sampler.
First cap the outer attempts, which bounds the number of inner sampling calls, and then cap each inner sampler.
The caps can be chosen polynomially in the stated parameters so that a union bound over the geometric tails makes the probability of any cap being reached at most $2^{-\kappa-1}$, where $2^{-\kappa}$ is the target error.
An arbitrary fixed termination rule on that event adds at most this probability to statistical distance. For the remaining error, use~\cref{eq:direct-simulation} with target error $2^{-\kappa-1}$.
With a uniform per-evaluation bound $\Rev$, the capped simulator also has polynomial total work in the stated parameters and $\Rev$.
The expected-work guarantee and the guarantee for the capped simulator are both stated for one evaluation. \Cref{thm:compiler-stateful} extends the expected-work guarantee to repeated evaluation on persistent state with an online simulator that runs the same operations.
The capped variant is not formalized (see \cref{sec:lean-formalization}).
\section{Formalization and extensions}\label{sec:formalization}
With the storage and single-evaluation compiler guarantees in place, we can describe precisely what is checked in Lean.
We then discuss extensions to adaptive leakage within a refresh, stateful circuits, and larger finite fields.
We prove and formalize the extension to stateful circuits in \cref{sec:stateful}. We sketch the extension to adaptivity within a refresh and discuss the extension to larger fields. Neither of these last two extensions is formalized.

\subsection{Lean formalization}\label{sec:lean-formalization}
Our formalization, written in Lean~4~\cite{DBLP:conf/cade/Moura021} with mathlib~\cite{DBLP:conf/cpp/X20}, covers the binary continual-refreshing bound (see \cref{thm:main}), the NAND compiler and its signed composition theorem with history-dependent labels (see \cref{thm:compiler-direct,thm:direct-composition}), the concrete expected-work simulator of \cref{sec:direct-simulator}, and continual stateful privacy against adaptively chosen public inputs (see \cref{thm:compiler-stateful}).
The source files are available in the \githubsub folder of the project repository. A separate note, \emph{Annotated Lean Reference Specifications}, is available at \path{lean/annotated-specifications.pdf}.

Lean checks that a proof establishes its formal statement. Whether that statement faithfully expresses the intended mathematical claim always remains for human readers to judge.
Simplicity of the model and statements is therefore particularly important.
We deliberately use deterministic strategies, with the adaptivity, locality, phase ordering, and erasures specified in the paper.
Randomized adversaries and randomized leakage functions are handled by the averaging reduction of \cref{sec:removing-randomness,sec:compiler-phase-randomness}: fixing the adversary's tape and the presampled auxiliary choices gives a deterministic strategy, and the uniform bound is then averaged over those choices.
This reduction, including its treatment of possibly infinite tapes, is a mathematical argument in the paper and is not part of the checked statements.
The protocol's own sampling records, including rejected proposals and retries, remain in the checked model.
The source interface is NAND-only and permits arbitrary fanout, repeated operands, zero gates, input wires designated directly as outputs, repeated output positions, and unused inputs.
The AND/OR/NOT conversion in \cref{sec:direct-source-model} is mathematical preprocessing outside that checked interface.

For the compiler, the formal development constructs the actual input-port refresh chains and proves their ordinary correctness, the $\Ipriv$ trusted-encoding count, $S=3t_0+\Ipub$, distinct gadget operands, and scope and incidence bounds of three.
The formal development also proves that the canonical experiment is exact and supplies the local approximation bounds for matrix refreshing and rectangular shortening, including full tapes of rejected proposals.
The formal proof of \cref{thm:direct-composition} follows \cref{sec:direct-composition}: energy trimming, the H\"older bound for incidence three, the joint expansion, tensor discrepancy, enumeration of guard answers under adaptive leakage, and inclusion--exclusion over marked sets.
Its explicit finite bound is $10^6\cdot S\cdot K^9\cdot q^{-1/27}$, where $q=2^n-1$ and $K=2^{2\lambda}$.
The formal local certificates allow separate mean energies of eight rather than the analytical bound of four in \cref{lem:compiler-local-interface}. The proof carries this larger allowance through trimming and absorbs it into the displayed absolute coefficient.
The construction and the exponents are unchanged.
The checked compiler theorem supplies the local certificates, exact canonical distribution, and incidence bounds for the actual compiled program, rather than assuming those properties.
Output encodings are hidden. The checked trusted decoder releases only the output tuple in the clear.

The simulator is fixed from the public source and supplied strategy before the private input and target error are quantified.
It receives the authorized output before answering leakage queries, runs the actual operations on zero private inputs, and discards its dummy output pairs.
The visible marginal of its counted execution, which includes every rejection and retry, is exactly this dummy-input view with the authorized output appended.
To keep the formal cost model simple, its counter records random-bit sampling, scalar arithmetic and tests, and explicitly charged preparation operations, including the corresponding costs of every rejected proposal and shortening retry.
Routine memory access and address bookkeeping are left to the conventional implementation analysis.
This keeps the formal development focused on the probabilistic and cryptographic arguments at the core of the security proof.
Writing $P:=\Ipriv$, Lean proves that the expected value of this counter is at most
\[
 W=P+k+16P\cdot(6n+1)+100S\cdot(n^2+2)^3,
\]
where $k$ counts designated output positions, including repetitions.
A uniform charge $\Rev$ for each adversary and leakage evaluation adds exactly $(8t_0+2\Ipub)\cdot\Rev$. Joint accounting can instead add the expected cost of tape-dependent additional computation without an independence assumption.
For $\kappa\ge1$, a checked conservative sufficient length is
\[
 n\ge12000\cdot\left(\kappa+\lambda+\ln(S+2)\right),\qquad n\ge2.
\]
Here $\ln$ is the natural logarithm used in Lean. The exact simulator marginal and work bound hold at every $n\ge2$, independently of this sufficient security threshold.

Each of the three public model files, \path{Refreshing.lean}, \path{Compiler.lean}, and \path{EfficientSimulation.lean}, has one bundled main theorem stated with \texttt{sorry}, Lean's placeholder for an omitted proof. Separate proof environments supply complete proofs and do not depend on those admitted declarations.
\path{Refreshing.lean} states its components directly in its main theorem, whereas \path{Compiler.lean} and \path{EfficientSimulation.lean} define named component propositions with their numerical bounds and then bundle them in their main theorems.
The public model files also admit needed construction certificates with \texttt{sorry}. Their complete proofs are checked separately by the Lean \texttt{comparator} tool, which checks the statements, definitions, and proof dependencies together.
Substantive proofs, proof-only definitions, and proof-only theorem statements are confined to \path{lean/Proofs/}. Elementary construction proofs may appear inline in the public definitions.
The repository describes the audit commands of the comparator and the permitted foundational axioms.
These results certify the stated claims within the probability model and the specified cost convention.
The results do not certify extracted machine code.
The strictly polynomial variant obtained by capping the same simulator is not formalized.
The extensions below to adaptivity within a phase and larger fields remain outside the formalized theorem statements.
The stateful development constructs the chronological unrolling, proves the retained-state and trusted-decoder semantics, and extends composition to causal history-dependent labels.
Its simulator is a function of an ideal oracle, independent of the initial state, and its complete view includes all public input/output records.
The formal oracle interface describes the chronological requests and authorized responses mathematically.
The counted stateful simulator uses the same cost convention and has the exact physical dummy-run marginal and expected counter bound obtained by replacing $P$ by the initial-state size and $k$ by the total number of output positions in the displayed bounds.
Adversary computations, public-history processing, and ideal-oracle evaluation are charged separately.

\subsection{Adaptivity within a refresh}\label{sec:within-round}
\Cref{thm:main} is stated for leakage that is adaptive across refreshes and nonadaptive within a refresh: the two leakage functions of a refresh are fixed before either answer of that refresh is revealed.
The proof can be extended to leakage that is also adaptive within a refresh.
In this stronger model, the adversary queries the parties' leakage functions one after another during the same refresh, choosing each function after seeing the earlier answers of that refresh, as long as each party leaks at most $\lambda$ bits in total.
We sketch the modifications. The theorems above and their formalization keep leakage nonadaptive within each phase.

\para{Answers and masks.}
Fix the adversary's random tape.
The interaction terminates after finitely many queries.
The queried party, the length of each answer, and whether to stop are determined by this tape and the preceding answers. Lengths, timing, and termination carry no additional observations.
The resulting binary transcript tree has depth at most $2\lambda$, and its leaves are prefix-free, so padding its answer strings to $2\lambda$ bits is injective.
Let $z$ be this padded complete answer, so that the answer alphabet still has at most $K:=2^{2\lambda}$ elements.
Fix the incoming history, the matrix $M$, and an answer $z$.
The part of $z$ that precedes an answer determines which leakage function produced that answer.
Set $\MaskL{z}(\ell,M):=1$ if, for every left answer in $z$, the function determined in this way outputs that answer on input $(\ell,M)$, and set it to zero otherwise. Define $\MaskR{z}(r,M)$ in the same way for the right party.
By induction along the sequence of answers, the old state $(\ell,r)$ produces $z$ exactly when $\MaskL{z}(\ell,M)\cdot \MaskR{z}(r,M)=1$.
Since every old state produces exactly one answer, the masks satisfy the partition identity in~\cref{eq:partition} as before.

\para{The rest of the proof.}
The local approximation in \cref{sec:joint-refresh} treats each answer separately and uses its two Boolean masks.
The approximation therefore provides the same per-answer nuclear, $L_2$, and energy estimates for complete interactive transcripts.
The history argument of \cref{sec:refresh-adaptivity} uses only the number of answers, at most $K$, and the fact that the true probability functions sum to one.
For private randomness in leakage functions, presample the auxiliary values for each possible query, indexed by the incoming history, the matrix, and the preceding answers within the refresh, and average the deterministic bounds.

\para{The compiler.}
Apply the same complete-transcript masks separately to the matrix and shortening phases, with total per-party budgets $\lambda_{\MatrixPhase}$ and $\lambda_{\ShortPhase}$ and at most $K=2^{2(\lambda_{\MatrixPhase}+\lambda_{\ShortPhase})}$ complete answers.
For shortening, each complete transcript still determines a left mask and a right mask.
The per-answer transport in \cref{lem:compiler-shortening-transport} in the \cref{sec:matrix-estimates} uses only their bounds in $[0,1]$, so it transports the chosen energies and $L_2$ error without requiring separate answer partitions on the two sides.
The proof of \cref{thm:direct-composition} uses these per-answer estimates and the normalization of the true complete-transcript probabilities.
Erasure before shortening remains essential, and the shortening budget covers every retry.
Auxiliary randomness is presampled as in \cref{sec:compiler-phase-randomness}, indexed by the gate and the preceding answers. Erased matrix state is unavailable during shortening.
These observations describe the extension. We do not give a separate full proof for interactive leakage.

\subsection{Stateful circuits}\label{sec:stateful}
We now extend \cref{thm:compiler-direct} to stateful circuits that are invoked repeatedly, with the input of each invocation chosen by the adversary after it has seen the leakage of the earlier ones.
This is the continual-leakage setting for computation: the total leakage grows without bound in the number of invocations.
A stateful circuit has registers whose contents persist between invocations.
Fix state, input, and output lengths $m_{\mathrm{state}}$, $m_{\mathrm{in}}$, and $m_{\mathrm{out}}$.
Starting from an initial state $\mathsf{st}_0\in\F_2^{m_{\mathrm{state}}}$, its transition circuit computes on every invocation $t\ge0$
\[
 (\mathsf{st}_{t+1},\vecy_t):=\Circuit(\mathsf{st}_t,\vecx_t),
\]
where $\Circuit:\F_2^{m_{\mathrm{state}}}\times\F_2^{m_{\mathrm{in}}}\to\F_2^{m_{\mathrm{state}}}\times\F_2^{m_{\mathrm{out}}}$ is a NAND source circuit as in \cref{sec:direct-source-model}, with $t_0$ gates and $I_{\mathrm{const}}$ public constants, $\mathsf{st}_t$ is the current private state, $\vecx_t\in\F_2^{m_{\mathrm{in}}}$ is the public input of invocation $t$, and $\vecy_t\in\F_2^{m_{\mathrm{out}}}$ is its public output.
The registers supply $\mathsf{st}_{t+1}$ to the next invocation. Every feedback cycle passes through a register.
See Ishai, Sahai, and Wagner~\cite[Section~2]{DBLP:conf/crypto/IshaiSW03} for this model and its privacy definition against probing attacks.

\para{The compiled stateful protocol.}
Fix the number $T\ge1$ of invocations.
The trusted encoder encodes the bits of $\mathsf{st}_0$ independently, as in \cref{sec:compiler-direct}. These are the only trusted encodings of the entire execution.
Invocation $t$ starts once the adversary has chosen $\vecx_t$.
It runs the operations of the execution schedule of \cref{sec:compiler-direct} for $\Circuit$, with the current state encodings as its private operand encodings: the public-input samplers for the bits of $\vecx_t$, the input-port refreshes, and the NAND gadgets.
At the end of the invocation, the trusted decoder reads the encodings of the output wires and releases $\vecy_t$. As before, it releases no shares and produces no leakage.
The encodings of the wires that carry $\mathsf{st}_{t+1}$ are retained and serve as source wires of invocation $t+1$: each is refreshed by its input-port chain immediately before every use, and an old copy is erased once all of its uses, including any output position, have finished.
All other encodings of invocation $t$ are erased as in the single-evaluation protocol.
Retained encodings are inactive between invocations and inaccessible to every leakage query.
The public constants are sampled once, at the start of the execution, and reused through their input-port chains.

Unrolling the $T$ invocations gives one NAND source circuit $\Circuit^{(T)}$ with private input $\mathsf{st}_0$, public inputs $\vecx_0,\ldots,\vecx_{T-1}$ and the constants, and designated outputs $\vecy_0,\ldots,\vecy_{T-1}$, in which the state wires of invocation $t$ feed the gates of invocation $t+1$.
The compiled stateful protocol is the compiled protocol of \cref{sec:compiler-direct} for $\Circuit^{(T)}$, except that the samplers for $\vecx_t$ run at the start of invocation $t$ instead of at the start of the execution, and the output wires of invocation $t$ are decoded at the end of that invocation instead of at the end of the execution.
Neither change affects the scopes: every operation reads or produces at most three wires, and every wire belongs to at most three scopes, by the argument given after the execution schedule in \cref{sec:compiler-direct}.
By~\cref{eq:direct-size}, the number of operations is
\begin{equation}\label{eq:stateful-size}
 S=3T\cdot t_0+T\cdot m_{\mathrm{in}}+I_{\mathrm{const}}.
\end{equation}

\para{Leakage and security.}
Leakage is as in \cref{sec:compiler-direct} for every operation of every invocation: each operation has its stated budget, each query reads only its party's data for the active operation, both functions of a phase are fixed before either answer is returned, and communication is private.
The adversary is adaptive across the whole execution.
At the start of invocation $t$, it chooses $\vecx_t$ as a function of its auxiliary input, its tape, all leakage answers so far, and $\vecy_0,\ldots,\vecy_{t-1}$. During the invocation it chooses the leakage functions of each phase as a function of the same data together with $\vecx_t$.
Its \emph{view} consists of all leakage answers, the inputs $\vecx_0,\ldots,\vecx_{T-1}$, and the outputs $\vecy_0,\ldots,\vecy_{T-1}$.
The \emph{ideal stateful circuit} with initial state $\mathsf{st}_0$ answers the queries $\vecx_0,\vecx_1,\ldots$ in turn with the outputs $\vecy_0,\vecy_1,\ldots$ of the transition circuit.
A \emph{simulator} interacts with the adversary in place of the compiled protocol and has oracle access to the ideal stateful circuit. It does not receive $\mathsf{st}_0$.

\begin{theorem}[Continual leakage for stateful circuits]\label{thm:compiler-stateful}
There exist absolute constants $C\ge1$ and $C'\ge1$ with the following property.
For every transition circuit, dimension $n\ge2$, pair of nonnegative integer phase budgets, number of invocations $T\ge1$, and admissible adaptive adversary, there exists a simulator, independent of the initial state, such that for every $\mathsf{st}_0$ the adversary's views in the compiled stateful protocol and in the interaction with the simulator are at statistical distance at most
\begin{equation}\label{eq:stateful-simulation}
 \min\left\{1,\ C\cdot S\cdot K^9\cdot(2^n-1)^{-1/27}\right\}
 \le C'\cdot S\cdot2^{18\lambda-n/27},
\end{equation}
with $S$ from~\cref{eq:stateful-size} and $K=2^{2\lambda}$.
The simulator answers every query when it is asked and uses $\vecy_t$ only at the end of invocation $t$.
Its expected primitive work is polynomial in the circuit and input sizes, $T$, and $n$, with $8T\cdot t_0+2(T\cdot m_{\mathrm{in}}+I_{\mathrm{const}})$ local-answer evaluations, adversary and leakage computation being charged separately as in \cref{sec:direct-simulator}.
For every fixed $\alpha<1/486$, choosing $\lambda\le\alpha\cdot n$ gives error $S\cdot2^{-\Omega(n)}$, so share length $n=O(\kappa+\lambda+\log(S+2))$ suffices for target distance $2^{-\kappa}$: the share length grows only logarithmically in the number of invocations, while the total leakage $2\lambda S$ grows linearly.
\end{theorem}

\begin{proof}
\emph{The simulator.}
The simulator encodes the all-zero initial state and, on receiving $\vecx_t$, runs invocation $t$ of the compiled protocol on its dummy state encodings against the adversary, answering each leakage query from this execution.
At the end of invocation $t$, it queries the ideal stateful circuit with $\vecx_t$ and returns its authorized response $\vecy_t$, without revealing the dummy output values or encodings.
It retains any encodings needed for the next state or subsequent uses.
An encoding may be erased only after its last required use has finished.
Its operations are those of the compiled protocol for $\Circuit^{(T)}$, so the expected-work bound and the count of local-answer evaluations follow from \cref{sec:direct-simulator} with the operation counts of~\cref{eq:stateful-size}.

\emph{The view is a function of the transcript.}
Fix the adversary's tape, its auxiliary input, and $\mathsf{st}_0$.
In both experiments, $\vecy_t$ is the output of the ideal stateful circuit on $\mathsf{st}_0$ and $\vecx_0,\ldots,\vecx_t$: in the real experiment because every refresh and gadget is correct, so that the decoder releases the values of the output wires in the clear, and in the simulated experiment by construction.
By induction on $t$, the inputs $\vecx_t$ and outputs $\vecy_t$ are therefore the same deterministic functions of the leakage answers preceding invocation $t$ in both experiments, and the view is the same deterministic function of the leakage transcript in both.
It suffices to bound the statistical distance between the two transcript distributions.

\emph{Two history-dependent labelings.}
With $\mathsf{st}_0$ fixed, the leakage functions of every phase and the input of every invocation are functions of the preceding formal history. Thus the true tables $\prob_{i,h,z}$ and the labels of the public wires are fixed for all formal histories, as \cref{sec:direct-composition-setup} requires. These tables and public labels are the same in both experiments.
The real execution carries the history-dependent labeling $s$ given by the plaintext values of $\Circuit^{(T)}$ on $\mathsf{st}_0$ and these public labels, and the simulated execution carries the labeling $s'$ given by the plaintext values on $0^{m_{\mathrm{state}}}$ and the same public labels.
Both are valid for every history, and they agree on the public wires.
The label of a wire of invocation $t$ depends only on $\mathsf{st}_0$ and $\vecx_0,\ldots,\vecx_t$, hence only on the answers preceding invocation $t$, which precede the operation producing the wire.
Decoding is not an operation and produces no answer, so it does not enter the transcript.

\emph{Composition.}
The canonical experiment of the compiled stateful protocol is exact by \cref{lem:compiler-canonical} in the \cref{sec:matrix-estimates} and the remark following it. Every computation gadget and input-port refresh has the local approximation and mean-energy bounds of \cref{lem:copy-local,lem:compiler-local-interface}. These bounds hold for every formal history and do not depend on the schedule.
Together with the scope and incidence bounds noted above, \cref{thm:direct-composition} applies to $s$ and $s'$ and gives~\cref{eq:stateful-simulation} by~\cref{eq:direct-composition-bound}, including for a view that contains the adversary's tape.
As in the proof of \cref{thm:compiler-direct}, the binary form and the parameter choice follow.
\end{proof}

The theorem keeps the assumptions of \cref{sec:compiler-direct}: a trusted encoder for the initial state, a trusted decoder at the end of every invocation, private channels, erasures, and leakage that is nonadaptive within a phase.
Its causal composition bound, chronological compiler, decoder correctness, oracle simulation, and expected counter bound are formalized in Lean under the cost convention of \cref{sec:lean-formalization}.

\subsection{General fields}
The matrix action and inner-product encoding are defined over any finite field.
Extending the quantitative security bounds requires adapting the local estimates and source compilation. We do not prove that extension here.
All quantitative theorems and their Lean specifications in this paper use the binary field $\F_2$.
\section*{Acknowledgments}
The work of Stefan Dziembowski was supported by the European Research Council (ERC) under the European Union's Horizon 2020 research and innovation programme (project PROCONTRA, grant agreement No.~885666).
The work of Sebastian Faust was supported by the European Research Council (ERC) under the European Union's Horizon Europe research and innovation programme (project CRYPTOLAYER, grant agreement No.~101044770).

\printbibliography[title={References}]
\appendix
\crefalias{section}{appendix}
\crefalias{subsection}{subappendix}
\crefalias{subsubsection}{subsubappendix}
\section{Proofs for the preliminaries}\label{sec:preliminaries-proofs}
We provide proofs and supporting arguments for the statements, definitions, and examples in \cref{sec:preliminaries}.
All table and function norms use the uniform measures specified there.
For a linear map $A$, its kernel is $\ker(A)=\{c:A\cdot c=0\}$.

\phantomsection\label{app:prelim-probability}
\para{Statistical distance.}
For probability distributions $P$ and $Q$ on a finite set $V$, the positive and negative parts of $P-Q$ have equal total mass.
Thus $\frac12\LoneNorm{P-Q}=\sum_{v:P(v)>Q(v)}(P(v)-Q(v))$, which is also the largest absolute difference between the acceptance probabilities of a Boolean test.
The maximum is attained by the test that outputs one exactly where $P(v)>Q(v)$.
Let $R$ be the common random tape, with distribution $\tau$, and let $P_r$ and $Q_r$ be the finite view distributions for each fixed tape $r$.
The statistical distance of the joint tape-and-view distributions is $\E_{R\sample\tau}{\StatDist{P_R}{Q_R}}$, also when $R$ is a possibly infinite bit tape.
Indeed, the maximizing test outputs one exactly where $P_R(v)>Q_R(v)$.

\subsection{Refresh correctness and encoding counts}\label{app:prelim-counting}
We check that a refresh preserves the encoded bit and count the encodings of each bit.
\para{Preservation of the encoded secret.}
Since $(L_{i-1}\cdot M_i)\cdot\left(M_i^{-1}\cdot R_{i-1}\right)=L_{i-1}\cdot R_{i-1}$, every refresh preserves the secret, and invertibility keeps both shares nonzero.

\para{Counting encodings.}
For every nonzero row $\ell$, exactly half of the $N$ binary columns satisfy $\ell\cdot r=0$ and half satisfy $\ell\cdot r=1$.
Indeed, $r\mapsto\ell\cdot r$ is a surjective linear functional on $\F_2^n$.
Removing the zero column leaves $N/2-1$ partners for secret zero and $N/2$ for secret one.
The same count holds with rows and columns exchanged.
Thus every nonzero share has the same number $m_s$ of nonzero partners for a given secret $s$, and
\[
 m_0=(N-2)/2,\qquad m_1=N/2,\qquad
 |\Omega_s|=q\cdot m_s,\qquad p_s=m_s/q.
\]
In particular, both $\mu_0$ and $\mu_1$ have uniform left and right marginals, and conditioning $\nu$ on $\Omega_s$ gives $\mu_s$.
Since $N\ge4$, the class probabilities satisfy
\[
 p_1\ge p_0=\frac{N-2}{2(N-1)}\ge\frac13.
\]

\begin{lemmarestatement}{lem:fresh}{Freshness}
For every $s\in\F_2$ and every $(\ell,r)\in\Omega_s$, if $M\sample\InvMatr_2^n$, then the refreshed pair $\left(\ell\cdot M,M^{-1}\cdot r\right)$ has distribution $\mu_s$.
\end{lemmarestatement}
\begin{proof}[Proof of \cref{lem:fresh}]
We show that the action $(\ell,r)\mapsto\left(\ell\cdot M,M^{-1}\cdot r\right)$ is transitive on $\Omega_s$.
For two states $(\ell,r)\in\Omega_s$ and $(\ell',r')\in\Omega_s$, it suffices to find an invertible $M$ with $M\cdot r'=r$ and $\ell\cdot M=\ell'$.
If $s\ne0$, map a basis of $\ker(\ell')$ to a basis of $\ker(\ell)$, and map $r'$ to $r$.
Appending the specified vectors to the kernel bases gives bases of the full column space because $\ell'\cdot r'=\ell\cdot r=s\ne0$.
If $s=0$, extend $r'$ and $r$ to bases of $\ker(\ell')$ and $\ker(\ell)$, respectively, and append vectors $v'$ and $v$ with $\ell'\cdot v'=\ell\cdot v=1$.
Map the first adapted basis to the second.
In both cases the resulting map is invertible, sends $r'$ to $r$, and makes $\ell\cdot M$ agree with $\ell'$ on a basis.
Since the action is transitive, for a fixed state the sets of matrices that take it to each state of $\Omega_s$ have equal cardinality: multiplication by a matrix taking one target state to another gives a bijection between the corresponding sets.
Thus a uniform matrix gives a uniform state of $\Omega_s$.
\end{proof}

The same freshness statement holds conditionally on the preceding execution because the next matrix is independent of that execution.
Conditioning on a leakage answer can instead change the matrix distribution, so freshness need not hold after that conditioning.

\subsection{Linear algebra for tables}\label{app:prelim-linear-algebra}
Using the orthonormal expansions recalled in \cref{lem:orthonormal-expansion}, we establish the minimum-cost characterization of their coefficient sum.
This characterization bounds the nuclear norm of arbitrary product representations, including the products of local approximations in \cref{lem:refresh-wire-change,clm:direct-marked-bound}, and it gives the minimum-cost representation in the proof of \cref{lem:disc}.

\subsubsection{Product tables}\label{app:prelim-products}
The following elementary identities for tables of the form $a\cdot b^{\transpose}$ are used throughout.
\begin{lemma}[Inner products and norms of product tables]\label{lem:la-products}
For functions $a$ and $c$ from $\UL$ to $\R$, and $b$ and $e$ from $\UR$ to $\R$,
\begin{align}
 \pairing{a\cdot b^{\transpose}}{c\cdot e^{\transpose}}
 &=\pairing{a}{c}\cdot\pairing{b}{e},
 \label{eq:product-table-pairing}\\
 \LtwoNorm{a\cdot b^{\transpose}}&=\LtwoNorm{a}\cdot\LtwoNorm{b}.
 \label{eq:product-table-L2}
\end{align}
In particular, every unit product table has $L_2$ norm one.
\end{lemma}
\begin{proof}[Proof of \cref{lem:la-products}]
Independence of $\ell$ and $r$ gives
\[
 \E_{\ell,r}{a(\ell)\cdot b(r)\cdot c(\ell)\cdot e(r)}
 =\E_\ell{a(\ell)\cdot c(\ell)}\cdot\E_r{b(r)\cdot e(r)}.
\]
Taking $c=a$ and $e=b$ in~\cref{eq:product-table-pairing} and then square roots proves~\cref{eq:product-table-L2}.
\end{proof}

\subsubsection{Minimum-cost factorizations}\label{app:prelim-factor-minimum}
We now prove that the nuclear norm is the smallest cost of a product representation, which is the form in which the security proofs use it.
\begin{definition}[Product cost]\label{def:product-cost}
For a finite product representation $F=\sum_j a_j\cdot b_j^{\transpose}$, its \emph{product cost} is $\sum_j\LtwoNorm{a_j}\cdot\LtwoNorm{b_j}$.
\end{definition}

\begin{lemma}[Minimum-cost characterization of the nuclear norm]\label{lem:nuclear-factor-characterization}
The nuclear norm of every real table $F:\UL\times\UR\to\R$ is its smallest product cost:
\begin{equation}\label{eq:nuclear-factor-cost}
 \NuclearNorm{F}
 =\min_{F=\sum_k a_k\cdot b_k^{\transpose}}\sum_k\LtwoNorm{a_k}\cdot\LtwoNorm{b_k}.
\end{equation}
The minimum ranges over all finite product representations, including the empty representation for the zero table.
It is attained by $a_j=\sqrt{\sigma_j}\cdot u_j$ and $b_j=\sqrt{\sigma_j}\cdot v_j$ from an orthonormal expansion of $F$.
\end{lemma}
\begin{proof}[Proof of \cref{lem:nuclear-factor-characterization}]
We verify the standard factorization characterization~\cite[Sec.~3, Lem.~1]{DBLP:conf/nips/SrebroRJ04} in our normalization.
Fix an orthonormal expansion of $F$ from \cref{lem:orthonormal-expansion} and put $W:=\sum_j u_j\cdot v_j^{\transpose}$.
\Cref{lem:la-products} gives $\pairing{F}{W}=\sum_j\sigma_j=\NuclearNorm{F}$.
For any left function $a$, Bessel's inequality (see~\cref{eq:bessel-inequality}) gives
\begin{equation}\label{eq:orthonormal-coefficient-bound}
 \sum_j\pairing{a}{u_j}^2\le\LtwoNorm{a}^2.
\end{equation}
The same inequality holds for the right family.
Consequently the Cauchy--Schwarz inequality (see \cref{lem:la-cauchy-schwarz}) gives, for any product table,
\[
 \left|\pairing{a\cdot b^{\transpose}}{W}\right|
 =\left|\sum_j\pairing{a}{u_j}\cdot\pairing{b}{v_j}\right|
 \le\LtwoNorm{a}\cdot\LtwoNorm{b}.
\]
Every finite representation therefore satisfies
\[
 \NuclearNorm{F}=\pairing{F}{W}
 \le\sum_k\LtwoNorm{a_k}\cdot\LtwoNorm{b_k}.
\]
The stated factors have $\LtwoNorm{a_j}^2=\LtwoNorm{b_j}^2=\sigma_j$, so they attain the minimum in~\cref{eq:nuclear-factor-cost}.
The empty representation gives the same conclusion when $F=0$.
\end{proof}

\begin{lemmarestatement}{lem:factor-minimum}{Basic nuclear-norm bounds}
The function $\NuclearNorm{\cdot}$ is a norm on the real tables on $\UL\times\UR$.
For every such table $F$,
\begin{equation*}\tag*{\labelcref*{eq:nuclear-dominates-ltwo}}
 \LtwoNorm{F}\le\NuclearNorm{F}.
\end{equation*}
For every integer $m\ge0$ and all functions $a_j:\UL\to\R$ and $b_j:\UR\to\R$, indexed by $j=1,\ldots,m$,
\begin{equation*}\tag*{\labelcref*{eq:nuclear-product-bound}}
 \NuclearNorm{\sum_j a_j\cdot b_j^{\transpose}}
 \le\sum_j\LtwoNorm{a_j}\cdot\LtwoNorm{b_j}.
\end{equation*}
For every pair of functions $a:\UL\to\R$ and $b:\UR\to\R$,
\begin{equation*}\tag*{\labelcref*{eq:nuclear-product-identity}}
 \NuclearNorm{a\cdot b^{\transpose}}=\LtwoNorm{a}\cdot\LtwoNorm{b}.
\end{equation*}
\end{lemmarestatement}
\begin{proof}[Proof of \cref{lem:factor-minimum}]
The product tables $u_j\cdot v_j^{\transpose}$ in an orthonormal expansion are themselves orthonormal by \cref{lem:la-products}.
Pythagoras' theorem (see~\cref{eq:orthonormal-sum-norm}) therefore gives $\LtwoNorm{F}^2=\sum_j\sigma_j^2$, and hence
\[
 \LtwoNorm{F}=\left(\sum_j\sigma_j^2\right)^{1/2}
 \le\sum_j\sigma_j=\NuclearNorm{F}.
\]
The inequality uses $\sigma_j\ge0$ and also holds for the empty expansion of the zero table.
\Cref{eq:nuclear-factor-cost} gives the arbitrary-representation bound~\cref{eq:nuclear-product-bound} and a representation attaining the minimum.
For a single product, \cref{eq:nuclear-factor-cost,eq:product-table-L2} give
\[
 \LtwoNorm{a}\cdot\LtwoNorm{b}=\LtwoNorm{a\cdot b^{\transpose}}
 \le\NuclearNorm{a\cdot b^{\transpose}}
 \le\LtwoNorm{a}\cdot\LtwoNorm{b},
\]
proving the product-table formula.
Concatenating representations attaining the minimum for two tables proves the triangle inequality for $\NuclearNorm{\cdot}$.
For real $\tau\ne0$, replacing $\sigma_j$ and $u_j$ in an orthonormal expansion of $F$ by $|\tau|\cdot\sigma_j$ and $\sgn{\tau}\cdot u_j$ gives an orthonormal expansion of $\tau F$, so $\NuclearNorm{\tau F}=|\tau|\cdot\NuclearNorm{F}$. The case $\tau=0$ is trivial.
The lower bound makes the norm positive on nonzero tables.
\end{proof}

\subsubsection{Largest singular coefficient}\label{app:prelim-largest-coefficient}
\begin{lemmarestatement}{lem:largest-singular-coefficient}{Largest singular coefficient}
For every real table $F:\UL\times\UR\to\R$ and all functions $u:\UL\to\R$ and $v:\UR\to\R$,
\begin{equation*}\tag*{\labelcref*{eq:largest-singular-coefficient}}
 \left|\pairing{F}{u\cdot v^{\transpose}}\right|
 \le\sigma_{\max}(F)\cdot\LtwoNorm{u}\cdot\LtwoNorm{v}.
\end{equation*}
For every nonzero $F$, there exist unit functions $u$ and $v$ attaining equality in~\cref{eq:largest-singular-coefficient}.
For every real table $F$,
\begin{equation*}\tag*{\labelcref*{eq:largest-singular-ltwo}}
 \sigma_{\max}(F)\le\LtwoNorm{F}.
\end{equation*}
\end{lemmarestatement}
\begin{proof}[Proof of \cref{lem:largest-singular-coefficient}]
Write $F=\sum_j\sigma_j\cdot u_j\cdot v_j^{\transpose}$.
For $F=0$ the claims hold immediately.
Otherwise, orthonormality and the Cauchy--Schwarz inequality (see \cref{lem:la-cauchy-schwarz}) give
\[
\begin{aligned}
 \left|\pairing{F}{u\cdot v^{\transpose}}\right|
 &\le\sigma_{\max}(F)\cdot
 \left(\sum_j\pairing{u}{u_j}^2\right)^{1/2}
 \cdot\left(\sum_j\pairing{v}{v_j}^2\right)^{1/2}\\
 &\le\sigma_{\max}(F)\cdot\LtwoNorm{u}\cdot\LtwoNorm{v}.
\end{aligned}
\]
The last inequality applies~\cref{eq:orthonormal-coefficient-bound} to each orthonormal family.
Choosing $u=u_j$ and $v=v_j$ at a largest coefficient attains equality.
Finally, $\sigma_{\max}(F)\le\left(\sum_j\sigma_j^2\right)^{1/2}=\LtwoNorm{F}$.
\end{proof}

\subsubsection{Pointwise energy bounds}\label{app:prelim-factor-covariance}
For every finite representation $F=\sum_j a_j\cdot b_j^{\transpose}$, with factor energies $\Phi_F$ and $\Psi_F$ from \cref{def:factor-energy}, the pointwise bound is
\begin{equation*}\tag*{\labelcref*{eq:factor-energy-entry-bound}}
 |F(\ell,r)|\le\sqrt{\Phi_F(\ell)\cdot\Psi_F(r)}
 \qquad(\ell\in\UL,\ r\in\UR).
\end{equation*}
\begin{proof}[Proof of \cref{eq:factor-energy-entry-bound}]
For any chosen finite factorization $F=\sum_j a_j\cdot b_j^{\transpose}$ and every $\ell$ and $r$, the Cauchy--Schwarz inequality (see \cref{lem:la-cauchy-schwarz}) in the factor index gives
\[
 |F(\ell,r)|
 =\left|\sum_j a_j(\ell)\cdot b_j(r)\right|
 \le\sqrt{\sum_j a_j(\ell)^2}\cdot\sqrt{\sum_j b_j(r)^2}
 =\sqrt{\Phi_F(\ell)\cdot\Psi_F(r)}.
\]
\end{proof}

\subsection{Binary signs and discrepancy}\label{app:prelim-sign-discrepancy}
Recall that $N:=2^n$, $q=N-1$, and $n\ge2$.
The discrepancy estimate follows by counting the signs of a nonzero binary linear form.
The local approximation proof uses the same cancellation of binary signs to control the consistency condition (see \cref{app:direct-approximation-proof}).

\begin{lemma}[The binary inner-product sign table]\label{lem:binary-sign-table}
On nonzero rows and columns, define
\[
 \Upsilon(\ell,r)=\begin{cases}1,&\ell\cdot r=0,\\-1,&\ell\cdot r=1,\end{cases}
 \qquad
 (\AvSign(v))(\ell)=\E_{r\sample \UR}{\Upsilon(\ell,r)\cdot v(r)}.
\]
For every real function $v:\UR\to\R$,
\begin{equation}\label{eq:binary-sign-L2}
 \LtwoNorm{\AvSign(v)}^2
 =\frac{N}{(N-1)^2}\cdot\LtwoNorm{v}^2
  -\frac1{N-1}\cdot
       \left(\E_{r\sample \UR}{v(r)}\right)^2.
\end{equation}
Consequently, for all real functions $u$ and $v$ on the respective share spaces,
\begin{equation}\label{eq:binary-sign-product-bound}
 \left|\pairing{\Upsilon}{u\cdot v^{\transpose}}\right|
 \le\frac{\sqrt N}{N-1}\cdot\LtwoNorm{u}\cdot\LtwoNorm{v}.
\end{equation}
\end{lemma}
\begin{proof}
For distinct nonzero columns $r$ and $r'$, the product $\Upsilon(\ell,r)\cdot\Upsilon(\ell,r')$ is $1$ or $-1$ according to the value of the nonzero linear form $\ell\cdot(r+r')$.
Among all $N$ rows its two values occur equally often. Removing the zero row removes one positive sign.
Thus its mean over nonzero $\ell$ is $-1/q$, whereas for $r=r'$ the mean is $1$.
Expanding the square gives
\[
 \begin{aligned}
 \LtwoNorm{\AvSign(v)}^2
 &=\frac1{q^2}\sum_r v(r)^2
  -\frac1{q^3}\sum_{r\ne r'}v(r)\cdot v(r')
 \\
 &=\frac{q+1}{q^2}\cdot\LtwoNorm{v}^2
  -\frac1q\cdot\left(\E_{r\sample \UR}{v(r)}\right)^2.
 \end{aligned}
\]
Since $q+1=N$, this is~\cref{eq:binary-sign-L2}.
Dropping its nonnegative last term and applying the Cauchy--Schwarz inequality (see \cref{lem:la-cauchy-schwarz}) to $\E_{\ell\sample \UL}{u(\ell)\cdot(\AvSign(v))(\ell)}$ proves~\cref{eq:binary-sign-product-bound}.
\end{proof}

\begin{table}[p]
\centering
\small
\renewcommand{\arraystretch}{1.08}
\begin{tabular}{@{}p{0.27\linewidth}p{0.56\linewidth}r@{}}
\toprule
Symbol & Meaning & Sec.\\
\midrule
\multicolumn{3}{@{}l}{\textit{Storage and refreshing (see \crefrange{sec:preliminaries}{sec:refreshing})}}\\
$n$, $N:=2^n$, $q:=N-1$ & share length in bits; number of $n$-bit strings; number of nonzero shares & \labelcref{sec:shares}\\
$\UL$, $\UR$ & nonzero rows and nonzero columns in $\F_2^n$ & \labelcref{sec:shares}\\
$\Omega_s$, $\mu_s$ & pairs of shares with inner product $s$; the uniform distribution on them & \labelcref{sec:shares}\\
$\nu$, $p_s$ & independent uniform nonzero shares; probability that their inner product is $s$ & \labelcref{sec:counting}\\
$(L,R)$, $(L',R')$ & old and refreshed shares in a single refresh & \labelcref{sec:refresh-joint-setup}\\
$T$, $M_i$ & number of refreshes; uniform invertible matrix of refresh $i$ & \labelcref{sec:shares}\\
$\lambda$, $K:=2^{2\lambda}$ & bits leaked per party per refresh; number of complete answers & \labelcref{sec:leakage}\\
$Z_i$, $H_i$ & answer of refresh $i$; history $(Z_1,\ldots,Z_i)$ & \labelcref{sec:leakage}\\
$\MaskL{z}$, $\MaskR{z}$ & left and right masks of the answer $z$ & \labelcref{sec:leakage}\\
$P_s^T$ & distribution of the complete transcript for secret $s$ & \labelcref{sec:security}\\
$\delta$, $\LoneNorm{P}$ & statistical distance; sum of absolute values of a finite signed weight function $P$ & \labelcref{sec:conventions}\\
$\LtwoNorm{F}$, $\NuclearNorm{F}$ & $L_2$ norm and nuclear norm of a table $F$ & \labelcref{sec:linear-algebra}\\
$\pairing{F}{W}$ & pairing of two tables & \labelcref{sec:linear-algebra}\\
$\Phi_F$, $\Psi_F$ & energies of a chosen product representation & \labelcref{sec:linear-algebra}\\
$\varphi$, $\project{F}$ & density of the refresh experiment; consistency projection of a table $F$ & \labelcref{sec:refresh-joint-setup}\\
$X_i=(L_i,R_i)$ & encoding after refresh $i$ & \labelcref{sec:shares}\\
$\prob_{i,h,z}$, $G_{i,h,z}$ & leakage probability function; its signed approximation & \labelcref{sec:copy-canonical}, \labelcref{sec:copy-local}\\
$\beta$, $\varepsilon$ & singular-coefficient threshold; per-answer $L_2$ approximation error & \labelcref{sec:copy-local}\\
$\sigma_{\max}(F)$, $\DiscBound$ & largest singular coefficient; one-encoding discrepancy bound & \labelcref{sec:linear-algebra}, \labelcref{sec:refresh-wire-change}\\
$\kappa$, $C$ & security parameter for target distance $2^{-\kappa}$; absolute constant & \labelcref{sec:field-size}, \labelcref{sec:conventions}\\
\midrule
\multicolumn{3}{@{}l}{\textit{Circuit compiler and the approximation lemma with retained local data}}\\
$t_0$, $S$ & source NAND gates after conversion; all $S=3t_0+\Ipub$ operations of the compiler & \labelcref{sec:compiler-direct}\\
$\Ipriv$, $\Ipub$, $k$ & private input wires; public or constant input wires; designated outputs & \labelcref{sec:direct-source-model}\\
$\vecx$, $\vecs$, $\vecy$ & public input; private input; authorized output & \labelcref{sec:direct-source-model}\\
$\lambda_{\MatrixPhase}$, $\lambda_{\ShortPhase}$ & per-party matrix and shortening budgets; $\lambda=\lambda_{\MatrixPhase}+\lambda_{\ShortPhase}$ & \labelcref{sec:direct-phase-model}\\
$S_i$ & scope of an operation: the encoding locations it reads or produces & \labelcref{sec:direct-source-model}\\
$\overline G$, $B$, $D_b$ & trimmed approximation; trimming threshold; encoding difference $\mu_b-\nu$ & \labelcref{sec:direct-composition}\\
$a$, $b$, $\OldData$, $d=|\OldData|$ & injective maps; set of old local data and its size & \labelcref{sec:joint-setup}\\
$\OldL$, $\OldR$ & old local data of the left and the right party & \labelcref{sec:joint-setup}\\
$m$, $N_m=2^m$, $q_m=N_m-1$ & long-share length; numbers of all and nonzero long shares & \labelcref{sec:compiler-gadget}, \labelcref{sec:joint-setup}\\
$U$, $V$; $A$, $R$ & long refreshed shares; short output shares & \labelcref{sec:compiler-gadget}\\
$H$ & shortening matrix with $A\cdot H=U$ and $R=H\cdot V$ & \labelcref{sec:compiler-gadget}\\
\bottomrule
\end{tabular}
\caption{Main notation.}
\label{tab:notation}
\end{table}

\subsubsection{Two-secret discrepancy}\label{app:prelim-discrepancy}
We express the difference between expectations under $\mu_0$ and $\mu_1$ using the sign table, then apply the Cauchy--Schwarz inequality (see \cref{lem:la-cauchy-schwarz}).
A minimum-cost product factorization (see~\cref{eq:nuclear-factor-cost}) extends the result to arbitrary real tables.

\begin{lemmarestatement}{lem:disc}{Two-secret discrepancy}
For every real table $F$,
\begin{equation*}\tag*{\labelcref*{eq:table-disc}}
 \left|\E_{(\ell,r)\sample \mu_0}{F(\ell,r)}-\E_{(\ell,r)\sample \mu_1}{F(\ell,r)}\right|
 \le\frac3{\sqrt{N-2}}\cdot\NuclearNorm{F}.
\end{equation*}
In particular, for real functions $u:\UL\to\R$ and $v:\UR\to\R$,
\begin{equation*}\tag*{\labelcref*{eq:disc}}
 \left|\E_{(\ell,r)\sample \mu_0}{u(\ell)\cdot v(r)}-\E_{(\ell,r)\sample \mu_1}{u(\ell)\cdot v(r)}\right|
 \le\frac3{\sqrt{N-2}}\cdot\LtwoNorm{u}\cdot\LtwoNorm{v}.
\end{equation*}
\end{lemmarestatement}
\begin{proof}[Proof of \cref{lem:disc}]
First consider a product table $u(\ell)\cdot v(r)$.
Subtracting the uniform mean from $v$ leaves its discrepancy unchanged, because the two secret distributions have the same left marginal, and does not increase its $L_2$ norm.
We may therefore assume $\E_{r\sample \UR}{v(r)}=0$.
For a fixed left share $\ell$, let
\[
 \GapFn(\ell)=\E_{R\sample \{r\in\UR:\ell\cdot r=0\}}{v(R)}
         -\E_{R\sample \{r\in\UR:\ell\cdot r=1\}}{v(R)}.
\]
The two classes have uniform probabilities $p_0$ and $p_1$.
Their weighted means sum to zero, while their difference weighted with signs is $(\AvSign(v))(\ell)$.
Since $p_0+p_1=1$, solving these two equations gives
\begin{equation}\label{eq:conditional-mean-gap}
 \GapFn(\ell)=\frac{(\AvSign(v))(\ell)}{2\cdot p_0\cdot p_1}.
\end{equation}
\Cref{eq:conditional-mean-gap,eq:binary-sign-L2} and \cref{eq:encoding-zero-probability,eq:encoding-one-probability} now imply
\begin{equation}\label{eq:conditional-mean-gap-L2}
 \LtwoNorm{\GapFn}
 \le\frac{2(N-1)}{(N-2)\sqrt N}\cdot\LtwoNorm{v}
 \le\frac3{\sqrt{N-2}}\cdot\LtwoNorm{v}.
\end{equation}
For the last inequality, use $N-1\le N$, $N/(N-2)\le2$, and $2\sqrt2\le3$.
Averaging over the common uniform left marginal and applying the Cauchy--Schwarz inequality (see \cref{lem:la-cauchy-schwarz}) with~\cref{eq:conditional-mean-gap-L2} gives
\[
 \begin{aligned}
 &\left|\E_{(\ell,r)\sample \mu_0}{u(\ell)\cdot v(r)}
       -\E_{(\ell,r)\sample \mu_1}{u(\ell)\cdot v(r)}\right|
 \\
 &\qquad=\left|\E_{\ell\sample \UL}{u(\ell)\cdot \GapFn(\ell)}\right|
 \le\frac3{\sqrt{N-2}}\cdot\LtwoNorm{u}\cdot\LtwoNorm{v}.
 \end{aligned}
\]
Finally, choose a minimum-cost representation $F=\sum_j a_j\cdot b_j^{\transpose}$ using~\cref{eq:nuclear-factor-cost}, and apply~\cref{eq:disc} to each term.
Since $\sum_j\LtwoNorm{a_j}\cdot\LtwoNorm{b_j}=\NuclearNorm{F}$, summing proves~\cref{eq:table-disc}.
\end{proof}

\section{The compiler approximation and shortening}\label{sec:matrix-estimates}
We first extend the refreshing approximation, proved in \cref{app:direct-approximation-proof}, to the retained local data of \cref{sec:joint-general}. We then specialize that extension to tensor encodings and show that rectangular shortening transports the approximation. Together, these three steps prove \cref{lem:compiler-local-interface}.
Recall that $\prob$ denotes a true leakage probability function, $G$ its signed approximation, and $\Phi$ and $\Psi$ the factor energies of a chosen product representation (see \cref{def:factor-energy}). The retained local data are the operand shares that the leakage may read in addition to the shares being refreshed.
All constants are absolute, and all unqualified table norms use the specified product of uniform measures.

\subsection{Extending the approximation to retained local data}\label{app:general-approximation-proof}
Recall the setup of \cref{sec:joint-setup}: the old local data $\OldL$ and $\OldR$ are sampled independently and uniformly from a set $\OldData$ of size $d$, and the injective maps $a$ and $b$ take them to nonzero $m$-bit shares.
The grouped coordinates are $x=(\OldL,U)$ and $y=(\OldR,V)$, with uniform product reference measure. The consistency projection keeps precisely the tuples with $U\cdot V=a(\OldL)\cdot b(\OldR)$.
The table $\prob_z$ is the conditional probability of one leakage answer $z$, extended by zero off these tuples.

\begin{lemmarestatement}{lem:joint-map}{Approximation of an answer with retained local data}
There exists an absolute constant $C>0$ such that the following holds for every formal incoming history, adversary tape, answer $z$, and threshold $\beta\in[d^{-1/3},1]$.
Take a probability-normalized orthonormal expansion (see \cref{lem:orthonormal-expansion})
$\prob_z=\sum_j\sigma_j\cdot u_j\cdot v_j^{\transpose}$ and put
\begin{equation*}\tag*{\labelcref*{eq:joint-direct-construction}}
 I:=\{j:\sigma_j>\beta\},\qquad
 \prob_z^{>\beta}:=\sum_{j\in I}\sigma_j\cdot u_j\cdot v_j^{\transpose}.
\end{equation*}
Then
\begin{align*}
 \NuclearNorm{\prob_z^{>\beta}}&\le2,\tag*{\labelcref*{eq:joint-map-budget}}\\
 \LtwoNorm{\prob_z^{>\beta}}&\le1.\tag*{\labelcref*{eq:joint-map-L2}}
\end{align*}
The balanced factors $\sqrt{\sigma_j}\cdot u_j$ and $\sqrt{\sigma_j}\cdot v_j$ of $\prob_z^{>\beta}$ (see \cref{def:balanced-factors}) have energies $\Phi_\beta$ and $\Psi_\beta$ satisfying, for every grouped row $x\in\TableRows$ and column $y\in\TableCols$,
\begin{align*}
 \Phi_\beta(x)&\le1/\beta,\tag*{\labelcref*{eq:joint-direct-energies}}\\
 \Psi_\beta(y)&\le1/\beta,\tag*{\labelcref*{eq:joint-direct-right-energy}}\\
 \E_{x\sample \TableRows}{\Phi_\beta(x)}&=\NuclearNorm{\prob_z^{>\beta}}\le2,\tag*{\labelcref*{eq:joint-direct-left-mean}}\\
 \E_{y\sample \TableCols}{\Psi_\beta(y)}&=\NuclearNorm{\prob_z^{>\beta}}\le2.\tag*{\labelcref*{eq:joint-direct-right-mean}}
\end{align*}
Finally,
\begin{equation*}\tag*{\labelcref*{eq:joint-map-error}}
 \LtwoNorm{\prob_z-2\cdot\project{\prob_z^{>\beta}}}\le C\cdot\beta^{1/3}.
\end{equation*}
The constant $C$ is independent of the dimension, threshold, answer, formal history, and admissible masks.
The construction depends on the masks and public maps, not on any plaintext labels.
\end{lemmarestatement}

\begin{proof}[Proof of \cref{lem:joint-map}]
We extend the proof of \cref{lem:refresh-rectangle}, keeping track of the two sizes $d=|\OldData|$ and $q_m=2^m-1$.
Throughout, $1\le d\le q_m$ and $d\cdot\beta^3\ge1$, where $q_m=N_m-1$ and $N_m=2^m$.
All averages use the uniform measures of \cref{sec:joint-setup}, unless full vector spaces are specified.

\para{Projection bounds.}
The row and column projection argument in \crefrange{eq:direct-left-unit-energy}{eq:direct-right-factor-energy} applies to any bounded table, so it applies to $\prob_z$ on $\TableRows\times\TableCols$.
Put $e_u(x):=\sum_{j\in I}u_j(x)^2$ and define $e_v$ similarly.
Every row and column of $\prob_z^{>\beta}$ has $L_2$ norm at most one, and
\begin{align}
 e_u&\le\beta^{-2},\nonumber\\
 e_v&\le\beta^{-2},\nonumber\\
 \Phi_\beta&\le\beta^{-1},\nonumber\\
 \Psi_\beta&\le\beta^{-1},\nonumber\\
 \E_{x\sample\TableRows}{e_u(x)^2}&\le |I|/\beta^2,\label{eq:general-left-projection-moment}\\
 \E_{y\sample\TableCols}{e_v(y)^2}&\le |I|/\beta^2,\label{eq:general-right-projection-moment}\\
 \LtwoNorm{\prob_z^{>\beta}}&\le1.\label{eq:general-projection-L2}
\end{align}
These statements also hold for an empty retained family.

\para{Gram variance on an injectively encoded set.}
For a function $g_{\OldL}(U)$ valued in a finite-dimensional real inner-product space, put
$\overline g_{\OldL}:=\E_U{g_{\OldL}(U)}$, $\widehat g_{\OldL}(U):=g_{\OldL}(U)-\overline g_{\OldL}$, and
$Y_M:=d^{-1}\sum_{\OldL} g_{\OldL}(a(\OldL)\cdot M)$.
Here $\OldL\in\OldData$ and $U\in\UL^{(m)}$.
For $\OldL\ne \OldL'$, the distinct nonzero vectors $a(\OldL)$ and $a(\OldL')$ are linearly independent over $\F_2$.
Their images under uniform invertible $M$ therefore form a uniform ordered pair of distinct nonzero rows.
Since each $\widehat g_{\OldL}$ has mean zero, deleting the diagonal from the sum over all pairs gives
\[
 \E_M{\pairing{\widehat g_{\OldL}(a(\OldL)\cdot M)}{\widehat g_{\OldL'}(a(\OldL')\cdot M)}}
 =-\frac1{q_m-1}\E_U{\pairing{\widehat g_{\OldL}(U)}{\widehat g_{\OldL'}(U)}}.
\]
Expanding the squared norm of the average and combining the diagonal and off-diagonal terms yields the counterpart of \cref{lem:refresh-Gram-variance}:
\begin{equation}\label{eq:general-Gram-variance}
\begin{aligned}
 \E_M{\VecNorm{Y_M-\E_{M'}{Y_{M'}}}^2}
 &=\frac{q_m}{d^2\cdot(q_m-1)}\sum_{\OldL}\E_U{\VecNorm{\widehat g_{\OldL}(U)}^2}\\
 &\quad-\frac1{d^2\cdot(q_m-1)}\E_U{\VecNorm{\sum_{\OldL}\widehat g_{\OldL}(U)}^2}\\
 &\le\frac{q_m}{d\cdot(q_m-1)}\E_{\OldL,U}{\VecNorm{g_{\OldL}(U)}^2}.
\end{aligned}
\end{equation}
For the inequality, discard the nonpositive term and use the fact that centering does not increase the second moment.
The column version follows from injectivity of $b$ and uniformity of $M^{-1}$.

\para{Nuclear norm.}
If $I$ is nonempty, set $\tau:=2\NuclearNorm{\prob_z^{>\beta}}$ and $W:=\sum_{j\in I}u_j\cdot v_j^{\transpose}$.
Then $\pairing{\prob_z}{W}=\tau/2$, $|I|\cdot\beta\le \tau/2$, and $\MaxNorm{W}\le\beta^{-2}$.
Define the Gram matrices of the substituted singular functions by
\[
 \Gram_{\LeftParty,M}:=\E_\OldL{u(\OldL,a(\OldL)\cdot M)\cdot u(\OldL,a(\OldL)\cdot M)^{\transpose}},
\]
where $u=(u_j)_{j\in I}$ is a coefficient column, and define $\Gram_{\RightParty,M}$ with $v(\OldR,M^{-1}\cdot b(\OldR))$.
As in \cref{lem:direct-nuclear}, their means are $\Id$.
Apply~\cref{eq:general-Gram-variance} to coefficient matrices with the ordinary square-sum norm $\EuclNorm{\cdot}$, the square root of the sum of the squared entries, without probability normalization.
Since $\EuclNorm{u\cdot u^{\transpose}}^2=e_u^2$, combining this with~\cref{eq:general-left-projection-moment,eq:general-right-projection-moment} gives, on each side,
\[
 \E_M{\EuclNorm{\Gram_{\LeftParty,M}-\Id}^2},\quad
 \E_M{\EuclNorm{\Gram_{\RightParty,M}-\Id}^2}
 \le\frac{q_m\cdot|I|}{d\cdot(q_m-1)\cdot\beta^2}.
\]
The Cauchy--Schwarz inequality (see \cref{lem:la-cauchy-schwarz}) applied to the two mask coefficient vectors, as in \cref{lem:direct-nuclear}, consequently gives
\[
 \E_{\OldL,\OldR,M}{\MaskL{z}(\OldL,M)\cdot \MaskR{z}(\OldR,M)\cdot W^M(\OldL,\OldR)}
 \le1+\sqrt{\frac{q_m\cdot|I|}{d\cdot(q_m-1)\cdot\beta^2}}.
\]
Using~\cref{eq:general-kernel-pairing,eq:general-encoding-gap}, we obtain
\[
 \tau\le1+\sqrt{\frac{q_m}{2d\cdot(q_m-1)\cdot\beta^3}}\cdot\sqrt \tau+\frac1{q_m\cdot\beta^2}
 \le2+\sqrt \tau.
\]
The last inequality uses $d\cdot\beta^3\ge1$, $d\le q_m$, $q_m\ge3$, and $\beta\le1$.
Thus $\tau\le4$. This also holds when $I$ is empty.
The balanced factors of $2\cdot\prob_z^{>\beta}$ have mean energies $\tau$ and pointwise energies at most $2/\beta$, while $\LtwoNorm{2\cdot\prob_z^{>\beta}}\le2$ follows from~\cref{eq:general-projection-L2}.

\para{Cancellation of the consistency sign.}
As in \cref{lem:consistency-sign}, the full binary sign matrix on pairs of $m$-bit vectors satisfies $\Upsilon^{\mathrm{full}}\cdot(\Upsilon^{\mathrm{full}})^{\transpose}=N_m^2\cdot\Id$.
The table $2J$ is its submatrix indexed by $(a(\OldL),U)$ and $(b(\OldR),V)$.
Injectivity makes these genuine subsets, so extending functions by zero and applying the Cauchy--Schwarz inequality (see \cref{lem:la-cauchy-schwarz}) bounds the unnormalized pairing by $N_m$ times the two Euclidean norms.
Each grouped domain now has $dq_m$ elements.
After converting to probability normalization, for every $f:\TableRows\to\R$ and $g:\TableCols\to\R$ we get
\begin{equation}\label{eq:general-consistency-sign}
 \left|\pairing{J}{f\cdot g^{\transpose}}\right|
 \le\frac{N_m}{2d\cdot q_m}\cdot\LtwoNorm{f}\cdot\LtwoNorm{g}
 \le\frac1d\cdot\LtwoNorm{f}\cdot\LtwoNorm{g}.
\end{equation}
Use the balanced representation of $\prob_z^{>\beta}$, with energies $\Phi_\beta$ and $\Psi_\beta$. Apply~\cref{eq:general-consistency-sign} to each product with the test factors inserted, and use the Cauchy--Schwarz inequality (see \cref{lem:la-cauchy-schwarz}) over the products.
For unit $L_2$ functions $f$ and $g$ the result is
$\left|\pairing{J\cdot \prob_z^{>\beta}}{f\cdot g^{\transpose}}\right|\le1/(d\cdot\beta)$.
Using~\cref{eq:largest-singular-coefficient} for the largest coefficient and expanding the residual as in~\cref{eq:direct-residual-identity}, we obtain, for $\Delta_z:=\prob_z-2\cdot\project{\prob_z^{>\beta}}$ and $t:=\LtwoNorm{\Delta_z}$,
\begin{align}
 \sigma_{\max}(\Delta_z)&\le\beta+2/(d\cdot\beta)\le3\beta,\label{eq:general-residual-estimates}\\
 \MaxNorm{\Delta_z}&\le3/\beta,\nonumber\\
 t^2&=\pairing{\prob_z}{\Delta_z}+4\cdot\pairing{\prob_z^{>\beta}}{J\cdot \prob_z^{>\beta}},\label{eq:general-residual-identity}\\
 \left|\pairing{\prob_z^{>\beta}}{J\cdot \prob_z^{>\beta}}\right|
 &\le2/(d\cdot\beta).\label{eq:general-residual-sign-pairing}
\end{align}
Every row and column of $\Delta_z$ has $L_2$ norm at most three, and $\Delta_z$ is supported on consistent tuples.

\para{Four cycles and collisions.}
We give the version of \cref{lem:four-cycle-refresh} needed for this smaller old-input domain.
For any consistent table $\Delta$, put $t:=\LtwoNorm{\Delta}$, $B_\infty:=\MaxNorm{\Delta}$, and let $B_2$ bound all row and column $L_2$ norms.
Extend $\Delta$ by zero whenever $U$ or $V$ is zero, obtaining $\widehat{\Delta}$.
The full-vector quantities satisfy $\widehat{t}^2=(q_m/N_m)^2\cdot t^2$ and $\FourCyc{\widehat{\Delta}}=(q_m/N_m)^4\cdot \FourCyc{\Delta}$.
The row and column norms remain at most $B_2$.
For a uniform unrestricted binary $m\times m$ matrix $A_0$, put
\[
 \widetilde\Delta_{A_0}(\OldL,(\OldR,V)):=
 \begin{cases}
 \sqrt{N_m}\cdot\widehat{\Delta}((\OldL,a(\OldL)\cdot A_0),(\OldR,V)),&A_0\cdot V=b(\OldR),\\
 0,&\text{otherwise}.
 \end{cases}
\]
The row measure is uniform on $\OldData$ and the column measure is uniform on $\OldData\times\F_2^{m\times1}$.
As in the refreshing proof, $\FourCyc{\widetilde\Delta_M}=\FourCyc{\Delta^M}$ for invertible $M$. Comparison with unrestricted matrices (see~\cref{eq:invertible-comparison}) applies to this nonnegative complete four-cycle and gives
\[
 \E_M{\FourCyc{\Delta^M}}\le4\E_{A_0}{\FourCyc{\widetilde\Delta_{A_0}}}.
\]

Expand the latter average and insert independent full rows $U_1$ and $U_2$.
The prescribed-row-and-column count (see \cref{lem:prescribed-entries}, in dimension $m$) applies to
$a(\OldL_i)\cdot A_0=U_i$ and $A_0\cdot V_j=b(\OldR_j)$.
The pairings required by that lemma are imposed by the support of $\widehat{\Delta}$.
Injectivity of $a$ makes row rank one equivalent to $\OldL_1=\OldL_2$, which forces $U_1=U_2$.
Column rank one means $V_1=V_2$, which forces $\OldR_1=\OldR_2$ by injectivity of $b$.
Thus the four rank contributions have exactly the form of~\cref{eq:joint-four-rank-expansion}, with the following coefficients:
\[
\begin{array}{c|cccc}
 (\text{row rank},\text{column rank})&(2,2)&(1,2)&(2,1)&(1,1)\\ \hline
 \text{coefficient}&16&4/d&4/d&2/d^2.
\end{array}
\]
For example, rank $(1,2)$ has weight $4N_m$ before the equality probabilities are included. $\OldL_1=\OldL_2$ costs $1/d$ and $U_1=U_2$ costs $1/N_m$.
Rank $(2,1)$ has the same cost with $V$ and $\OldR$ exchanged. Rank $(1,1)$ has weight $2N_m^2$ and equality probability $1/(d^2N_m^2)$.

For completeness, the rank $(2,2)$ term is the full four-cycle restricted to $\OldL_1\ne \OldL_2$ and $V_1\ne V_2$, multiplied by $16$.
Inclusion--exclusion subtracts two nonnegative complete four-cycles of sliced tables.
The intersection adds at most $16B_\infty^2\widehat{t}^2/(d\cdot N_m)$: its equality probability is $1/(d\cdot N_m)$, each entry retains its full product marginal, and
$|e_1e_2e_3e_4|\le(B_\infty^2/4)\sum_i e_i^2$.
Hence this term is at most $16\FourCyc{\widehat{\Delta}}+16B_\infty^2\widehat{t}^2/(d\cdot N_m)$.
Each mixed-rank term is a nonnegative product of two squared entries. Dropping its remaining inequality restriction bounds it by $4B_2^2\widehat{t}^2/d$.
The rank $(1,1)$ term is at most $2B_\infty^2\widehat{t}^2/d^2$.
Multiplying by four, using $N_m>d$ and $q_m/N_m\le1$, gives
\begin{equation}\label{eq:general-four-cycle}
 \E_M{\FourCyc{\Delta^M}}
 \le64\FourCyc{\Delta}+\frac{32B_2^2}{d}t^2+
 \frac{72B_\infty^2}{d^2}t^2.
\end{equation}

\para{Completing the error bound.}
We apply~\cref{eq:general-four-cycle} to our residual $\Delta_z$, with $t:=\LtwoNorm{\Delta_z}$, $B_2=3$, and $B_\infty\le3/\beta$.
By~\cref{eq:four-cycle-norms} and \cref{eq:general-residual-estimates},
$\FourCyc{\Delta_z}\le\sigma_{\max}(\Delta_z)^2\cdot t^2\le9\beta^2\cdot t^2$.
Since $d\cdot\beta^3\ge1$ and $\beta\le1$, both $d^{-1}$ and $d^{-2}\cdot\beta^{-2}$ are at most $\beta^2$.
\Cref{eq:general-four-cycle} therefore yields $\E_M{\FourCyc{\Delta_z^M}}\le C\cdot\beta^2\cdot t^2$.
Applying the test bound~\cref{eq:four-cycle-test}, followed by Jensen's inequality over $M$, gives
\begin{equation}\label{eq:general-mask-residual}
 \left|\E_{\OldL,\OldR,M}{\MaskL{z}(\OldL,M)\cdot \MaskR{z}(\OldR,M)\cdot\Delta_z^M(\OldL,\OldR)}\right|
 \le C\cdot\beta^{1/2}\cdot t^{1/2}.
\end{equation}
Replacing $p_s^{(m)}$ by $1/2$ in~\cref{eq:general-kernel-pairing} incurs error at most
$\MaxNorm{\Delta_z}/(2q_m)\le3/(2d\cdot\beta)$.
Combining~\cref{eq:general-mask-residual,eq:general-residual-identity,eq:general-residual-sign-pairing} gives~\cref{eq:general-error-bootstrap}:
\begin{equation*}\tag*{\labelcref*{eq:general-error-bootstrap}}
 t^2\le C\cdot\beta^{1/2}\cdot t^{1/2}+C/(d\cdot\beta)
 \le C\cdot\beta^{1/2}\cdot t^{1/2}+C\cdot\beta^2.
\end{equation*}
If the second term on the right of~\cref{eq:general-error-bootstrap} is at least $t^2/2$, then $t\le C\cdot\beta\le C\cdot\beta^{1/3}$.
Otherwise, the first is at least $t^2/2$, and division by $t^{1/2}$ gives $t\le C\cdot\beta^{1/3}$. The case $t=0$ is immediate.
This includes $d=1$, $\beta=1$, and an empty retained family.
\end{proof}

\subsection{Tensor encodings for the matrix phase of a circuit gate}\label{sec:joint-tensor-instance}
The approximation lemma of \cref{sec:joint-masks} (see \cref{lem:joint-map}) also covers the matrix phase of a computation gadget.
All estimates below are per answer. Their later use enumerates the relevant answers explicitly.

\para{The matrix phase of a circuit gate.}\label{sec:joint-tensor}
Fix a NAND computation gadget with two distinct operand encoding locations.
Take $m=n^2+1$, $d=q^2$, and $\OldData=(\{0,1\}^n\setminus\{0^n\})^2$, with
\[
 \OldL=(L_1,L_2),\quad \OldR=(R_1,R_2),\qquad
 a(\OldL)=(L_1\otimes L_2,1),\quad b(\OldR)=\binom{R_1\otimes R_2}{1}.
\]
The tensor product lists all pairwise coordinate products, so
$a(\OldL)\cdot b(\OldR)=1+(L_1\cdot R_1)\cdot(L_2\cdot R_2)$.
Over $\F_2$, a nonzero rank-one tensor uniquely determines its two nonzero factors: any nonzero row or column determines one factor, and its entries determine the other.
Appending a fixed coordinate preserves injectivity.
For operand bits $s_1$ and $s_2$, put
\[
 \LongViewDist{s_1,s_2}:=
 \mu_{s_1}^{(n)}\otimes\mu_{s_2}^{(n)}\otimes\mu_{1+s_1\cdot s_2}^{(m)}.
\]

\begin{corollary}[Tensor instance]\label{cor:joint-tensor}
For every answer $z$ and every formal incoming history, and for $q^{-2/3}\le\beta\le1$, the construction of \cref{lem:joint-map} gives a common table $2\cdot\prob_z^{>\beta}$ with nuclear norm at most $4$ and balanced energies $2\cdot\Phi_\beta$ and $2\cdot\Psi_\beta$, bounded pointwise by $2/\beta$ and in mean by~$4$.
For every $(s_1,s_2)\in\F_2^2$,
\begin{equation}\label{eq:joint-tensor-error}
 \left(\E_{\omega\sample\LongViewDist{s_1,s_2}}{|\prob_z\left(\omega\right)-2\cdot\prob_z^{>\beta}\left(\omega\right)|^2}\right)^{1/2}
 \le C\cdot\beta^{1/3}.
\end{equation}
All constants are absolute and uniform in the masks and formal history.
\end{corollary}
\begin{proof}
Each fixed-bit encoding distribution has density at most three relative to independent uniform nonzero shares by \cref{eq:encoding-probability-lower}, so the three-encoding distribution $\LongViewDist{s_1,s_2}$ has density at most $27$.
On a valid local labeling, $\prob_z-2\cdot\prob_z^{>\beta}=\prob_z-2\cdot\project{\prob_z^{>\beta}}$. Apply~\cref{eq:joint-map-error} and take square roots of this density bound.
The admissible threshold range is $d^{-1/3}\le\beta\le1$ with $d=q^2$.
\end{proof}

The compiler shortens the long output back to length $n$. \Cref{sec:compiler-shortening} transports both the chosen factor energies and the $L_2$ error in~\cref{eq:joint-tensor-error}.

\subsection{Shortening by a random rectangular matrix}\label{sec:compiler-shortening}
The shortener of \cref{sec:compiler-gadget} applies to any nonzero shares $U\in\F_2^{1\times m}$ and $V\in\F_2^{m\times1}$, with $m\ge n\ge2$.
It samples a nonzero row $A$ uniformly, samples a full-row-rank matrix $H\in\F_2^{n\times m}$ uniformly subject to $A\cdot H=U$, and sets $R:=H\cdot V$.
An attempt succeeds when $R\ne0$. Otherwise, the parties repeat with fresh $A$ and $H$ and the same $U$ and $V$.
All attempts belong to one shortening phase and share its leakage budget.

For the circuit gadget, put $\OldL:=(L_1,L_2)$, $\OldR:=(R_1,R_2)$, and $m:=n^2+1$.
The NAND gadget forms the long shares
\[
 a(\OldL)=(L_1\otimes L_2,1),\qquad b(\OldR)=\binom{R_1\otimes R_2}{1}.
\]
Thus the long and short encodings both represent
$s=1+(L_1\cdot R_1)\cdot(L_2\cdot R_2)$.
This output bit can be either zero or one. Fixing the NAND gadget's final coordinate does not restrict the bit represented by the shortener.
Recall that $N=2^n$, and let $\FullRank_{n,m}$ be the finite set of full-row-rank $n\times m$ binary matrices.
The matrix from the preceding refresh has been erased before this phase.

The sampler in \cref{sec:direct-simulator} produces the required uniform matrix $H$ in expected polynomial time.
The local sampling tape may be included in left leakage. Its conditional distribution will be averaged into the left mask below.

\begin{lemma}[Exact output distribution of shortening]\label{lem:compiler-shortening-freshness}
Let $m\ge n\ge2$, let $s\in\F_2$, and fix any nonzero long shares $U\in\F_2^{1\times m}$ and $V\in\F_2^{m\times1}$ with $U\cdot V=s$.
Let $p_s^{\mathrm{succ}}$ be the probability that one shortening attempt succeeds.
Then
\begin{equation}\label{eq:compiler-acceptance-probability}
 1-p_s^{\mathrm{succ}}=
 \begin{cases}
 \left(2^{m-n}-1\right)/\left(2^{m-1}-1\right),&s=0,\\
 0,&s=1.
 \end{cases}
\end{equation}
In particular, $p_s^{\mathrm{succ}}\ge1/2$ and $1-p_s^{\mathrm{succ}}\le2/N$.
Conditional on success, the output has distribution $\mu_s^{(n)}$, independently of the particular fixed pair of shares $U$ and $V$.
The procedure that retries until success terminates almost surely, takes at most two attempts in expectation, and has exactly the same output distribution $\mu_s^{(n)}$.

If instead $(U,V)\sample \mu_s^{(m)}$, then, conditional on first-attempt success and any fixed output $(A,R)\in\Omega_s$, the preceding variables have the following distribution:
\begin{equation}\label{eq:compiler-inverse-shortening}
 H\sample \FullRank_{n,m},\qquad
 V\sample \{v:H\cdot v=R\},\qquad
 U:=A\cdot H.
\end{equation}
The fiber in the middle has exactly $2^{m-n}$ elements, all nonzero.
These statements concern encoding distributions before conditioning on leakage answers.
\end{lemma}
\begin{proof}
For fixed nonzero $A$, the number of matrices $H\in\FullRank_{n,m}$ with $A\cdot H=U$ is the same for every nonzero $U$.
Indeed, right multiplication by an invertible $m\times m$ matrix maps one such set bijectively to any other.
This number is also independent of $A$, by invertible changes of the $n$ rows.
Consequently every admissible pair $(A,H)$ in one attempt has the same probability.

The kernel of $H$ is a uniformly random $(m-n)$-dimensional subspace of $\ker(U)$.
To see this, the row space of $H$ is a uniformly random $n$-dimensional subspace containing $U$: every such subspace has the same number of ordered bases, and each basis gives exactly one nonzero coordinate row $A$ for $U$.
Taking annihilators gives the assertion about $\ker(H)$.
If $s=1$, the vector $V$ is outside $\ker(U)$, so it cannot lie in $\ker(H)$.
If $s=0$, symmetry among the $2^{m-1}-1$ nonzero vectors of $\ker(U)$ gives
\[
 \Prb{H\cdot V=0}
 =\frac{2^{m-n}-1}{2^{m-1}-1}.
\]
This proves~\cref{eq:compiler-acceptance-probability} and its bounds.

For exact output uniformity, an invertible $n\times n$ matrix $B$ sends an admissible attempt $(A,H)$ to $\left(A\cdot B,B^{-1}\cdot H\right)$, preserving its probability and changing its output to $\left(A\cdot B,B^{-1}\cdot R\right)$.
The action on nonzero pairs of inner product $s$ is transitive by the argument of \cref{lem:fresh}.
All successful outputs therefore have equal probability.
Each retry uses an independent attempt with the same positive success probability, proving almost-sure termination, the expected-attempt bound, and exact freshness of the final output.

Finally, suppose the long pair is uniform.
Fix a valid short output $(A,R)$.
For every $H\in\FullRank_{n,m}$ there exist exactly $2^{m-n}$ choices of $V$ with $H\cdot V=R$. Each is nonzero because $R\ne0$.
Putting $U=A\cdot H$ gives a nonzero share and $U\cdot V=A\cdot R=s$.
Every such choice has the same forward probability: the long pair is uniform, $A$ is uniform, and the number of eligible matrices for $A$ and $U$ is constant.
These choices are precisely all the successful preimages of $(A,R)$.
Counting them proves~\cref{eq:compiler-inverse-shortening}.
\end{proof}

\para{Conditional leakage probability functions.}
Fix a gate and an incoming history.
For a matrix-phase answer $\zeta$, let $\prob_\zeta\left((\OldL,U),(\OldR,V)\right)$ be its conditional probability given the old operands and the refreshed long pair.
On valid long tuples, $\prob_\zeta\ge0$ and $\sum_\zeta \prob_\zeta=1$.
For fixed $\zeta$, the two shortening leakage functions are chosen before either shortening answer is revealed.
Write that answer as $z=(z_{\LeftParty},z_{\RightParty})$.
On a first-attempt success, their masks have the form
\[
 \MaskL{\zeta z_{\LeftParty}}(\OldL,A,H),\qquad
 \MaskR{\zeta z_{\RightParty}}(\OldR,V,H).
\]
Here the left mask averages over the local sampling tape conditional on $A$, $H$, and $U=A\cdot H$. Hence it takes values in $[0,1]$.
The sampler's finite rejection records have countable support, and the bounded mask makes this conditional average an absolutely convergent weighted sum.
The right mask is Boolean, or likewise averaged if the implementation uses additional independent right randomness.
The successful acknowledgment is fixed on this branch.
The two masks depend on separate local variables once $H$ is fixed.
Erasure is essential: the shortening views retain the old operands and long shares, but not the preceding matrix or its sampling tape.
Thus the conditional averaging of the new sampling tape does not also depend on the other party's inputs or on hidden matrix-phase state.

\begin{definition}[Shortening operator]\label{def:shortening-operator}
For every real table $G:(\OldData\times\UL^{(m)})\times(\OldData\times\UR^{(m)})\to\R$, define a table $\cS_{\zeta,z}(G)$ on $(\OldData\times\UL)\times(\OldData\times\UR)$ as follows.
In the expectation below, sample $H\sample \FullRank_{n,m}$ and, conditional on $H$, sample $V$ uniformly from $\{v:H\cdot v=R\}$:
\begin{align*}
 (\cS_{\zeta,z}(G))((\OldL,A),(\OldR,R))
 =\E_{H,V}{\begin{aligned}&G((\OldL,A\cdot H),(\OldR,V))\\[-1mm]
 &{}\cdot \MaskL{\zeta z_{\LeftParty}}(\OldL,A,H)
 \cdot \MaskR{\zeta z_{\RightParty}}(\OldR,V,H)\end{aligned}}.
\end{align*}
For nonzero $A$ and $R$, every variable in this formula belongs to its prescribed ambient share space, whether or not the operand and output plaintexts satisfy the gate equation.
\end{definition}
On valid local triples, the leakage probability function conditioned on success of the first attempt is
\begin{equation}\label{eq:compiler-first-attempt-answer}
 \widehat{\prob}_{\zeta,z}=\cS_{\zeta,z}\left(\prob_\zeta\right).
\end{equation}
This follows directly from the reverse description in~\cref{eq:compiler-inverse-shortening}. No density or conditioning factor is needed.

The operator $\cS_{\zeta,z}$ is an average over $H$ of products: for fixed $H$, the left mask depends on the left data and the right mask on the right data.
It therefore carries a product representation of $G$ to a product representation of $\cS_{\zeta,z}(G)$, and Jensen's inequality keeps the energies and the $L_2$ error under control.

\begin{lemma}[Transport through shortening]\label{lem:compiler-shortening-transport}
Fix any gate, formal incoming history, and adversary tape.
For every operand plaintext pair $(s_1,s_2)\in\F_2^2$, put $s:=1+s_1\cdot s_2$ and
\[
 \LongViewDist{s_1,s_2}:=\mu_{s_1}^{(n)}\otimes\mu_{s_2}^{(n)}\otimes\mu_s^{(m)},\qquad
 \ShortViewDist{s_1,s_2}:=\mu_{s_1}^{(n)}\otimes\mu_{s_2}^{(n)}\otimes\mu_s^{(n)}.
\]
Fix a matrix answer $\zeta$, a shortening answer $z$, a threshold $\beta\in[q^{-2/3},1]$, where $q=2^n-1$, and a real number $\varepsilon\ge0$.
Suppose the real table $G_\zeta=\sum_t a_t\cdot b_t^{\transpose}$ has chosen factor energies $\Phi$ and $\Psi$ such that, for every $x\in\TableRows$ and $y\in\TableCols$,
\begin{align}
 \Phi(x)&\le2/\beta,\label{eq:compiler-long-error-hypothesis}\\
 \Psi(y)&\le2/\beta,\label{eq:compiler-long-right-energy}\\
 \E_{x\sample\TableRows}{\Phi(x)}&\le4,\label{eq:compiler-long-left-mean}\\
 \E_{y\sample\TableCols}{\Psi(y)}&\le4,\label{eq:compiler-long-right-mean}\\
 \left(\E_{\omega\sample\LongViewDist{s_1,s_2}}{|\prob_\zeta(\omega)-G_\zeta(\omega)|^2}\right)^{1/2}
 &\le\varepsilon\quad\text{for every }(s_1,s_2)\in\F_2^2.\label{eq:compiler-long-L2-hypothesis}
\end{align}
Then $\widetilde G_{\zeta,z}:=\cS_{\zeta,z}(G_\zeta)$ has a product representation whose energies satisfy, for every $(\OldL,A)\in\OldData\times\UL$ and $(\OldR,R)\in\OldData\times\UR$,
\begin{align}
 \Phi_{\mathrm{out}}(\OldL,A)&\le2/\beta,\label{eq:compiler-shortening-energies}\\
 \Psi_{\mathrm{out}}(\OldR,R)&\le2/\beta,\label{eq:compiler-output-right-energy}\\
 \E_{(\OldL,A)\sample\OldData\times\UL}{\Phi_{\mathrm{out}}(\OldL,A)}&\le4,\label{eq:compiler-output-left-mean}\\
 \E_{(\OldR,R)\sample\OldData\times\UR}{\Psi_{\mathrm{out}}(\OldR,R)}&\le4,\label{eq:compiler-output-right-mean}\\
 \NuclearNorm{\widetilde G_{\zeta,z}}&\le4.\label{eq:compiler-output-nuclear}
\end{align}
Let $\prob^{\mathrm{out}}_{\zeta,z}$ be the actual complete-gadget leakage probability function of the answer $(\zeta,z)$, including all retries, conditional on the operand encodings and a short output encoding with the prescribed output bit.
Then, for every operand plaintext pair $(s_1,s_2)\in\{0,1\}^2$,
\begin{equation}\label{eq:compiler-shortening-L2}
 \left(\E_{\omega\sample\ShortViewDist{s_1,s_2}}
 {|\prob^{\mathrm{out}}_{\zeta,z}\left(\omega\right)-\widetilde G_{\zeta,z}(\omega)|^2}\right)^{1/2}
 \le\varepsilon+4/2^n.
\end{equation}
Under any assignment of the three wire labels, including labels violating the gate equation, the $L_2$ norm of $\widetilde G_{\zeta,z}$ is at most $C_0:=4\cdot\sqrt{27}$.
The transported representation need not be balanced. Its two mean energy bounds are separate inequalities.
\end{lemma}
\begin{proof}
For fixed $H\in\FullRank_{n,m}$, define
\[
 a_t^H(\OldL,A):=a_t(\OldL,A\cdot H)\cdot \MaskL{\zeta z_{\LeftParty}}(\OldL,A,H),
\]
\[
 b_t^H(\OldR,R):=\E_{V\sample \{v:H\cdot v=R\}}
 {b_t(\OldR,V)\cdot \MaskR{\zeta z_{\RightParty}}(\OldR,V,H)}.
\]
Use the factors $|\FullRank_{n,m}|^{-1/2}\cdot a_t^H$ and $|\FullRank_{n,m}|^{-1/2}\cdot b_t^H$, indexed by $(H,t)$.
The sum of their products is exactly $\widetilde G_{\zeta,z}$.
Since the masks lie in $[0,1]$, Jensen's inequality on the right fibers and~\cref{eq:compiler-long-error-hypothesis,eq:compiler-long-right-energy} give
\begin{align}
 \Phi_{\mathrm{out}}(\OldL,A)&\le\E_{H\sample \FullRank_{n,m}}{\Phi(\OldL,A\cdot H)}\le2/\beta,
 \label{eq:compiler-shortening-left-energy}\\
 \Psi_{\mathrm{out}}(\OldR,R)&\le\E_{H\sample \FullRank_{n,m}}
 {\E_{V\sample \{v:H\cdot v=R\}}{\Psi(\OldR,V)}}\le2/\beta.
 \label{eq:compiler-shortening-right-energy}
\end{align}
Average~\cref{eq:compiler-shortening-left-energy} over independent uniform $\OldL$ and $A$.
For each fixed nonzero $A$, right multiplication by invertible $m\times m$ matrices preserves the uniform distribution on $\FullRank_{n,m}$ and acts transitively on nonzero long rows, so $A\cdot H$ is uniform among them and~\cref{eq:compiler-long-left-mean} gives~\cref{eq:compiler-output-left-mean}.
For~\cref{eq:compiler-shortening-right-energy}, sample uniform nonzero $R$ and then a uniform member $V$ of its fiber.
Averaged over uniform full-row-rank $H$, the distribution of this nonzero $V$ is invariant under all invertible changes of the $m$ coordinates and therefore uniform.
The old data $\OldR$ remain independent, so~\cref{eq:compiler-long-right-mean} gives~\cref{eq:compiler-output-right-mean}.
These averaging facts preserve the original energies before masks are imposed. Masks can only reduce the bounds.
The Cauchy--Schwarz inequality (see \cref{lem:la-cauchy-schwarz}) over $(H,t)$ and the product-representation bound~\cref{eq:nuclear-product-bound} now yield
\[
 \NuclearNorm{\widetilde G_{\zeta,z}}
 \le\sqrt{\E_{(\OldL,A)}{\Phi_{\mathrm{out}}(\OldL,A)}\cdot\E_{(\OldR,R)}{\Psi_{\mathrm{out}}(\OldR,R)}}\le4,
\]
which is~\cref{eq:compiler-output-nuclear}.
By~\cref{eq:nuclear-dominates-ltwo}, the uniform $L_2$ norm is at most the nuclear norm.
Every fixed three-wire labeling has density at most $27$ relative to independent uniform shares, proving the arbitrary-label bound $C_0$.

For the error, put $e_\zeta:=\prob_\zeta-G_\zeta$ on the full long table.
The operator $\cS_{\zeta,z}$ is a positive average with total mask weight at most one, so the weighted Cauchy--Schwarz inequality (see \cref{lem:la-cauchy-schwarz}) gives
\[
 |\cS_{\zeta,z}(e_\zeta)|^2
 \le\cS_{\zeta,z}(|e_\zeta|^2).
\]
Average over $\omega\sample\ShortViewDist{s_1,s_2}$ and use the reverse description in~\cref{eq:compiler-inverse-shortening}.
Removing the masks only increases the nonnegative summand and leaves the long distribution $\LongViewDist{s_1,s_2}$.
Thus, using~\cref{eq:compiler-first-attempt-answer,eq:compiler-long-L2-hypothesis},
\begin{equation}\label{eq:compiler-first-attempt-L2}
 \E_{\omega\sample\ShortViewDist{s_1,s_2}}
 {|\widehat{\prob}_{\zeta,z}\left(\omega\right)-\widetilde G_{\zeta,z}(\omega)|^2}
 \le\E_{\omega\sample\LongViewDist{s_1,s_2}}{|\prob_\zeta\left(\omega\right)-G_\zeta(\omega)|^2}\le\varepsilon^2.
\end{equation}
Conditioning on first-attempt success preserves this bound: this event has the same probability $p_s^{\mathrm{succ}}$ for every long pair encoding $s$, so it leaves the old operands, long pair, and matrix answer distributed as before.

Finally, retain the actual retries, whose leakage may inspect all attempts within the same phase budget.
For fixed old operands, the final short encoding is uniform $\mu_s^{(n)}$ both on first-attempt success and on failure followed by fresh attempts.
Consequently, conditional on any valid old and short encodings, the complete answer distribution has the pointwise mixture
\begin{equation*}
 \prob^{\mathrm{out}}=p_s^{\mathrm{succ}}\cdot\widehat{\prob}
       +(1-p_s^{\mathrm{succ}})\cdot \prob^{\mathrm{retry}}.
\end{equation*}
The components are probability distributions on the complete answer alphabet, including the matrix answer $\zeta$.
If failure has probability zero, its unused component may be arbitrary.
The sum of absolute differences between two probability distributions is at most two. Hence, in particular, for each individual answer,
\begin{equation}\label{eq:compiler-retry-answer-error}
 |\prob^{\mathrm{out}}_{\zeta,z}-\widehat{\prob}_{\zeta,z}|
 \le 2(1-p_s^{\mathrm{succ}})\le4/2^n.
\end{equation}
Combining~\cref{eq:compiler-first-attempt-L2,eq:compiler-retry-answer-error} with the $L_2$ triangle inequality proves~\cref{eq:compiler-shortening-L2}.
The proof is per answer throughout, so no factor of the answer-alphabet size appears.
\end{proof}

\begin{lemma}[Exact canonical execution]\label{lem:compiler-canonical}
Fix the public input and all secret input values, hence every wire's plaintext value.
The joint distribution of all generated wire encodings and leakage answers is exactly the distribution generated by the following canonical experiment: sample every distinct wire encoding independently from its corresponding $\mu_s$, then generate successive adaptive answers from the conditional leakage probability functions of the actual gadgets given their incident wire encodings.
\end{lemma}
\begin{proof}
Fix an incoming history and the complete input encodings of one operation.
For a computation gadget, \cref{lem:compiler-shortening-freshness} shows that its final output encoding has distribution $\mu_s^{(n)}$, independently of this information.
This statement holds for every refreshed long pair and includes all retries.
For an input-port refresh, the same freshness follows from \cref{lem:fresh}, since its matrix is sampled independently of the preceding execution.
A public-input sampler draws an independent encoding of its prescribed public bit.
In every case, generate the actual transition by first sampling that fresh output and then sampling its complete answer conditional on the old and new encodings.
This reproduces exactly the original transition, including any dependence of the answer on the new encoding.

Apply this identity successively in topological order, using the actual adaptive leakage functions selected by the incoming history.
Since each new encoding is fresh independently of the entire preceding execution, the wire samples may equivalently be drawn before any leakage answers are generated.
A location read by multiple operations has one sample, reused at every such read.
The initial input encodings already have this independent product distribution.
The constructed experiment therefore reproduces the joint encoding-and-answer distribution.
Marginalizing the hidden encodings gives the leakage transcript. Appending the trusted decoder's prescribed output gives the compiler view.
\end{proof}

The lemma extends to labels that are chosen during the execution, where the public bit of a sampler, and hence the plaintext values of the wires that later operations produce, may be functions of the leakage answers preceding the operation. This is the setting of the history-dependent labelings in \cref{sec:direct-composition-setup} and of the stateful protocol in \cref{sec:stateful}.
The transition argument above is unchanged, since the fresh output of an operation is sampled from $\mu_b$ for the bit $b$ that its incoming history prescribes.
Applying it in topological order gives the canonical experiment of \cref{sec:direct-composition-setup}, which samples each wire encoding immediately before its producing operation and whose transcript probabilities are given by~\cref{eq:direct-canonical} with the induced labeling.
Only the final step, which moves all wire samples before the first answer, requires the labels to be fixed in advance. The composition proof does not use that step.

We return to the gate notation of \cref{sec:compiler-local}: $S_i$ is the set of three wires incident to gate $i$, $X_{S_i}$ is their tuple of encodings, $\UL^{S_i}$ and $\UR^{S_i}$ are the sets of triples of nonzero left and of nonzero right shares indexed by $S_i$, with their product uniform measures, and $h$ is the incoming answer history.
Complete gate answers belong to the alphabet $\cZ$ of size $K$ (see \cref{sec:direct-phase-model}). A labeling is valid when its wire bits satisfy every gate operation.

For a local labeling $s$, write $\LtwoNormAt{F}{s}:=\left(\E_{X_{S_i}\sample\bigotimes_{w\in S_i}\mu_{s(w)}}{|F(X_{S_i})|^2}\right)^{1/2}$.
\begin{lemmarestatement}{lem:compiler-local-interface}{A gate with its inputs and output retained}
There exists an exact canonical experiment with conditional leakage probability functions $\prob_{i,h,z}\left(X_{S_i}\right)$, defined on valid local labelings, with $\prob_{i,h,z}\ge0$ and $\sum_z \prob_{i,h,z}=1$.
Its joint distribution of generated wire encodings and answers equals that of the actual execution.
There exists an absolute constant $C>0$ such that, for every threshold $\beta\in[q^{-2/3},1]$, computation operation $i$, formal incoming history $h\in\cZ^{i-1}$, and answer $z\in\cZ$, there exists a real table $G_{i,h,z}^{\beta}$, common to all wire labelings, with a product representation whose left and right energies $\Phi$ and $\Psi$ (the sums of squared factors; see \cref{def:factor-energy}) satisfy, for all grouped shares $x\in\UL^{S_i}$ and $y\in\UR^{S_i}$,
\begin{align*}
 \NuclearNorm{G_{i,h,z}^{\beta}}&\le4,\tag*{\labelcref*{eq:compiler-local-interface}}\\
 \Phi(x)&\le2/\beta,\tag*{\labelcref*{eq:compiler-local-left-energy}}\\
 \Psi(y)&\le2/\beta,\tag*{\labelcref*{eq:compiler-local-right-energy}}\\
 \E_{x\sample \UL^{S_i}}{\Phi(x)}&\le4,\tag*{\labelcref*{eq:compiler-local-left-mean}}\\
 \E_{y\sample \UR^{S_i}}{\Psi(y)}&\le4,\tag*{\labelcref*{eq:compiler-local-right-mean}}\\
 \LtwoNormAt{\prob_{i,h,z}-G_{i,h,z}^{\beta}}{s}
 &\le C\cdot\beta^{1/3}\quad\text{for every valid local labeling }s,\tag*{\labelcref*{eq:compiler-local-error}}\\
 \LtwoNormAt{G_{i,h,z}^{\beta}}{s}
 &\le C_0\quad\text{for every local labeling }s.\tag*{\labelcref*{eq:compiler-local-rms}}
\end{align*}
Here $C_0:=4\cdot\sqrt{27}$, and $C$ is independent of the dimension, threshold, answer, formal history, and admissible leakage functions.
\end{lemmarestatement}

\begin{proof}[Proof of \cref{lem:compiler-local-interface}]
Use the true complete-answer functions $\prob^{\mathrm{out}}_{\zeta,z}$ from \cref{lem:compiler-shortening-transport}. They are nonnegative and sum to one on every valid local labeling.
\Cref{lem:compiler-canonical} gives the exact joint encoding-and-answer distribution, whose answer marginal is the leakage transcript.

Fix a gate, a formal incoming history, and $q^{-2/3}\le\beta\le1$.
Apply \cref{cor:joint-tensor} to each matrix-phase answer $\zeta$, with $d=q^2$ and $m=n^2+1$.
It supplies the global approximation $2\cdot\prob_\zeta^{>\beta}$, where the superscript keeps the singular coefficients exceeding $\beta$ as in~\cref{eq:joint-direct-construction}, with nuclear norm at most four, balanced pointwise energies at most $2/\beta$, both mean energies at most four, and valid-label $L_2$ error $C\cdot\beta^{1/3}$ (see~\cref{eq:joint-tensor-error}).
For each shortening answer apply \cref{lem:compiler-shortening-transport} to this representation.
The transported table satisfies the separate energy bounds, the nuclear bound, and the arbitrary-label $L_2$ bound by~\crefrange{eq:compiler-shortening-energies}{eq:compiler-output-nuclear} and the last assertion of that lemma.
By~\cref{eq:compiler-shortening-L2}, its valid-label error is at most
\[
 C\cdot\beta^{1/3}+4/2^n\le C'\cdot\beta^{1/3},
\]
since $\beta\ge q^{-2/3}$ and $2^{-n}\le q^{-2/9}\le\beta^{1/3}$.
Relabel the complete answer $(\zeta,z)$ as one answer of the gadget.
All constructions depend only on the public maps, masks, threshold, and formal incoming history. The same tables serve every plaintext labeling.
The constants are absolute and uniform in all these choices.
The transported representation is the chosen representation in the interface and is not assumed to be balanced.
\end{proof}

\section{An error in the Goldwasser--Rothblum proof}\label{app:gr}

We describe an error in Goldwasser and Rothblum's leakage-resilient computation proof~\cite{Goldwasser2015}.
Their construction uses ciphertext banks to protect arbitrary circuits for an unbounded number of executions in the model where only computation leaks.
All claim, lemma, figure, and page references to their work below are to the journal version, \emph{SIAM Journal on Computing} \textbf{44}(5), 1480--1549 (2015).
The main issue is Claim~6.24, an auxiliary statement used in the security argument for continual refreshing.
We give a counterexample to that claim and briefly record other issues in the printed argument and procedures.
All the errors and gaps discussed in this appendix were identified with the help of ChatGPT.

We contacted Goldwasser and Rothblum, who confirmed the gap.
They told us that they had known about it for some time and had a fix, which had not yet been published at the time of our correspondence.

\subsection{Claim 6.24 and its hypotheses}

We first recall Claim~6.24 and the notions it uses.
All arithmetic in this appendix is over $\F_2$.
The symbols below are those of~\cite{Goldwasser2015} and are local to this appendix. Several of them, including $K$, $\kappa$, $L$, $\ell$, $m$, and $d$, denote different objects here than in the rest of the paper.
Write $K\in\F_2^\kappa$ for a column key and $M\in\F_2^{\kappa\times m}$ for a matrix.
The relevant notion of \emph{independence up to orthogonality} (IuO) is defined in~\cite[Definitions~6.18--6.19, p.~1523]{Goldwasser2015}.
\begin{definition}[Independence up to orthogonality]\label{def:iuo}
A distribution $\cV$ on $\F_2^\kappa\times\F_2^{\kappa\times m}$ is IuO if there exist distributions $\mu_K$ on $\F_2^\kappa$ and $\mu_M$ on $\F_2^{\kappa\times m}$ such that one obtains $\cV$ by sampling $K\sample\mu_K$ and then sampling $M\sample\mu_M$ conditioned on $K^{\transpose}\cdot M=0$.
The conditioning event must have positive probability for every key in the support of $\mu_K$.
\end{definition}
Its independently drawn variant is $\mu_K\otimes\mu_M$, with the orthogonality condition omitted.
The underlying distributions need not in general be the actual marginals of $\cV$. Our counterexample uses the actual marginals and therefore also applies to that more restrictive reading.

For a random variable $Z$ taking values in a finite set $\cZ$, let
\[
 H_\infty(Z):=-\log_2\max_{z\in\cZ}\Prb{Z=z}.
\]
Consider the positive integer parameters $a$, $\kappa$, and $m$ of~\cite[Lemma~6.21]{Goldwasser2015}, where $m\ge\kappa$ and $a$ is the number of matrix pieces.
The piece width is $\ell=\kappa/10$, and the per-piece leakage bound is $\lambda=\kappa/(20\cdot a)$. The parameters are chosen so these are integers.
A matrix piece is $M\cdot L$, where $L\in\F_2^{m\times\ell}$ has rank $\ell$. Thus a piece consists of $\ell$ linear combinations of the columns of $M$.
The adversary may obtain bounded leakage separately from the key and from the matrix piece.
One conclusion of Claim~6.24~\cite[pp.~1526--1527]{Goldwasser2015} is the following: if $\cV$ is IuO, every matrix in its support has rank $\kappa-1$, and
\begin{equation}\label{eq:gr-claim-entropy}
 H_\infty(K,M)\ge\kappa+(\kappa-1)\cdot2\kappa-0.15\kappa,
\end{equation}
then the joint distributions of the key and leakage transcript $W$ under $\cV$ and under $\mu_K\otimes\mu_M$ satisfy the following bound.
Here $\cL_{\cV}(K,W)$ denotes the joint distribution in the experiment with key--matrix distribution $\cV$, and the same leakage strategy is used in both experiments:
\begin{equation}\label{eq:gr-claimed-distance}
 \StatDist{\cL_{\cV}(K,W)}{\cL_{\mu_K\otimes\mu_M}(K,W)}
 \le8\ell\cdot2^{-0.05\kappa}.
\end{equation}
In particular, the conclusion includes disclosure of the entire key after the bounded leakage.
We take $m=2\kappa$, so the entropy threshold is exactly the one printed in the claim.

\subsection{A counterexample with one matrix entry}

The claim says that, for a distribution meeting these hypotheses, the key and leakage transcript are nearly indistinguishable from those obtained with an independently drawn key and matrix.
To refute the bound, we seek a relation that can be tested from the disclosed key and bounded matrix leakage and holds with noticeably different probabilities in the two experiments.
We therefore build a distribution in which one matrix entry copies one key bit: a single leaked bit then exposes the relation once the key is disclosed, while the copy costs only a constant number of bits of joint min-entropy.
The following construction is defined for every integer $\kappa\ge3$, with $m=2\kappa$. Below we specialize to parameters satisfying the claim's integrality and entropy requirements.
Index vector coordinates and matrix rows by $0,\ldots,\kappa-1$, and matrix columns by $1,\ldots,m$.
Put
\[
 \cK:=\{k\in\F_2^\kappa:k[0]=1\}.
\]
For $k\in\cK$, define
\begin{equation}\label{eq:gr-marker-fiber}
 \cF_k:=\left\{M\in\F_2^{\kappa\times m}:
 \rank{M}=\kappa-1,\quad
 k^{\transpose}\cdot M=0,\quad M[2,1]=k[1]\right\}.
\end{equation}
Sample $K$ uniformly from $\cK$ and, conditional on $K=k$, sample $M$ uniformly from $\cF_k$.
Call the resulting joint distribution $\cD$ and its actual marginals $\mu_K$ and $\mu_M$.
The entry $M[2,1]$ is a marker: it copies the key bit $K[1]$.
This equality is not forced by orthogonality alone.
We verify that $\cD$ nevertheless satisfies the hypotheses of Claim~6.24.

\para{Rank and entropy.}
Put $d:=\kappa-1$.
For every $k\in\cK$, the linear functional $v\mapsto v[2]$ is nonzero on the $d$-dimensional hyperplane $k^\perp$.
Otherwise $e_2$ would lie in $(k^\perp)^\perp=\spn\{k\}$, contradicting $k[0]=1$.
Hence exactly $2^{d-1}$ choices for the first column meet the marker constraint.
Choose the remaining $m-1$ columns independently and uniformly in $k^\perp$.
If they do not span $k^\perp$, some nonzero linear functional on that hyperplane vanishes on all of them.
A union bound therefore gives
\[
 \Prb{\text{the last $m-1$ columns do not span }k^\perp}
 \le(2^d-1)2^{-(m-1)}<2^{-\kappa}.
\]
If these columns span the hyperplane, the whole matrix has rank $d$, regardless of the first column.
Consequently,
\[
 |\cF_k|>2^{dm-1}(1-2^{-\kappa}),
\]
so these sets are nonempty and every supported pair has probability $2^{-(\kappa-1)}/|\cF_k|$.
It follows that
\begin{equation}\label{eq:gr-marker-entropy}
 \begin{aligned}
 H_\infty(K,M)
 &>\kappa+dm-2+\log_2(1-2^{-\kappa})\\
 &>\kappa+dm-3.
 \end{aligned}
\end{equation}
Thus the marker costs only a constant number of bits of joint min-entropy.

\para{Independence up to orthogonality.}
To sample the matrix marginal $\mu_M$, first choose $U$ uniformly from $\cK$ and then choose $M$ uniformly from $\cF_U$.
The matrix has rank $\kappa-1$, and the unique nonzero vector in its left kernel is $U$.
For any independently chosen $k\in\cK$, therefore,
\[
 k^{\transpose}\cdot M=0\quad\Longleftrightarrow\quad U=k.
\]
This event has probability $2^{-(\kappa-1)}$, independent of $k$.
Conditional on it, $M$ is uniform on $\cF_k$.
Sampling from the actual marginals and imposing orthogonality therefore recovers $\cD$ exactly.
This proves the IuO property. Since the acceptance probability is constant, the example also works if one conditions the entire product distribution on orthogonality.
The marker is compatible with this property because each matrix in the support determines its key, the unique nonzero vector in its left kernel.
Conditioning the common matrix distribution on orthogonality to a chosen key selects that key's fiber.
The relative weights inside the fiber already enforce the marker, so this conditioning does not remove it.

\para{One leaked bit distinguishes the two distributions.}
Let $L$ select the first $\ell$ columns of $M$, so that $L$ has rank $\ell$, and request only
\[
 W:=M[2,1].
\]
This is one bit of leakage from the allowed matrix piece $M\cdot L$. No key leakage is needed before the final disclosure.
Under $\cD$, one always has $W=K[1]$.
Under $\mu_K\otimes\mu_M$, one has $W=U[1]$, where $U$ is uniform on $\cK$ and independent of $K$.
Thus $W$ is a fair bit independent of $K$ in the second experiment.
The test $W=K[1]$ accepts with probabilities one and $1/2$, respectively. Indeed, for every fixed key the conditional transcript distributions have distance $1/2$.
Hence
\begin{equation}\label{eq:gr-marker-distance}
 \StatDist{\cL_{\cD}(K,W)}{\cL_{\mu_K\otimes\mu_M}(K,W)}=\frac12.
\end{equation}
For a concrete choice, take
\[
 \kappa=1000,\qquad m=2000,\qquad a=1,\qquad
 \ell=100,\qquad\lambda=50.
\]
The entropy bound in \cref{eq:gr-marker-entropy} exceeds the threshold in \cref{eq:gr-claim-entropy}, and the one-bit query is within the leakage allowance.
But \cref{eq:gr-claimed-distance} would bound the distance by $800\cdot2^{-50}<10^{-12}$, contradicting \cref{eq:gr-marker-distance}.

\subsection{The failed inference and its scope}

The proof on p.~1527 treats conditioning the key on the next observed matrix column as adding only its orthogonality equation.
Independence up to orthogonality and high joint entropy do not justify this step.
At the entropy threshold used here, a single exact relation can remain even though probability is spread over many key--matrix pairs.
In \cref{eq:gr-marker-fiber}, observing the first column also reveals $K[1]=M[2,1]$, an additional constraint that survives the entropy requirement.
Thus Claim~6.24 is false as stated, and its use in the proof of Lemma~6.21 needs a different justification.
The counterexample does not show that an honestly initialized bank reaches this distribution, and it does not by itself refute the fresh-matrix statement of Lemma~6.21 or give an attack on the complete compiler.

Repairing the security argument is quite involved and requires techniques such as Fourier analysis.
We do not present the repair here.

\subsection{Other issues, briefly}
Besides Claim~6.24, several other places in the printed argument and procedures need correction or further justification, and we record them briefly.

\para{Posterior distributions and conditioning.}
The exact assertions in Lemma~5.3 (pp.~1503--1504) that the actual posterior key and ciphertext marginals are independent of the plaintexts, and that conditioning their product recovers the joint posterior, do not hold in general.
For example, a rare ciphertext-side leakage event can impose a plaintext-dependent equation on the key.
This concerns an exact identity and does not itself contradict an averaged negligible-error security guarantee.
The collision-conditioning argument around Claim~6.28 (pp.~1529--1532) also needs justification, since conditioning on equal transcripts can reweight the starting key--matrix pair.
The joint sampling and locality steps in Propositions~6.11 and~6.13 (pp.~1518--1519) need further justification as well. We do not claim that those propositions or Claim~6.28 are false.

\para{Printed procedures and the permutation simulator.}
The unrestricted redraw in \textsf{BankGenRand} (Figure~3, p.~1509) can change the ciphertext coordinate required to equal one, leaving the encoding domain.
Figure~6 (p.~1514) computes a refreshed key but returns the old key, which can change the decrypted bit if the return line is executed literally.
The all-zero simulator used for Lemma~7.2 and Proposition~7.3 (pp.~1535--1537) is inconsistent with an observable linear relation satisfied by the real permutation input.
Finally, the row constructed by \textsf{SafeNAND} (Figure~10, p.~1540) contains ciphertexts outside the domain of the refresh it invokes. The printed procedure can produce incorrect outputs and allows one bit of leakage to reveal its mask, contrary to its stated local simulation guarantee.
These issues concern the procedures and guarantees as printed and are distinct from an end-to-end secret-recovery attack.
\end{document}